\documentclass[12pt]{report}
\usepackage{axodraw,epsfig,cite,feynarts,amsmath,amssymb,wrapfig,psfrag}

\input paperdef

\evensidemargin \oddsidemargin
\begin{document}

\thispagestyle{empty}
\setcounter{page}{0}
\def\thefootnote{\fnsymbol{footnote}}

\begin{flushright}
CERN--PH--TH/2004--139\\
LMU 09/04\\
hep-ph/0407244 \\
\end{flushright}

\vspace{1cm}

\begin{center}

{\large\sc {\bf MSSM Higgs Physics at Higher Orders}}


 
\vspace{1cm}

{\sc 
S.~Heinemeyer$^{1,2}$%
\footnote{email: Sven.Heinemeyer@cern.ch}%
}

\vspace*{1cm}

{\sl
$^1$CERN, Theory Division, Dept.\ of Physics, 
1211 Geneva 23, Switzerland

\vspace{0.4cm}

$^2$Institut f\"ur theoretische Elementarteilchenphysik,
LMU M\"unchen, Theresienstr.\ 37, D-80333 M\"unchen, Germany

}

\end{center}

\vspace*{1cm}

\BC
{\bf Abstract}
\EC
Various aspects of the Higgs boson phenomenology of the Minimal
Supersymmetric Standard Model (MSSM) are reviewed. Emphasis is put on
the effects of higher-order corrections. The masses and couplings are
discussed in the MSSM with real and complex parameters. Higher-order
corrections to Higgs boson production channels at a prospective
$e^+e^-$~linear collider are investigated. 
Corrections to Higgs boson decays to SM fermions and their
phenomenological implications for hadron and lepton colliders are
explored. 


\def\thefootnote{\arabic{footnote}}
\setcounter{footnote}{0}



\newpage
\thispagestyle{empty}
\mbox{}
\setcounter{page}{0}
\newpage
\tableofcontents
\newpage

\newpage

\chapter{Introduction}
\label{chap:intro}

The search for the lightest Higgs boson is a crucial test of 
Supersymmetry (SUSY)~\cite{mssm} which can be performed with the
present and the 
next generation of accelerators. The prediction of a relatively
light Higgs boson is common to all supersymmetric models whose
couplings remain in the perturbative regime up to a very high energy
scale~\cite{susylighthiggs}.
Within the Minimal Supersymmetric Standard Model (MSSM), where all
parameters are assumed to be real (rMSSM), the mass of the lightest
$\cp$-even Higgs boson, $\mh$, is 
bounded from above by about $\mh \lsim 140 \gev$%
\footnote{
This value has been obtained with the top-quark mass value of 
$\mt^{\rm exp} = 178.0 \gev$~\cite{mtopexpnew} that has become
available while finalizing this report. Most results in this 
report, however, have been derived with the former value 
$\mt^{\rm exp,old} = 174.3 \gev$~\cite{mtopexpold}. 
The use of the new value is always indicated. The main effect of the
higher $\mt^{\rm exp}$ value is an increase in the $\mh$ prediction by
$\sim 3 \gev$~\cite{mhiggslong,tbexcl}.
}%
~\cite{mhiggslong,mhiggsAEC}, 
including radiative corrections at
the \onel~\cite{ERZ,mhiggsf1lA,mhiggsf1lB,mhiggsf1lC} 
and at the \twol\ level~%
\cite{mhiggslong,mhiggsAEC,mhiggsEP1b,mhiggsRG1a,mhiggsRG1,mhiggsRG2,mhiggsletter,mhiggslle,mhiggsEP0,mhiggsEP1,mhiggsEP3,mhiggsEP2,mhiggsEP4,mhiggsFD2,mhiggsEP4b,mhiggsEP5}.
In the case where also complex parameters are allowed (cMSSM) the
evaluations of the Higgs boson sector are less
advanced~\cite{mhiggsCPXgen,mhiggsCPXEP,mhiggsCPXRG1,mhiggsCPXRG2,mhiggsCPXsn,mhiggsCPXFD1,mhiggsCPXFDproc,mhiggsCPXFD2}, 
but the upper bound of $\sim 140 \gev$ still holds.
This places the lightest MSSM Higgs boson in the reach of the
currently operating Tevatron (depending on the luminosity
performance), the LHC, and a prospective future $e^+e^-$ linear
collider (LC). 

While at the Tevatron one can at most hope for the discovery of the Higgs
boson and possibly a crude mass measurement~\cite{TevHiggsReport}, at
the LHC already a high-precision determination of $\mh$ down to 
$\de\mh^{\rm exp} = 100 - 200 \mev$~\cite{LHCHiggs} as well as some coupling
and total width measurements~\cite{HcoupLHCSM} 
seem to be feasible. At the LC eventually, $\mh$ can be determined at
the $50 \mev$ level~\cite{teslatdr}. The mass and width measurements
are summarized in 
\refta{tab:masswidth} for a Higgs boson with $\mh \sim 120 \gev$. 
The various production cross
sections~\cite{teslatdr,eennHexp}, see \refta{tab:xs}, as well as the
branching ratios to SM fermions and gauge
bosons~\cite{teslatdr,HiggsBRBrient,HiggsBRBarklow}, see
\refta{tab:br},  can be determined 
with high precision down to a few per cent at the LC. Even the
trilinear Higgs boson couplings seem to be in reach~\cite{hhh}.

\begin{table}[htb!]
\begin{center}
\renewcommand{\arraystretch}{1.5}
\begin{tabular}{|l|c|c|} \hline\hline
collider & $\de\mh$ [GeV] & $\de\Ga_h/\Ga_h$ \\ \hline \hline
Tevatron & \order{2} & -- \\ \hline
LHC      & $0.2$ &  $40\%$ \\ \hline
LC       & $0.05$  & $6\%$ \\ \hline \hline
\end{tabular}
\renewcommand{\arraystretch}{1.0}
\end{center}
\caption{Expected precisions for the measurement of the lightest Higgs
boson mass and width at the Tevatron, the LHC and the
LC~\cite{TevHiggsReport,LHCHiggs,teslatdr}.
}
\label{tab:masswidth}
\end{table}

\begin{table}[htb!]
\begin{center}
\renewcommand{\arraystretch}{1.5}
\begin{tabular}{|c|l|c|} \hline\hline
collider & decay mode & $\de\,\si/\si$ \\ \hline \hline
LC   & $e^+e^- \to Z^* \to Zh$     & 1.5\% \\ \hline
LC   & $e^+e^- \to \bar\nu \nu W^+W^- \to \bar\nu \nu h$ &  2\% \\ \hline\hline
\end{tabular}
\renewcommand{\arraystretch}{1.0}
\end{center}
\caption{Expected precisions for the measurement of Higgs production
cross sections at the LC~\cite{teslatdr,eennHexp}.}
\label{tab:xs}
\end{table}

\begin{table}[htb!]
\begin{center}
\renewcommand{\arraystretch}{1.5}
\begin{tabular}{|l|c|c|} \hline\hline
decay mode & $\de\br/\br$ ($\sqrt{s} = 500 \gev$) &
                        $\de\br/\br$ ($\sqrt{s} = 1 \tev$) \\ \hline \hline
$\hbb$     &  1.5\% & 1.5\% \\ \hline
$\htautau$ &  4.5\% & 2\%   \\ \hline
$\hcc$     &  6\%   & --    \\ \hline
$\hgg$     &  4\%   & 2.5\% \\ \hline
$\hWW$     &  3\%   & -- \\ \hline \hline
\end{tabular}
\renewcommand{\arraystretch}{1.0}
\end{center}
\caption{Expected precisions for the measurement of Higgs branching
  ratios at the LC for Higgs boson masses of $\sim 120 \gev$. For the
  analysis at $\sqrt{s} = 1 \tev$ not all channels have been
  investigated~\cite{teslatdr,HiggsBRBrient,HiggsBRBarklow}. }
\label{tab:br}
\end{table}

These expected accuracies make it mandatory to have a corresponding
precision of the theoretical predictions in terms of the relevant SUSY
parameters at hand. In this report we review several recent
theoretical evaluations of higher-order corrections to the Higgs boson
masses in the MSSM, to the dominant production cross sections at the
LC, and to the decays to SM fermions, which are relevant for
measurements at the Tevatron, the LHC, and the LC.


\newpage

\chapter{The Higgs boson sector of the MSSM with real parameters}
\label{chap:mhrMSSM}

In this section we will concentrate on the corrections in the Higgs
boson sector in the MSSM with real parameters (rMSSM). 
The corresponding case with complex parameters
(cMSSM) is treated in \refse{chap:mhcMSSM}. An introduction to other
ascpects of MSSM phenomenology can be found in \citere{mssm}. 


\section{The Higgs boson sector at tree-level}

Contrary to the Standard Model (SM), in the MSSM two Higgs doublets
are required.
The  Higgs potential~\cite{hhg}
\BEA
V &=& m_{1}^2 |\cHe|^2 + m_{2}^2 |\cHz|^2 
      - m_{12}^2 (\epsilon_{ab} \cHe^a\cHz^b + \hc)  \non \\
  & & + \frac{1}{8}(g_1^2+g_2^2) \left[ |\cHe|^2 - |\cHz|^2 \right]^2
        + \frac{1}{2} g_2^2|\cHe^{\dag} \cHz|^2~,
\label{higgspot}
\EEA
contains $m_1, m_2, m_{12}$ as soft SUSY breaking parameters;
$g, g'$ are the $SU(2)$ and $U(1)$ gauge couplings, and 
$\epsilon_{12} = -1$.

The doublet fields $H_1$ and $H_2$ are decomposed  in the following way:
\BEA
\cHe &=& \VL \cHe^0 \\[0.5ex] \cHe^- \VR \; = \; \VL v_1 
        + \frac{1}{\sqrt2}(\phi_1^0 + i\chi_1^0) \\[0.5ex] -\phi_1^- \VR~,  
        \non \\
\cHz &=& \VL \cHz^+ \\[0.5ex] \cHz^0 \VR \; = \; \VL \phi_2^+ \\[0.5ex] 
        v_2 + \frac{1}{\sqrt2}(\phi_2^0 + i\chi_2^0) \VR~.
\label{higgsfeldunrot}
\EEA
The potential (\ref{higgspot}) can be described with the help of two  
independent parameters (besides $g$ and $g'$): 
$\Tb = v_2/v_1$ and $M_A^2 = -m_{12}^2(\Tb+\CTb)$,
where $M_A$ is the mass of the $\cp$-odd Higg boson~$A$.

The diagonalization of the bilinear part of the Higgs potential,
i.e.\ of the Higgs mass matrices, is performed via the orthogonal
transformations 
\BEA
\label{hHdiag}
\VL H^0 \\[0.5ex] h^0 \VR &=& \ML \Ca & \Sa \\[0.5ex] -\Sa & \Ca \MR 
\VL \phi_1^0 \\[0.5ex] \phi_2^0~, \VR  \\
\label{AGdiag}
\VL G^0 \\[0.5ex] A^0 \VR &=& \ML \Cb & \Sbe \\[0.5ex] -\Sbe & \Cb \MR 
\VL \chi_1^0 \\[0.5ex] \chi_2^0 \VR~,  \\
\label{Hpmdiag}
\VL G^{\pm} \\[0.5ex] H^{\pm} \VR &=& \ML \Cb & \Sbe \\[0.5ex] -\Sbe & 
\Cb \MR \VL \phi_1^{\pm} \\[0.5ex] \phi_2^{\pm} \VR~.
\EEA
The mixing angle $\al$ is determined through
\BE
\al = {\rm arctan}\KKL 
  \frac{-(\MA^2 + \MZ^2) \Sbe \Cb}
       {\MZ^2 \CQb + \MA^2 \SQb - m^2_{h,{\rm tree}}} \KKR~, ~~
 -\frac{\pi}{2} < \al < 0~.
\label{alphaborn}
\end{equation}

One gets the following Higgs spectrum:
\BEA
\mbox{2 neutral bosons},\, {\cal CP} = +1 &:& h, H \non \\
\mbox{1 neutral boson},\, {\cal CP} = -1  &:& A \non \\
\mbox{2 charged bosons}                   &:& H^+, H^- \non \\
\mbox{3 unphysical Goldstone bosons}      &:& G, G^+, G^- .
\EEA

At tree level the mass matrix of the neutral $\cp$-even Higgs bosons
is given in the $\Pe$-$\Pz$-basis 
in terms of $\MZ$, $\MA$, and $\Tb$ by
\BEA
M_{\rm Higgs}^{2, {\rm tree}} &=& \ML \mpe^2 & \mpez^2 \\ 
                           \mpez^2 & \mpz^2 \MR \non\\
&=& \ML \MA^2 \SQb + \MZ^2 \CQb & -(\MA^2 + \MZ^2) \Sbe \Cb \\
    -(\MA^2 + \MZ^2) \Sbe \Cb & \MA^2 \CQb + \MZ^2 \SQb \MR,
\label{higgsmassmatrixtree}
\EEA
which by diagonalization according to \refeq{hHdiag} yields the
tree-level Higgs boson masses
\BE
M_{\rm Higgs}^{2, {\rm tree}} 
   \stackrel{\al}{\longrightarrow}
   \ML m_{H,{\rm tree}}^2 & 0 \\ 0 &  m_{h,{\rm tree}}^2 \MR~.
\end{equation}
The charged Higgs boson mass is given by
\BE
\label{rMSSM:mHp}
\mHp^2 = \MA^2 + \MW^2~.
\end{equation}
The masses of the gauge bosons are given in analogy to the SM:
\BE
M_W^2 = \frac{1}{2} g_2^2 (v_1^2+v_2^2) ;\qquad
M_Z^2 = \frac{1}{2}(g_1^2+g_2^2)(v_1^2+v_2^2) ;\qquad M_\gamma=0.
\end{equation}


\section{The scalar quark sector}
\label{sec:squark}

The squark mass term of the MSSM Lagrangian is given by
\BE
{\cal L}_{m_{\tilde{f}}} = -\frac{1}{2} 
   \Big( \tilde{f}_L^{\dag},\tilde{f}_R^{\dag} \Big)\; {\bf Z} \; 
   \VL \tilde{f}_L \\[0.5ex] \tilde{f}_R \VR~,
\end{equation}
where
\BE
\renewcommand{\arraystretch}{1.5}
{\bf Z} = \ML M_{\tilde{Q}}^2 + M_Z^2\CZb(I_3^f-Q_f\sw^2) + m_f^2 
    & m_f (A_f - \mu \{\CTb;\Tb\}) \\
    m_f (A_f - \mu \{\cot\be;\tan\be\}) 
    & M_{\tilde{Q}'}^2 + M_Z^2 \CZb Q_f\sw^2 + m_f^2 \VR ,
\label{squarkmassenmatrix}
\end{equation}
and $\{\CTb;\Tb\}$ corresponds to $\{u;d\}$-type squarks.
The soft SUSY breaking term $M_{\tilde{Q}'}$ is given by:
\BEA
M_{\tilde{Q}'} &=& \left\{ \begin{array}{cl} 
M_{\tilde{U}} & \quad\mbox{for $u$-type squarks} \\
M_{\tilde{D}} & \quad\mbox{for $d$-type squarks}
\end{array} \right. .
\EEA

In order to diagonalize the mass matrix and to determine the physical mass
eigenstates the following rotation has to be performed:
\BE
\VL \sfe \\ \sfz \VR = \ML \costf & \sintf \\ -\sintf & \costf \MR 
                       \VL \sfl \\ \sfr \VR .
\label{squarkrotation}
\end{equation}
The mixing angle $\tsf$ is given for $\Tb > 1$ by:
\BEA
\costf &=& \sqrt{
    \frac{(m_{\tilde{f}_R}^2 - m_{\tilde{f}_1}^2)^2}
          {m_f^2 \, (A_f - \mu \{\CTb ; \Tb \})^2 
    + (m_{\tilde{f}_R}^2-m_{\tilde{f}_1}^2)^2} } \\
\sintf &=& \mp\; {\rm sgn} \Big[ 
    A_f - \mu \{\CTb ; \Tb \} \Big] \non \\ 
& & \times \; \sqrt{ 
    \frac{m_f^2 \, (A_f - \mu \{\CTb ; \Tb \})^2}
    {m_f^2 \, (A_f - \mu \{\CTb ; \Tb \})^2 
    + (m_{\tilde{f}_R}^2 - m_{\tilde{f}_1}^2)^2} }~. 
\label{stt}
\EEA
The negative sign in (\ref{stt}) corresponds to $u$-type squarks, the
positive sign 
to $d$-type ones. 
$m_{\tilde{f}_R}^2 =M_{\tilde{Q}'}^2  + M_Z^2 \CZb Q_f\sw^2 + m_f^2 $
denotes the lower right entry in the 
squark mass matrix~(\ref{squarkmassenmatrix}).
The masses are given by the eigenvalues of the mass matrix:
\BEA
\label{Squarkmasse} 
m_{\tilde{f}_{1,2}}^2 &=& \edz \KKL M_{\tilde{Q}}^2 + M_{\tilde{Q}'}^2 \KKR
  + \frac{1}{2} M_Z^2 \CZb I^f_3 + m_f^2 \\
& & \left\{ \begin{array}{l} \displaystyle \pm\; \frac{c_f}{2}\, 
    \sqrt{ \Big[ M_{\tilde{Q}}^2 - M_{\tilde{Q}'}^2 + 
                 M_Z^2 \CZb (I^f_3-2Q_f\sw^2)\Big]^2 
    + 4 m_f^2 \Big( A_u - \mu \CTb \Big)^2} \non \\[2ex]
    \displaystyle \pm\; \frac{c_f}{2}\, 
    \sqrt{ \Big[ M_{\tilde{Q}}^2 - M_{\tilde{Q}'}^2 + 
                 M_Z^2 \CZb (I^f_3-2Q_f\sw^2)\Big]^2 
    + 4 m_f^2 \Big( A_d - \mu \Tb \Big)^2} 
    \end{array} \right. \non \\
c_f &=& {\rm sgn} \KKL M_{\tilde{Q}}^2 - M_{\tilde{Q}'}^2 + 
              M_Z^2 \CZb (I^f_3-2Q_f\sw^2) \KKR \non
\EEA
for $u$-type and $d$-type squarks, respectively.
For most of our discussions 
we make the choice 
\BE
M_{\tilde{Q}} = M_{\tilde{Q}'} =: \msq \equiv \msusy~.
\end{equation}
Since the non-diagonal entry of the mass matrix
\refeq{squarkmassenmatrix} is proportional to the fermion mass, 
mixing becomes particularly important for scalar tops ($\Sf = \Stop$), 
in the case of $\Tb \gg 1$ also for scalar bottoms ($\Sf = \Sbot$).

\bigskip
Furthermore it is possible to express the squark mass matrix
in terms of the physical masses $\msfe, \msfz$ and the mixing angle $\tsf$:
\BE
{\bf Z} = \ML \cosQtf \msfe^2 + \sinQtf \msfz^2 & 
              \sintf \costf (\msfe^2 - \msfz^2) \\
              \sintf \costf (\msfe^2 - \msfz^2) &
              \sinQtf \msfe^2 + \cosQtf \msfz^2 
          \MR~.
\label{smm:physparam}
\end{equation}
$A_f$ can be written as follows:
\BE
A_f = \frac{\sintf \costf (\msfe^2 - \msfz^2)}{\mf} + \mu \{\CTb; \Tb\}.
\label{eq:af}
\end{equation}

Since the most relevant squarks for the MSSM Higgs boson sector are
the $\Stop$~and $\Sbot$~particles, here we explicitly list 
their mass matrices in the basis of the gauge eigenstates 
$\StopL, \StopR$ and $\SbotL, \SbotR$:
\BEA
\label{stopmassmatrix}
{\cal M}^2_{\Stop} &=&
  \ML \MstL^2 + \mt^2 + \CZb (\edz - \frac{2}{3} \sw^2) \MZ^2 &
      \mt \Xt \\
      \mt \Xt &
      \MstR^2 + \mt^2 + \frac{2}{3} \CZb \sw^2 \MZ^2 
  \MR, \\
&& \non \\
\label{sbotmassmatrix}
{\cal M}^2_{\Sbot} &=&
  \ML \MsbL^2 + \mb^2 + \CZb (-\edz + \frac{1}{3} \sw^2) \MZ^2 &
      \mb \Xb \\
      \mb \Xb &
      \MsbR^2 + \mb^2 - \frac{1}{3} \CZb \sw^2 \MZ^2 
  \MR,
\EEA
where 
\BE
\mt \Xt = \mt (\At - \mu \CTb) , \quad
\mb\, \Xb = \mb\, (\Ab - \mu \Tb) .
\label{eq:Xtb}
\end{equation}
Here $\At$ denotes the trilinear Higgs--stop coupling, $\Ab$ denotes the
Higgs--sbottom coupling, and $\mu$ is the Higgs mixing parameter.
SU(2) gauge invariance requires the relation
\BE
\MstL = \MsbL .
\end{equation}


\section{Corrections in the Feynman-diagrammatic approach}

\subsection{Renormalization}
\label{subsec:renrMSSM}

In order to calculate the higher-order corrections to the Higgs boson
masses and effective mixing angle, the renormalized Higgs boson
self-energies are needed. The parameters appearing in the Higgs
potential, see \refeq{higgspot}, are renormalized as follows:
\begin{align}
\label{rMSSM:PhysParamRenorm}
  \MZ^2 &\to \MZ^2 + \dMZsq,  & \tadh &\to \tadh +
  \dtadh, \\ 
  \MW^2 &\to \MW^2 + \dMWsq,  & \tadH &\to \tadH +
  \dtadH, \notag \\ 
  M_{\rm Higgs}^2 &\to M_{\rm Higgs}^2 + \de M_{\rm Higgs}^2, & 
  \tanb &\to \tanb (1+\dtanb). \notag \\
  \mHp^2 &\to \mHp^2 + \de\mHp^2 \notag
\end{align}
$M_{\rm Higgs}^2$ denotes the tree-level Higgs boson mass matrix given
in \refeq{higgsmassmatrixtree}. $\tadh$ and $\tadH$ are the tree-level
tadpoles, i.e.\ the terms linear in $h$ and $H$ in the Higgs potential.

The field renormalization matrices of both Higgs multiplets
can be set up symmetrically, 
\begin{align}
\label{rMSSM:higgsfeldren}
  \begin{pmatrix} h \\[.5em] H \end{pmatrix} \to
  \begin{pmatrix}
    1+\tfrac{1}{2} \dZ{hh} & \tfrac{1}{2} \dZ{hH} \\[.5em]
    \tfrac{1}{2} \dZ{hH} & 1+\tfrac{1}{2} \dZ{HH} 
  \end{pmatrix} \cdot
  \begin{pmatrix} h \\[.5em] H \end{pmatrix}~,
\end{align}
and for the charged Higgs boson
\BE
\label{rMSSM:ZHp}
H^\pm \to H^\pm (1 + \dZ{H^-H^+})~.
\end{equation}

\noindent
For the mass counter term matrices we use the definitions
\begin{align}
  \delta M_{\rm Higgs}^2 =
  \begin{pmatrix}
    \dmhsq  & \dmhHsq \\[.5em]
    \dmhHsq & \dmHsq  
  \end{pmatrix}~.
\end{align}
The renormalized self-energies, $\hSi(p^2)$, can now be expressed
through the unrenormalized self-energies, $\Si(p^2)$, the field
renormalization constants and the mass counter terms.
This reads for the $\cp$-even part,
\begin{subequations}
\label{rMSSM:renses_higgssector}
\begin{align}
\ser{hh}(p^2)  &= \se{hh}(p^2) + \dZ{hh} (p^2-\mhtree^2) - \dmhsq, \\
\ser{hH}(p^2)  &= \se{hH}(p^2) + \dZ{hH}
(p^2-\tfrac{1}{2}(\mhtree^2+\mHtree^2)) - \dmhHsq, \\ 
\ser{HH}(p^2)  &= \se{HH}(p^2) + \dZ{HH} (p^2-\mHtree^2) - \dmHsq~,
\end{align}
\end{subequations}
and for the charged Higgs boson
\BE
\label{rMSSM:SEHp}
\ser{H^-H^+}(p^2) = \se{H^-H^+}(p^2) + \dZ{H^-H^+} (p^2 - \mHp^2)
                                     - \de\mHp^2~.
\end{equation}

Inserting the renormalization transformation into the Higgs mass terms
leads to expressions for their counter terms which consequently depend
on the other counter terms introduced in~(\ref{rMSSM:PhysParamRenorm}). 

For the $\cp$-even part of the Higgs sectors, these counter terms are:
\begin{subequations}
\label{rMSSM:HiggsMassenCTs}
\begin{align}
\dmhsq &= \de\MA^2 \cos^2(\alpha-\beta) + \delta \MZ^2 \sin^2(\alpha+\beta) \\
&\quad + \tfrac{e}{2 \MZ \sw \cw} (\dtadH \cos(\alpha-\beta)
\sin^2(\alpha-\beta) + \dtadh \sin(\alpha-\beta)
(1+\cos^2(\alpha-\beta))) \notag \\ 
&\quad + \dtanb \sinb \cosb (\MA^2 \sin 2 (\alpha-\beta) + \MZ^2 \sin
2 (\alpha+\beta)), \notag \\ 
\dmhHsq &= \tfrac{1}{2} (\de\MA^2 \sin 2(\alpha-\beta) - \dMZsq \sin
2(\alpha+\beta)) \\ 
&\quad + \tfrac{e}{2 \MZ \sw \cw} (\dtadH \sin^3(\alpha-\beta) -
\dtadh \cos^3(\alpha-\beta)) \notag \\ 
&\quad - \dtanb \sinb \cosb (\MA^2 \cos 2 (\alpha-\beta) + \MZ^2 \cos
2 (\alpha+\beta)), \notag \\ 
\dmHsq &= \de\MA^2 \sin^2(\alpha-\beta) + \dMZsq \cos^2(\alpha+\beta) \\
&\quad - \tfrac{e}{2 \MZ \sw \cw} (\dtadH \cos(\alpha-\beta)
(1+\sin^2(\alpha-\beta)) + \dtadh \sin(\alpha-\beta)
\cos^2(\alpha-\beta)) \notag \\ 
&\quad - \dtanb \sinb \cosb (\MA^2 \sin 2 (\alpha-\beta) + \MZ^2 \sin
2 (\alpha+\beta))~. \notag 
\end{align}
\end{subequations}
For the charged Higgs boson it reads
\BE
\label{rMSSM:dmHp}
\de\mHp^2 = \de\MA^2 + \de\MW^2~.
\end{equation}

\bigskip
For the field renormalization we chose to give each Higgs doublet one
renormalization constant,
\begin{align}
\label{rMSSM:HiggsDublettFeldren}
  \cHe \to (1 + \tfrac{1}{2} \dZ{\cHe}) \cHe, \quad
  \cHz \to (1 + \tfrac{1}{2} \dZ{\cHz}) \cHz~.
\end{align}
This leads to the following expressions for the various field
renormalization constants in \refeq{rMSSM:higgsfeldren}:
\begin{subequations}
\label{rMSSM:FeldrenI_H1H2}
\begin{align}
  \dZ{hh} &= \sinasq \dZ{\cHe} + \cosasq \dZ{\cHz}, \\[.2em]
  \dZ{hH} &= \sina \cosa (\dZ{\cHz} - \dZ{\cHe}), \\[.2em]
  \dZ{HH} &= \cosasq \dZ{\cHe} + \sinasq \dZ{\cHz}, \\[.2em]
  \dZ{H^-H^+} &= \sinbsq \dZ{\cHe} + \cosbsq \dZ{\cHz}~.
\end{align}
\end{subequations}
The counter term for $\tb$ can be expressed in terms of the vaccuum
expectation values as
\begin{equation}
\de\tb = \frac{1}{2} \KL \dZ{\cHz} - \dZ{\cHe} \KR +
\frac{\de v_2}{v_2} - \frac{\de v_1}{v_1}~,
\end{equation}
where the $\de v_i$ are the renormalization constants of the $v_i$:
\begin{equation}
v_1 \to \KL 1 + \dZ{\cHe} \KR \KL v_1 + \de v_1 \KR, \quad
v_2 \to \KL 1 + \dZ{\cHz} \KR \KL v_2 + \de v_2 \KR~.
\end{equation}

The renormalization conditions are fixed by an appropriate
renormalization scheme. For the mass counter terms on-shell conditions
are used:
\begin{align}
\label{rMSSM:mass_osdefinition}
  \dMZsq = \re \se{ZZ}(\MZ^2), \quad \dMWsq = \re \se{WW}(\MW^2),
  \quad \de\MA^2 = \re \se{AA}(\MA^2). 
\end{align}
Here $\Si$ denotes the transverse part of the self-energy. 
Since the tadpole coefficients are chosen to vanish in all orders,
their counter terms follow from $T_{\{h,H\}} + \de T_{\{h,H\}} = 0$: 
\begin{align}
  \dtadh = -{\tadh}, \quad \dtadH = -{\tadH}~. 
\end{align}
For the remaining renormalization constants for $\de\tb$, $\dZ{\cHe}$
and $\dZ{\cHz}$ several choices are possible, see the discussion in 
\refse{subsec:recentHO}. As will be shown there, the most convenient
choice is a \drbar\ renormalization of $\de\tb$, $\dZ{\cHe}$
and $\dZ{\cHz}$, 
\begin{subequations}
\label{rMSSM:deltaZHiggsTB}
\begin{align}
  \dtanb &= \dtanb^{\drbarm} 
       \; = \; -\ed{2\CZa} \KKL \re \Sip_{hh}(\mhtree^2) - 
                             \re \Sip_{HH}(\mHtree^2) \KKR^{\rm div}, \\[.5em]
  \dZ{\cHe} &= \dZ{\cHe}^{\drbarm}
       \; = \; - \KKL \re \Sip_{HH \; |\al = 0} \KKR^{\rm div}, \\[.5em]
  \dZ{\cHz} &= \dZ{\cHz}^{\drbarm} 
       \; = \; - \KKL \re \Sip_{hh \; |\al = 0} \KKR^{\rm div}~.
\end{align}
\end{subequations}
The corresponding renormalization scale, $\mudim$, is set to 
$\mudim = \mt$ in all numerical evaluations.


\subsection{The concept of higher order corrections in the
  Feynman-diagrammatric approach} 
\label{subsec:FDconcept}

In the Feynman diagrammatic (FD) approach the higher-order corrected 
$\cp$-even Higgs boson masses in the rMSSM are derived by finding the
poles of the $(h,H)$-propagator 
matrix. The inverse of this matrix is given by
\BE
\left(\Delta_{\rm Higgs}\right)^{-1}
= - i \ML p^2 -  \mHtree^2 + \hSi_{HH}(p^2) &  \hSi_{hH}(p^2) \\
     \hSi_{hH}(p^2) & p^2 -  \mhtree^2 + \hSi_{hh}(p^2) \MR~.
\label{higgsmassmatrixnondiag}
\end{equation}
Determining the poles of the matrix $\Delta_{\rm Higgs}$ in
\refeq{higgsmassmatrixnondiag} is equivalent to solving
the equation
\begin{equation}
\left[p^2 - \mhtree^2 + \hSi_{hh}(p^2) \right]
\left[p^2 - \mHtree^2 + \hSi_{HH}(p^2) \right] -
\left[\hSi_{hH}(p^2)\right]^2 = 0\,.
\label{eq:proppole}
\end{equation}

The status of the available results for the self-energy contributions to
\refeq{higgsmassmatrixnondiag} can be summarized as follows. For the
one-loop part, the complete result within the MSSM is 
known~\cite{ERZ,mhiggsf1lA,mhiggsf1lB,mhiggsf1lC}. The by far dominant
one-loop contribution is the \order{\alt} term due to top and stop 
loops ($\alt \equiv h_t^2 / (4 \pi)$, $h_t$ being the 
superpotential top coupling).
Concerning the two-loop
effects, their computation is quite advanced and has now reached a
stage such that all the presumably dominant
contributions are known. They include the strong corrections, usually
indicated as \order{\alt\als}, and Yukawa corrections, \order{\alt^2},
to the dominant one-loop \order{\alt} term, as well as the strong
corrections to the bottom/sbottom one-loop \order{\alb} term ($\alb
\equiv h_b^2 / (4\pi)$), i.e.\ the \order{\alb\als} contribution. The
latter can be relevant for large values of $\Tb$. Presently, the
\order{\alt\als}~\cite{mhiggsEP1b,mhiggsletter,mhiggslong,mhiggsEP0,mhiggsEP1},
\order{\alt^2}~\cite{mhiggsEP1b,mhiggsEP3,mhiggsEP2} and the
\order{\alb\als}~\cite{mhiggsEP4,mhiggsFD2} contributions to the self-energies
are known for vanishing external momenta.  In the (s)bottom
corrections the all-order resummation of the $\Tb$-enhanced terms,
\order{\alb(\als\tb)^n}, is also performed \cite{deltamb1,deltamb}.
Recently the \order{\alt\alb} and \order{\alb^2} corrections
became available~\cite{mhiggsEP4b}. Most recently the
\order{\alb\al_\tau} corrections have been evaluated, which are,
however, completely negligible~\cite{mhiggsWN}. Finally a ``full'' two-loop
effective potential 
calculation (including even the momentum dependence for the leading
pieces) has been published~\cite{mhiggsEP5}. However, the latter results
have been obtained using a certain renormalization in which all
quantities, including SM gauge boson masses and couplings, are \drbar\
parameters. This makes them not usable
in the approach and evaluations presented here.

\bigskip
The charged Higgs boson mass is obtained by solving the equation
\BE
\label{rMSSM:mHpHO}
p^2 - \mHp^2 - \ser{H^-H^+}(p^2) = 0~.
\end{equation}
The charged Higgs boson self-energy is known at the one-loop
level~\cite{chargedmhiggs,markusPhD}. 
We will not explore the corrections to the charged Higgs boson mass
further in this report. For a detailed analysis, see \citere{markusPhD}.


\subsection{The $\aeff$-approximation}
\label{subsec:aeff}


The dominant contributions for the Higgs boson self-energies can be
obtained by setting $p^2=0$. Approximating the 
renormalized Higgs boson self-energies by
\BE
\hSi(p^2) \to \hSi(0) \equiv \hSi
\label{zeroexternalmomentum}
\end{equation}
yields the
Higgs boson masses by re-diagonalizing the dressed
mass matrix
\BE
\label{deltaalpha}
M_{\rm Higgs}^{2} 
 = \ML \mH^2 - \hSiH & -\hSihH \\
       -\hSihH & \mh^2 - \hSih \MR
   \stackrel{\De\al}{\longrightarrow}
   \ML \MH^2 & 0 \\ 0 &  \Mh^2 \MR ,
\end{equation}
where $\Mh$ and $\MH$ are the corresponding
higher-order-corrected Higgs boson
masses. The rotation matrix in the transformation~(\ref{deltaalpha}) 
reads:
\BE
D(\De\al) = \ML \CDea & -\SDea \\ \SDea & \CDea \MR~.
\label{rotmatrix}
\end{equation}
The angle $\De\al$ is related to the renormalized
self-energies and masses through the eigenvector equation
\BEA
&& 
\ML \mH^2 - \hSiH - \Mh^2 & -\hSihH \\
    -\hSihH & \mh^2 - \hSih - \Mh^2 \MR
\VL -\SDea \\ \CDea \VR~=~0 
\EEA
which yields
\BE
\frac{\hSihH}{\Mh^2 - \mH^2 + \hSiH}~=~\TDea~.
\label{tandeltaalpha}
\end{equation}
The second eigenvector equation leads to:
\BE
\frac{-\hSihH}{\MH^2 - \mh^2 + \hSih}~=~\TDea~.
\label{tandeltaalpha2}
\end{equation}
Using the relations
\BE
D(\aeff) = D(\al)\;D(\De\al)
\end{equation}
and
\BE
\ML \hSiH & \hSihH \\ \hSihH & \hSih \MR = 
D^{-1}(\al) \ML \hSi_{\Pe} & \hSi_{\PePz} \\ 
                \hSi_{\PePz} & \hSi_{\Pz} \MR D(\al)
\end{equation}
it is obvious that $\aeff = (\al + \De\al$) is exactly the angle that
diagonalizes the higher-order corrected Higgs boson mass matrix in the
$\Pe,\Pz$-basis:
\BEA
\label{higgsmassmatrixPhi1Phi2}
&&
\ML \mpe^2 - \hSi_{\Pe} & \mpez^2 - \hSi_{\PePz} \\ 
    \mpez^2 - \hSi_{\PePz} & \mpz^2 - \hSi_{\Pz} \MR 
   \stackrel{\aeff}{\longrightarrow}
   \ML \MH^2 & 0 \\ 0 &  \Mh^2 \MR~ \non \\
&& \Big\downarrow~~\al \\
&&
\ML \mH^2 - \hSiH & - \hSihH \\ 
    - \hSihH & \mh^2 - \hSih \MR 
 \stackrel{\De\al}{\longrightarrow}
   \ML \MH^2 & 0 \\ 0 &  \Mh^2 \MR . \non
\EEA
The angle $\aeff$ can be obtained from 
\BE
\aeff = {\rm arctan}\KKL 
  \frac{-(\MA^2 + \MZ^2) \Sbe \Cb - \hSi_{\PePz}}
       {\MZ^2 \CQb + \MA^2 \SQb - \hSi_{\Pe} - \mh^2} \KKR~, ~~
 -\frac{\pi}{2} < \aeff < \frac{\pi}{2}~.
\end{equation}


\subsection{Recently calculated higher-order corrections}
\label{subsec:recentHO}

In order to discuss the impact of recent improvements in the MSSM Higgs
sector we will make use of the program
\fhto~\cite{feynhiggs,feynhiggs1.2}, which 
is a Fortran code for the evaluation
of the neutral $\cp$-even Higgs sector of the MSSM including
higher-order corrections to the renormalized Higgs boson
self-energies. 
The code comprises all existing higher-order corrections (except of
the results of \citere{mhiggsEP5}, which have been obtained in a pure
\drbar\ renormalization scheme, see \refse{subsec:FDconcept}). This
includes the well known full
\onel\ corrections~\cite{ERZ,mhiggsf1lA,mhiggsf1lB,mhiggsf1lC}, the
two-loop leading, momentum-independent, \order{\alt\als} correction
in the $t/\Stop$~sector~\cite{mhiggsletter,mhiggslong,mhiggsEP1}, 
the two-loop leading logarithmic corrections at
\order{\alt^2}~\cite{mhiggsRG1,mhiggsRG2} and all further corrections
discussed below. By two-loop
momentum-independent corrections here and hereafter we mean the
two-loop contributions to Higgs boson self-energies evaluated at zero
external momenta. At the one-loop level, the momentum-independent
contributions are the dominant part of the self-energy corrections, that, 
in principle, should be 
evaluated at external momenta squared equal to the poles of the
$h,H$-propagator matrix, \refeq{higgsmassmatrixnondiag}.

With the implementation of the latest results obtained in the MSSM Higgs
sector, \fh\ 
allows the presently most precise prediction of the masses of the
$\cp$-even Higgs bosons and the corresponding mixing angle.
The latest version of \fhto\ can be obtained from 
{\tt www.feynhiggs.de}.


\subsubsection{Hybrid renormalization scheme at the \onel\ level}
\label{subsec:newren}

\begin{figure}[htb!]
\newcommand{\psfragtextscale}{0.75}
\psfrag{0}[][][\psfragtextscale]{0}
\psfrag{50}[][][\psfragtextscale]{50}
\psfrag{60}[][][\psfragtextscale]{60}
\psfrag{70}[][][\psfragtextscale]{70}
\psfrag{80}[][][\psfragtextscale]{80}
\psfrag{90}[][][\psfragtextscale]{90}
\psfrag{100}[][][\psfragtextscale]{100}
\psfrag{110}[][][\psfragtextscale]{110}
\psfrag{120}[][][\psfragtextscale]{120}
\psfrag{130}[][][\psfragtextscale]{130}
\psfrag{140}[][][\psfragtextscale]{140}
\psfrag{200}[][][\psfragtextscale]{200}
\psfrag{300}[][][\psfragtextscale]{300}
\psfrag{400}[][][\psfragtextscale]{400}
\psfrag{500}[][][\psfragtextscale]{500}
\psfrag{600}[][][\psfragtextscale]{600}
\psfrag{700}[][][\psfragtextscale]{700}
\psfrag{800}[][][\psfragtextscale]{800}
\psfrag{900}[][][\psfragtextscale]{900}
\psfrag{1000}[][][\psfragtextscale]{1000}
\psfrag{MA0}[][][\psfragtextscale]{$M_A [\mathrm{GeV}]$}
\psfrag{Mh0}[][][\psfragtextscale]{$M_h [\mathrm{GeV}]$}
\begin{center}
\mbox{
\epsfig{figure=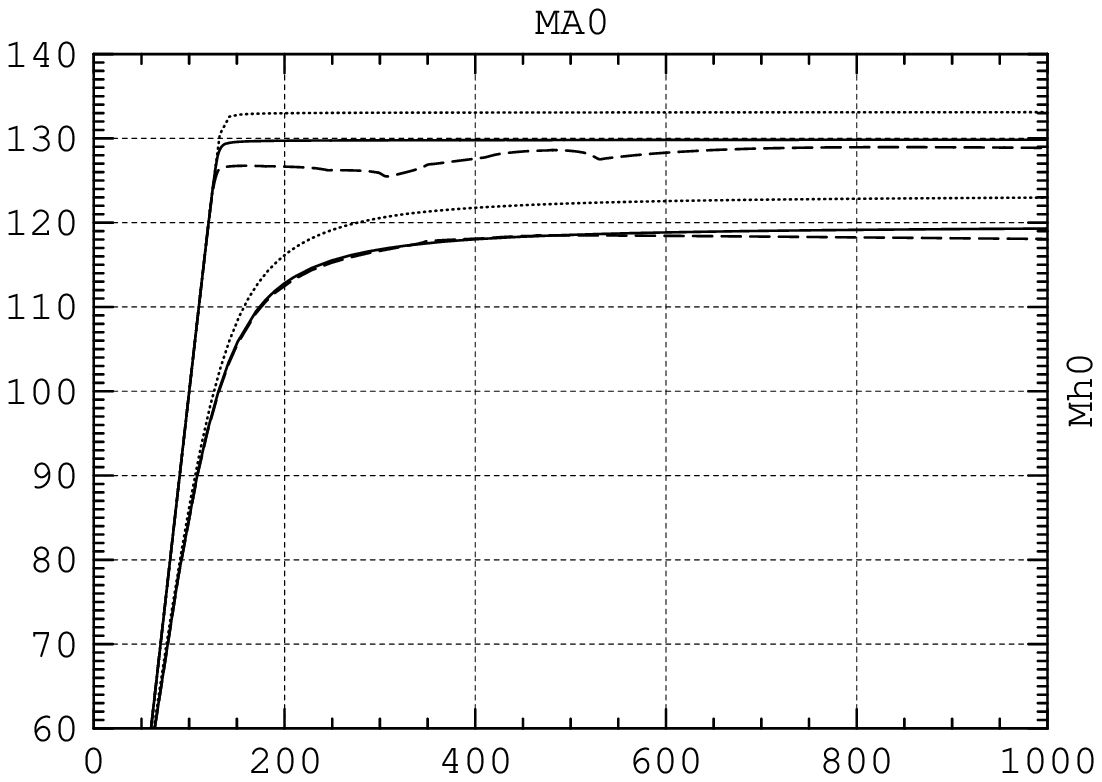,width=7cm,height=5.5cm} 
\hspace{1em}
\epsfig{figure=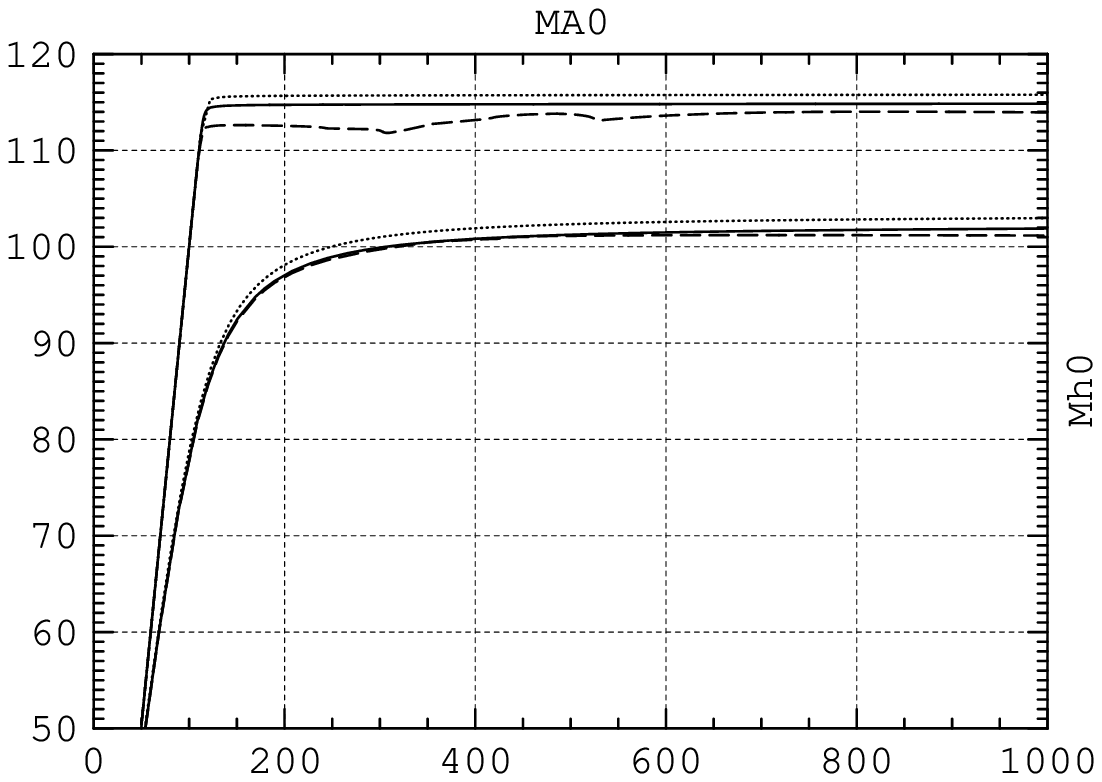,width=7cm,height=5.5cm}}
\end{center}
\caption{
The lightest Higgs boson mass is shown as a function of $\MA$ in the
$\mhmax$ and the no-mixing scenario~\cite{benchmark}, see the Appendix.
$\tb$ has been set to $\tb = 3$ (lower curves) and $\tb = 50$ (upper
curves). 
The hybrid \drbar/on-shell scheme (solid lines) is compared with the
original on-shell (type-1) renormalization (dashed).
The dotted lines indicate the effect of the ``subleading''
\order{\alt^2} correction. 
}
\label{fig:newrenorm}
\end{figure}

\fh\ is based on the FD approach with on-shell renormalization 
conditions~\cite{mhiggslong}. This means in particular that all the 
masses in the FD result are the physical ones, i.e.\ they correspond to
physical observables. Since \refeq{eq:proppole} is solved
iteratively, the result for $\mh$ and $\mH$ contains a dependence on
the field-renormalization constants of $h$ and $H$, which is 
formally of higher-order. Accordingly, there is some freedom in choosing 
appropriate renormalization conditions for fixing the
field-renormalization constants (this can also be interpreted as
affecting the 
renormalization of $\tb$). Different renormalization conditions have
been considered in the literature, e.g.\ ($\hSip$ denotes the derivative 
with respect to the external momentum squared):
\begin{enumerate}
\item
on-shell renormalization for $\hSi_Z, \hSi_A,
\hSip_A, \hSi_{AZ}$, and  
$\de v_1/v_1 = \de v_2/v_2$~\cite{mhiggsf1lC}
\item
on-shell renormalization for 
$\hSi_Z, \hSi_A, \hSi_{AZ}$, and 
$\de v_i = \de v_{i, {\rm div}}, i = 1,2$~\cite{mhiggsf1lB}
\item
on-shell renormalization for $\hSi_Z, \hSi_A$~,
\drbar\ renormalization (employing dimensional reduction~\cite{dred})
for $\dZ{\cHe}, \dZ{\cHz}$ and $\tb$~\cite{feynhiggs1.2}, see also the
discussion in \refse{subsec:renrMSSM}.
\end{enumerate}
The original full \onel\ evaluations of the Higgs boson
self-energies~\cite{mhiggsf1lB,mhiggsf1lC}  
were based on type-1 renormalization conditions, thus requiring the
derivative of the $A$~boson self-energy. In \citere{feynhiggs1.2} a
hybrid \drbar/on-shell scheme, type 3, has been proposed. The choice of a 
\drbar\ definition
for $\de Z_h, \de Z_H$ and $\tb$ requires to specify a renormalization
scale $Q^2$ at which these parameters are defined, which is commonly
chosen to be $\mt$. Variation of this scale gives some indication of
the size of 
unknown higher-order corrections, see \refse{subsec:muvariation}. These new
renormalization conditions lead to a more stable behavior around
thresholds, e.g.\ $\MA= 2\, \mt$, and avoid unphysically large
contributions in certain regions of the MSSM parameter space%
\footnote{
A more detailed discussion can be found in
\citeres{feynhiggs1.2,mhiggsrenorm}; see also \citere{ayresdoink}.
}%
~. This effect is demonstrated in \reffi{fig:newrenorm}.

\subsubsection{Two-loop \order{\alt^2} corrections}
\label{subsec:nonlogaltsq}


Recently, the two-loop \order{\alt^2} corrections in the limit of
zero external momentum became available, first only for the lightest
eigenvalue, $\mh$, and in the limit $\MA \gg \MZ$~\cite{mhiggsEP3},
then for all the entries of the Higgs propagator matrix for arbitrary
values of $\MA$~\cite{mhiggsEP2}. These corrections were obtained in the
effective-potential approach, that allows to construct the Higgs
boson self-energies, at zero external momenta, by taking the relevant
derivatives of the field-dependent potential.  In this procedure it
is important, in order to make contact with the physical $\MA$, to
compute the effective potential as a function of both $\cp$-even and
$\cp$-odd fields, as emphasized in \citere{mhiggsEP1}. In the
evaluation of the \order{\alt^2} corrections, the specification of a
renormalization prescription for the Higgs mixing parameter $\mu$ is also
required and it has been chosen as \drbar.
In \fhto, which includes the two-loop \order{\alt^2} corrections, 
the corresponding renormalization scale is fixed to be the same as for 
$\de Z_h, \de Z_H$ and $\tb$. 

\begin{figure}[htb!]
\begin{center}
\epsfig{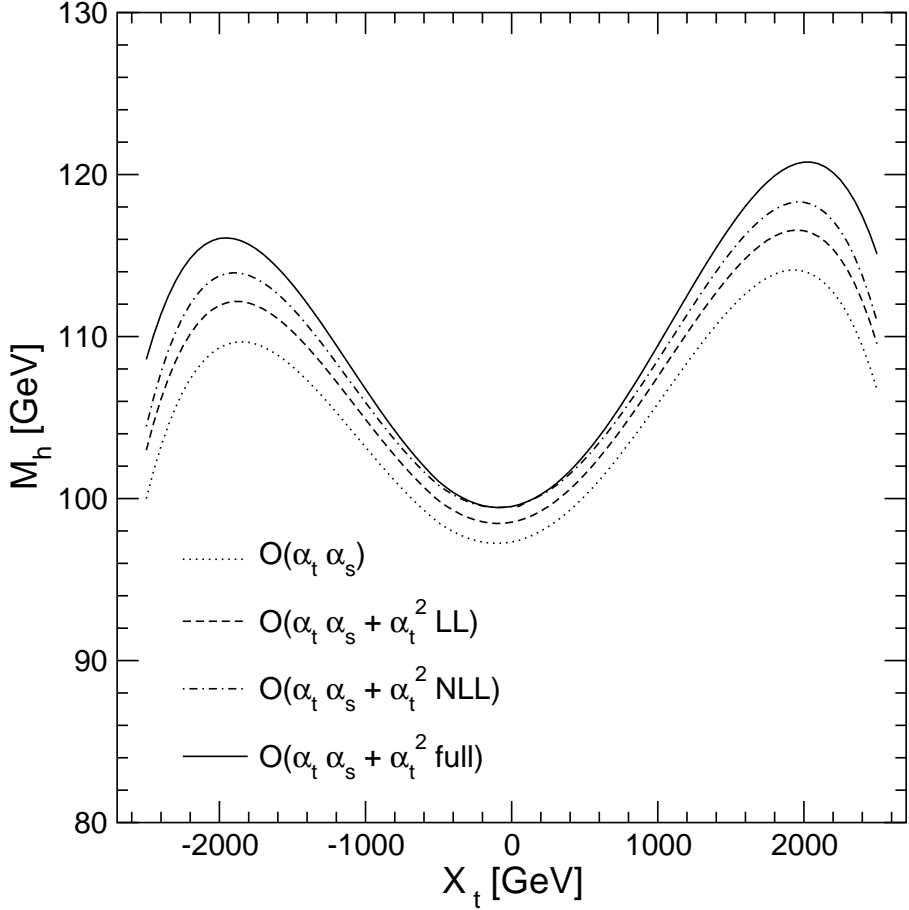}
\end{center}
\caption[]{ Two-loop corrected
$\Mh$ as a function of $\Xt$ in various steps of approximation.
The relevant MSSM parameters are chosen as $\tb=3\,,\, 
\mt^{\rm pole} = 174.3 \gev$,
$\MstL = \MstR = \MA = \mu = 1 \tev$ and $\mgl = 800 \gev$.
The meaning of the different curves is explained in the text.
}
\label{fig:RGmeth}
\end{figure}
 
The availability of the complete result for the momentum-independent
part of the \order{\alt^2} corrections allows to judge the quality
of results that incorporate only the logarithmic
contributions~\cite{mhiggsRG1,mhiggsRG2}.  
In \reffi{fig:RGmeth} (see also \reffi{fig:newrenorm}) we plot the
two-loop corrected $\mh$ as a 
function of the stop mixing parameter, $\Xt$.
For simplicity, the soft SUSY breaking parameters in the
diagonal entries of the stop mass matrix, $\MstL, \,\MstR$, are chosen
to be equal, $\MstL = \MstR = \msusy$. For the numerical analysis
$\msusy$ as well as
$\MA$ and $\mu$, are chosen to be all equal to $1 \tev$, while the
gluino mass is $\mgl = 800 \gev$, and $\tb = 3$. If not otherwise stated, in
all plots below we choose the trilinear couplings in the stop and
sbottom sectors equal to each other, $\Ab = \At$, and set 
$M_2 = \msusy$, where $M_2$ is the SU(2) gaugino mass parameter (the U(1)
gaugino mass parameter is obtained via the GUT relation, 
$M_1 = 5/3\, \sw^2/\cw^2\, M_2$).

The solid and dotted lines in \reffi{fig:RGmeth} 
are computed with and without the inclusion of the full
\order{\alt^2} corrections, while the dashed and dot--dashed ones are
obtained including only the logarithmic contributions. In particular,
the dashed curve corresponds to the result of \cite{mhiggsRG1},
obtained with the one-loop
renormalization group method. The dot--dashed curve,
instead, corresponds to the leading and next-to-leading logarithmic
terms. For a more detailed discussion see \citere{mhiggsAEC}.
From \reffi{fig:RGmeth} it can also be seen that, for small $\Xt$, the
full \order{\alt^2} result is very well reproduced by the logarithmic
approximation, once the next-to-leading terms are correctly taken
into account. On the other hand, when $\Xt$ is large there are
significant differences, amounting to several GeV, between the
logarithmic approximation and the full result. Such differences are
due to non-logarithmic terms that scale like powers of $\Xt/\msusy$. 
It should be noted that for more general choices of the
MSSM parameters the renormalization group method becomes rather
involved (see e.g.\ \citere{espnav}, where the case of a large splitting
between $\MstL$ and $\MstR$ is discussed), and a suitable
next-to-leading logarithmic
approximation to the full result is much more difficult to devise.


\subsubsection{Two-loop sbottom corrections}
\label{subsec:abass}

Due to the smallness of the bottom mass, the \order{\alb} \onel\
corrections to the Higgs boson self-energies can be numerically
non-negligible only for $\tb \gg 1$ and sizable values of the $\mu$
parameter.  In fact, at the classical level $h_b / h_t = (\mb / \mt)
\tb$, thus $\tb \gg 1$ is needed in order to have $\alb \sim \alt$
in spite of $\mb \ll \mt$.  In contrast to the \order{\alt}
corrections where both top and stop loops give sizable contributions,
in the case of
the \order{\alb} corrections the numerically dominant contributions
come from sbottom loops: those coming from bottom loops are
always suppressed by the small value of the bottom mass. A sizable
value of $\mu$ is then required to have sizable sbottom--Higgs scalar
interactions in the large-$\tb$ limit.

The relation between the bottom-quark mass and the Yukawa coupling
$h_b$, which controls also the interaction between the Higgs fields and
the sbottom squarks, reads at lowest order $\mb =h_b v_1 /\sqrt{2}$. 
This relation is affected at \onel\ order by large radiative
corrections \cite{deltamb1,deltamb} (see also \citere{deltamb2}),
proportional to $h_b v_2$,  
in general giving rise to $\tb$-enhanced contributions.
These terms proportional to $v_2$, often indicated as threshold
corrections to the bottom mass, are generated either by
gluino--sbottom \onel\ diagrams, resulting in \order{\alb\al_s}
corrections to the Higgs masses, or by chargino--stop loops, giving
\order{\alb\alt} corrections.  Because the $\tb$-enhanced
contributions can be numerically relevant, an accurate determination
of $h_b$ from the experimental value of the bottom mass requires a
resummation of such effects to all orders in the perturbative
expansion, as described in \citere{deltamb}.

The leading effects are included in the effective Lagrangian
formalism developed in \citere{deltamb}.
Numerically this is by far the dominant part of the
contributions from the sbottom sector. The effective Lagrangian is
given by
\BEA
\cL = \frac{g}{2\MW} \frac{\mbms}{1 + \dmb} \Bigg[ 
&& \tb\; A \, i \, \bar b \ga_5 b 
   + \wz \, V_{tb} \, \tb \; H^+ \bar{t}_L b_R \non \\
&+& \KL \frac{\Sa}{\Cb} - \dmb \frac{\Ca}{\Sbe} \KR h \bar{b}_L b_R 
                                                               \non \\
&-& \KL \frac{\Ca}{\Cb} + \dmb \frac{\Sa}{\Sbe} \KR H \bar{b}_L b_R
    \Bigg] + {\rm h.c.}~.
\label{effL}
\EEA
Here $\mbms$ denotes the running bottom quark mass including SM QCD
corrections. In the numerical evaluations we choose 
$\mbms = \mbms(\mt) \approx 2.74 \gev$. 
Furthermore, $\dmb$ is given at \order{\als} by
\BE
\dmb = \frac{2\als}{3\,\pi} \, \mgl \, \mu \, \tb \,
                    \times \, I(\msbe, \msbz, \mgl) ,
\label{def:dmb}
\end{equation}
where $I$ is given by
\BE
I(a, b, c) = \ed{(a^2 - b^2)(b^2 - c^2)(a^2 - c^2)} \,
             \KL a^2 b^2 \log\frac{a^2}{b^2} +
                 b^2 c^2 \log\frac{b^2}{c^2} +
                 c^2 a^2 \log\frac{c^2}{a^2} \KR .
\end{equation}
The large $\Sbot-\gl$~loops are resummed to all orders of
$(\als\tb)^n$ via the inclusion of $\dmb$~\cite{deltamb}.
The prefactor $1/(1 + \dmb)$ in \refeq{effL} arises from the
resummation of the leading corrections to all orders. They are due to
the shift in the relation of the bottom Yukawa coupling to the
physical $b$~quark mass, coming from the loop induced coupling of the
$\phi_1$ to the bottom quarks. The 
additional terms $\sim \dmb$ in the $h\bar b b$ and $H\bar b b$
couplings arise from the mixing and coupling of the ``other'' Higgs
boson, $H$ and $h$, respectively, to the $b$~quarks.

\begin{figure}[ht!]
\vspace{2.5em}
\begin{center}
\mbox{
\psfig{figure=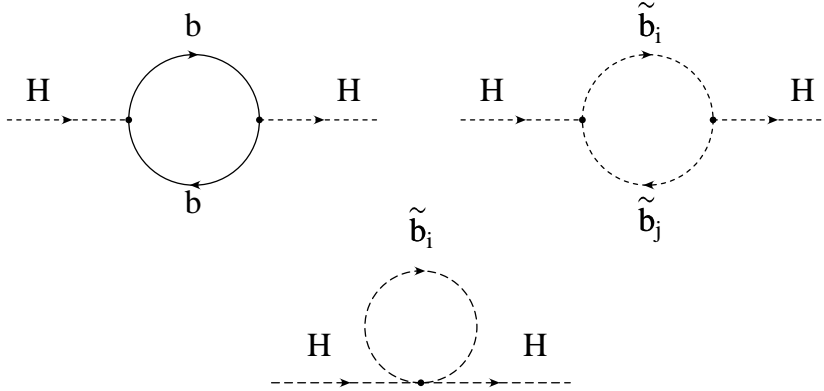,width=12cm,bbllx=150pt,bblly=630pt,
                                      bburx=450pt,bbury=730pt}}
\end{center}
\caption[]{
Generic Feynman diagrams for the $b/\Sbot$ contributions to Higgs
boson self-energies (H = $h, H, A$). 
}
\label{fig:fdol}
\end{figure}

\begin{figure}[ht!]
\begin{center}
\mbox{
\psfig{figure=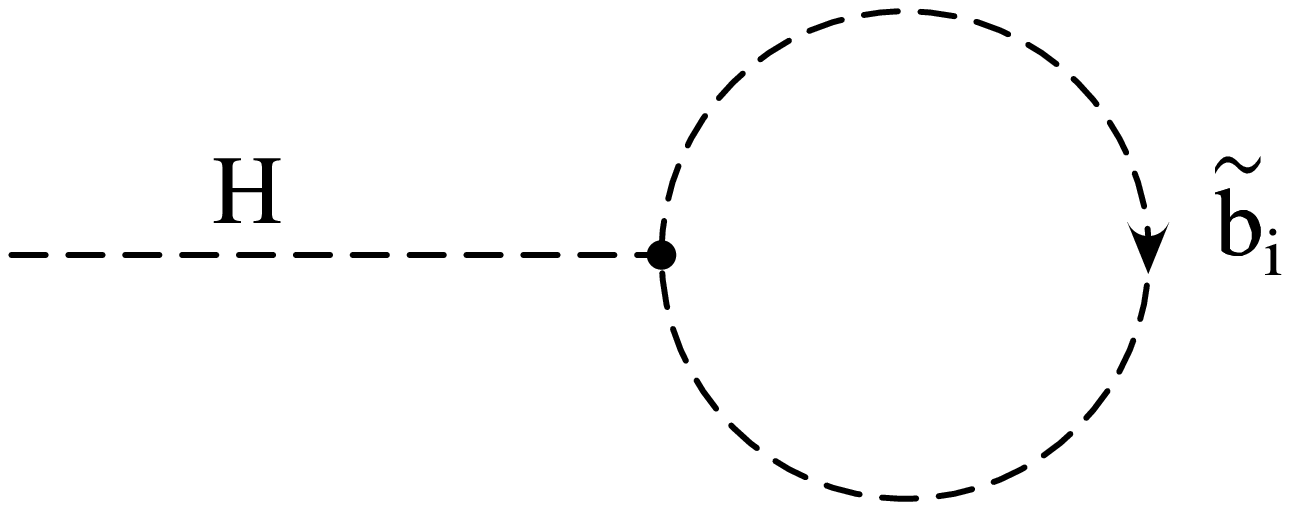,width=3.5cm,bbllx=150pt,bblly=370pt,
                                      bburx=450pt,bbury=470pt}
\hspace{2cm}
\psfig{figure=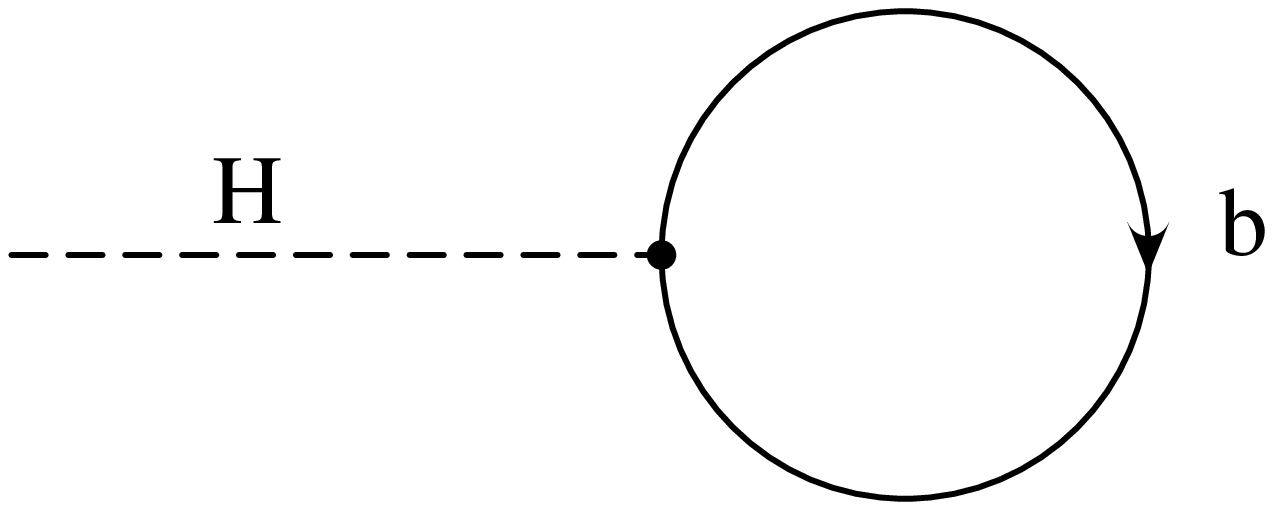,width=3.5cm,bbllx=150pt,bblly=370pt,
                                      bburx=450pt,bbury=470pt}
}
\end{center}
\caption[]{
Generic Feynman diagrams for the $b/\Sbot$ contributions to Higgs
tadpoles (H = $h, H$).
}
\label{fig:fdol_tp}
\end{figure}

\smallskip
The $b/\Sbot$ corrections to the renormalized Higgs boson
self-energies in \refeq{higgsmassmatrixnondiag} are given at the
\onel\ level by the Feynman diagrams shown in \reffis{fig:fdol},
\ref{fig:fdol_tp}. 

The $\dmb$ corrections included in the effective Lagrangian,
\refeq{effL}, enter the \onel\ diagrams in the following ways:
\begin{itemize}
\item
The factor $(\mb \Sa/\Cb)$ in the $hb\bar b$ coupling changes to 
\BE
\mb \frac{\Sa}{\Cb} \to 
 \frac{\mb}{1 + \dmb} \KL \frac{\Sa}{\Cb} - \dmb \frac{\Ca}{\Sbe} \KR .
\end{equation} 
The corresponding factor $(-\mb \Ca/\Cb)$ in the $Hb\bar b$ coupling is
shifted to
\BE
-\mb \frac{\Ca}{\Cb} \to 
 -\frac{\mb}{1 + \dmb} \KL \frac{\Ca}{\Cb} + \dmb \frac{\Sa}{\Sbe} \KR .
\end{equation} 
Finally, the $Ab\bar b$ coupling receives the factor
\BE
\mb \to \frac{\mb}{1 + \dmb} .
\end{equation}
It should be noted that the $b$~quark masses appearing in the bottom
propagators do not receive a correction.
\item
The $b$~quark masses in the $\Sbot$~mass matrix (\ref{sbotmassmatrix})
are due to Yukawa couplings from the scalar bottoms to the Higgs
doublets. Thus they receive the shift
\BE
\mb \to \frac{\mb}{1 + \dmb} .
\label{yukshift}
\end{equation}
\item
For the same reason the $\mb$~factors in the Higgs-$\Sbot\bar{\Sbot}$
couplings receive the same shift as given in \refeq{yukshift}.
\end{itemize}
Using the above changes in the \onel\ corrections includes
the leading $b/\Sbot$~corrections beyond \onel, resummed to all
orders.

Recently the complete \twol\, momentum-independent,
\order{\alb\al_s} corrections (which are not included in the
\order{\alb(\als\tb)^n} resummation) have been
computed~\cite{mhiggsEP4,mhiggsFD2}.  The result obtained makes use of an
appropriate choice of renormalization conditions on the relevant
parameters that allows to disentangle the genuine \twol\ effects
from the large threshold corrections to the bottom mass, and also ensures
the decoupling of heavy gluinos. 

\begin{figure}[htb!]
\begin{center}
\epsfig{figure=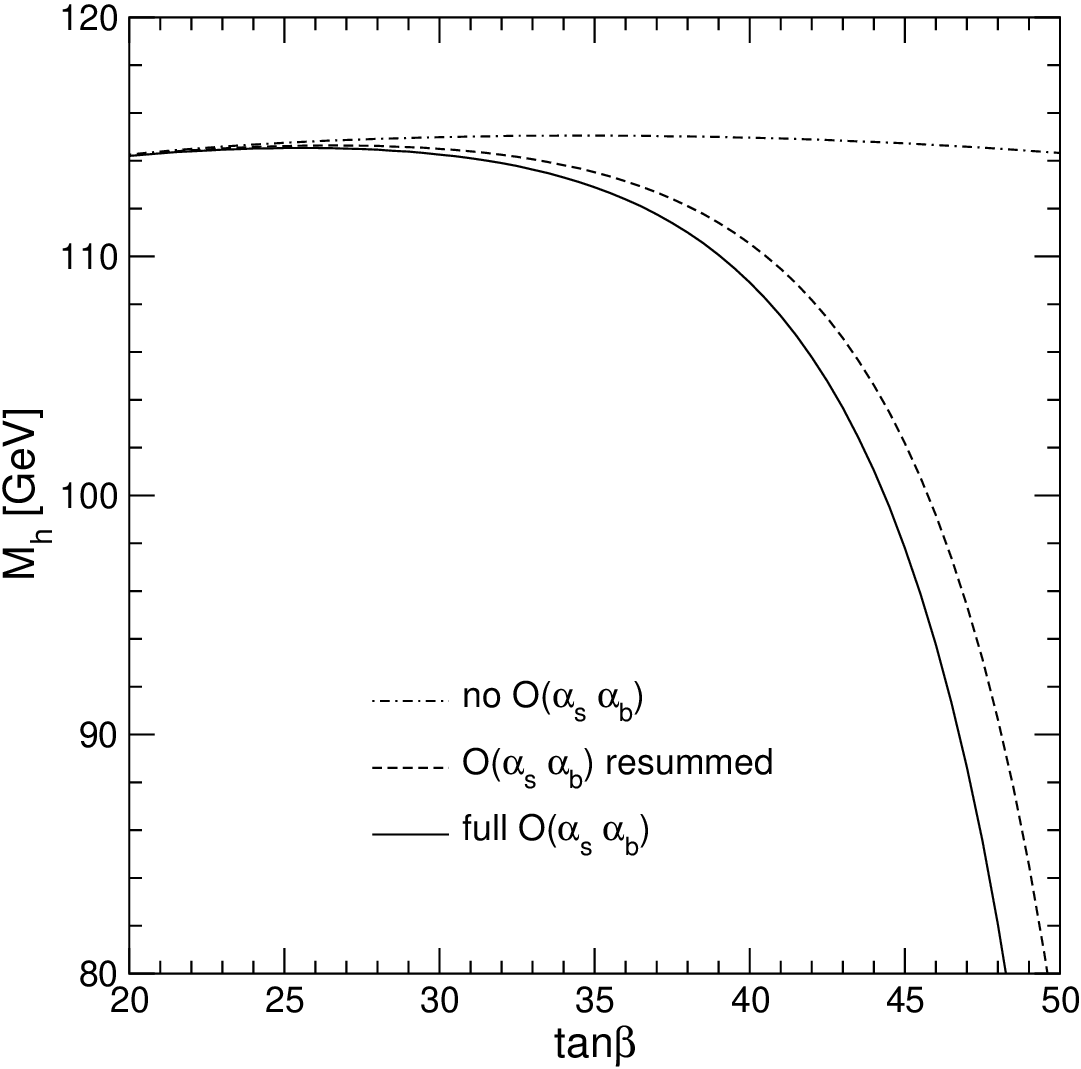,width=12cm}
\end{center}
\caption[]{ The result for the lightest $\cp$-even Higgs-boson mass in 
the MSSM, $\Mh$, as obtained with the program \fh\ is shown 
as a function of $\tb$ for $\MA = 120$ GeV, $\mu = -1 \tev$, 
$\MstL = \MstR = \MsbR = \mgl = 1$ TeV, $\At = \Ab = 2 \tev$. 
The meaning of the different curves is explained in the text.}
\label{fig:sbottom}
\end{figure}

To appreciate the importance of the various sbottom contributions, we
plot in Fig.~\ref{fig:sbottom} the light Higgs mass $\Mh$ as a
function of $\tb$. The SM running bottom mass computed at the top mass
scale, $\mb(\mt) = 2.74$ GeV, is used in order to account for the
universal large QCD corrections.  The relevant MSSM parameters are
chosen as $\MA = 120 \gev$, $\mu = -1 \tev$, $\MstL = \MstR = \MsbR =
\mgl = 1 \tev$, $\At = \Ab = 2 \tev$. The dot--dashed curve
in Fig.~\ref{fig:sbottom} includes the full \onel\ contribution as
well as the \twol\ \order{\alt\als + \alt^2} corrections (the
latter being approximately $\tb$-independent when $\tb$ is
large). The dashed curve includes also the resummation of the
$\tb$-enhanced threshold effects in the relation between $h_b$ and
$\mb$. Finally, the solid curve includes in addition the complete
\order{\alb\al_s} \twol\ corrections of \citere{mhiggsEP4}. In the
last two curves, the steep dependence of $\mh$ on $\tb$ when the
latter is large is driven by the sbottom contributions. One sees that
the $\tb$-enhanced threshold effects account for the bulk of
the sbottom contributions beyond \onel . The genuine
\order{\alb\al_s} \twol\ corrections can still shift $\mh$ by
a few GeV for very large values of $\tb$ and $\mu$.%


\subsection{Implications for $\tb$ exclusion bounds}
\label{sec:phenoimp}

The improved knowledge of the \twol\ contributions to the Higgs boson
self-energies results in a very precise prediction for the Higgs boson 
masses and mixing angle with interesting implications for MSSM parameter
space analyses. In this section an important consequence of the
various corrections is presented: the implications on the upper limit
on $\mh$ within the MSSM and on the corresponding limit on $\tb$ arising
from confronting the upper bound on $\mh$ with the lower limit from
Higgs searches, see also the discussion in \citere{tbexcl}. 

The theoretical upper bound on the lightest Higgs boson mass as a
function of $\tb$ can be combined with the results from direct
searches at LEP to constrain $\tb$. The diagonalization of the
tree-level mass matrix, \refeq{higgsmassmatrixtree}, yields a value for
$\mhtree$ that is maximal when $\MA \gg \MZ$, in which case 
$\mhtree^2 \simeq \MZ^2\, \CQZb$, 
which vanishes for $\tb=1$. 
Radiative corrections significantly increase the light
Higgs boson mass compared to its tree-level value, but still $\mh$ is
minimized for values of $\tb$ around one. Thus, in principle, the region
of low $\tb$ can be probed experimentally via the search for the
lightest MSSM Higgs boson~\cite{LEPHiggs}. If the remaining MSSM
parameters are tuned in such a way to obtain the maximal value of $\Mh$
as a function of $\tb$ (for reasonable values of $\msusy$ and taking
into account the experimental uncertainties of $\mt$ and the
other SM input parameters as well as the theoretical uncertainties from
unknown higher-order corrections), the experimental lower bound on
$\Mh$ can be used to obtain exclusion limits for $\tb$.
While in general a detailed investigation of a variety of different
possible production and decay modes is necessary in order to determine
whether a particular point of the MSSM parameter space can be excluded
via the Higgs searches or not, the situation simplifies considerably in
the region of small $\tb$ values. In this parameter region the lightest
$\cp$-even Higgs boson of the MSSM couples to the $Z$~boson with
SM-like strength, and its decay into a $b\bar b$ pair is not significantly
suppressed. Thus, within good approximation, constraints on $\tb$ can 
be obtained in this parameter region by confronting the exclusion bound
on the SM Higgs boson with the upper limit on $\Mh$ within the MSSM.
We use this approach below in order to discuss the implications of the
new $\Mh$ evaluation on $\tb$ exclusion bounds.

Concerning the upper bound on $\Mh$ within the MSSM,
the \onel\ corrections contribute positively to $\Mh^2$. The
\twol\ effects of \order{\alt\als} and \order{\alt^2}, on the other
hand, enter with
competing signs, the former reducing $\Mh^2$ while the latter give a
(smaller) positive contribution.  The actual bound that can be derived depends
sensitively on the precise value of the top-quark mass, because the
dominant \onel\ contribution to $\Mh^2$, as well as the
\twol\ \order{\alt\als} term, scale as $\mt^4$.  Furthermore, a
large top mass amplifies the relative importance of the \twol\
\order{\alt^2} correction, because of the additional $\mt^2$ factor.

In order to discuss restrictions on the MSSM parameter space it has
become customary in the recent years to refer to so-called benchmark
scenarios of MSSM parameters~\cite{benchmark,sps}. The $\mhmax$ benchmark
scenario~\cite{benchmark} has been designed such that for fixed values
of $\mt$ and $\msusy$ the predicted value of the lightest $\cp$-even
Higgs boson mass is maximized for each value of $\MA$ and $\tb$. The
value of the 
top-quark mass had been fixed to its experimental central value, 
$\mt = 174.3 \gev$, while the SUSY parameters are taken as in the
$\mhmax$ scenario, see the Appendix.
Furthermore, $\MA$ has been fixed to $\MA = 1 \tev$.

\begin{figure}[htb!]
\begin{center}
\epsfig{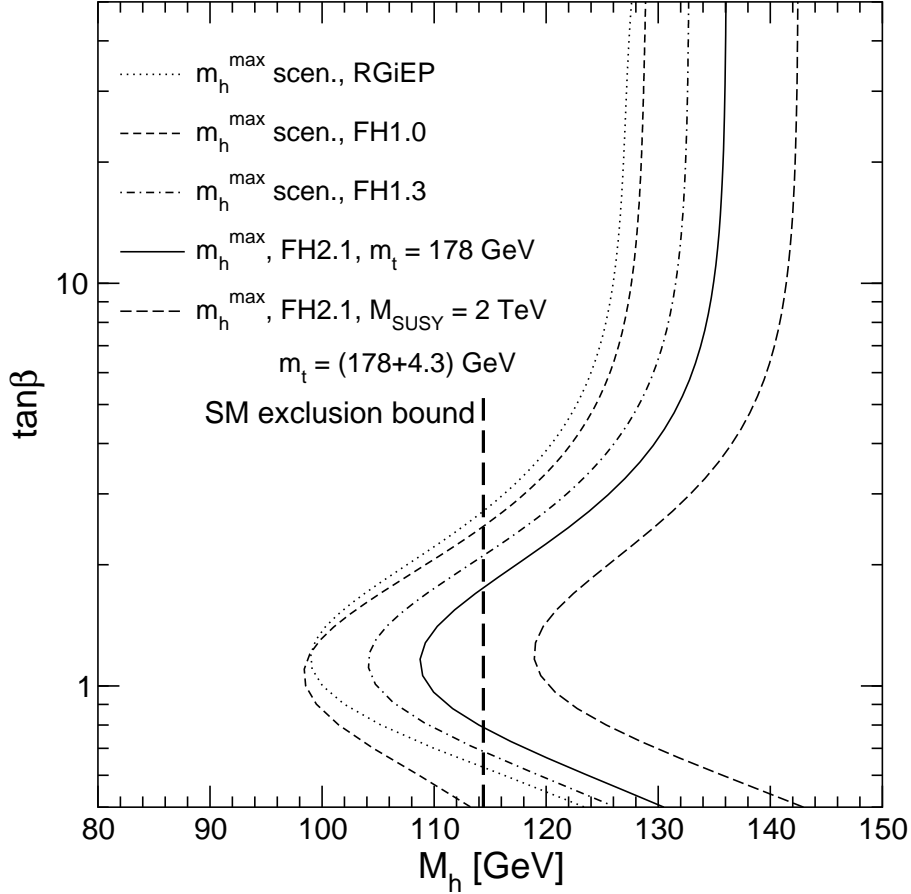}
\end{center}
\vspace{-1em}
\caption[]{The result for the lightest $\cp$-even Higgs boson mass, $\Mh$,
as a function of $\tb$ in the $\mhmax$~scenario. The
dotted curve has been obtained with a renormalization-group improved
effective potential
method, using the code {\em subhpoledm}.  The short-dashed curve
corresponds to the result obtained with \fh1.0, while the dash-dotted
curve shows the  
result of \fh1.3, where the improvements described in
\refse{subsec:recentHO} are included.
The full curve shows the result from \fh2.1, using the top quark mass
of $\mt = 178.0 \gev$. 
The dot-dashed curve, also 
obtained with \fh2.1, employs $\mt = 182.3 \gev$ and $\msusy = 2
\tev$. The vertical long-dashed line corresponds to the LEP exclusion
bound for the SM Higgs boson of $114.4 \gev$.  }
\label{fig:mhtb}
\end{figure}

In \reffi{fig:mhtb} we plot $\Mh$ as a function of $\tb$ in the
$\mhmax$~scenario. The dashed and dot-dashed curves correspond to the
result obtained with the previous (used for the LEP evaluations so
far~\cite{LEPHiggs}) and the more advanced version (where the improvements
described in \refse{subsec:recentHO} are included, and which will be
used for the final 
LEP evaluations~\cite{arnulf}) of \fh, respectively. The two
versions, \fh1.0 and \fh1.3, differ by the recent improvements obtained
in the MSSM Higgs sector which are described in \refse{subsec:recentHO}.
Also shown as a solid line is the result obtained with \fh2.1 (which
for the $\mhmax$~scenario should yield quantitatively the same result
as \fh1.3), but with $\mt$ set to the recently obtained new
experimental value, $\mt^{\rm exp} = 178.0 \gev$~\cite{mtopexpnew}.
For comparison, also the result obtained with a renormalization-group
improved effective potential method is indicated. The dotted curve in
Fig.~\ref{fig:mhtb} corresponds to the code 
{\em subhpoledm}~\cite{mhiggsRG1,bse,deltamb} in the 
$\mhmax$~scenario, for a $\Stop$~mixing parameter $\Xt^{\msbarm} =
\sqrt{6}\,\msusy$~\cite{mhiggsRG1,bse} (see also the Appendix).
It deviates from the 
result of \fh1.0 by typically not more than 1~GeV for $\tb \geq 1$.
The LEP exclusion bound for the mass of a SM-like Higgs~\cite{LEPHiggsSM},
$\MHSM \ge 114.4 \gev$, is shown in the figure as a vertical long--dashed 
line. As can be seen from the figure, the improvements on the
theoretical prediction described in \refse{subsec:recentHO}, in particular
the inclusion of the complete momentum-independent \order{\alt^2}
corrections into \fh, gives rise to a significant increase in the upper
bound on $\Mh$ as a function of $\tb$. Comparison of this prediction 
with the exclusion bound on a SM-like Higgs shows that the lower limit
on $\tb$ is considerably weakened. Also the new top mass value, see
the solid line, shifting the resulting $\Mh$ values upwards, weakens
the $\tb$ exclusion bound.

Concerning the interpretation of the results shown in
Fig.~\ref{fig:mhtb}, it should be kept in mind that within the (pure)
$\mhmax$ 
benchmark scenario $\mt$ and $\msusy$ are kept fixed, and no theoretical
uncertainties from unknown higher-order corrections are taken into
account. In order to arrive at a more general exclusion bound on $\tb$
that is not restricted to a particular benchmark scenario, the impact of
the parametric and higher-order uncertainties in the prediction for
$\Mh$ has to be considered~\cite{tbexcl}. In order to demonstrate in 
particular the dependence of the $\tb$ exclusion bound on the chosen
value of the top pole mass, besides the solid line obtained with 
$\mt = 178.0 \gev$, the dot--dashed curve in Fig.~\ref{fig:mhtb}
shows the result obtained with \fh2.1 where the top-quark mass has been
increased by one standard deviation, $\sigma_{\mt} = 4.3\, \gev$, to
$\mt = 182.3 \gev$, and $\msusy$ has been changed from 1~TeV to 2~TeV.
It can be seen that in this more general scenario no lower limit on
$\tb$ from the LEP Higgs searches can be obtained. 

Constraints from the Higgs searches at LEP do of course play an
important role in regions of the MSSM parameter space where the
parameters are such that $\Mh$ does not reach its maximum value. Also in
this case, however, the remaining theoretical uncertainties from unknown
higher-order corrections (see \refse{sec:hocorr} below) have to be
taken into account in order to obtain conservative exclusion limits.


\section{Comparison with the RG approach}
\label{sec:RGcomp}

The diagrammatic \twol\ computation of the
dominant contributions at $\oaas$ to the neutral $\cp$-even Higgs boson
masses~\cite{mhiggsletter} had been obtained first in the
on-shell scheme, was subsequently combined~\cite{mhiggslong}
with the complete
diagrammatic \onel\ on-shell result of \citere{mhiggsf1lC}.
Within the RG approach~\cite{mhiggsRG1,mhiggsRG2}, on the
other hand, the \twol\ results was expressed in
terms of the top-quark mass in the \msbar\ scheme.

Besides a sizable shift in the upper bound of $\Mh$ of about $5 \gev$,
apparent deviations 
between the explicit diagrammatic \twol\ calculation and the
results of the RG computation were observed in the dependence
of $\Mh$ on the stop-mixing parameter $\Xt$. While the value of
$\Xt$ that maximizes the lightest $\cp$-even Higgs mass is
$(\Xt)_{\rm max}\simeq \pm\sqrt{6}\,\ms \approx \pm 2.4\,\ms$ ($\ms$
denotes the soft SUSY-breaking parameter in the $\Stop$~and
$\Sbot$~sector, $\ms \equiv \MstL = \MstR = \MsbL = \MsbR$) in the
RG results,
the corresponding on-shell \twol\ diagrammatic computation found
a maximal value for $\Mh$ at $(\Xt)_{\rm max}\approx2\ms$.%
\footnote{
A local maximum for $\Mh$ is also found for
$\Xt\simeq -2\ms$, although the corresponding value of $\Mh$ at
$\Xt\simeq +2\ms$ is significantly
larger~\cite{mhiggslong,mhiggslle}.
}
Moreover, the
RG result is symmetric under $\Xt\to -\Xt$ and has a (local) minimum
at $\Xt=0$.  In contrast, the \twol\ diagrammatic
computation yields $\Mh$
values for positive and negative $\Xt$ that differ significantly from
each other and the local minimum in $\Mh$ is shifted slightly
away from $\Xt= 0$~\cite{mhiggslle}.  

In this section, it is 
shown that this apparent discrepancy is caused by the different
renormalization schemes employed in the two approaches, leading to
differences in the leading-logarithmic contributions.   
It will furthermore be shown that in
the analytic approximation for the leading $\mt^4$ corrections, the dominant
numerical contribution of the new genuine non-logarithmic \twol\
contributions of the FD result~\cite{mhiggsletter,mhiggslong}
can be absorbed into an
effective \onel\ expression by choosing an appropriate scale for
the running top-quark mass in different terms of the expression.


\subsection{Approximation formulas for $\Mh$}
\label{subsec:mhapprox}

Within the RG approach the following approximation for $\Mh$ has been
obtained~\cite{mhiggsRG1a,mhiggsRG1,mhiggsRG2,bse}, taking into
account terms up to $\oaas$: 
\BEA
\lefteqn{
\Mh^2 = \mh^{2,{\rm tree}} +
  \frac{3}{2} \frac{G_F \sqrt{2}}{\pi^2} \mtms^4 \left\{
  - \ln\left(\frac{\mtms^2}{\Msbarii} \right) +
    \frac{\Xtbarii}{\Msbarii}
  \left(1 - \frac{1}{12} \frac{\Xtbarii}{\Msbarii} \right)
  \right\} \non } \\
&& {} - 3 \frac{G_F \sqrt{2}}{\pi^2} \frac{\alpha_s}{\pi} \mtms^4
  \left \{\ln^2 \left(\frac{\mtms^2}{\Msbarii} \right) +
    \left[\frac{2}{3} - 2 \frac{\Xtbarii}{\Msbarii}
          \left(1 - \frac{1}{12} \frac{\Xtbarii}{\Msbarii} \right) \right]
    \ln\left(\frac{\mtms^2}{\Msbarii} \right) \right \},
\label{eq:mhrg}
\EEA
where we have introduced the notation $\Msbar, \Xtbar$ to emphasize that
the corresponding quantities are \msbar\ parameters, which are
evaluated at the scale $\mu = \ms$:
\BE
\label{eq:MsXtbar}
\Msbar \equiv \ms^{\MS}(\ms), \quad
\Xtbar \equiv \Xt^{\MS}(\ms),
\end{equation}
and
$\mtms \equiv \mtsmreg(\mt)$ is the \msbar\ top mass
\BE
\label{mtmsbar}
 \mtms = \mtms(\mt) \approx \frac{\Mt}{1 + \frac{4}{3\,\pi}\als(\mt)} ~.
\end{equation}
$\Mt$ denotes the top quark pole mass.
Since in this section only the \order{\al\als} corrections are
investigated, we do not include here the contributions of
\order{\al_t}, see however \refse{subsec:missing3L}.

Concerning the FD result,
we consider the dominant \onel\ and \twol\ terms and we
focus on the case $\MA \gg \MZ$, for which
the result for $\Mh^2$ can be expressed in a particularly compact
form~\cite{mhiggslle}
\BE
\label{eq:mhdiagos}
\Mh^2 = \mh^{2,{\rm tree}} + \mh^{2,{\alpha}} +
\mh^{2,{\alpha\alpha_s}},
\end{equation}
and neglect the non-leading terms of ${\cal O}(\MZ^2/\MA^2)$.
Assuming that $\ms \gg \Mt$ and neglecting the
non-leading
terms of ${\cal O}(\Mt/\ms)$ and ${\cal O}(\MZ^2/\Mt^2)$, one obtains
the following simple result for the \onel\ and \twol\ contributions
\BEA
\mh^{2,{\alpha}} &=&
  \frac{3}{2} \frac{G_F \sqrt{2}}{\pi^2} \Mt^4 \left\{
  - \ln\left(\frac{\Mt^2}{\ms^2} \right)
  + \frac{\Xt^2}{\ms^2}
  \left(1 - \frac{1}{12} \frac{\Xt^2}{\ms^2} \right)
  \right\}, 
\label{eq:mh1ldiagos} \\
\mh^{2,{\alpha\alpha_s}} &=&
  - 3 \frac{G_F \sqrt{2}}{\pi^2} \frac{\alpha_s}{\pi} \Mt^4
  \left\{\ln^2 \left(\frac{\Mt^2}{\ms^2} \right)
  - \left(2 + \frac{\Xt^2}{\ms^2} \right)
  \ln\left(\frac{\Mt^2}{\ms^2} \right)
- \frac{\Xt}{\ms} \left(2 - \frac{1}{4} \frac{\Xt^3}{\ms^3} \right)
  \right\} . \non \\
&& \label{eq:mh2ldiagos}
\EEA
The corresponding formulae, in which terms up to ${\cal O}(\Mt^4/\ms^4)$
are kept, can be found in \citere{bse}.

In \refeqsto{eq:mh1ldiagos}{eq:mh2ldiagos} the parameters
$\Mt$, $\ms$, $\Xt$ are on-shell quantities.
Using \refeq{mtmsbar}, the on-shell result for $\Mh^2$,
\eqs{eq:mhdiagos}{eq:mh2ldiagos}, can easily be rewritten
in terms of the running top-quark mass $\mtms$.
While this reparameterization does not change the
form of the \onel\ result, it induces an extra contribution at
$\oaas$. Keeping 
again only terms that are not
suppressed by powers of $\mtms/\ms$, 
the resulting expressions read
\BEA
\mh^{2,{\alpha}} &=&
  \frac{3}{2} \frac{G_F \sqrt{2}}{\pi^2} \mtms^4 \Biggl\{
  - \ln\left(\frac{\mtms^2}{\ms^2} \right) +
  \frac{\Xt^2}{\ms^2}
  \left(1 - \frac{1}{12} \frac{\Xt^2}{\ms^2} \right)
  \Biggr\} , 
  \label{eq:mh1ldiagosmtms} \\
\mh^{2,{\alpha\alpha_s}} &=&
  - 3 \frac{G_F \sqrt{2}}{\pi^2} \frac{\alpha_s}{\pi} \mtms^4
  \Biggl\{\ln^2 \left(\frac{\mtms^2}{\ms^2} \right) +
  \left(\frac{2}{3} - \frac{\Xt^2}{\ms^2} \right)
  \ln \left(\frac{\mtms^2}{\ms^2} \right)\non  \\
&& {} + \frac{4}{3} - 2 \frac{\Xt}{\ms} - \frac{8}{3} \frac{\Xt^2}{\ms^2}
   + \frac{17}{36} \frac{\Xt^4}{\ms^4} \Biggr\} ,
\label{eq:mh2ldiagosmtms}
\EEA
in accordance with the formulae given in \citere{mhiggslle}.

We now compare the diagrammatic result expressed in terms of the parameters
$\mtms$, $\ms$, $\Xt$,
\refeqsto{eq:mh1ldiagosmtms}{eq:mh2ldiagosmtms},
with the RG result of \refeq{eq:mhrg}, which is given
in terms of the \msbar\ parameters $\mtms$, $\Msbar$, $\Xtbar$
\refeq{eq:MsXtbar}.  While the $\Xt$-independent logarithmic terms
are the same in both the diagrammatic
and RG results, the corresponding logarithmic terms at two loops that
are proportional to powers of $\Xt$ and $\Xtbar$, respectively,
are different.  Furthermore, \refeq{eq:mh2ldiagosmtms} does not
contain a logarithmic term proportional to $\Xt^4$,
while the corresponding  term proportional to $\Xtbariv$ appears in
\refeq{eq:mhrg}.
To check whether these results are consistent with each other, one
must relate the 
on-shell and \msbar\ definitions of the parameters $\ms$ and $\Xt$.

Finally, we note that the non-logarithmic terms contained in
\refeq{eq:mh2ldiagosmtms}
correspond to genuine \twol\ contributions
that are not present in the RG result of \refeq{eq:mhrg}.
They can be interpreted as a \twol\
finite threshold correction to the quartic Higgs self-coupling in the
RG approach.  In particular, note that
\refeq{eq:mh2ldiagosmtms} contains a term that is
linear in $\Xt$.
This is the main source of the asymmetry in the \twol\ corrected
Higgs mass under $\Xt\to -\Xt$ obtained by the diagrammatic method.
The non-logarithmic terms in \refeq{eq:mh2ldiagosmtms} give rise to
a numerically significant increase of the
maximal value of $\Mh$ of about 5~GeV in this approximation.


\subsection{On-shell and $\overline{\rm MS}$ definitions
of $\ms$ and $\Xt$}

Since the parameters $p = \{\mste^2, \mstz^2, \tst, \mt\}$ of the $t$--$\Stop$
sector are renormalized differently in different schemes,
the parameters $\ms$ and $\Xt$ also have a different meaning in these
schemes. The relation between these parameters in the
\msbar\ and in the on-shell scheme have been derived in
\citeres{bse,reconcA} and read in leading order in $\mt/\ms$:
\BEA
\Msbarii &=& \ms^{2, \OS}
 - \frac{8}{3} \frac{\alpha_s}{\pi} \ms^2 \,,
\label{eq:msms} \\
\Xtbar &=& \Xt^{\OS} \frac{\Mt}{\mtms(\ms)} +
  \frac{8}{3} \frac{\alpha_s}{\pi} \ms \,.
\label{eq:xtms}
\EEA
As previously noted, it is not necessary to specify the definition of
the parameters that appear in the ${\cal O}(\alpha_s)$ terms as long
as higher orders are neglected. Thus, we
use the generic symbol $\ms^2$ in the ${\cal O}(\alpha_s)$ terms of
\eqs{eq:msms}{eq:xtms}.  The corresponding results including terms
up to ${\cal O}\left(\mt^4/\ms^4\right)$ can be found in \citere{bse}.

Finally, the evaluation of the ratio $\Mt/\mtms(\ms)$ is needed.
In leading order in $\mt/\ms$ it is given by~\cite{bse}:
\BE
\mtms(\ms) =
  \mtms \left[1 +
\frac{\alpha_s}{\pi}
    \ln\left(\frac{\mt^2}{\ms^2}\right)
   +  \frac{\alpha_s}{3\pi} \frac{\Xt}{\ms} \right]\,,
\label{eq:mtsusy}
\end{equation}
where $\mtms\equiv \mtsmreg(\mt)$ is given in terms of $\Mt$ by
\refeq{mtmsbar}.
Note that the term in \refeq{eq:mtsusy} that is proportional to $\Xt$ is a
threshold correction due to the supersymmetry-breaking stop-mixing
effect. Inserting the result of \refeq{eq:mtsusy} into \refeq{eq:xtms} yields
\BE \label{eq:xtrelsusylead}
\Xtbar  = \Xt^{\OS} + \frac{\alpha_s}{3 \pi} \ms
   \left[8 + \frac{4\Xt}{\ms} - \frac{\Xt^2}{\ms^2} - \frac{3\Xt}{\ms}
\ln\left(\frac{\mt^2}{\ms^2}\right)
   \right]\,.
\end{equation}
It is interesting to note that $\Xtbar\neq 0$ when $\Xt^{\OS}= 0$.
Moreover, it is clear from \refeq{eq:xtrelsusylead} that the relation
between $\Xt$ defined in the on-shell and the \msbar\ schemes
includes a leading logarithmic effect, which has to be taken into
account in a comparison of the leading logarithmic contributions
in the RG and the \twol\ diagrammatic results.

A remark on the regularization scheme is
in order here.  In the effective field theory, the running top-quark mass
at scales below $\ms$ is the SM running coupling
of \refeq{mtmsbar}, which
is calculated in dimensional regularization.
This is matched to the running top-quark mass as computed in the
full supersymmetric theory.
One could argue that the appropriate regularization
scheme for the latter should be dimensional reduction
(DRED)~\cite{dred}, which is usually applied in loop calculations in
supersymmetry.\footnote{In order
to obtain the corresponding DRED result, one simply has to replace the
term $4 \alpha_s/3 \pi$ in the denominator of
\refeq{mtmsbar} by $5 \alpha_s/3 \pi$.}
The result of such a change would be to modify slightly the \twol\
non-logarithmic contribution to $\Mh$ that is proportional to
powers of $\Xt$.
Of course, the physical Higgs mass is independent of any scheme.
One is free to re-express \refeqsto{eq:mh1ldiagos}{eq:mh2ldiagos}
(which depend on the on-shell parameters $\Mt$, $\ms$, $\Xt$)
in terms of parameters defined in any other scheme.  In
this paper, we find \msbar--renormalization via DREG to be
the most convenient scheme
for the comparison of the diagrammatic and RG results for $\Mh$.


\subsection{Comparing the RG and the Feynman diagrammatic results}
\label{subsec:rgcomp}

In order to directly compare the \twol\ diagrammatic and RG results,
we must convert from on-shell to \msbar\ parameters.
Inserting \refeqsto{eq:msms}{eq:xtrelsusylead} into
\refeqsto{eq:mh1ldiagosmtms}{eq:mh2ldiagosmtms}, one finds
\BEA
\mh^{2,{\alpha}} &=&
  \frac{3}{2} \frac{G_F \sqrt{2}}{\pi^2} \mtms^4 \left\{
  - \ln\left(\frac{\mtms^2}{\Msbarii} \right)
  + \frac{\Xtbarii}{\Msbarii}
  \left(1 - \frac{1}{12} \frac{\Xtbarii}{\Msbarii} \right)\right\}\,,
\label{eq:MSbarDREG1} \\
\mh^{2,{\alpha\alpha_s}} &=&
  - 3 \frac{G_F \sqrt{2}}{\pi^2} \frac{\alpha_s}{\pi} \mtms^4
  \Biggl\{\ln^2 \left(\frac{\mtms^2}{\Msbarii} \right) +
    \left[\frac{2}{3} - 2 \frac{\Xtbarii}{\Msbarii}
          \left(1 - \frac{1}{12} \frac{\Xtbarii}{\Msbarii} \right) \right]
    \ln\left(\frac{\mtms^2}{\Msbarii} \right)  \non \\
&& {} + \frac{\Xtbar}{\Msbar} \left(\frac{2}{3} - \frac{7}{9}
   \frac{\Xtbarii}{\Msbarii} + \frac{1}{36}\frac{\Xtbariii}{\Msbariii}
+ \frac{1}{18} \frac{\Xtbariv}{\Msbariv} \right)
   \Biggr\} + {\cal O}\left(\frac{\mtms}{\Msbar}\right)\,.
\label{eq:MSbarDREG2}
\EEA

Comparing \refeq{eq:MSbarDREG2} with \refeq{eq:mhrg} shows that
the logarithmic contributions of the diagrammatic result expressed in
terms of the \msbar\ parameters $\mtms$, $\Msbar$, $\Xtbar$ agree with
the logarithmic contributions obtained by the RG
approach. The differences in the logarithmic terms
observed in the comparison of \refeqsto{eq:mh1ldiagosmtms}{eq:mh2ldiagosmtms}
with \refeq{eq:mhrg} have thus been traced to the different renormalization
schemes applied in the respective
calculations. The fact that the logarithmic contributions obtained
within the two approaches agree after a proper rewriting of the
parameters of the stop sector is an important
consistency check of the calculations.  In addition to the
logarithmic contributions, \refeq{eq:MSbarDREG2} also
contains non-logarithmic contributions, which are
numerically sizable.

\begin{figure}[htb!]
\begin{center}
\epsfig{figure=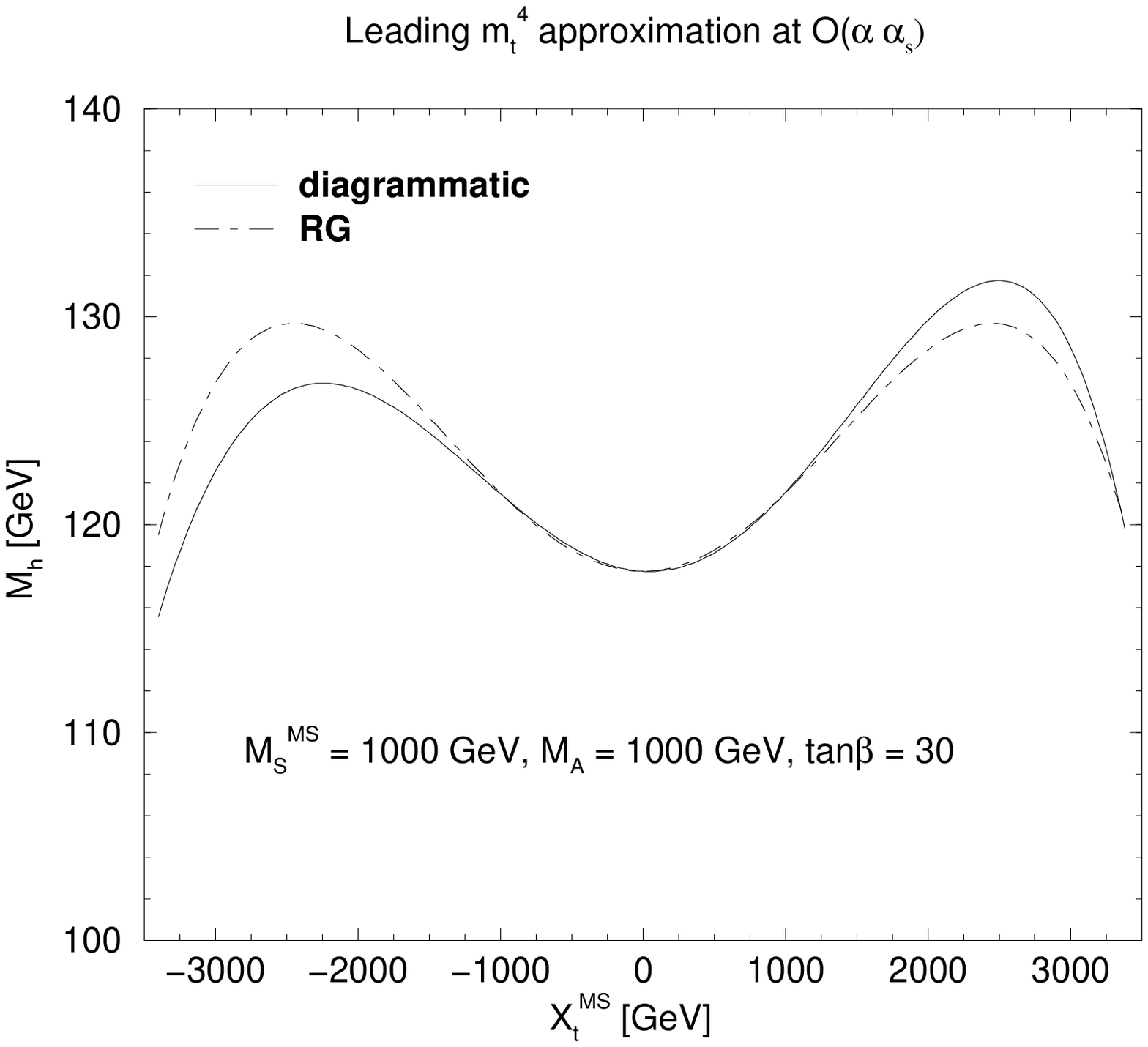,width=14cm,height=10cm}
\end{center}
\caption{Comparison of the diagrammatic \twol\
$\oaas$ result for $\Mh$, to leading
order in $\mtms/\Msbar$ (\refeqsto{eq:MSbarDREG1}{eq:MSbarDREG2})
with the RG result of \refeq{eq:mhrg}.
Note that the latter omits the
\onel\ threshold corrections due to stop mixing in the
evaluation of $\mtms(\ms)$.  Since this quantity enters in the
definition of $\Xtbar$, see \refeq{eq:xtms}, the meaning of
$\Xt^{\rm MS}$ plotted along the $x$-axis is slightly different
for the diagrammatic curve, where $\Xt^{\rm MS}=\Xtbar$, and the
RG curve, where 
$\Xt^{\rm MS}=\Xtbar\,[1+(\alpha_s/3\pi)(\Xt/\ms)]$. See text for further
details. The graph is plotted for 
$\Msbar=\MA=(\mgl^2+\mtms^2)^{1/2}= 1 \tev$.
} 
\label{fig:msbar1}
\end{figure}

In \reffi{fig:msbar1}, 
we compare the diagrammatic result for $\Mh$
in the leading $\mt^4$ approximation to the results obtained 
by RG techniques.
While the diagrammatic result expressed in terms of $\mtms$, $\Msbar$,
$\Xtbar$ agrees well with the RG result in the region of no mixing in
the stop sector, sizable deviations occur for large mixing.
In particular, the non-logarithmic contributions give rise
to an asymmetry under the change of sign of the parameter $\Xtbar$,
while the RG result is symmetric under $\Xtbar\to -\Xtbar$. In the
approximation considered here, the maximal value for $\Mh$ in
the diagrammatic result lies about 3~GeV higher than the maximal value
of the RG result. The differences are slightly larger for smaller
$\tb$ values.
In addition, as previously noted, the
maximal-mixing point $(\Xtbar)_{\rm max}$
(where the radiatively corrected value of $\Mh$ is maximal)
is equal to its \onel\ value, $(\Xtbar)_{\rm max}\simeq \pm\sqrt{6}\ms$, 
in the RG result of \refeq{eq:mhrg},
while it is shifted in the \twol\ diagrammatic result.
However, \reffi{fig:msbar1} illustrates that the shift in 
$(\Xtbar)_{\rm max}$ from its \onel\ value, while significant in the \twol\
on-shell diagrammatic result, is largely diminished when the latter is
re-expressed in terms of \msbar\ parameters.

The differences between the diagrammatic and RG results shown
in \reffi{fig:msbar1} can be attributed to
non-negligible non-logarithmic terms proportional to powers of $\Xt$.
Clearly, the RG technique can be improved to incorporate these terms.
In \citere{mhiggsRG2}, it was shown that the leading \twol\ contributions
to $\Mh^2$ given by the RG result of \refeq{eq:mhrg} could be absorbed
into an effective \onel\ expression.
This was accomplished by considering separately the
$\Xt$--independent leading double
logarithmic term (the ``no-mixing'' contribution) and the
leading single logarithmic term that is proportional to powers of
$\Xtbar$ (the ``mixing'' contribution) at $\oaas$.
Both terms can be reproduced by an effective \onel\ expression, where
$\mtms$ in \refeq{eq:MSbarDREG1}, which appears in the
no-mixing and mixing contributions, is replaced by the
{\it running} top-quark mass evaluated at the scales $\mu_t$ and
$\mu_{\Stop}$, respectively:
\BE
\mbox{no mixing: } \mu_t \equiv (\mtms \Msbar)^{1/2}\,, \qquad
\mbox{mixing: } \mu_{\Stop} \equiv \Msbar\,.
\label{eq:miximpr}
\end{equation}
That is, at $\oaas$, the leading double logarithmic term
is precisely reproduced by the single-logarithmic term at
$\oa$, by replacing $\mtms$ with $\mtms(\mu_t)$,
while the leading single logarithmic term at two loops proportional to
powers of $\Xtbar$ is
precisely reproduced by the corresponding non-logarithmic terms
proportional to
$\Xtbar$ by replacing $\mtms$ with $\mtms(\Msbar)$.

\begin{figure}[htb!]
\begin{center}
\epsfig{figure=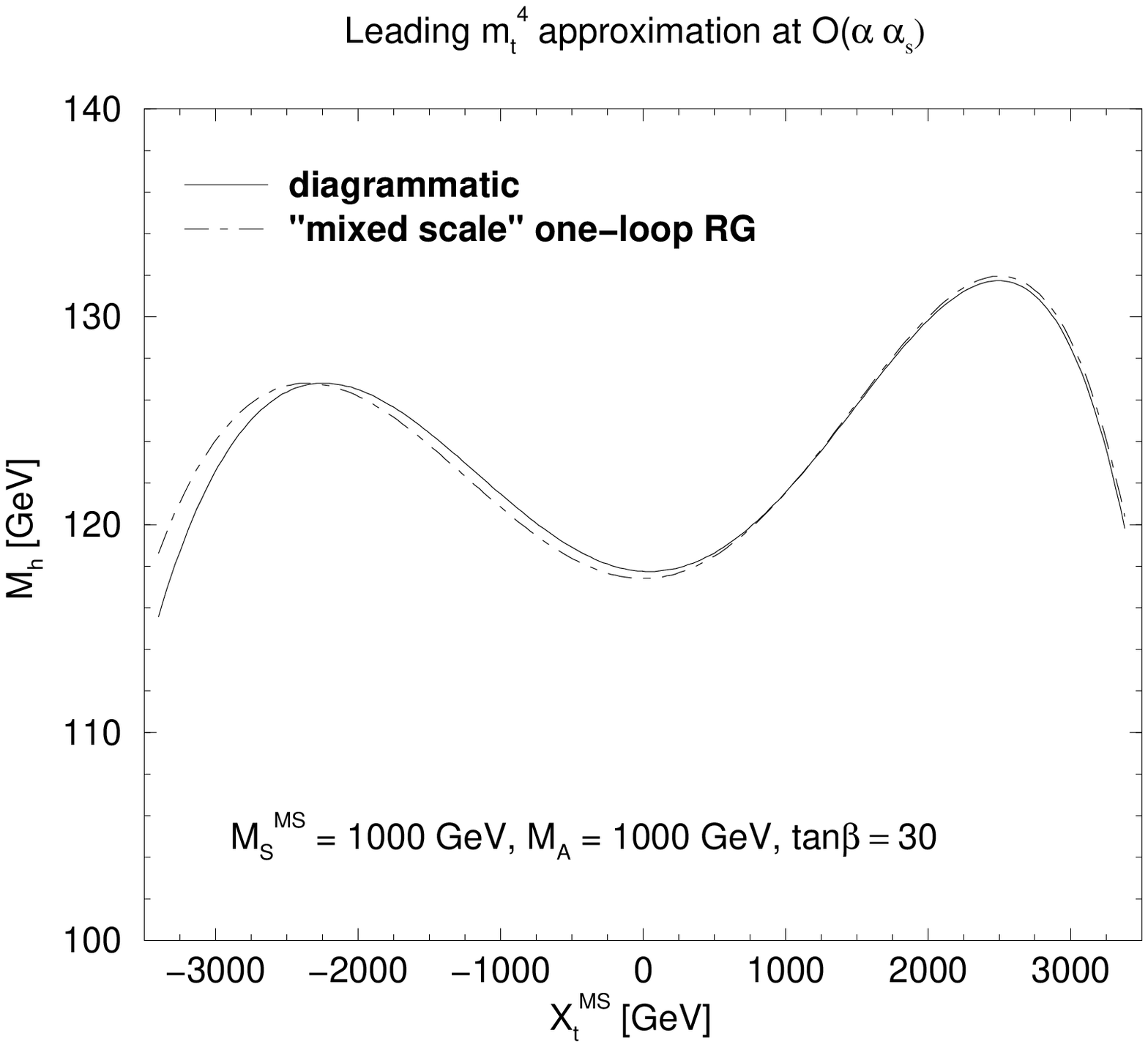,width=14cm,height=10cm}
\end{center}
\caption{Comparison of the diagrammatic \twol\
$\oaas$ result for $\Mh$, to leading
order in $\mtms/\Msbar$ (\refeqsto{eq:MSbarDREG1}{eq:MSbarDREG2})
with the ``mixed-scale'' \onel\ RG result
of \refeq{eq:mh1lmixed}.  Note that the latter now
includes the threshold corrections due to stop mixing in the
evaluation of $\mtms(\ms)$ in contrast to the RG results
depicted in \reffi{fig:msbar1}.  ``Mixed-scale'' indicates that in
the no-mixing and mixing contributions to the \onel\ Higgs mass,
the running top quark mass is evaluated at different scales
according to \refeq{eq:miximpr}. See text for further details.  The
graph is plotted for
$\Msbar=\MA=(\mgl^2+\mtms^2)^{1/2}= 1 \tev$.}
\label{fig:impr1}
\end{figure}

Applying the same procedure to \refeq{eq:MSbarDREG2} and rewriting it in
terms of the running top-quark
mass at the corresponding scales as specified in \refeq{eq:miximpr},
we obtain
\BEA
\mh^{2,{\alpha}} &=&
  \frac{3}{2} \frac{G_F \sqrt{2}}{\pi^2} \left\{
- \mtms^4(\mu_t)\ln\left(\frac{\mtms^2(\mu_t)}{\Msbarii}\right)
+ \mtms^4(\ms)
  \frac{\Xtbarii}{\Msbarii}
  \left(1 -  \frac{\Xtbarii}{12\Msbarii} \right)
  \right\} , 
  \label{eq:mh1lmixed} \\
\mh^{2,{\alpha\alpha_s}} &=&
  - 3 \frac{G_F \sqrt{2}}{\pi^2} \frac{\alpha_s}{\pi} \mt^4
  \left\{\frac{1}{6} \ln\left(\frac{\mt^2}{\ms^2} \right) +
  \frac{\Xt}{\ms} \left(\frac{2}{3} - \frac{1}{9} \frac{\Xt^2}{\ms^2} +
  \frac{1}{36} \frac{\Xt^3}{\ms^3} \right)
   \right\} \,.
\label{eq:impr}
\EEA
Indeed, the $\Xt$--independent leading double logarithmic term
and the leading single logarithmic term that is
proportional to powers of $\Xtbar$ have disappeared from the
\twol\ expression \refeq{eq:impr}, having been absorbed into an
effective \onel\ result (\ref{eq:mh1lmixed}), (denoted henceforth
as the ``mixed-scale'' \onel\ RG result).  Among the terms that
remain in \refeq{eq:impr}, there is a subleading \onel\ logarithm at
two loops which is a remnant of the no-mixing contribution.  But,
note that the magnitude of the coefficient ($1/6$) has been
reduced from the corresponding coefficients that appear in
\refeqsto{eq:mh2ldiagos}{eq:MSbarDREG2} ($-2$ and $2/3$,
respectively). In addition, the remaining leftover \twol\
non-logarithmic terms are also numerically insignificant. We
conclude that the ``mixed-scale'' \onel\ RG result provides a
very good approximation to the leading $\mt^4$ corrections to $\Mh^2$
at \order{\als\alt}, in which the most significant
\twol\ terms have been absorbed into an effective \onel\
expression.

To illustrate this result, we compare in
\reffi{fig:impr1} the diagrammatic \twol\ result
expressed in terms of \msbar\ parameters
(\refeqsto{eq:MSbarDREG1}{eq:MSbarDREG2}) with the
``mixed-scale'' \onel\ RG result of \refeq{eq:mh1lmixed}
as a function of $\Xtbar$.  
The difference
between the solid and dashed lines of \reffi{fig:impr1} is precisely
equal to the leftover \twol\ term given by \refeq{eq:impr}, which
is seen to be numerically small.
Hence, within the simplifying framework under consideration
({\it i.e.}, only leading $t$--$\Stop$ sector-contributions are taken
into account assuming a simplified stop squared-mass
matrix (\ref{stopmassmatrix}), with
$\Msbar$, $\MA\gg\mtms$ and $\mgl = \msq$), we see that the
``mixed-scale'' \onel\ result for $\Mh$ provides a very good
approximation to a more complete \twol\ result for all values of
$\Xtbar$.\footnote{Strictly speaking, the analytic approximations 
discussed here break down when $\mtms\Xtbar\sim\Msbarii$.  Thus, one does
not expect an accurate result for the corresponding formulae when
$\Xtbar$ is too large
\cite{mhiggsRG1a,mhiggsRG1,mhiggsRG2,mhiggslle}.  
In practice, one should not trust the accuracy of the analytic formulae
once $\Xtbar>(\Xtbar)_{\rm max}$.}


\section{Missing higher-order corrections and parametric uncertainties}
\label{sec:hocorr}

The prediction for $\Mh$ in the MSSM is affected by two kinds of
uncertainties, parametric uncertainties from the experimental errors of
the input parameters and uncertainties from unknown higher-order
corrections.


\subsection{Parametric uncertainties}
\label{subsec:paraunc}

Currently the parametric uncertainties dominate over
those from unknown higher-order corrections, as the present
experimental error of the top-quark mass of about 
$\pm 4 \gev$~\cite{mtopexpnew} induces an
uncertainty of $\De\Mh \approx \pm 4 \gev$~\cite{tbexcl}. However, at the next
generation of colliders $\mt$ will be measured with a much higher
precision, reaching the level of about 0.1~GeV at an $e^+e^-$
LC~\cite{teslatdr}. Thus, the $\mt$-induced parametric error will be
drastically reduced, see also \citeres{deltamt,lhclc}. 

Besides $\mt$, the other SM input parameters whose experimental errors
can be relevant for the prediction of $\Mh$ are $\MW$, $\als$, and
$\mb$. The $W$~boson mass $\MW$ mainly
enters via the reparameterization of the electromagnetic coupling
$\alpha(0)$ in terms of the Fermi constant $\gf$, 
\BE
\alpha(0) = \frac{\sqrt{2} \, \gf}{\pi}  
\MW^2 \left(1 - \frac{\MW^2}{\MZ^2}\right) 
\frac{1}{1 + \Delta r}~,
\label{eq:Deltar}
\end{equation}
where the quantity $\Delta r$ summarizes the radiative corrections.

The present experimental error of the $W$~boson mass of $34 \mev$ leads to
a parametric theoretical uncertainty of $\Mh$ below $0.1 \gev$. In view
of the prospective improvements in the experimental accuracy of $\MW$
the parametric uncertainty induced by $\MW$ will be substantially
smaller than the one induced by $\mt$ even for $\de\mt = 0.1 \gev$. 

The current experimental error of the strong coupling constant, 
$\de\als(\MZ) = \pm 0.002$~\cite{pdg}, induces a parametric
theoretical uncertainty of $\Mh$ of about $0.3 \gev$. Since a future
improvement of the error of $\als(\MZ)$ by about a factor of two can be
envisaged~\cite{alsdet,ewpo:gigaz1}, the parametric uncertainty
induced by $\mt$ 
will dominate over the one induced by $\als(\MZ)$ down to the level of 
$\de\mt = 0.1$--$0.2 \gev$.
The effect of the experimental uncertainties in $\mb$ are negligible,
once the proper resummation of the leading
effects~\cite{deltamb1,deltamb} is taken into account.

Also the (so far unknown) SUSY parameters have a large impact on the
value of $\Mh$. Most prominent here are the parameters of the
$\Stop$~sector. However, the induced uncertainty will depend strongly
on the future experimental uncertainty, and can thus currently not
very well estimated. Therefore these uncertainties are not further
investigated here.


\subsection{Estimating missing two-loop corrections}
\label{sec:misstwol}

Given  our present knowledge of the \twol\ contributions to the
Higgs boson self-energies, see \refse{subsec:recentHO}, the
theoretical accuracy reached in the  
prediction for the $\cp$-even Higgs boson masses is quite advanced. 
However, obtaining a complete \twol\ result for the 
Higgs boson masses and 
mixing angle requires additional contributions that are not yet
available%
\footnote{
Concerning the ``full'' two-loop effective potential
calculation~\cite{mhiggsEP5}, see \refse{subsec:FDconcept}. 
}%
. 
In this and the following sections we discuss the possible effect of
the missing \twol\  
corrections, and we estimate the size of the higher-order (i.e.\ three-loop) 
contributions.

\bigskip
It is customary to separate the corrections to the Higgs boson
self-energies into two parts: i) the momentum-independent part,
namely the contributions to the self-energies evaluated at zero
external momenta, which can also be computed in the effective
potential approach; ii) the momentum-dependent corrections, i.e.\ the
effects induced by the dependence on the external momenta of the
self-energies that are required to determine the poles of the
$(h,H)$-propagator matrix.  

All the presently available \twol\ contributions are computed at zero
external momentum, and moreover they are obtained in the so-called
gaugeless limit, namely by switching off the electroweak gauge
interactions (with the already discussed exception in
\citere{mhiggsEP5}, see \refse{subsec:FDconcept}). 
The two approximations are in fact related, since the 
leading Yukawa corrections are obtained by neglecting both the momentum 
dependence and the gauge interactions. In order to systematically
improve the result beyond the approximation of the leading Yukawa terms,
both effects from the gauge interactions and the momentum dependence
should be taken into account.

To try to estimate, although in a very rough way,
the importance of the various contributions we look at their 
relative size in the \onel\ part. There, in the effective
potential part, the effect of the
\order{\alt} corrections typically amounts to an increase in $\Mh$ of
40--60 GeV, depending on the choice of the MSSM parameters, whereas
the corrections due to the electroweak (D-term) Higgs--squark
interactions usually decrease $\Mh$ by less than 5 GeV
\cite{mhiggsf1lA}.  Instead, the purely electroweak gauge
corrections to $\Mh$, namely those coming from Higgs, gauge boson and
chargino or neutralino loops \cite{mhiggsf1lC}, are typically quite
small at \onel\, and can reach at most 5 GeV in specific regions of
the parameter space (namely for large values of $\mu$ and $M_2$). 
Concerning the effects induced by the dependence on the external
momentum, as a general rule we expect them to be more
relevant in the determination of the heaviest eigenvalue $\MH$ of
the Higgs boson mass matrix, and when $\MA$ is larger than $\MZ$, see
\refse{subsec:thresholds}. 
Indeed, only in this case the self-energies are evaluated at external
momenta comparable to or larger than the masses circulating in the
dominant loops. In addition, if $\MA$ is much larger than $\MZ$, the
relative importance of these corrections decreases, since the
tree-level value of $\mH$ grows with $\MA$.  In fact,
the effect of the \onel\ momentum-dependent corrections
on $\Mh$ amounts generally to less than 2~GeV.

Assuming that the relative size of the \twol\ contributions follows
a pattern similar to the \onel\ part, we estimate that the \twol\
diagrams involving D-term interactions should induce a variation in
$\Mh$ of at most 1--2 GeV, while we expect those with pure gauge
electroweak interactions to contribute to $\Mh$ not very
significantly, probably of the order of 1 GeV or less. Given the
smallness of the \onel\ contribution it seems quite unlikely that
the effect of the momentum-dependent part of the \order{\alt \als}
corrections to $\Mh$, which should be the largest among this type of
\twol\ contributions, could be larger than 1 GeV.  As already said,
the situation can, in principle, be different for the heavier
Higgs boson mass. 
The momentum-dependent corrections turn out to be more relevant in
processes where the $H$~boson appears as an external particle, see
\citeres{markusPhD,eennH} and \refse{subsec:thresholds}.


\subsection{Effects of the variation of the renormalization scale}
\label{subsec:muvariation}

\begin{figure}[htb!]
\newcommand{\psfragtextscale}{0.75}
\psfrag{5x-1}[][][\psfragtextscale]{0.5}
\psfrag{1x0}[][][\psfragtextscale]{1}
\psfrag{5x0}[][][\psfragtextscale]{5}
\psfrag{1x1}[][][\psfragtextscale]{10}
\psfrag{5x1}[][][\psfragtextscale]{50}
\psfrag{0}[][][\psfragtextscale]{0}
\psfrag{60}[][][\psfragtextscale]{60}
\psfrag{70}[][][\psfragtextscale]{70}
\psfrag{80}[][][\psfragtextscale]{80}
\psfrag{90}[][][\psfragtextscale]{90}
\psfrag{100}[][][\psfragtextscale]{100}
\psfrag{110}[][][\psfragtextscale]{110}
\psfrag{120}[][][\psfragtextscale]{120}
\psfrag{130}[][][\psfragtextscale]{130}
\psfrag{140}[][][\psfragtextscale]{140}
\psfrag{150}[][][\psfragtextscale]{150}
\psfrag{500}[][][\psfragtextscale]{500}
\psfrag{1000}[][][\psfragtextscale]{1000}
\psfrag{1500}[][][\psfragtextscale]{1500}
\psfrag{TB}[][][\psfragtextscale]{$\tb$}
\psfrag{MA0}[][][\psfragtextscale]{$M_A [\mathrm{GeV}]$}
\psfrag{Mh0}[][][\psfragtextscale]{$M_h [\mathrm{GeV}]$}
\begin{center}
\epsfig{figure=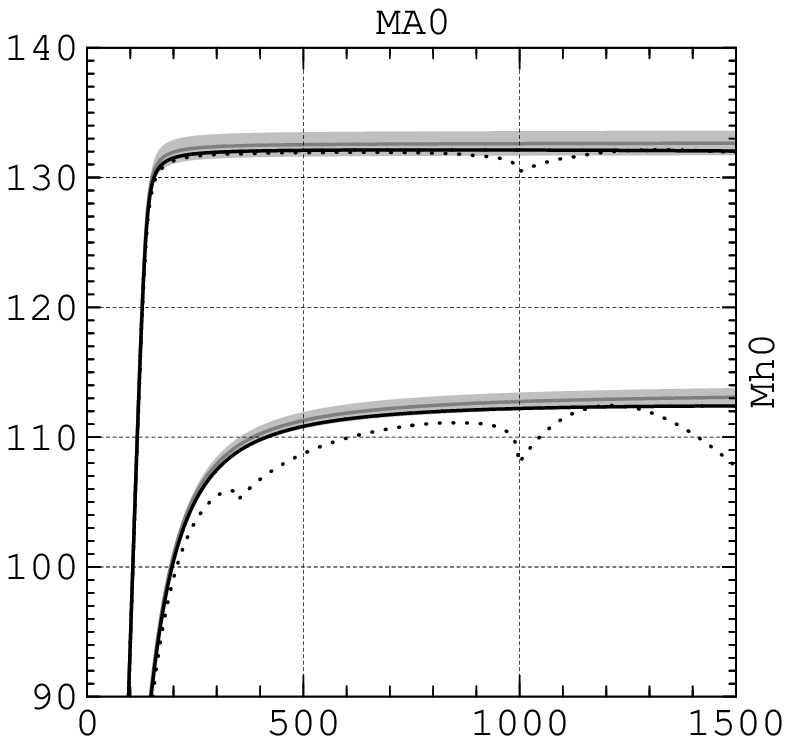,width=7cm}
\epsfig{figure=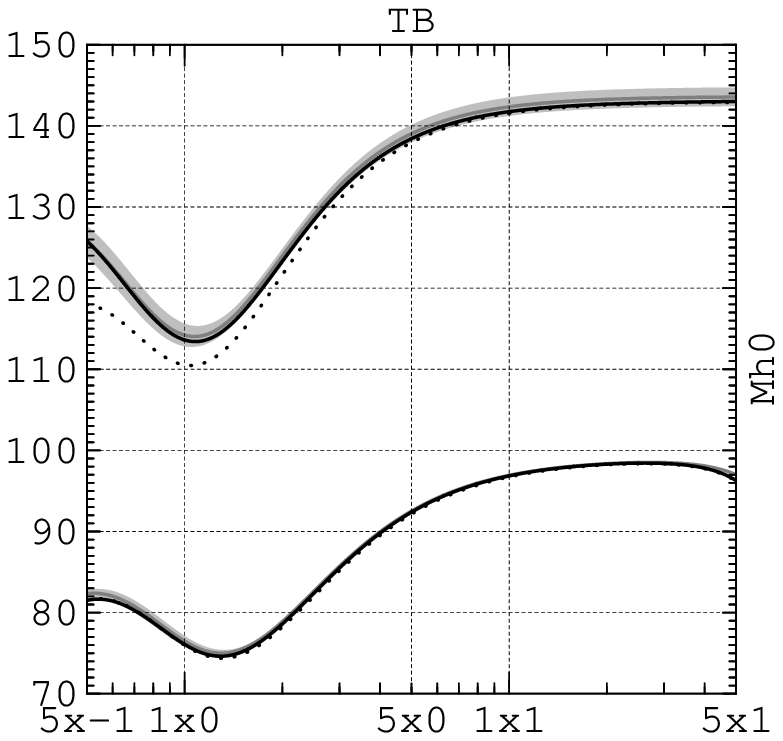,width=7cm}
\end{center}
\caption{ 
The renormalization scale dependence of $\Mh$  introduced via
the \drbar\ definition of $\tb$ and the Higgs field
renormalization constants is shown as a function of $\MA$ (left plot)
and $\tb$ (right). The lower curves correspond to $\tb = 2$ (left) and
$\MA = 100 \gev$ (right). For the upper curves we have set $\tb = 20$
(left) and $\MA = 500 \gev$ (right). 
$\mu_{\drbarm}$ has been varied from $\mt/2$ to
$2\mt$. The other parameters are chosen according to the $\mhmax$
scenario, see the Appendix.
The dotted line corresponds to the full on-shell scheme, see
\refse{subsec:recentHO}. 
}
\label{fig:muvariation}
\end{figure}

Another way of estimating the uncertainties of the kind discussed above
is to investigate the renormalization scale dependence introduced via
the \drbar\ definition of $\tb$, $\mu$, and the Higgs 
field-renormalization constants~\cite{mhiggsrenorm}, see
\refse{subsec:recentHO}.  
The variation of the scale parameter between $0.5 \mt$ 
and $2 \mt$ is shown in \reffi{fig:muvariation}. It gives rise to a
shift in $\Mh$ of about  $\pm 1.5 \gev$. This intrinsic error is 
in accordance with the estimates in \refse{sec:misstwol}.

\subsection{Estimate of the uncertainties from unknown three-loop
corrections}
\label{subsec:missing3L}

Even in the case that a complete \twol\ computation of the MSSM
Higgs masses is achieved, non-negligible uncertainties will remain,
due to the effect of higher-order corrections. Although a three-loop
computation of the Higgs masses is not available so far, it is
possible to give at least a rough estimate for the size of these 
unknown contributions.

A first estimate can be obtained by varying the renormalization scheme
in which the parameters entering the \twol\ part of the corrections
are expressed. In fact, the resulting difference in the numerical
results amounts formally to a three-loop effect. Since the
\order{\alt \als + \alt^2} corrections are particularly sensitive to
the value of the top mass, we compare the predictions for $\Mh$
obtained using in the \twol\ corrections either the top pole mass,
$\mt^{\rm pole} = 174.3 \gev$, or the SM running top mass
$\overline{m}_t$, expressed in the \msbar\ renormalization scheme,
i.e.
\BE \mtms \;\;\equiv \;\;\mt(\mt)^{\msbarm}_{\SM}
\;\;\simeq \;\; \frac{\mt^{\rm pole}}{1+ 4\,\als(\mt)/3 \pi
-\alt(\mt)/2\pi} ~.
\label{mtrun}
\end{equation}
%
\begin{figure}[htb!]
\begin{center}
\epsfig{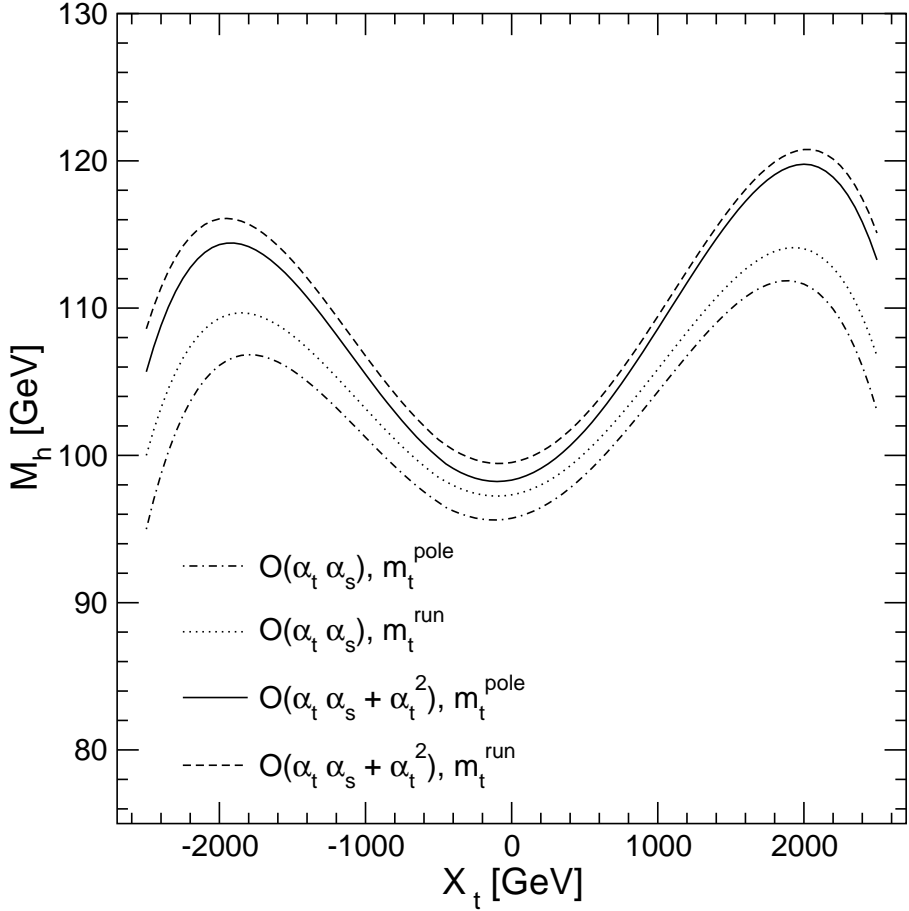}
\end{center}
\caption[]{$\Mh$ as a function of $\Xt$, using either $\mt^{\rm pole}$
or $\overline{m}_t$ in the \twol\ corrections.  The relevant MSSM
parameters are chosen as $\tb=3\,,\, \msusy = \MA = \mu = 1  \tev$
and $\mgl = 800 \gev$. For the different lines see text.} 
\label{fig:mtpolevsms}
\end{figure}

Inserting appropriate values for the SM running couplings $\als$ and
$\alt$ we find $\overline{m}_t = 168.6 \gev$.  In
Fig.~\ref{fig:mtpolevsms} we show the effect of changing the
renormalization scheme for $\mt$ in the \twol\ part of the
corrections. The relevant MSSM parameters are chosen as in
Fig.~\ref{fig:RGmeth}, i.e.~$ \tb = 3\,,\;\msusy = \MA = \mu = 1 \tev$
and $\mgl = 800 \gev$. The dot--dashed and dotted curves show the
\order{\alt \als} predictions for $\Mh$ obtained using $\mt^{\rm
pole}$ or $\overline{m}_t$, respectively, in the \twol\
corrections.  The solid and dashed curves, instead, show the
corresponding \order{\alt \als + \alt^2} predictions for $\Mh$.  The
difference in the two latter curves induced by the shift in $\mt$,
which should give an indication of the size of the unknown three-loop
corrections, is of the order of 1--1.5 GeV. However, as can be seen
from the figure, the effect of the shift in $\mt$ partially cancels
between the \order{\alt \als} and \order{\alt^2} corrections, and there
is no guarantee that such a compensating effect will appear again in the
three-loop corrections.

An alternative way of estimating the typical size of the leading
three-loop corrections makes use of the renormalization group
approach.  If all the supersymmetric particles (including the $\cp$-odd
Higgs boson $A$) have the same mass $\msusy$, and $\beta = \pi/2$, the
effective 
theory at scales below $\msusy$ is just the SM, with the role of the
Higgs doublet played by the doublet that gives mass to the up-type
quarks. In this simplified case, it is easy to  
apply the techniques of \citeres{mhiggsRG1,mhiggsRG2} in order to obtain
the leading logarithmic corrections to $\Mh$ up to three loops (see also
\citere{ahoang}). 
Considering, for further simplification, the case of zero stop mixing, 
we find
\BE
\left(\Delta \Mh^2\right)^{\rm LL} = 
\frac{3\,\alt\,\overline{m}_t^2}{\pi} \,t\,
\left[ \,1 + \left(\frac{3}{8}\,\alt - 2\,\als\right)\,\frac{t}{\pi}
+\left(\frac{23}{6}\,\als^2 - \frac{5}{4}\,\als\alt - 
\frac{33}{64}\,\alt^2 \right) \,\frac{t^2}{\pi^2}\, + ... \,\right]\;,
\end{equation}
where $t = \log\,(\msusy^2/\overline{m}_t^2)$, $\overline{m}_t$ is
defined in \refeq{mtrun}, $\alt$ and $\als$ have to be
interpreted as SM running quantities computed at the scale $Q =
\overline{m}_t$, and the ellipses stand for higher loop contributions.
It can be checked that, for $\msusy = 1 \tev$, the effect of the
three-loop leading logarithmic terms amounts to an increase in $\Mh$ of
the order of 1--1.5 GeV. If $\msusy$ is pushed to larger values, the
relative importance of the higher-order logarithmic corrections
obviously increases. In that case, it becomes necessary to resum the
logarithmic corrections to all orders, by solving the
appropriate renormalization group equations numerically.
Since it is unlikely that a complete three-loop
diagrammatic computation of the MSSM Higgs boson masses will be
available in the near future, it will probably be necessary to combine
different approaches
(e.g.\ diagrammatic, effective potential, and renormalization group), in
order to improve the accuracy of the theoretical predictions up to the
level required to compare with the experimental results expected at
the next generation of colliders.

\smallskip
To summarize this discussion, the uncertainty in the prediction
for the lightest $\cp$-even Higgs boson arising from not yet
calculated three-loop and even
higher-order corrections can conservatively be estimated 
to be 1--2 GeV. From the various missing \twol\ corrections an
uncertainty of less than 1--2 GeV is expected.
However, it is extremely unlikely that all these effects would
coherently sum up, with no partial compensation among them. 
Therefore we believe that a realistic estimate of the uncertainty from
unknown higher-order corrections in the theoretical prediction for
the lightest Higgs boson mass should not exceed $3 \gev$.


\section{Outlook}

The current precision in the theory prediction of $\Mh$, with an
uncertainty of about $3 \gev$ from missing higher-order corrections
and at least $4 \gev$ from parametric uncertainties can certainly not
match the anticipated experimental precision at the LHC 
($\De\mh \approx 200 \mev$) or at the LC ($\De\mh \approx 50 \mev$).

The intrinsic uncertainties from unknown higher-order corrections
could be reduced to the level of $0.5 \gev$ or below, if a full
two-loop calculation (including a proper renormalization) and
leading/subleading three-loop (and possibly leading four-loop)
calculations are available. On the other hand, the top quark mass 
will be measured extremely precisely at the LC~\cite{mtatLC}, bringing
down the parametric error induced by $\mt$ to the level of 
$0.1 \gev$. Also the other parametric uncertainties from SM parameters
are expected to be reduced to this level.

If both the intrinsic and the parametric error will reduce as
anticipated, the mass of the lightest Higgs boson $\Mh$ can serve as
an extremely accurate electroweak precision observable. It will be
possible to use it to constrain unknown SUSY parameters, especially
from the scalar top sector, see e.g.\ \citeres{mondplot,MAdet,sfittino}.

\newpage

\chapter{The Higgs boson sector of the cMSSM}
\label{chap:mhcMSSM}

The radiative corrections to the Higgs boson masses in the $\cp$ conserving 
MSSM (rMSSM) are meanwhile quite advanced, as has been described in
detail in \refse{chap:mhrMSSM}. 
Comparisons of different methods, see e.g.\ \refse{sec:RGcomp}, showing
agreement where expected, have lead to deeper insight into the radiative
corrections in the MSSM Higgs sector and thus to the confidence that
the higher-order contribution, although being large, are under
control.

In the case of the MSSM with complex parameters (cMSSM) the 
higher-order corrections have yet been restricted, after the first more
general investigations~\cite{mhiggsCPXgen}, to evaluations in the EP
approach~\cite{mhiggsCPXEP} and to the RG improved \onel\ EP
method~\cite{mhiggsCPXRG1,mhiggsCPXRG2}. These results have been
restricted to the corrections coming from the (s)fermion sector and
some leading logarithmic corrections from the gaugino sector. A more
complete calculation has been attempted in \citere{mhiggsCPXsn}.
More recently the leading \onel\ corrections have also been evaluated
in the FD method, using the on-shell renormalization
scheme~\cite{mhiggsCPXFD1}. Some
preliminary results including the full \onel\ evaluation have been
presented in \citere{mhiggsCPXFDproc}. The full \onel\ result
including a detailed phenomenological analysis can be found in
\citere{mhiggsCPXFD2,markusPhD}. 

In this chapter we describe the full \onel\ calculation in the FD approach
using the on-shell renormalization scheme (\MSbar\ renormalization for
$\tb$ and the field renormalizations). Following
\citere{mhiggsCPXFD1}, we present all analytical details of the
calculation. The results are then briefly analyzed in view of the corrections
coming from the non-(s)fermion sectors and the effects of the
non-vanishing external momentum that can only be incorporated
completely in the FD approach. For a more detailed analysis, see
\citere{mhiggsCPXFD2}. 
All results are incorporated into the public Fortran code
\fhto~~\cite{feynhiggs2.1}.


\section{Tree-level relations and on-shell renormalization scheme}

In this section we review the tree-level structure of of the MSSM,
however with an emphasis on the $\cp$-violating parameters and their
effects. 

\subsection{The scalar quark sector in the cMSSM}

The mass matrix of two squarks of the same flavor, $\sql$ and $\sqr$,
is given by 
\begin{align}
M_{\sq} =
\begin{pmatrix}
        M_L^2 + \mq^2 & \mq \; \Xq^* \\
        \mq \; \Xq    & M_R^2 + \mq^2
\end{pmatrix}
\label{squarkmassmatrix}
\end{align}
with
\BEA
M_L^2 &=& M_{\tilde Q}^2 + \MZ^2 \CZb (I_3^q - Q_q \sw^2) \non \\
M_R^2 &=& M_{\tilde Q'}^2 + \MZ^2 \CZb Q_q \sw^2 \\ \non
\Xq &=& A_q - \mu^* \{\CTb, \tb\} ,
\label{squarksoftSUSYbreaking}
\EEA
where $\{\CTb, \tb\}$ applies for up- and down-type squarks, respectively.
In the Higgs and scalar quark sector of the cMSSM $N_q + 1$ phases are
present, one for each $A_q$ and one
for $\mu$, i.e.\ $N_q + 1$ new parameters appear. As an abbreviation it
will be used
\BE
\phi_q = {\rm arg}\KL \Xq \KR .
\end{equation}
As an independent parameter one can trade $\arg(\Aq) = \varphi_{\Aq}$ for
$\phi_q$. \\
The squark mass eigenstates are obtained by the rotation
\BE
\VL \sqe \\ \sqz \VR = \matr{U}_{\sq} \VL \sql \\ \sqr \VR
\end{equation}
with
\BE
\matr{U}_{\sq} = \ML \ctq & \stq \\ -\stq^* & \ctq \MR
               ,\quad
\matr{U}^{\sq+} \matr{U}^{\sq} = \id~,
\end{equation}
The mass eigenvalues are given by
\BE
m_{\tilde q_{1,2}}^2 = \mq^2
  + \edz \KKL M_L^2 + M_R^2
           \mp \sqrt{( M_L^2 - M_R^2)^2 + 4 \mq^2 |\Xq|^2}~\KKR ,
\end{equation}
and are independent of the phase of $\Xq$.%
\footnote{
In the limit of having all parameters real, this definition differs
slightly from the one given in \refse{sec:squark}. 
}%
~The elements of the mixing matrix $\matr{U}$ can be calculated as
\begin{align}
  c_{\tilde{q}} =
  \frac{\sqrt{M_L^2 + \mq^2 - m_{\tilde{q}_2}^2}}
       {\sqrt{m_{\tilde{q}_1}^2 - m_{\tilde{q}_2}^2}}, \quad
  s_{\tilde{q}} =
  \frac{\Xq^*}
       {\sqrt{M_L^2 + \mq^2 - m_{\tilde{q}_2}^2} 
        \sqrt{m_{\tilde{q}_1}^2 - m_{\tilde{q}_2}^2}}.
\end{align}
The parameter $\ctq \equiv \costq$ is real, whereas 
$\stq \equiv e^{-i\phi_q}\,\sintq$ 
can be complex with the phase
\BE
\phi_{\stq} = - \phi_q = {\rm arg}\KL \Xq^* \KR ~.
\end{equation}


\subsection{The chargino/neutralino sector in the cMSSM}

The physical masses of the charginos can be determined from the matrix
\begin{align}
  \matr{X} =
  \begin{pmatrix}
    \MTwo & \sqrt{2} \sinb \MW \\
    \sqrt{2} \cosb \MW & \mu
  \end{pmatrix},
\end{align}
which contains the soft breaking term $\MTwo$ and the Higgs mass term $\mu$,
both of which may take on complex values in the cMSSM.
The rotation to the physical chargino mass eigenstates is done by transforming
the original Wino and Higgsino fields with the help of two unitary 2$\times$2
matrices $\matr{U}$ and $\matr{V}$,
\begin{align}
\label{eq:charginotransform}
  \begin{pmatrix} \tilde{\chi}_1^+ \\ \tilde{\chi}_2^+ \end{pmatrix} =
  \matr{V}
  \begin{pmatrix} \tilde{W}^+ \\ \tilde{H}_1^+ \end{pmatrix}, \quad
  \begin{pmatrix} \tilde{\chi}_1^- \\ \tilde{\chi}_2^- \end{pmatrix} =
  \matr{U}
  \begin{pmatrix} \tilde{W}^- \\ \tilde{H}_2^- \end{pmatrix}
\end{align}
These definitions lead to the diagonal mass matrix
\begin{align}
  \begin{pmatrix} m_{\tilde{\chi}^\pm_1} & 0 \\ 0 &
    m_{\tilde{\chi}^\pm_2} \end{pmatrix} = 
  \matr{U}^* \, \matr{X} \, \matr{V}^+.
\end{align}
From this relation, it becomes clear that the chargino masses
$m_{\tilde{\chi}^\pm_1}$ and
$m_{\tilde{\chi}^\pm_2}$ can be determined as the (real and positive)
singular values of $\matr{X}$.
The singular value decomposition of $\matr{X}$
also yields results for $\matr{U}$ and~$\matr{V}$.

A similar procedure is used for the determination of the neutralino masses and
mixing matrix, which can both be calculated from the mass matrix
\begin{align}
  \matr{Y} =
  \begin{pmatrix}
    \MOne                  & 0                & -\MZ \, \sw \cosb
    & \MZ \, \sw \sinb \\ 
    0                      & \MTwo            & \quad \MZ \, \cw \cosb
    & \MZ \, \cw \sinb \\ 
    -\MZ \, \sw \cosb      & \MZ \, \cw \cosb & 0
    & -\mu             \\ 
    \quad \MZ \, \sw \sinb & \MZ \, \cw \sinb & -\mu                    & 0
  \end{pmatrix}.
\end{align}
This symmetric matrix contains the additional complex soft-breaking
parameter $\MOne$. 
Its diagonalization is achieved by a transformation starting from the
original bino/wino/Higgsino basis,
\begin{align}
  \begin{pmatrix} \tilde{\chi}^0_1 \\[.5em] \tilde{\chi}^0_2 \\[.5em]
    \tilde{\chi}^0_3 \\[.5em] \tilde{\chi}^0_4 \end{pmatrix} = 
  \matr{N}
  \begin{pmatrix} \tilde{B}^0 \\[.5em] \tilde{W}^0 \\[.5em] 
  \tilde{H}_2^0 \\[.5em]  \tilde{H}_1^0 \end{pmatrix}, \quad 
  \begin{pmatrix}
    m_{\tilde{\chi}^0_1} & 0 & 0 & 0 \\[.5em]
    0 & m_{\tilde{\chi}^0_2} & 0 & 0 \\[.5em]
    0 & 0 & m_{\tilde{\chi}^0_3} & 0 \\[.5em]
    0 & 0 & 0 & m_{\tilde{\chi}^0_4}
  \end{pmatrix} =
  \matr{N}^* \, \matr{Y} \, \matr{N}^+.
\end{align}
The unitary 4$\times$4 matrix $\matr{N}$ and the physical neutralino
masses again result 
from a numerical singular value decomposition of $\matr{Y}$.
The symmetry of $\matr{Y}$ permits the non-trivial condition of
using only one 
matrix $\matr{N}$ for its diagonalization, in contrast to the chargino
case shown above.


\subsection{The cMSSM Higgs potential}

The Higgs potential $\VHiggs$ contains the real soft breaking terms
$m_1^2$ and $m_2^2$, the potentially complex soft breaking parameter
$m_{12}^2$ and the $U(1)$ and $SU(2)$ coupling constants $g_1$ and $g_2$: 

\begin{align}
\label{eq:higgspotential}
  \VHiggs &= m_1^2 H_{1i}^* H_{1i} + m_2^2 H_{2i}^* H_{2i} +
  \epsilon_{ij} (m_{12}^2 H_{1i} H_{2j} + {m_{12}^2}^* H_{1i}^*
  H_{2j}^*) \notag \\[.5em]
  &\quad +\tfrac{1}{8}(g_1^2+g_2^2)(H_{1i}^* H_{1i} - H_{2i}^* H_{2i})^2 
    + \tfrac{1}{2} g_2^2 | H_{1i}^* H_{2i} |^2. 
\end{align}

\noindent
The indices $\{i,j\}=\{1,2\}$ refer to the respective Higgs doublet
component, and $\epsilon_{12}=+1$. 
The Higgs doublets are decomposed in the following way,
\begin{align}
\label{eq:higgsdoublets}
\cHe = \begin{pmatrix} H_{11} \\ H_{12} \end{pmatrix} &=
\begin{pmatrix} v_1 + \tfrac{1}{\sqrt{2}} (\phi_1-i \chi_1) \\
  -\phi^-_1 \end{pmatrix}, \notag \\ 
\cHz = \begin{pmatrix} H_{21} \\ H_{22} \end{pmatrix} &= e^{i \xi}
\begin{pmatrix} \phi^+_2 \\ v_2 + \tfrac{1}{\sqrt{2}} (\phi_2+i
  \chi_2) \end{pmatrix}. 
\end{align}
As compared to the real case, see \refeq{higgsfeldunrot}, 
besides the vacuum expectation values $v_1$ and $v_2$,
\refeq{eq:higgsdoublets} introduces a real, as yet undetermined,
possible new phase $\xi$ between the two Higgs doublets. 
Using this decomposition, $\VHiggs$ can be rearranged in powers of the fields,
\begin{align}
\VHiggs &=  \cdots - T_{\phi_1} \phi_1 - T_{\phi_2} \phi_2 -
        T_{\chi_1} \chi_1 - T_{\chi_2} \chi_2 \notag \\ 
        &+ \tfrac{1}{2} \begin{pmatrix} \phi_1,\phi_2,\chi_1,\chi_2
        \end{pmatrix} 
\matr{M}_\mathrm{n}
\begin{pmatrix} \phi_1 \\ \phi_2 \\ \chi_1 \\ \chi_2 \end{pmatrix} +
\begin{pmatrix} \phi^-_1,\phi^-_2  \end{pmatrix} \matr{M}_\mathrm{c}
\begin{pmatrix} \phi^+_1 \\ \phi^+_2  \end{pmatrix} + \cdots,
\end{align}
where the coefficients of the linear terms are the tadpoles and
those of the quadratic terms are the mass matrices
$\matr{M}_\mathrm{n}$ and $\matr{M}_\mathrm{c}$. 
The tadpole coefficients read
\begin{subequations}
\label{eq:phichi_tadpoles}
\begin{align}
\label{eq:phi1_tadpole}
T_{\phi_1} &= -\sqrt{2} (m_1^2 v_1 + \cos \xi' |m_{12}^2| v_2 +
\tfrac{1}{4}(g_1^2 + g_2^2)(v_1^2 - v_2^2) v_1), \\ 
\label{eq:phi2_tadpole}
T_{\phi_2} &= -\sqrt{2} (m_2^2 v_2 + \cos \xi' |m_{12}^2| v_1 -
\tfrac{1}{4}(g_1^2 + g_2^2)(v_1^2 - v_2^2) v_2), \\ 
\label{eq:chi1_tadpole}
T_{\chi_1} &= -\sqrt{2} \sin \xi' |m_{12}^2| v_1 = - T_{\chi_2},
\end{align}
\end{subequations}
with $\xi' := \xi + \arg(m_{12}^2)$.

\smallskip
The real, symmetric 4$\times$4-matrix $\matr{M}_\mathrm{n}$ and the
hermitian 2$\times$2-matrix $\matr{M}_\mathrm{c}$ contain the
following elements, 
\begin{subequations}

\begin{align}
\matr{M}_\mathrm{n} &=
\begin{pmatrix} \matr{M}_\phi & \matr{M}_{\phi \chi} \\[.5em] 
                \matr{M}_{\phi\chi} & \matr{M}_{\chi} 
\end{pmatrix}, 
\end{align}

\begin{align}
\label{eq:MphiKomponenten}
\matr{M}_\phi  &= 
\begin{pmatrix}
m_1^2 + \tfrac{1}{4}(g_1^2+g_2^2)(3 v_1^2 - v_2^2) &
\cos \xi' |m_{12}^2| - \tfrac{1}{2}(g_1^2+g_2^2) v_1 v_2 \\[.5em]
\cos \xi' |m_{12}^2| - \tfrac{1}{2}(g_1^2+g_2^2) v_1 v_2 &
m_2^2 + \tfrac{1}{4}(g_1^2+g_2^2)(3 v_2^2 - v_1^2)
\end{pmatrix}, \\[.5em]
\matr{M}_{\phi \chi} &= 
\begin{pmatrix}
0 &
-\sin \xi' |m_{12}^2| \\[.5em]
 \sin \xi' |m_{12}^2| &
0
\end{pmatrix}, \\[.5em]
\matr{M}_\chi &= 
\begin{pmatrix}
m_1^2 + \tfrac{1}{4}(g_1^2+g_2^2)(v_1^2 - v_2^2) &
\cos \xi' |m_{12}^2| \\[.5em]
\cos \xi' |m_{12}^2| &
m_2^2 + \tfrac{1}{4}(g_1^2+g_2^2)(v_2^2 - v_1^2)
\end{pmatrix}, \\[.5em]
\label{eq:massen_phipm}
\matr{M}_\mathrm{c} &= 
\begin{pmatrix}
m_1^2 + \tfrac{1}{4} g_1^2 (v_1^2 - v_2^2) + \tfrac{1}{4} g_2^2 (v_1^2
+ v_2^2) & 
e^{i \xi'} |m_{12}^2| - \tfrac{1}{2} g_2^2 v_1 v_2 \\[.5em]
e^{-i \xi'} |m_{12}^2| - \tfrac{1}{2} g_2^2 v_1 v_2 &
m_2^2 + \tfrac{1}{4} g_1^2 (v_2^2 - v_1^2) + \tfrac{1}{4} g_2^2 (v_1^2 + v_2^2)
\end{pmatrix}.
\end{align}
\end{subequations}
The non-vanishing elements of $\matr{M}_{\phi \chi}$ lead to
$\cp$-violating mixing terms in the Higgs potential between the $\cp$-even
fields $\phi_1$ and $\phi_2$ and the $\cp$-odd fields $\chi_1$ and
$\chi_2$ if $\xi' \ne 0$. 
The physically relevant mass eigenstates in lowest order follow from a
unitary transformation of the original fields, 
\begin{align}
\label{eq:RotateToMassES}
\begin{pmatrix} h \\ H \\ A \\ G \end{pmatrix} = \matr{U}_{\mathrm{n}(0)} \cdot
\begin{pmatrix} \phi_1 \\ \phi_2 \\ \chi_1 \\ \chi_2 \end{pmatrix}, \quad
\begin{pmatrix} H^\pm \\ G^\pm \end{pmatrix} = \matr{U}_{\mathrm{c}(0)} \cdot
\begin{pmatrix} \phi_1^\pm \\ \phi_2^\pm \end{pmatrix},
\end{align}
such that the resulting mass matrices 
\BE
\matr{M}_\mathrm{n}^{\rm diag} = \matr{U}_{\mathrm{n}(0)}
\matr{M}_\mathrm{n} \matr{U}_{\mathrm{n}(0)}^+ 
\quad {\rm and} \quad
\matr{M}_\mathrm{c}^{\rm diag} = \matr{U}_{\mathrm{c}(0)} \matr{M}_\mathrm{c}
\matr{U}_{\mathrm{c}(0)}^+ 
\end{equation}
of the new fields will be diagonal. 
The new fields correspond to the three neutral Higgs bosons $h$, $H$
and $A$, the charged pair $H^\pm$ and the Goldstone bosons $G$ and
$G^\pm$. 

The lowest-order mixing matrices can be determined from
the eigenvectors of $\matr{M}_\mathrm{n}$ and $\matr{M}_\mathrm{c}$,
calculated under the additional condition that the tadpole
coefficients~(\ref{eq:phichi_tadpoles}) must vanish in order that
$v_1$ and $v_2$ are indeed stationary points of the Higgs potential. 
This automatically requires $\xi'=0$, which in turn leads to a
vanishing matrix $\matr{M}_{\phi \chi}$ and a real, symmetric matrix
$\matr{M}_\mathrm{c}$. 
Therefore, no $\cp$-violation occurs in the Higgs potential at the lowest
order, and the corresponding mixing matrices can be parametrized by
real mixing angles as 
\begin{align}
\label{eq:MixMatrixO0}
  \matr{U}_{\mathrm{n}(0)} =
  \begin{pmatrix}
        - \sina & \cosa &                 0 &           0 \\
    \quad \cosa & \sina &                 0 &           0 \\
              0 &     0 &     - \sin \betan & \cos \betan \\
              0 &     0 & \quad \cos \betan & \sin \betan
  \end{pmatrix}, \quad
  \matr{U}_{\mathrm{c}(0)} =
  \begin{pmatrix}
        - \sin \betac & \cos \betac \\
    \quad \cos \betac & \sin \betac
  \end{pmatrix}.
\end{align}
The mixing angles $\alpha$, $\betan$ and $\betac$ can be determined
from the requirement that this transformation will result in diagonal
mass matrices for the physical fields. 


\subsection{Higgs mass terms and tadpoles}

The terms in $\VHiggs$ that are linear or quadratic in the physical
fields are denoted as follows, 
\begin{align}
\label{eq:PhysTadMass}
\VHiggs &= \mathrm{const.} - \tadh \cdot h - \tadH \cdot H -\tadA
\cdot A - \tadG \cdot G \notag \\[.5em]
&\quad + \tfrac{1}{2}
\begin{pmatrix}
  h,H,A,G
\end{pmatrix} \cdot
\begin{pmatrix}
  \mh^2  & \mhH^2 & \mhA^2 & \mhG^2 \\
  \mhH^2 & \mH^2  & \mHA^2 & \mHG^2 \\
  \mhA^2 & \mHA^2 & \mA^2  & \mAG^2 \\
  \mhG^2 & \mHG^2 & \mAG^2 & \mG^2
\end{pmatrix} \cdot
\begin{pmatrix}
  h \\ H \\ A \\ G
\end{pmatrix} + \\[.5em]
&\quad +
\begin{pmatrix}
  H^-,G^-
\end{pmatrix} \cdot
\begin{pmatrix}
  \mHpm^2  & \mHmGp^2 \\
  \mGmHp^2 & \mGpm^2
\end{pmatrix} \cdot
\begin{pmatrix}
  H^+ \\ G^+
\end{pmatrix} + \cdots \notag.
\end{align}
In order to 
perform a simple renormalization procedure, the parameters in
$\VHiggs$ have to be expressed in term of physical parameters.
In total, $\VHiggs$ contains eight independent real parameters: $v_1$,
$v_2$, $g_1^2$, $g_2^2$, $m_1^2$, $m_2^2$, $|m_{12}^2|$ and
$\arg(m_{12}^2)$ (or $\xi'$). 
The coupling constants $g_1$ and $g_2$ can be replaced by the electric
coupling constant $e$ and the weak mixing angle $\theta_\mathrm{w}$
($\sw \equiv \sin \theta_\mathrm{w}$, $\cw \equiv \cos \theta_\mathrm{w}$), 
\begin{align} 
\label{eq:kopplungskonstanten}
  e = g_1 \, \cw = g_2 \, \sw,
\end{align}
while the $Z$~boson mass $\MZ$ and $\tb$ substitute for $v_1$ and $v_2$:
\begin{align}
\label{eq:tanb_mz}
\tb = \frac{v_2}{v_1}, \; \MZ^2 = \tfrac{1}{2}(g_1^2 + g_2^2)(v_1^2 + v_2^2).
\end{align}
The $W$~boson mass is then given by
\BE
\MW = \edz g_2^2 (v_1^2 + v_2^2).
\end{equation}
The tadpole coefficients in the physical basis follow from the
original ones~(\ref{eq:phichi_tadpoles}) by applying the
transformation~(\ref{eq:MixMatrixO0}), 
\begin{subequations}
\label{eq:tadpoles}
\begin{align}
\label{eq:tadpoleH}
\tadH &= \sqrt{2} [-m_1^2 v_1 \cosa - m_2^2 v_2 \sina - \cos \xi'
|m_{12}^2| (v_1 \sina + v_2 \cosa) \\ 
&\quad - \tfrac{1}{4}(g_1^2 + g_2^2)(v_1^2 - v_2^2)(v_1 \cosa - v_2
\sina)], \notag \\ 
\label{eq:tadpoleh}
\tadh &= \sqrt{2} [+m_1^2 v_1 \sina - m_2^2 v_2 \cosa - \cos \xi'
|m_{12}^2| (v_1 \cosa - v_2 \sina) \\ 
&\quad + \tfrac{1}{4}(g_1^2 + g_2^2)(v_1^2 - v_2^2)(v_1 \sina + v_2
\cosa)], \notag \\ 
\label{eq:tadpoleA}
\tadA &= \sqrt{2} \sin \xi' |m_{12}^2| (v_1 \cos \betan + v_2 \sin \betan), \\
\label{eq:tadpoleG} 
\tadG &= -\tan(\beta-\betan) \tadA.
\end{align}
\end{subequations}
Due to the linear dependence of $\tadG$ on $\tadA$,
\refeq{eq:tadpoles} provides only three replacements for the
original parameters. 
Typically, the remaining parameter is replaced by either the mass of
the neutral $A$-boson, $\mA$, or the mass of the charged pair, $\mHpm$. 
Their expressions in terms of the original parameters are given by
\begin{subequations}
\label{eq:massen_A_Hp}
\begin{align}
\label{eq:masse_A}
\mA^2 &= m_1^2 \sin^2 \betan + m_2^2 \cos^2 \betan - \sin 2 \betan
\cos \xi' |m_{12}^2| \notag \\ 
&\quad - \cos 2 \betan \tfrac{1}{4}(g_1^2 + g_2^2)(v_1^2 - v_2^2), \\
\label{eq:masse_Hp}
\mHpm^2 &= m_1^2 \sin^2 \betac + m_2^2 \cos^2 \betac - \sin 2 \betac
\cos \xi' |m_{12}^2| \notag \\ 
&\quad - \cos 2 \betac \tfrac{1}{4}(g_1^2 + g_2^2)(v_1^2 - v_2^2) +
\tfrac{1}{2} g_2^2 (v_1 \cos \betac + v_2 \sin \betac)^2. 
\end{align}
\end{subequations}
If $\betan = \betac$ (what will be shown below) the relation between
$\mA^2$ and $\mHpm^2$ becomes
\BE
\mHpm^2 = \mA^2 + \MW^2~.
\end{equation}

\noindent
Using (\ref{eq:tadpoles}a-c) and (\ref{eq:masse_A}) or
(\ref{eq:masse_Hp}), all of $m_1^2$, $m_2^2$, $|m_{12}^2|$ and $\xi'$
can be substituted by $\tadh$, $\tadH$, $\tadA$ and ($\mA$ or $\mHpm$). 
In summary, this leads to the following replacements:
\begin{align}
v_1 &\to \frac{\wz\,\Cb\,\sw\,\cw\,\MZ}{e}, \\
v_2 &\to \frac{\wz\,\Sbe\,\sw\,\cw\,\MZ}{e}, \\
g_1 &\to e/\cw, \\
g_2 &\to e/\sw, \\ 
m_1^2 &\to     
-\ed{8} \Big[-4 \mA^2  + 2 (2 \mA^2  + \MZ^2 ) \CZb 
  + \MZ^2 \cos(4 \be - 2 \betan)
  + \MZ^2 \cos(2 \betan) \\
&+ \frac{e \tadh \cos\betan}{2 \cw \sw \MZ} 
   \KL \Cb \cos\betan \Sa + \Sbe (\Ca \cos\betan + 2 \Sa \sin\betan) \KR
                         \notag \\
&+ \frac{e \tadH \cos\betan}{2 \cw \sw \MZ}
   \KL \cos \betan \Sa \Sbe - \Ca (\Cb \cos\betan + 2 \Sbe \sin\betan) \KR
\Big] /\KL\cos^2(\be - \betan)\KR, \notag \\
m_2^2 &\to \ed{8} \Big[4 \mA^2  + 2 (2 \mA^2  + \MZ^2 ) \CZb 
  + \MZ^2 \cos(4 \be - 2 \betan)
  + \MZ^2 \cos(2 \betan) \\
&- \frac{e \tadh \sin\betan}{2 \cw \sw \MZ} 
   \KL \Cb \sin\betan \Sa + \Ca (\Sbe \sin\betan + 2 \Cb \cos\betan) \KR
                         \notag \\
&- \frac{e \tadH \sin\betan}{4 \cw \sw \MZ}
   \KL 2 \Sa \Sbe \sin\betan + \Cb (3 \sin(\al-\betan) + \sin(\al+\betan) \KR
\Big] /\KL\cos^2(\be - \betan)\KR, \notag \\
|m_{12}^2| &\to \sqrt{\KL f_m^2 + f_s^2 \KR} \\
 &{} f_m = \Big[ \edz \mA^2 \SZb 
         - \frac{e \tadh}{4 \cw \sw \MZ} 
           \KL \cos(\be+\al) + \Cba \cos(2 \betan) \KR \notag \\
 &{}~~~~~~~- \frac{e \tadH}{4 \cw \sw \MZ}
           \KL \sin(\be+\al) + \Sba \cos(2 \betan) \KR \Big]
         /\KL\cos^2(\be - \betan)\KR, \notag \\
 &{} f_s = \frac{e \tadA}{2 \sw \cw \MZ 
                          \KL \Cb \cos\betan + \Sbe \sin\betan \KR},
                                                             \notag \\
\sin\xi' &\to f_s/\sqrt{f_m^2 + f_s^2}, \\
\cos\xi' &\to f_m/\sqrt{f_m^2 + f_s^2}~.
\end{align}
The resulting physical mass terms are given either in terms of $\mA$
or $\mHpm$, depending on which parameter leads to more compact
expressions. 

The charged Higgs sector contains, apart from $\mHpm^2$, the mass terms
\begin{subequations}
\label{eq:fullmass1}
\begin{align}
\label{eq:mHmGp2}
\mHmGp^2 &= - \mHpm^2 \tan(\beta-\betac) \\
&\quad - \tfrac{e}{2 \MZ \sw \cw} \tadH
\sin(\alpha-\betac)/\cos(\beta-\betac) \notag \\ 
&\quad - \tfrac{e}{2 \MZ \sw \cw} \tadh
\cos(\alpha-\betac)/\cos(\beta-\betac) \notag \\ 
&\quad - \tfrac{e}{2 \MZ \sw \cw} i \tadA/\cos(\beta-\betan), \notag \\
\mGmHp^2 &= (\mHmGp^2)^*, \\
\mGpm^2 &= \mHpm^2 \tan^2(\beta-\betac) \notag \\
&\quad - \tfrac{e}{2 \MZ \sw \cw} \tadH \cos(\alpha + \beta - 2
\betac)/\cos^2(\beta-\betac) \\ 
&\quad + \tfrac{e}{2 \MZ \sw \cw} \tadh \sin(\alpha + \beta - 2
\betac)/\cos^2(\beta-\betac), 
\end{align}
\end{subequations}
where the star denotes a complex conjugation.

The neutral mass matrix is more easily parametrized by $\mA$, as can
be seen from the 2$\times$2 sub-matrix of the $A$~and $G$~boson: 
\begin{subequations}
\begin{align}
\label{eq:mAG2}
\mAG^2 &= - \mA^2 \tan(\beta-\betan) \\
&\quad - \tfrac{e}{2 \MZ \sw \cw} \tadH
\sin(\alpha-\betan)/\cos(\beta-\betan) \notag \\ 
&\quad - \tfrac{e}{2 \MZ \sw \cw} \tadh
\cos(\alpha-\betan)/\cos(\beta-\betan), \notag \\ 
\mG^2 &= \mA^2 \tan^2(\beta-\betan) \\
&\quad - \tfrac{e}{2 \MZ \sw \cw} \tadH \cos(\alpha + \beta - 2
\betan)/\cos^2(\beta-\betan) \notag \\ 
&\quad + \tfrac{e}{2 \MZ \sw \cw} \tadh \sin(\alpha + \beta - 2
\betan)/\cos^2(\beta-\betan). 
\end{align}
\end{subequations}
The $\cp$-violating mixing terms connecting the $h$-/$H$- and the
$A$-/$G$-sector are 
\begin{subequations}
\begin{align}
\mhA^2 &= \tfrac{e}{2 \MZ \sw \cw} \tadA
\sin(\alpha-\betan)/\cos(\beta-\betan), \\ 
\mhG^2 &= \tfrac{e}{2 \MZ \sw \cw} \tadA
\cos(\alpha-\betan)/\cos(\beta-\betan), \\ 
\mHA^2 &= - \mhG^2, \\
\mHG^2 &= \tfrac{e}{2 \MZ \sw \cw} \tadA
\sin(\alpha-\betan)/\cos(\beta-\betan). 
\end{align}
\end{subequations}
Finally, the mass terms of the $\cp$-even $h$~and $H$~bosons are:
\begin{subequations}
\label{eq:fullmass2}
\begin{align}
\label{eq:masse_h_voll}
\mh^2 &= \MZ^2 \sin^2(\alpha+\beta) \\
&\quad + \mA^2 \cos^2(\alpha-\beta)/\cos^2(\beta-\betan) \notag \\
&\quad + \tfrac{e}{2 \MZ \sw \cw} \tadH \cos(\alpha-\beta)
\sin^2(\alpha-\betan)/\cos^2(\beta-\betan) \notag \\ 
&\quad + \tfrac{e}{2 \MZ \sw \cw} \tadh \tfrac{1}{2}
\sin(\alpha-\betan) (\cos(2 \alpha - \beta - \betan) + 3 \cos(\beta -
\betan))/\cos^2(\beta-\betan), \notag \\ 
\label{eq:masse_hH_voll}
\mhH^2 &= - \MZ^2 \sin(\alpha+\beta) \cos(\alpha+\beta) \\
&\quad + \mA^2 \sin(\alpha-\beta)
\cos(\alpha-\beta)/\cos^2(\beta-\betan) \notag \\ 
&\quad + \tfrac{e}{2 \MZ \sw \cw} \tadH \sin(\alpha-\beta)
\sin^2(\alpha-\betan)/\cos^2(\beta-\betan) \notag \\ 
&\quad - \tfrac{e}{2 \MZ \sw \cw} \tadh \cos(\alpha-\beta)
\cos^2(\alpha-\betan)/\cos^2(\beta-\betan), \notag \\ 
\label{eq:masse_H_voll}
\mH^2 &= \MZ^2 \cos^2(\alpha+\beta) \\
&\quad + \mA^2 \sin^2(\alpha-\beta)/\cos^2(\beta-\betan) \notag \\
&\quad + \tfrac{e}{2 \MZ \sw \cw} \tadH \tfrac{1}{2}
\cos(\alpha-\betan) (\cos(2 \alpha - \beta - \betan) - 3 \cos(\beta -
\betan))/\cos^2(\beta-\betan) \notag \\ 
&\quad - \tfrac{e}{2 \MZ \sw \cw} \tadh \sin(\alpha-\beta)
\cos^2(\alpha-\betan)/\cos^2(\beta-\betan). \notag 
\end{align}
\end{subequations}


\subsection{Masses and mixing angles in lowest order}
\label{sec:MassenMischungenTR}

The masses and mixing angles in lowest order follow from
eqs.~(\ref{eq:fullmass1})-(\ref{eq:fullmass2}) and the additional
requirement that both the tadpole coefficients $T_{\{h,H,A\}}$ and all
non-diagonal entries of the mass matrices must vanish. 
From (\ref{eq:mHmGp2}) and (\ref{eq:mAG2}), it immediately follows that
\begin{align}
\label{eq:equalbetas}
  \betac = \betan = \beta,
\end{align}
which in turn determines the tree-level value of $\al = \alpha_{(0)}$
(up to a sign) from 
(\ref{eq:masse_hH_voll}) as (which is equivalent to \refeq{alphaborn}
in the rMSSM)  
\begin{align}
\label{eq:Tan2Alpha}
  \tan 2 \alpha_{(0)} = \tan 2 \beta \frac{\mA^2+\MZ^2}{\mA^2-\MZ^2},
\quad -\frac{\pi}{2} \le \al_{(0)} \le 0~.
\end{align}
The Higgs masses $\mhtr$ and $\mHtr$ are 
the eigenvalues of their 2$\times$2 mass matrix with
entries~(\ref{eq:fullmass2}, $\alpha$ set to zero), 
\begin{subequations}
\label{eq:mh_mH_tree}
\begin{align}
  \{\mhtr^2,\mHtr^2\} = \tfrac{1}{2} \left(\mA^2+\MZ^2 \mp
  \sqrt{(\mA^2 + \MZ^2)^2 - 4 \mA^2 \MZ^2 \cos^2 2 \beta} \right). 
\end{align}
\end{subequations}
Finally, combining eqs.~(\ref{eq:massen_A_Hp})
and~(\ref{eq:equalbetas}) relates the remaining masses $\mA$ and
$\mHpm$ with each other, 
\begin{align}
\label{eq:MA0_MHp_Relation}
  \mHpm^2 = \mA^2 + \cw^2 \MZ^2 = \mA^2 + \MW^2.
\end{align}
Specifying one Higgs boson mass as an input parameter therefore
unambiguously determines the other ones. 
Since the $\cp$-odd $A$~boson will --due to the $\cp$-violating mixing in
the neutral Higgs sector-- no longer be an eigenstate in higher orders,
the charged Higgs mass $\mHpm$ will be used as input parameter in the
cMSSM.


\subsection{Renormalization of the Higgs potential}
\label{sec:RenormVHiggs}

In this and the following sections we will focus one the \onel\
corrections to the cMSSM Higgs sector. However, \twol\ corrections
(taken from the rMSSM, see
\citeres{mhiggsletter,mhiggslong,mhiggsAEC}) will be included 
in our numerical evaluation, see \refse{sec:cMSSMnumeval}.

In order to calculate the first-order corrections to the Higgs boson
masses and effective mixing angles, the counter terms for the mass and
tadpole coefficients in the Higgs potential are needed. 
Apart from these, counter terms are needed for several other
parameters which appear in the Higgs potential: 
\begin{align}
\label{eq:PhysParamRenorm}
  \MZ^2 &\rightarrow \MZ^2 + \dMZsq,  & \tadh &\rightarrow \tadh + \dtadh, \\
  \MW^2 &\rightarrow \MW^2 + \dMWsq,  & \tadH &\rightarrow \tadH +
  \dtadH, \notag \\ 
  \matr{M}_\mathrm{n} &\rightarrow \matr{M}_\mathrm{n} + \delta
  \matr{M}_\mathrm{n}, & 
  \tadA &\rightarrow \tadA + \dtadA, \notag \\
  \matr{M}_\mathrm{c} &\rightarrow \matr{M}_\mathrm{c} + \delta
  \matr{M}_\mathrm{c}, & 
  \tanb &\rightarrow \tanb (1+\dtanb). \notag
\end{align}
These definitions explain why the
expressions~(\ref{eq:fullmass1}-\ref{eq:fullmass2}) for the Higgs
masses must differentiate between the mixing angles $\betan$ and
$\betac$ (which, like $\alpha$, are not renormalized) and the
parameter $\beta$ (which is). 
This distinction is necessary to arrive at the following 
expressions for the counter terms. 

The field renormalization matrices of both Higgs multiplets
can be set up symmetrically, 
\begin{subequations}
\label{eq:higgsfeldren}
\begin{align}
  \begin{pmatrix} h \\[.5em] H \\[.5em] A \\[.5em] G \end{pmatrix} \rightarrow
  \begin{pmatrix}
    1+\tfrac{1}{2} \dZ{hh} & \tfrac{1}{2} \dZ{hH} & \tfrac{1}{2}
    \dZ{hA} & \tfrac{1}{2} \dZ{hG} \\[.5em]
    \tfrac{1}{2} \dZ{hH} & 1+\tfrac{1}{2} \dZ{HH} & \tfrac{1}{2}
    \dZ{HA} & \tfrac{1}{2} \dZ{HG} \\[.5em] 
    \tfrac{1}{2} \dZ{hA} & \tfrac{1}{2} \dZ{HA} & 1+\tfrac{1}{2}
    \dZ{AA} & \tfrac{1}{2} \dZ{AG} \\[.5em] 
    \tfrac{1}{2} \dZ{hG} & \tfrac{1}{2} \dZ{HG} & \tfrac{1}{2} \dZ{AG}
    & 1+\tfrac{1}{2} \dZ{GG} 
  \end{pmatrix} \cdot
  \begin{pmatrix} h \\[.5em] H \\[.5em] A \\[.5em] G \end{pmatrix}
\end{align}
and
\begin{align}
\label{eq:dZHpGp}
  \begin{pmatrix} H^+ \\[.5em] G^+ \end{pmatrix} &\rightarrow
  \begin{pmatrix}
    1 + \tfrac{1}{2} \dZ{H^+}     &     \tfrac{1}{2} \dZ{H^- G^+} \\[.5em]
        \tfrac{1}{2} \dZ{G^- H^+} & 1 + \tfrac{1}{2} \dZ{G^+}
  \end{pmatrix} \cdot
  \begin{pmatrix} H^+ \\[.5em] G^+ \end{pmatrix}, \\[.5em]
\label{eq:dZHmGm}
  \begin{pmatrix} H^- \\[.5em] G^- \end{pmatrix} &\rightarrow
  \begin{pmatrix}
    1 + \tfrac{1}{2} \dZ{H^+}^*   &     \tfrac{1}{2} \dZ{G^- H^+} \\[.5em]
        \tfrac{1}{2} \dZ{H^- G^+} & 1 + \tfrac{1}{2} \dZ{G^+}^*
  \end{pmatrix} \cdot
  \begin{pmatrix} H^- \\[.5em] G^- \end{pmatrix}.
\end{align}
\end{subequations}

\noindent
For the mass counter term matrices we use the definitions
\begin{align}
  \delta \matr{M}_\mathrm{n} =
  \begin{pmatrix}
    \dmhsq  & \dmhHsq & \dmhAsq & \dmhGsq \\[.5em]
    \dmhHsq & \dmHsq  & \dmHAsq & \dmHGsq \\[.5em]
    \dmhAsq & \dmHAsq & \dmAsq  & \dmAGsq \\[.5em]
    \dmhGsq & \dmHGsq & \dmAGsq & \dmGsq
  \end{pmatrix}, \quad
  \delta \matr{M}_\mathrm{c} =
  \begin{pmatrix}
    \dmHpmsq  & \dmHmGpsq \\[.5em]
    \dmGmHpsq & \dmGpmsq
  \end{pmatrix} .
\end{align}
The renormalized self-energies, $\hSi(p^2)$, can now be expressed
through the unrenormalized self-energies, $\Si(p^2)$, the field
renormalization constants, and the mass counter terms.
This reads for the $\cp$-even part,
\begin{subequations}
\label{eq:renses_higgssector}
\begin{align}
\label{renSEhh}
\ser{hh}(p^2)  &= \se{hh}(p^2) + \dZ{hh} (p^2-\mhtr^2) - \dmhsq, \\
\label{renSEhH}
\ser{hH}(p^2)  &= \se{hH}(p^2) + \dZ{hH}
(p^2-\tfrac{1}{2}(\mhtr^2+\mHtr^2)) - \dmhHsq, \\ 
\label{renSEHH}
\ser{HH}(p^2)  &= \se{HH}(p^2) + \dZ{HH} (p^2-\mHtr^2) - \dmHsq,
\end{align}
the $\cp$-odd part,
\begin{align}
\ser{AA}(p^2)  &= \se{AA}(p^2) + \dZ{AA} (p^2 - \mAtr^2) - \dmAsq, \\
\ser{AG}(p^2)  &= \se{AG}(p^2) + \dZ{AG} (p^2 - \tfrac{1}{2}\mAtr^2) -
\dmAGsq, \\ 
\ser{GG}(p^2)  &= \se{GG}(p^2) + \dZ{GG} p^2 - \dmGsq,
\end{align}
the $\cp$-violating self energies,
\begin{align}
\ser{hA}(p^2)  &= \se{hA}(p^2) + \dZ{hA}
(p^2-\tfrac{1}{2}(\mhtr^2+\mAtr^2)) - \dmhAsq, \\ 
\ser{hG}(p^2)  &= \se{hG}(p^2) + \dZ{hG} (p^2-\tfrac{1}{2} \mhtr^2) -
\dmhGsq, \\ 
\ser{HA}(p^2)  &= \se{HA}(p^2) + \dZ{HA}
(p^2-\tfrac{1}{2}(\mHtr^2+\mAtr^2)) - \dmHAsq, \\ 
\ser{HG}(p^2)  &= \se{HG}(p^2) + \dZ{HG} (p^2-\tfrac{1}{2} \mHtr^2) - \dmHGsq,
\end{align}
and finally for the charged sector:
\begin{align}
\label{renSEHp}
\ser{H^- H^+}(p^2)  &= \se{H^- H^+}(p^2) + \dZ{H^- H^+} (p^2 -
\mHpmtr^2) - \dmHpmsq, \\ 
\ser{H^- G^+}(p^2)  &= \se{H^- G^+}(p^2) + \dZ{H^- G^+} (p^2 -
\tfrac{1}{2} \mHpmtr^2) - \dmHmGpsq, \\ 
\ser{G^- H^+}(p^2)  &= \ser{H^- G^+}^*(p^2), \\
\label{renSEGp}
\ser{G^- G^+}(p^2)  &= \se{G^- G^+}(p^2) + \dZ{G^- G^+} p^2 - \dmGpmsq.
\end{align}
\end{subequations}
It follows from the definition of the field renormalization
matrices~(\ref{eq:dZHpGp}) and~(\ref{eq:dZHmGm}) that $\dZ{H^- H^+} =
2 \, \re \dZ{H^+}$ and $\dZ{G^- G^+} = 2 \, \re \dZ{G^+}$ are real
quantities. 

Inserting the renormalization transformation into the Higgs mass terms
leads to expressions for their counter terms which consequently depend
on the other counter terms introduced in~(\ref{eq:PhysParamRenorm}). 
Since the counter terms themselves are of first order, the zero-order
equalities $T_{\{h,H,A\}} = 0$ and $\betan = \betac = \beta$ can
afterwards be used to simplify these expressions. 

For the $\cp$-even part of the Higgs sectors, these counter terms are:
\begin{subequations}
\label{eq:HiggsMassenCTs}
\begin{align}
\dmhsq &= \dmAsq \cos^2(\alpha-\beta) + \delta \MZ^2 \sin^2(\alpha+\beta) \\
&\quad + \tfrac{e}{2 \MZ \sw \cw} (\dtadH \cos(\alpha-\beta)
\sin^2(\alpha-\beta) + \dtadh \sin(\alpha-\beta)
(1+\cos^2(\alpha-\beta))) \notag \\ 
&\quad + \dtanb \sinb \cosb (\mA^2 \sin 2 (\alpha-\beta) + \MZ^2 \sin
2 (\alpha+\beta)), \notag \\ 
\dmhHsq &= \tfrac{1}{2} (\dmAsq \sin 2(\alpha-\beta) - \dMZsq \sin
2(\alpha+\beta)) \\ 
&\quad + \tfrac{e}{2 \MZ \sw \cw} (\dtadH \sin^3(\alpha-\beta) -
\dtadh \cos^3(\alpha-\beta)) \notag \\ 
&\quad - \dtanb \sinb \cosb (\mA^2 \cos 2 (\alpha-\beta) + \MZ^2 \cos
2 (\alpha+\beta)), \notag \\ 
\dmHsq &= \dmAsq \sin^2(\alpha-\beta) + \dMZsq \cos^2(\alpha+\beta) \\
&\quad - \tfrac{e}{2 \MZ \sw \cw} (\dtadH \cos(\alpha-\beta)
(1+\sin^2(\alpha-\beta)) + \dtadh \sin(\alpha-\beta)
\cos^2(\alpha-\beta)) \notag \\ 
&\quad - \dtanb \sinb \cosb (\mA^2 \sin 2 (\alpha-\beta) + \MZ^2 \sin
2 (\alpha+\beta)), \notag 
\end{align}
while for the $\cp$-odd part they follow as
\begin{align}
\dmAGsq &= \tfrac{e}{2 \MZ \sw \cw} (-\dtadH \sin(\alpha-\beta) -
\dtadh \cos(\alpha-\beta)) - \dtanb \mA^2 \sinb \cosb, \\ 
\dmGsq &= \tfrac{e}{2 \MZ \sw \cw} (-\dtadH \cos(\alpha-\beta) +
\dtadh \sin(\alpha-\beta)), 
\end{align}
for the $\cp$-violating mixing as
\begin{align}
\dmhAsq &= + \tfrac{e}{2 \MZ \sw \cw} \dtadA \sin(\alpha-\beta), \\
\dmhGsq &= + \tfrac{e}{2 \MZ \sw \cw} \dtadA \cos(\alpha-\beta), \\
\dmHAsq &= - \dmhGsq, \\
\dmHGsq &= + \tfrac{e}{2 \MZ \sw \cw} \dtadA \sin(\alpha-\beta),
\end{align}
and for the charged Higgs bosons as
\begin{align}
\dmHmGpsq &= \tfrac{e}{2 \MZ \sw \cw} (-\dtadH \sin(\alpha-\beta) -
\dtadh \cos(\alpha-\beta) - i \, \dtadA), \\ 
&\quad - \dtanb \mHpm^2 \sinb \cosb \notag, \\
\dmGmHpsq &= (\dmHmGpsq)^*, \\
\dmGpmsq &= \tfrac{e}{2 \MZ \sw \cw} (-\dtadH \cos(\alpha-\beta) +
\dtadh \sin(\alpha-\beta)). 
\end{align}
\end{subequations}
Note that neither $\dmAsq$ nor $\dmHpmsq$ are listed here, since one
of these masses can be a free input parameter whose definition 
depends on the renormalization.
However, from (\ref{eq:MA0_MHp_Relation}) the relation
\begin{align}
  \dmHpmsq = \dmAsq + \dMWsq
\label{massct}
\end{align}
can be derived between them, which, being generally valid, is used to
replace $\dmAsq$ in other expressions. 

\bigskip
For the field renormalization we chose to give each Higgs doublet one
renormalization constant,
\begin{align}
\label{eq:HiggsDublettFeldren}
  \cHe \rightarrow (1 + \tfrac{1}{2} \dZ{\cHe}) \cHe, \quad
  \cHz \rightarrow (1 + \tfrac{1}{2} \dZ{\cHz}) \cHz.
\end{align}
This leads to the following expressions for the various field
renormalization constants:
\begin{subequations}
\label{eq:FeldrenI_H1H2}
\begin{align}
  \dZ{hh} &= \sinasq \dZ{\cHe} + \cosasq \dZ{\cHz}, \\[.2em]
  \dZ{AA} &= \sinbsq \dZ{\cHe} + \cosbsq \dZ{\cHz}, \\[.2em]
  \dZ{hH} &= \sina \cosa (\dZ{\cHz} - \dZ{\cHe}), \\[.2em]
  \dZ{AG} &= \sinb \cosb (\dZ{\cHz} - \dZ{\cHe}), \\[.2em]
  \dZ{HH} &= \cosasq \dZ{\cHe} + \sinasq \dZ{\cHz}, \\[.2em]
  \dZ{GG} &= \cosbsq \dZ{\cHe} + \sinbsq \dZ{\cHz}, \\[.2em]
  \dZ{H^- H^+} &= \sinbsq \dZ{\cHe} + \cosbsq \dZ{\cHz}, \\[.2em]
  \dZ{H^- G^+} = \dZ{G^- H^+} &= \sinb \cosb (\dZ{\cHz} - \dZ{\cHe}), \\[.2em]
  \dZ{G^- G^+} &= \cosbsq \dZ{\cHe} + \sinbsq \dZ{\cHz} ~.
\end{align}
\end{subequations}
For the field renormalization constants of the $\cp$-violating
self-energies it follows,
\begin{align}
  \dZ{hA} = \dZ{hG} = \dZ{HA} = \dZ{HG} = 0,
\end{align}
which corresponds to the fact that the $\cp$-violating self-energies
do not possess divergences depending on the external momentum.


\subsection{Hybrid on-shell/\MSbar\ renormalization}

Up to now, the counter terms for $\MZ$, $\MW$, $\mHpm$, $\tanb$ and
$T_{\{h,H,A\}}$ as well as the field renormalization constants are
undetermined. 
For the mass counter terms, an on-shell definition is appropriate,
\begin{align}
\label{eq:mass_osdefinition}
  \dMZsq = \re \se{ZZ}(\MZ^2), \quad \dMWsq = \re \se{WW}(\MW^2),
  \quad \dmHpmsq = \re \se{H^- H^+}(\mHpm^2). 
\end{align}
Here $\Si$ denotes the transverse part of the self-energy. 
Since the tadpole coefficients are chosen to vanish in all orders,
their counter
terms follow from $T_{\{h,H,A\}(1)} + \de T_{\{h,H,A\}} = 0$: 
\begin{align}
  \dtadh = -{\tadh}_{(1)}, \quad \dtadH = -{\tadH}_{(1)}, \quad \dtadA
  = -{\tadA}_{(1)}. 
\end{align}

Concerning the field and $\tb$ renormalization, we adopt the
\drbar\ scheme (see also \citere{feynhiggs1.2}),
\begin{subequations}
\label{eq:deltaZHiggsTB}
\begin{align}
  \dtanb &= \dtanb^{\drbarm} 
       \; = \; -\ed{2\CZa} \KKL \re \Sip_{hh}(\mhtr^2) - 
                             \re \Sip_{HH}(\mHtr^2) \KKR^{\rm div}, \\[.5em]
  \dZ{\cHe} &= \dZ{\cHe}^{\drbarm}
       \; = \; - \KKL \re \Sip_{HH \; |\al = 0} \KKR^{\rm div}, \\[.5em]
  \dZ{\cHz} &= \dZ{\cHz}^{\drbarm} 
       \; = \; - \KKL \re \Sip_{hh \; |\al = 0} \KKR^{\rm div}, 
\end{align}
\end{subequations}
i.e.\ only the divergent parts of the renormalization constants in
\refeqs{eq:deltaZHiggsTB} are taken into account.  As renormalization
scale we have chosen $\mu_{\drbarm} = \mt$.


\section{Higher-order corrections}
\label{sec:higherorderhiggs}

\subsection{Calculation of the renormalized self-energies}
\label{subsec:SEcalc}

In order to obtain the higher-order corrections in the cMSSM Higgs
sector the renormalized self-energies \refeqs{renSEhh} --(\ref{renSEGp}) 
have to be evaluated. A renormalized self-energy can be decomposed as
\BE
\hSi(p^2) = \hSi^{(1)}(p^2) + \hSi^{(2)}(p^2) + \ldots~,
\end{equation}
where $\hSi^{(i)}$ denotes the contribution at the $i$-loop order. 
In this section we present in detail the full \onel\ contribution to
$\hSi(p^2)$, i.e.\ $\hSi^{(1)}(p^2)$ in the cMSSM. However, for the
numerical evaluation in \refse{sec:cMSSMnumeval}, also corrections
beyond the one-loop level (from the rMSSM) are taken into account.

\begin{figure}[htb!]
\input psfrag_definitionen
\begin{center}
\input{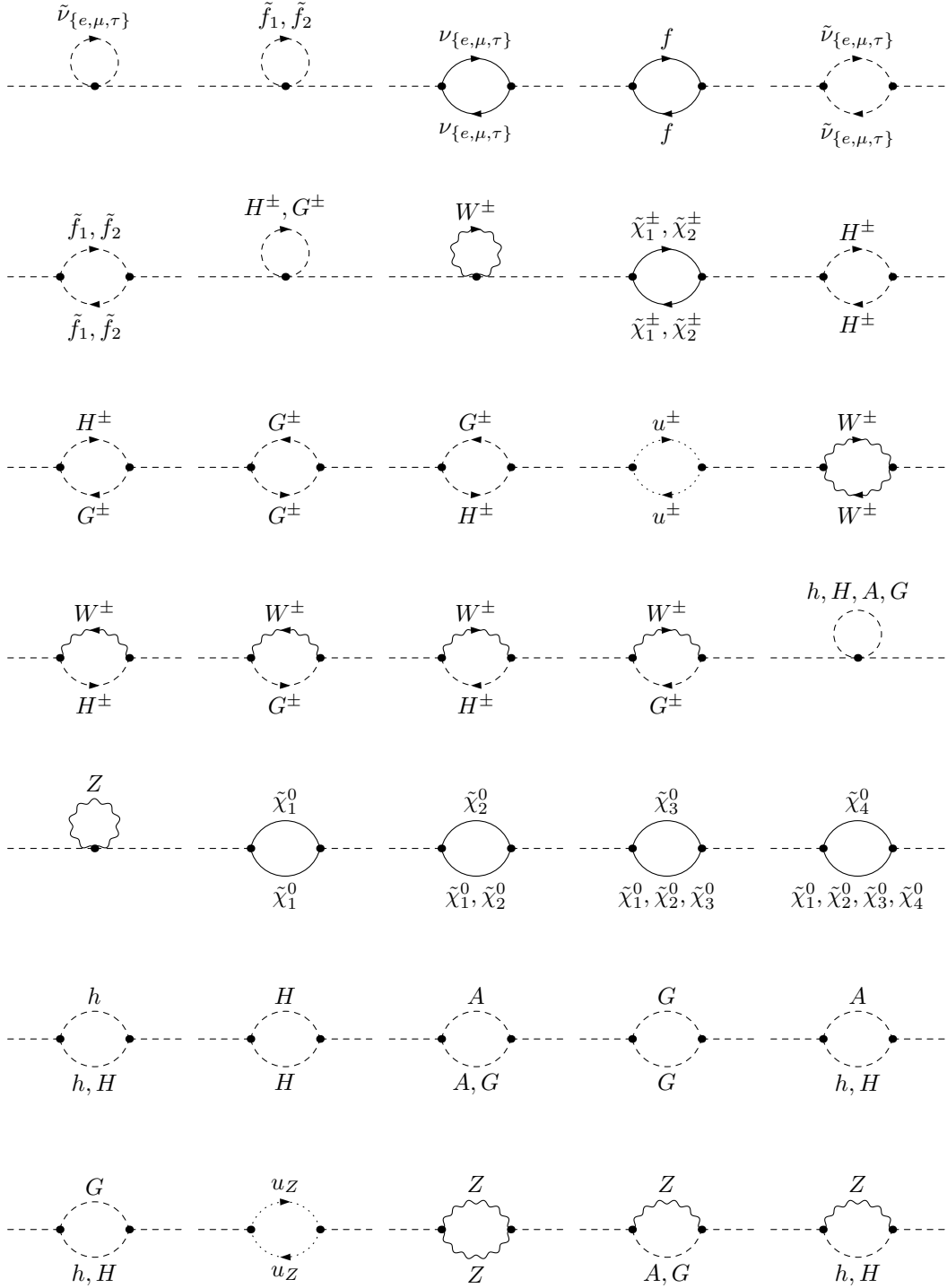}
\caption{Generic Feynman diagrams for $h$, $H$, $A$, $G$ self-energies 
($f$ = \{$e$, $\mu$, $\tau$, $d$, $s$, $b$, $u$, $c$, $t$\} ).
Corresponding diagrams for the $Z$~boson self-energy are obtained by
replacing the external Higgs boson by a $Z$~boson.}
\label{fig:fdSEn}
\end{center}
\end{figure}

The generic Feynman diagrams for the \onel\ contribution to the Higgs
and gauge boson self-energies are shown in \reffis{fig:fdSEn},
\ref{fig:fdSEc}. The \onel\ tadpole diagrams entering via the
renormalization are generically depicted in \reffi{fig:fdTP}. 

The diagrams and corresponding amplitudes have been obtained with the
program \fa~\cite{feynarts,fa-mssm} and further evaluated with
\fc~\cite{formcalc}. As a regularization scheme differential
regularization~\cite{cdr} has been used, which has been shown to be
equivalent to 
dimensional reduction~\cite{dred} at the \onel\ level~\cite{formcalc}. 
Thus, the employed regularization preserves SUSY.

\setlength{\unitlength}{0.099mm}

\setlength{\unitlength}{0.096mm}
\begin{figure}[htb!]
\input psfrag_definitionen
\begin{center}
\input{plots/fdSEc}
\caption{Generic Feynman diagrams for $H^\pm$, $G^\pm$ self-energies 
($l$ = \{$e$, $\mu$, $\tau$\}, 
 $d$ = \{$d$, $s$, $b$\},
 $u$ = \{$u$, $c$, $t$\} ).
Corresponding diagrams for the $W$~boson self-energy are obtained by
replacing the external Higgs boson by a $W$~boson.}
\label{fig:fdSEc}
\end{center}
\end{figure}

\setlength{\unitlength}{0.099mm}
\begin{figure}[htb!]
\input psfrag_definitionen
\begin{center}
\input{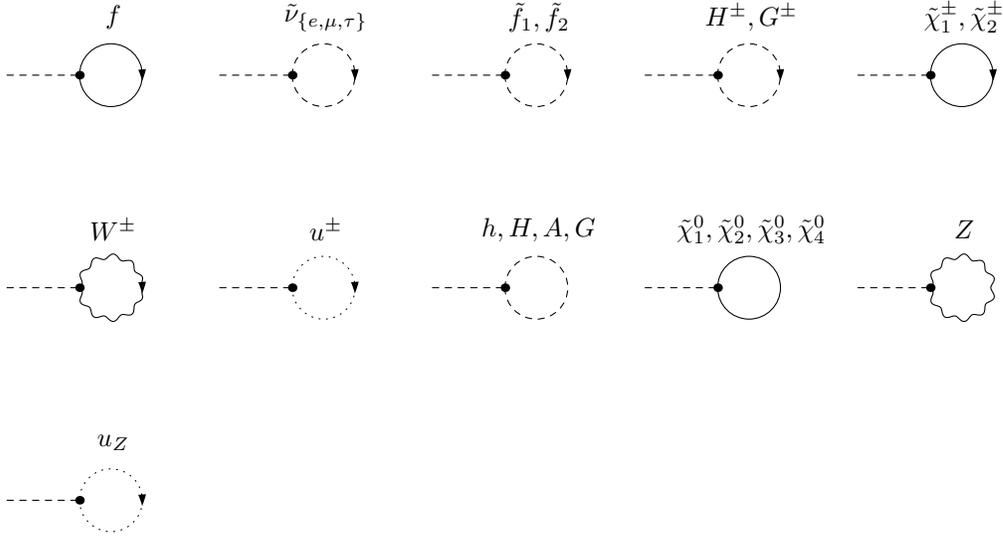}
\caption{Generic Feynman diagrams for $h$, $H$, $A$, tadpoles
($f$ = \{$e$, $\mu$, $\tau$, $d$, $s$, $b$, $u$, $c$, $t$\}).}
\label{fig:fdTP}
\end{center}
\end{figure}


\subsection{Masses and mixing at higher orders}
\label{subsec:massmix}

The masses of particles in a multiplet are determined by the poles of
their propagator matrix. 
In higher orders, self-energy terms appear in this matrix. Its
inverse is given in the case of the three physical neutral Higgs
bosons by 
\begin{align}
\label{eq:invprophiggs}
  \hat{\Gamma}_{hHA}(p^2) &= p^2 \id - \matr{M}_{\mathrm{n}}(p^2), \\
  \matr{M}_{\mathrm{n}}(p^2) &=
  \begin{pmatrix}
    \mhtr^2 - \ser{hh}(p^2) & - \ser{hH}(p^2) & - \ser{hA}(p^2) \\
    - \ser{hH}(p^2) & \mHtr^2 - \ser{HH}(p^2) & - \ser{HA}(p^2) \\
    - \ser{hA}(p^2) & - \ser{HA}(p^2) & \mAtr^2 - \ser{AA}(p^2)
  \end{pmatrix}. \notag
\end{align}
The mixing terms between the
Goldstone boson $G$ and physical Higgs bosons will not be considered
in the actual calculations of this article, since the size of these
mixing self-energies is absolutely negligible compared to the size of
the self-energies containing only physical Higgs bosons. 
 
The loop corrected pole masses correspond to the roots of $\det
\hat{\Gamma}_{hHA}(p^2)$. 
A full calculation therefore involves solving
\begin{align}
\label{eq:deteq1}
   \det(p^2 \id - \matr{M}_{\mathrm{n}}(p^2))=0
\end{align}
for all its roots. %
The $\cp$-even Higgs bosons $h$ and $H$ and the $\cp$-odd boson $A$ mix
to form new mass eigenstates $H_1$, $H_2$ and $H_3$ with
\BE
\mHe \le \mHz \le \mHd~.
\end{equation}

A simpler approximation for calculating the Higgs masses consists of
setting $p^2=0$ in~(\ref{eq:deteq1}). 
This ``$p^2=0$-approximation'' identifies the masses with the eigenvalues of
$\matr{M}_{\mathrm{n}}(0)$ instead of the true pole masses and is
mainly useful for comparisons with effective-potential calculations
and the determination of effective mixing in higher orders. 

Another simple approximation consists of choosing the $p^2$ values as follows,
\BEA
\hSi_{hh}(p^2) &\to& \hSi_{hh}(\mhtr^2) \non \\
\label{p2onshell}
\hSi_{HH}(p^2) &\to& \hSi_{hh}(\mHtr^2) \\
\hSi_{hH}(p^2) &\to& \hSi_{hH}((\mhtr^2 + \mHtr^2)/2) ~. \non 
\EEA
This ``$p^2$=on-shell'' approximation removes all dependencies
from the field renormalization constants. It results in Higgs boson
masses much closer to the true pole masses than the ``$p^2=0$''
approximation, see \refse{subsec:thresholds}.

In each of these approximations, the mass eigenstates $H_1, H_2, H_3$
are connected to $h$, $H$ and $A$ by an orthogonal 
transformation matrix $\matr{U}_{\mathrm{n}}$ which diagonalizes
$\matr{M}_{\mathrm{n}}$: 
\begin{align}
  \begin{pmatrix} H_1 \\ H_2 \\ H_3 \end{pmatrix} &=
  \matr{U}_{\mathrm{n}} \cdot \begin{pmatrix} h \\ H \\ A
  \end{pmatrix}, \quad 
  \matr{U}_{\mathrm{n}} =
  \begin{pmatrix}
    u_{11} & u_{12} & u_{13} \\
    u_{21} & u_{22} & u_{23} \\
    u_{31} & u_{32} & u_{33}
  \end{pmatrix}, \notag \\[.5em]
  \begin{pmatrix}
    \mHe^2 & 0 & 0 \\ 0 & \mHz^2 & 0 \\ 0 & 0 & \mHd^2
  \end{pmatrix} &= 
  \matr{U}_{\mathrm{n}} \matr{M}_{\mathrm{n}}
  \matr{U}_{\mathrm{n}}^+. 
\end{align}
The elements of $\matr{U}_{\mathrm{n}}$ are used in the following to
quantify the extent of $\cp$-violation. 
For example, $u_{13}^2$ can be understood as the $\cp$-odd part in $H_1$,
while $u_{11}^2+u_{12}^2$ make up the $\cp$-even part. 
The unitarity of $\matr{U}_{\mathrm{n}}$ ensures that both parts
add up to 1. 


\subsection{The Higgs boson couplings}

The leading corrections in the neutral MSSM Higgs boson sector are
taken into account by the Higgs boson self-energies at vanishing
external momentum (or in the ``$p^2$=on-shell''
approximation). The matrix $\matr{U}_{\mathrm{n}}$ then also provides
the leading  
corrections to the neutral Higgs boson couplings to SM gauge bosons
and fermions, see e.g.\ \citere{mhiggsCPXRG1}.

Taking complex phases into account, all three neutral Higgs bosons
contain a $\cp$-even part, thus all three Higgs bosons can
couple to two gauge bosons, $VV = ZZ, W^+W^-$. The couplings normalized
to the SM values are given by
\begin{align}
g_{H_i VV} &= u_{i 1} \Sba + u_{i 2} \Cba~.
\end{align}
The coupling of two Higgs bosons to a $Z$~boson, normalized to
the SM value, is given by
\begin{align}
g_{H_iH_jZ} &= u_{i 3} \KL u_{j 1} \Cba - u_{j 2} \Sba \KR \notag \\
            &- u_{j 3} \KL u_{i 1} \Cba - u_{i 2} \Sba \KR~.
\end{align}
The Bose symmetry that forbids any anti-symmetric derivative coupling
of a vector particle to two identical real scalar fields is respected,
$g_{H_iH_iV} = 0$.

Concerning the decay to light SM fermions, 
the decay width of the $H_i$ can be obtained from the SM decay
width of the Higgs boson by multiplying it with
\BE
\KKL \KL g_{H_iff}^\rmS \KR^2 + \KL g_{H_iff}^\rmP \KR^2 \KKR,
\end{equation}
with
\begin{align}
 g_{H_iuu}^\rmS &= (u_{i 1}\Ca + u_{i 2} \Sa)/\sbe, \quad
& g_{H_iuu}^\rmP = u_{i 3} \; \cbe/\sbe \\
 g_{H_idd}^\rmS &= (-u_{i 1} \Sa + u_{i 2} \Ca)/\cbe, \quad
& g_{H_idd}^\rmP = u_{i 3} \; \sbe/\cbe
\end{align}
for up- and down-type quarks, respectively.
For more details, see \citere{feynhiggs2.1}.


%


\section{Phenomenological implications}
\label{sec:cMSSMnumeval}

In this section we briefly describe some of the phenomenological
implications of a complete \onel\ evaluation of the cMSSM Higgs
sector. More details can be found in \citere{mhiggsCPXFD2}. 

The higher-order
corrected Higgs boson sector has been evaluated with the help of the
Fortran code \fhto~\cite{feynhiggs,feynhiggs1.2,feynhiggs2.1}. 
The code includes the full \onel\ calculation, see
\refse{subsec:SEcalc}. Furthermore, the \twol\ corrections are taken
over from the rMSSM~\cite{mhiggslong,mhiggsAEC,mhiggsEP5} and for the
$b/\Sbot$~sector~\cite{deltamb1,deltamb} from the cMSSM.
The code can be obtained from the \fh\ home page:
{\tt www.feynhiggs.de} .

For a more detailed phenomenological analysis,
constraints on $\cp$-violating parameters from experimental bounds,
e.g.\ on electric
dipole moments (EDMs), have to be taken into account~\cite{pdg}.
However, in our analysis below we only take non-zero phases for
$\At = \Ab$ and $M_2, M_1$, which are not severely restricted from EDM
bounds. However, our analysis is confined to
$\varphi_\mu = 0$, since this is the most restricted phase, see e.g.\
\citere{plehnix} and
references therein. 

The numerical analysis given below has been performed in the
the following scenario (if not indicated otherwise):
\BEA
&& \msusy = 500 \gev, \; |\At| = |\Ab| = |\Atau| = 1000 \gev, \;
   \phiab = \phiatau = 0, \; \non \\
&& \mu = 1000 \gev, \; M_2 = 500 \gev, \; M_1 = 250 \gev \non \\
&& \MHp = 150 \gev, \; \tb = 4, 15,  \; \mudim = \mt~,
\label{parameters}
\EEA
where the parameter under investigation has been varied.

Larger effects of the $\cp$-violating phases are observed for smaller
values of $\mHp$. Values of $\mHp$ and $\tb$ as given in
\refeq{parameters} could already be
challenged by the Higgs search performed at
LEP~\cite{LEPHiggsSM,LEPHiggs} (depending on the other
parameters). It has been shown, however, that within the cMSSM the
obtained limits on $\mh$ cannot be taken over directly to the complex
case~\cite{cpx,CPXOPAL}. Therefore, we do not apply the bounds of
\citeres{LEPHiggsSM,LEPHiggs}, but one should be aware that
effects for experimentally not excluded parameters might be slightly
weaker than shown below. 


\subsection{Dependence on the gaugino phases}

\begin{figure}[htb!]
\input psfrag_definitionen
\psfrag{phiM2}[][]{$\varphi_{M_2}/\pi$}
\vspace{-1em}
\begin{center}
\epsfig{figure=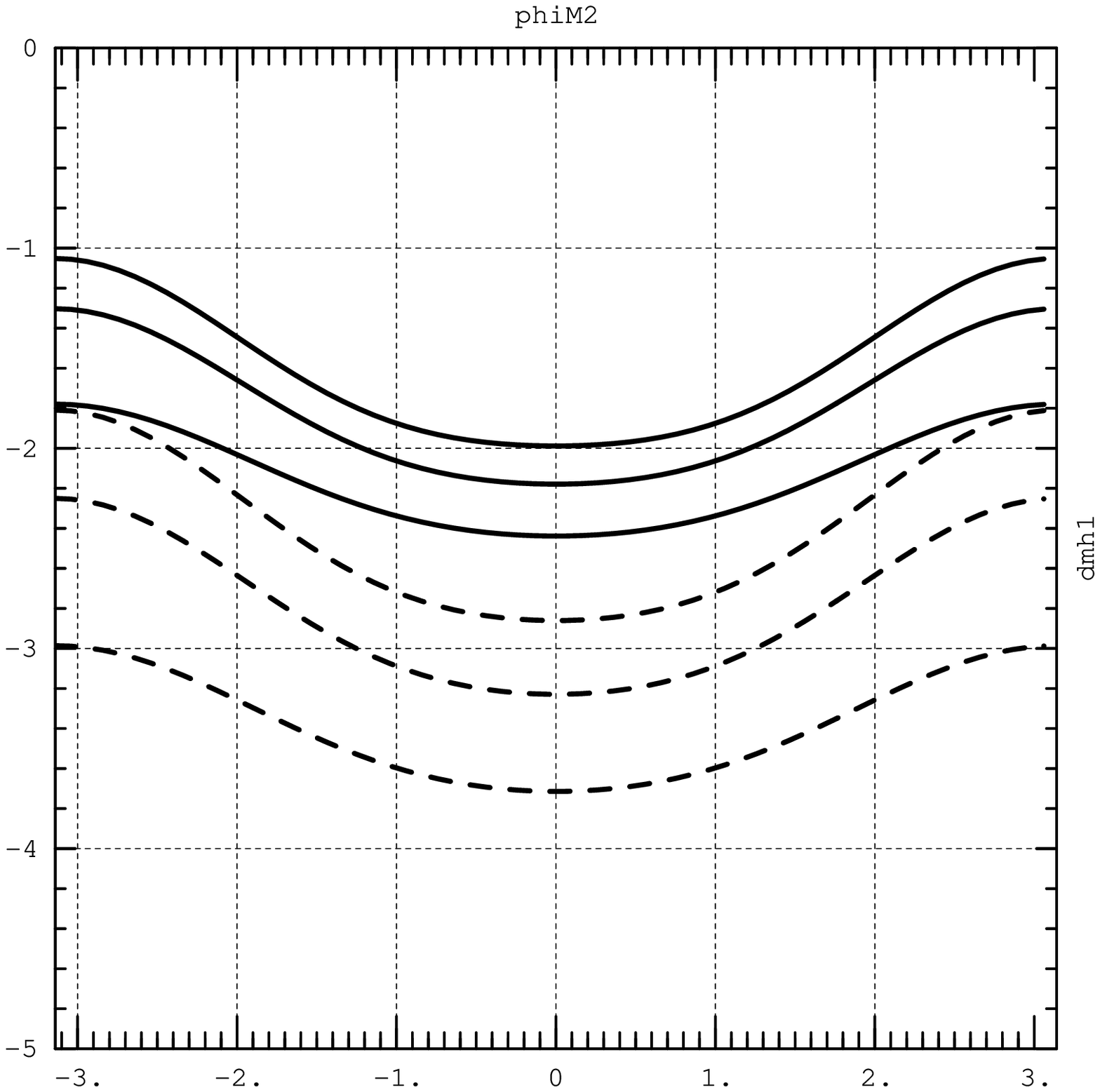,width=7.9cm,height=11cm}
\hspace{1em}  
\epsfig{figure=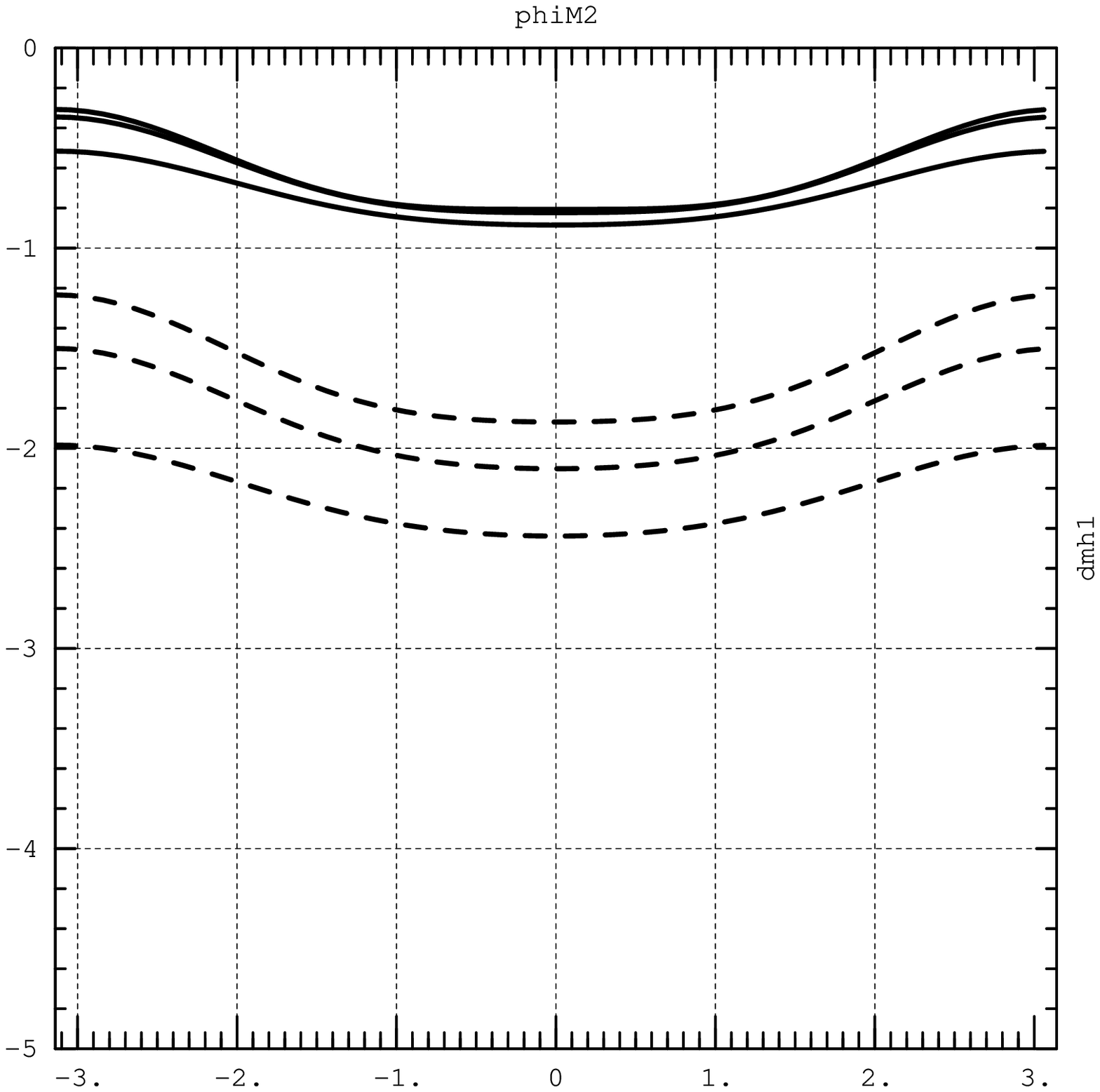,width=7.9cm,height=11cm} 
\end{center}
\vspace{-5em}
\caption{
$\De\mHe$ := (all sectors) -- ($f/\Sf$ sector) is shown as a function of
$\phiMz$ for $p^2 \neq 0$ (solid) and $p^2 = 0$ (dashed) and for 
$\tb = 4$ (left) and $\tb = 15$ (right). 
$|M_2|$ is chosen as $500, 1000, 2000 \gev$.
Only \onel\ corrections are taken into account.
}
\label{fig1c}
\end{figure}

First we analyze the dependence on the gaugino phases $\phiMe$
and $\phiMz$. In \reffi{fig1c} the dependence of the
lightest cMSSM Higgs boson mass on $\phiMz$ is shown. 
$\De\mHe$ := (all sectors) -- ($f/\Sf$ sector) is evaluated at the pure
\onel\ level for three
different values of $|M_2|$, $|M_2| = 500, 1000, 2000 \gev$ from 
the upper to the lowest line.
The other parameters
are chosen as in \refeq{parameters}. The result including the full
momentum dependence is given by the solid lines, while the $p^2 = 0$
approximation is shown as dashed lines. In the left plot we have chosen 
$\tb = 4$, in the right one $\tb = 15$. 
For the low $\tb$ value the effects from the gaugino (and Higgs)
sector are about $2 \gev$ if 
the external momentum is not neglected, and about $3 \gev$ if the
external momentum is neglected as e.g.\ in the effective potential
approach. The effect coming from varying the gaugino phase $\phiMz$
itself is of \order{1 \gev}. Both types of effects become smaller for
larger $\tb$ values. However, being in the ballpark of 1--$2 \gev$ the
effects from the gaugino sector are non-negligible and have to be
taken into account in a precision analysis.

\begin{figure}[hb!]
\input psfrag_definitionen
\psfrag{phiM1}[][]{$\varphi_{M_1}/\pi$}
\vspace{-3em}
\begin{center}
\epsfig{figure=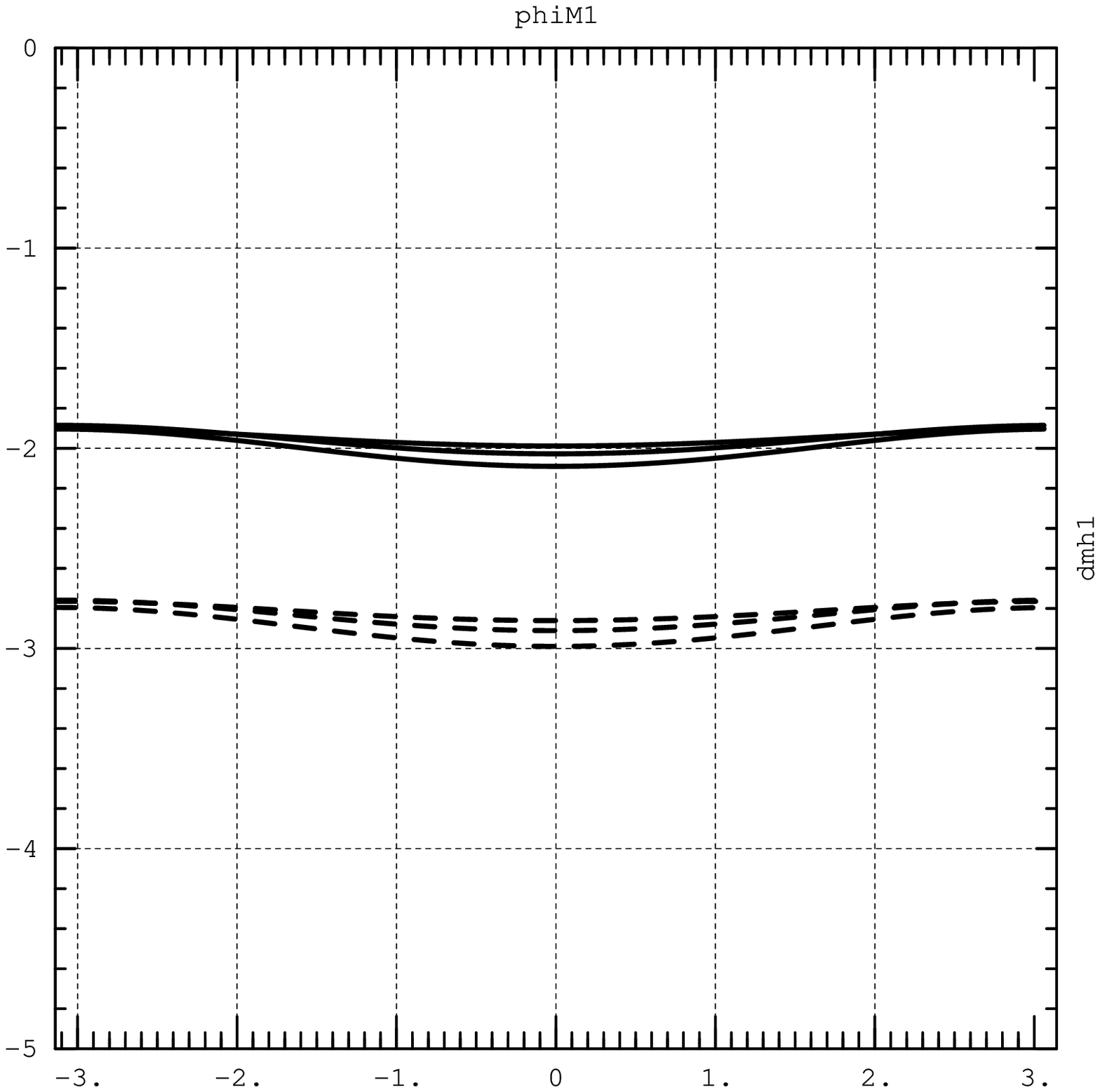,width=7.9cm,height=11cm}
\hspace{1em}  
\epsfig{figure=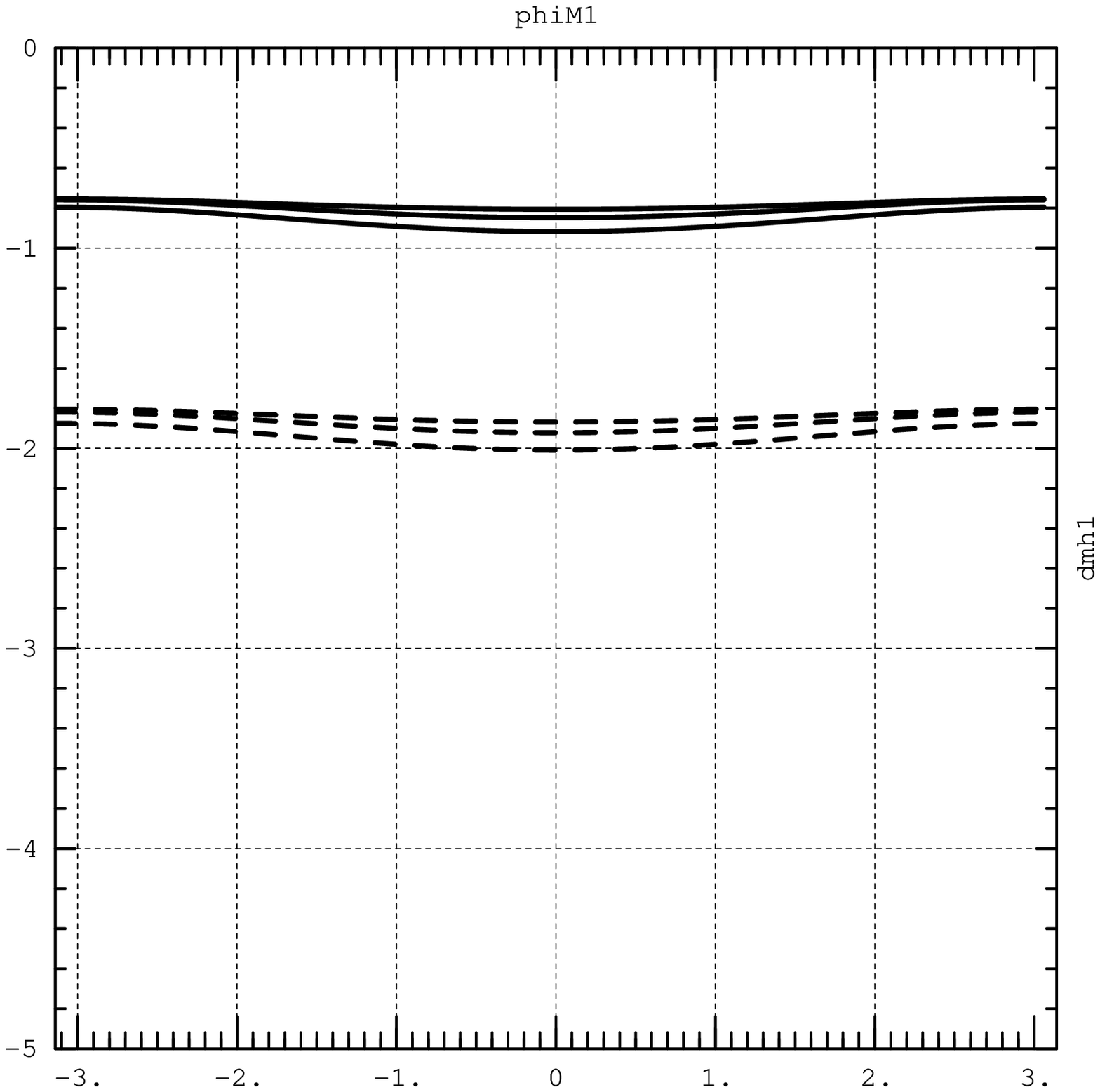,width=7.9cm,height=11cm}
\end{center}
\vspace{-5em}
\caption{
$\De\mHe$ := (all sectors) -- ($f/\Sf$ sector) is shown as a function of
$\phiMe$ for $p^2 \neq 0$ (solid) and $p^2 = 0$ (dashed) and for 
$\tb = 4$ (left) and $\tb = 15$ (right).
$|M_2|$ is chosen as $250, 500, 1000 \gev$.
Only \onel\ corrections are taken into account.
}
\label{fig1d}
\end{figure}

We now turn to the effects from varying $\phiMe$ as shown in
\reffi{fig1d}. The parameters are as in \reffi{fig1c}, but with 
$M_2 = 500 \gev$, and $|M_1| = 250, 500, 1000 \gev$ from 
the most upper to the most lower line. 
The size of the effects from the gaugino sector is of course the same
as in \reffi{fig1c}. However, the dependence on $\phiMe$ is much
smaller, being at \order{200 \mev}. 
Aiming to match the anticipated LC accuracy of 
$\de\mHe^{\rm exp} = 50 \mev$ even these relatively small corrections
have to be taken into account.

%


\subsection{Threshold effects for heavy Higgs bosons}
\label{subsec:thresholds}

\begin{figure}[htb!]
\input psfrag_definitionen
\psfrag{dm32}[][]{$\Delta m_{32}/\gev$}
\vspace{-3em}
\begin{center}
\epsfig{figure=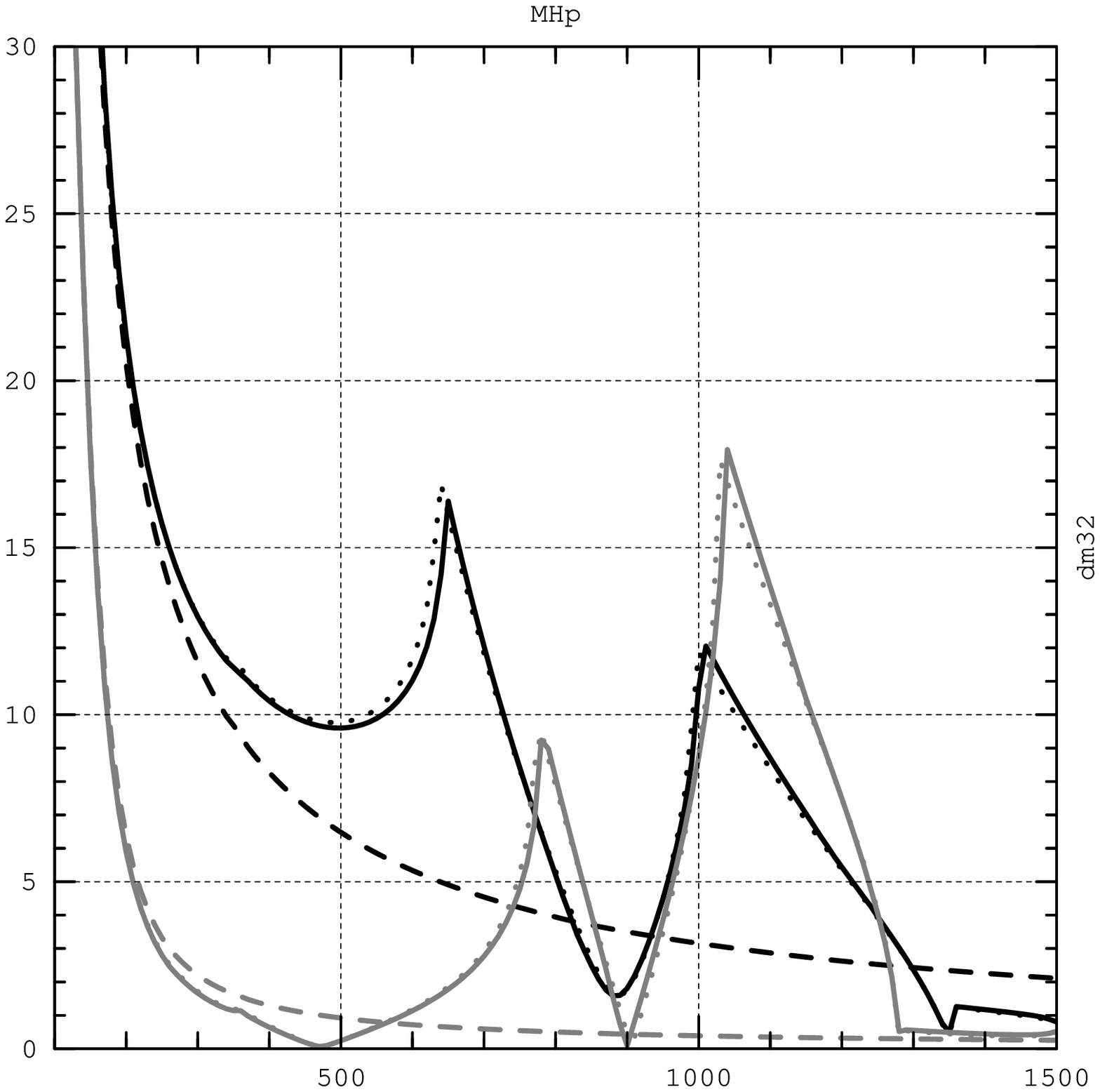,width=7.9cm,height=11cm}
\hspace{1em}  
\epsfig{figure=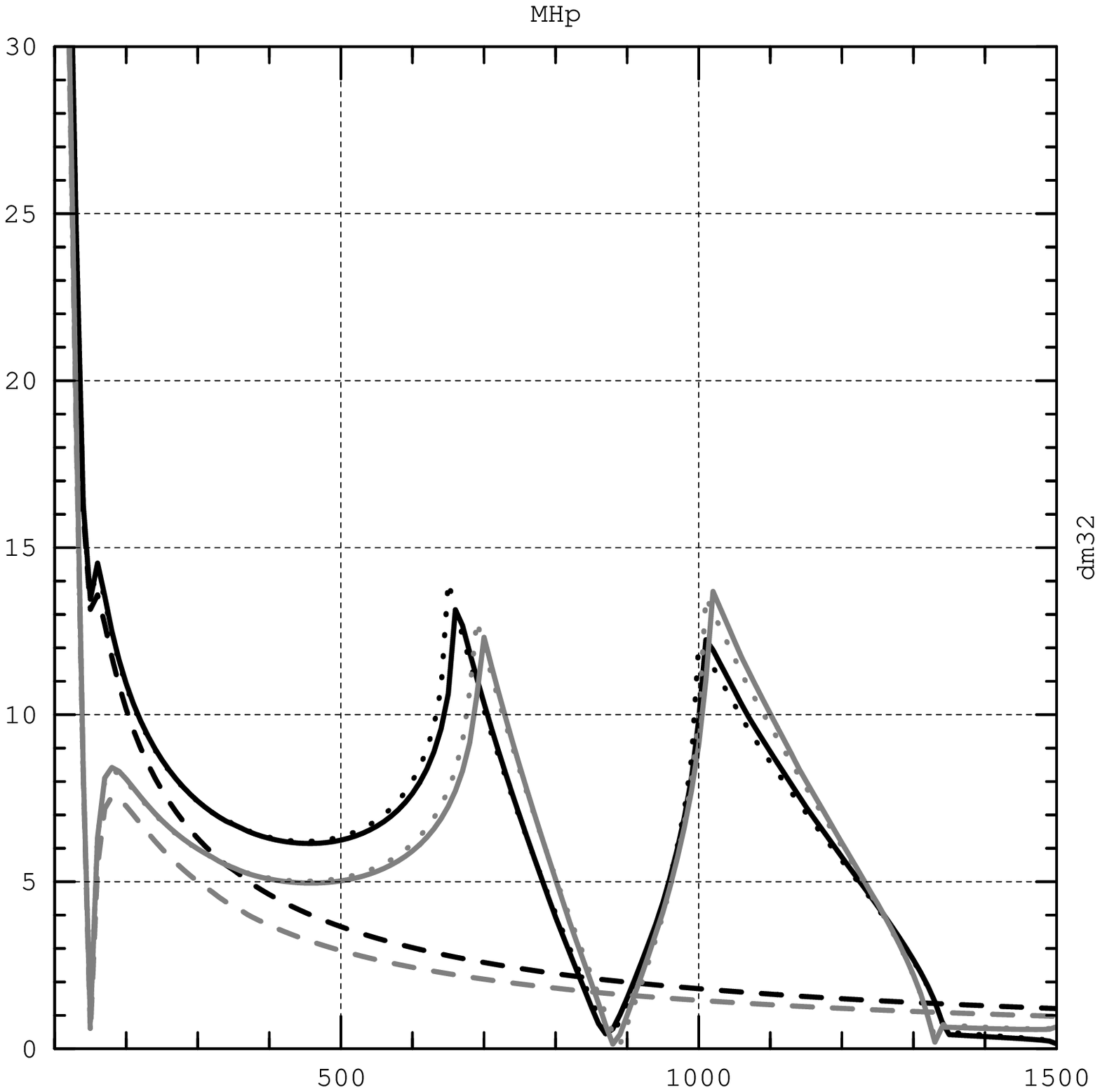,width=7.9cm,height=11cm} \\[-2em]
\epsfig{figure=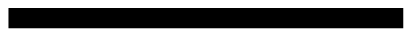,width=1cm} all sectors, \quad
\epsfig{figure=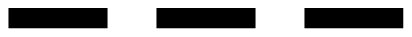,width=1cm} all sectors, $p^2 = 0$, \quad
\epsfig{figure=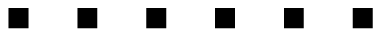,width=1cm} all sectors, $p^2 =$ on-shell
approx. 
\end{center}
\vspace{-1em}
\caption{
$\De m_{32} = \mHd - \mHz$ is shown as a function of $\mHp$ for 
$\tb = 4$ (left) and $\tb = 15$ (right) for two different values of
  $\phiat$, $\phiat = \pi/2$ (black) and $\phiat = 0$ (gray).
}
\label{fig2a}
\end{figure}

In this subsection the threshold effects on the masses of the heavy
neutral Higgs bosons are analyzed. In \reffi{fig2a} the mass
difference $\De m_{32} = \mHd - \mHz$ is shown as a function of $\mHp$ for 
$\tb = 4$ (left) and $\tb = 15$ (right) for two different values of
$\phiat$, $\phiat = \pi/2$ (black) and $\phiat = 0$ (gray).
In general it can be observed that the mass differences are larger in
the case of non-vanishing complex phases%
\footnote{
However, in \citere{mhiggsCPXFD2} it is shown that all mass
differences that appear for complex parameters can also be realized
(for other parameter combinations) in the rMSSM.
}%
.
The dashed curves are evaluated with the $p^2 = 0$ approximation. The
mass difference monotonuously decreases with increasing $\mHp$. The full
calculation shown in the solid curves, however, exhibits strong
threshold effects coming from the scalar top quarks in the Higgs boson
self-energies (e.g. the second and sixth diagram in \reffi{fig:fdSEn}). 
Their effects can be larger than $\sim 10 \gev$ and
thus can be more relevant than the effect induced by the complex phases
for $\At$ and $\Ab$. Thus,
neglecting the external momentum can lead to large uncertainties in
the calculation of the heavy Higgs boson masses. On the other hand, it
turns out that the $p^2 =$~on-shell approximation, see \refeq{p2onshell}, 
shown as dotted lines, gives a rather
good approximation to the full result. The remaining deviations stay
below the level of $1 \gev$.
For LC precisions for the heavier Higgs boson of \order{1 \gev} these
corrections should be taken into account.
It should be kept in mind that the inclusion of the finite widths of
the scalar top quarks will change the results close to the
threshold peaks. These effects will have to be taken into account in order
to obtain a reliable result at the thresholds.

\newpage

\chapter{Higgs boson production at the LC}

Controlling the Higgs boson production properties at the per-cent
level is mandatory to perform high-precision measurements at the
LC. The Higgs 
boson production cross sections themselves will be measured at
\order{1\%}~\cite{teslatdr,eennHexp}. In order to be able to exploit
these measurements the experimental accuracy has to be matched with a
corresponding theoretical precision. 
At the LC, the possible channels for neutral Higgs-boson production are
the production via $Z$-boson exchange (Higgs-strahlung),
\BEA
e^+e^- &\to& Z \to Z \{h,H\}\,, \non \\
e^+e^- &\to& Z \to A \{h,H\}\,,
\label{eehZhA}
\EEA
and the $WW$-fusion channel,
\BE
e^+e^- \to \bar\nu_e W^+ \; \nu_e W^- \to \bar\nu_e \nu_e \{h,H,A\}\,.
\label{eennH}
\end{equation}
Charged Higgs bosons can be produced in pairs,
\BE
e^+e^- \to \{\ga, Z\} \to H^+H^-
\label{eeHpHm}
\end{equation}
or singly,
\BEA
\label{eeWH}
e^+e^- &\to& \{\ga, Z\} \to W^+H^- \, , \\
\label{eeneH}
e^+e^- &\to& \bar\nu_e W^+ \; e^- \{\ga,Z\} \to \bar\nu_e e^- H^+ 
\EEA
(and also with the charged conjugated processes).
We do not discuss here the possibility of Higgs-particle
production by brems\-strahlung off heavy quarks (e.g. $e^+e^- \to
\bar{b}b\{ h,H,A\} $, which can be significant for large $\tb$~\cite{hbb}).

The Higgs-strahlung processes, \refeq{eehZhA}, are possible at the
tree-level. The full MSSM \onel\ corrections can be found
in \citeres{hprod2,hprodboxes,hprodf1l}. The leading \twol\
corrections have been incorporated in \citere{eehZhA}.
While the $WW$~fusion channel, \refeq{eennH}, for the $\cp$-even Higgs
bosons are possible already at the tree-level~\cite{eennHtree0,hprod},
the $\cp$-odd Higgs 
boson can only be produced at the loop level. Within the SM recently
the full \onel\ corrections to the $WW$~fusion channel have been
obtained by two groups~\cite{eennHSMf1l} (see also \citere{eennHSMf1lB}
for a partial analytical calculation). In the MSSM so far only
the corrections from all~\cite{eennH} or third
generation~\cite{eennHWiener} fermion and sfermion loops are
available. The $\cp$-odd channel has been evaluated including 
3- and 4-point function \onel\ corrections~\cite{eennA}. 
The pair production of the charged Higgs bosons, \refeq{eeHpHm}, is
possible at the tree-level, the full \onel\ corrections in the MSSM
can be found in \citere{eeHpHm}. 
The charged Higgs boson production in association with a $W$~boson,
\refeq{eeWH}, is mediated via loop diagrams, where the full \onel\
corrections can be found in \citeres{eeWH1,eeWH2}. 
Finally the single Higgs production in the gauge boson fusion channel,
\refeq{eeneH}, possible only at the loop level, has been evaluated
including the full \onel\ SM fermion and scalar fermion corrections in
\citere{eeenH}. 

In the following subsections we will review the leading \twol\
corrections to the Higgs-strahlung process, the (s)fermion one-loop
corrections to the $WW$~fusion channel and the (s)fermion one-loop
contribution to the single charged Higgs production in gauge boson
fusion. The main phenomenological consequences are also discussed.


\section{Corrections to the Higgs-strahlung channel}
\label{sec:Higgsstrahlung}

The most promising channels for the production of the $\cp$-even
neutral MSSM Higgs bosons in the first phase of a LC
are the Higgs-strahlung processes~\cite{hprod},
\BE 
e^+e^- \to Z\,H_i ~, 
\label{eetohZ}
\end{equation}
($H_{1,2} = h,H$) and the associated production of a scalar and a
pseudoscalar Higgs boson,  
\BE
e^+e^- \to A\,H_i~.  
\label{eetohA}
\end{equation} 
We review the computation of the MSSM predictions for the cross
sections of both 
channels in the Feynman-diagrammatic (FD) approach using the on-shell
renormalization scheme. 
We take into account the complete set of \onel\ contributions,
thereby keeping the full dependence on all kinematical variables.
The one-loop contributions consist of the corrections to the Higgs- 
and gauge-boson propagators, where the former contain the dominant 
electroweak \onel\ corrections of
\order{\alt}, and of the contributions to the 3-point and 4-point vertex
functions~\cite{hprod2,hprodboxes,hprodf1l}%
\footnote{
Only photonic corrections to the $Ze^+e^-$ vertex are
omitted. These virtual
IR-divergent photonic corrections constitute, together with
real-photon bremsstrahlung, the initial-state QED corrections, which
are conventionally treated separately and are the same as for the
SM Higgs-boson production.
}%
. 
We combine the complete \onel\ result with the dominant
\twol\ QCD corrections of 
\order{\alt\als}~\cite{mhiggsletter,mhiggslong} and further
sub-dominant corrections.  In this way the currently most
accurate results for the cross sections are obtained.  

Furthermore
we show analytically that the Higgs-boson propagator corrections with
neglected momentum dependence can be absorbed into the tree-level
coupling using the effective mixing angle from the neutral $\cp$-even
Higgs-boson sector. We  compare our results for the
cross sections with the approximation in which only the corrections to 
the effective mixing angle, evaluated within the
renormalization-group-improved one-loop 
effective potential approach, are taken 
into account.
For most parts of the MSSM parameter space we find
agreement of the two approaches of better than 10\% for the highest
LEP energies, while for $\sqrt{s} = 500 \gev$ the difference can reach
25\%.


\subsection{Cross sections for Higgs production in $e^+e^-$ collisions}
\label{sec:form}

\subsubsection{Classification of radiative corrections}
\label{subsec:rcXsec}

The set of diagrams taken into account for Higgs-strahlung \eetohZ\ is
schematically shown in \reffi{fig:diagrams}, where a) is the tree-level
diagram. The shaded blobs
summarize the loops with all possible virtual particles, except
photons in the $Ze^+e^-$ vertex corrections, see above.  
More details can be found in 
\citeres{hprod2,hprodboxes,hprodf1l}.  An analogous set has been evaluated
for the second process \eetohA.

\setlength{\unitlength}{1bp}
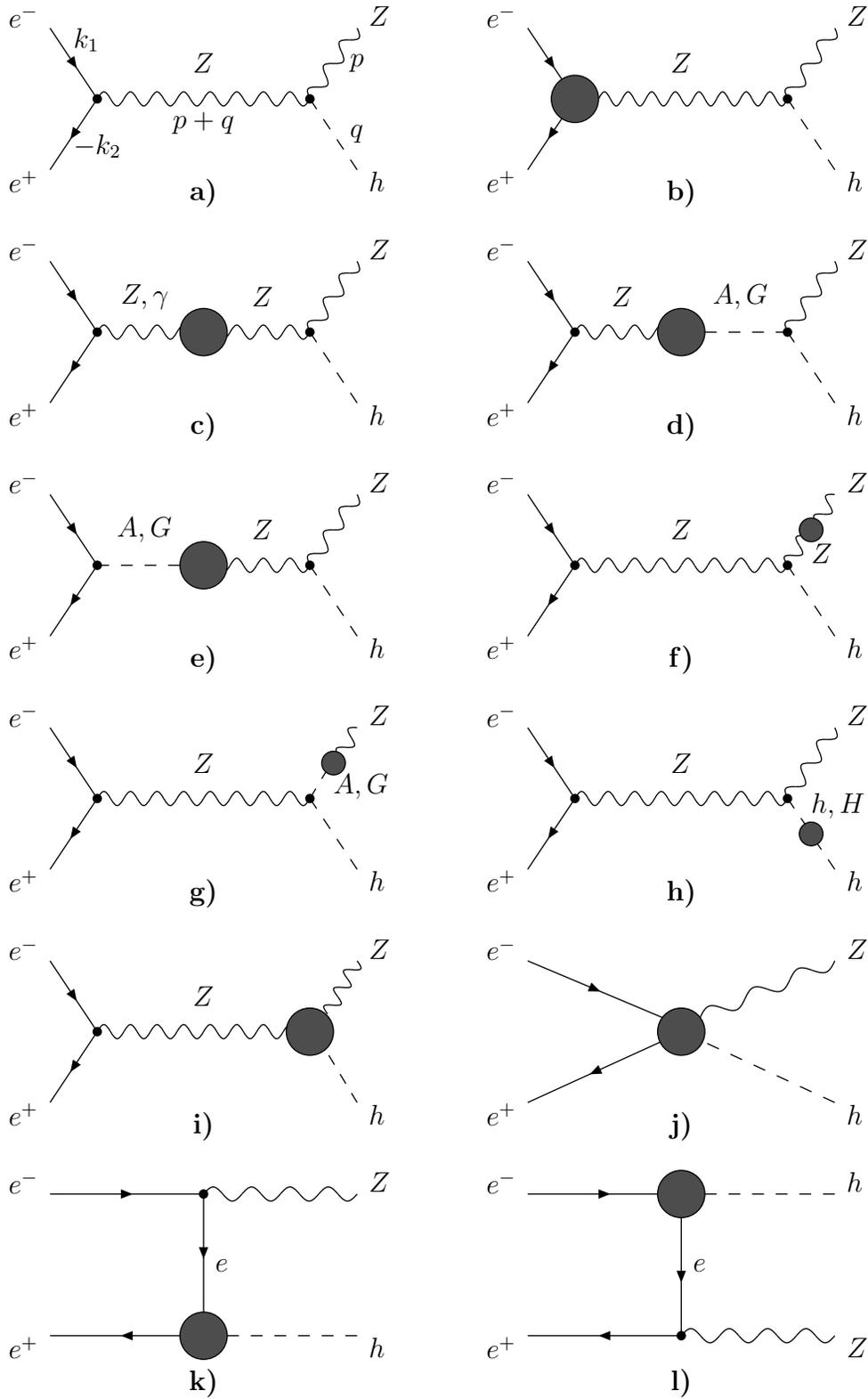
\begin{figure}[ht!]
\begin{center}
\begin{tabular}{lp{1cm}r}
\begin{picture}(150,80)(0,0)
\ArrowLine(30,40)(10,10)
\Text(5,5)[r]{\mbox{$e^+$}}
\Text(20,20)[l]{\mbox{$-k_2$}}
\ArrowLine(10,70)(30,40)
\Text(5,75)[r]{\mbox{$e^-$}}
\Text(20,65)[l]{\mbox{$k_1$}}
\Vertex(30,40){2}
\Photon(30,40)(120,40){3}{8}
\Text(75,55)[c]{\mbox{$Z$}}
\Text(75,30)[c]{\mbox{$p+q$}}
\Vertex(120,40){2}
\DashLine(120,40)(140,10){5}
\Text(145,5)[l]{\mbox{$h$}}
\Text(137,25)[l]{\mbox{$q$}}
\Photon(120,40)(140,70){3}{3}
\Text(145,75)[l]{\mbox{$Z$}}
\Text(137,55)[l]{\mbox{$p$}}
\Text(75,0)[c]{\mbox{\bf a)}}
\end{picture}
&&
\begin{picture}(150,80)(0,0)
\ArrowLine(30,40)(10,10)
\Text(5,5)[r]{\mbox{$e^+$}}
\ArrowLine(10,70)(30,40)
\Text(5,75)[r]{\mbox{$e^-$}}
\GCirc(30,40){10}{0.3}                                                      
\Photon(40,40)(120,40){3}{8}
\Text(75,55)[c]{\mbox{$Z$}}
\Vertex(120,40){2}
\DashLine(120,40)(140,10){5}
\Text(145,5)[l]{\mbox{$h$}}
\Photon(120,40)(140,70){3}{3}
\Text(145,75)[l]{\mbox{$Z$}}
\Text(75,0)[c]{\mbox{\bf b)}}
\end{picture}
\\
\\
\begin{picture}(150,80)(0,0)
\ArrowLine(30,40)(10,10)
\Text(5,5)[r]{\mbox{$e^+$}}
\ArrowLine(10,70)(30,40)
\Text(5,75)[r]{\mbox{$e^-$}}
\Vertex(30,40){2}
\Photon(30,40)(120,40){3}{8}
\GCirc(75,40){10}{0.3}                                                      
\Text(50,55)[c]{\mbox{$Z,\gamma$}}
\Text(100,55)[c]{\mbox{$Z$}}
\Vertex(120,40){2}
\DashLine(120,40)(140,10){5}
\Text(145,5)[l]{\mbox{$h$}}
\Photon(120,40)(140,70){3}{3}
\Text(145,75)[l]{\mbox{$Z$}}
\Text(75,0)[c]{\mbox{\bf c)}}
\end{picture}
&&
\begin{picture}(150,80)(0,0)
\ArrowLine(30,40)(10,10)
\Text(5,5)[r]{\mbox{$e^+$}}
\ArrowLine(10,70)(30,40)
\Text(5,75)[r]{\mbox{$e^-$}}
\Vertex(30,40){2}
\Photon(30,40)(75,40){3}{4}
\Text(50,55)[c]{\mbox{$Z$}}
\GCirc(75,40){10}{0.3}                                                      
\DashLine(85,40)(120,40){5}
\Text(100,55)[c]{\mbox{$A,G$}}
\Vertex(120,40){2}
\DashLine(120,40)(140,10){5}
\Text(145,5)[l]{\mbox{$h$}}
\Photon(120,40)(140,70){3}{3}
\Text(145,75)[l]{\mbox{$Z$}}
\Text(75,0)[c]{\mbox{\bf d)}}
\end{picture}
\\
\\
\begin{picture}(150,80)(0,0)
\ArrowLine(30,40)(10,10)
\Text(5,5)[r]{\mbox{$e^+$}}
\ArrowLine(10,70)(30,40)
\Text(5,75)[r]{\mbox{$e^-$}}
\Vertex(30,40){2}
\Photon(75,40)(120,40){3}{4}
\Text(100,55)[c]{\mbox{$Z$}}
\GCirc(75,40){10}{0.3}                                                      
\DashLine(30,40)(65,40){5}
\Text(50,55)[c]{\mbox{$A,G$}}
\Vertex(120,40){2}
\DashLine(120,40)(140,10){5}
\Text(145,5)[l]{\mbox{$h$}}
\Photon(120,40)(140,70){3}{3}
\Text(145,75)[l]{\mbox{$Z$}}
\Text(75,0)[c]{\mbox{\bf e)}}
\end{picture}
&&
\begin{picture}(150,80)(0,0)
\ArrowLine(30,40)(10,10)
\Text(5,5)[r]{\mbox{$e^+$}}
\ArrowLine(10,70)(30,40)
\Text(5,75)[r]{\mbox{$e^-$}}
\Vertex(30,40){2}
\Photon(30,40)(120,40){3}{8}
\Text(75,55)[c]{\mbox{$Z$}}
\Vertex(120,40){2}
\DashLine(120,40)(140,10){5}
\Text(145,5)[l]{\mbox{$h$}}
\Photon(120,40)(127.5,52.5){3}{1.5}
\Photon(132.5,57.5)(140,70){3}{1.5}
\GCirc(130,55){5}{0.3}                                                      
\Text(145,75)[l]{\mbox{$Z$}}
\Text(130,45)[l]{\mbox{$Z$}}
\Text(75,0)[c]{\mbox{\bf f)}}
\end{picture}
\\
\\
\begin{picture}(150,80)(0,0)
\ArrowLine(30,40)(10,10)
\Text(5,5)[r]{\mbox{$e^+$}}
\ArrowLine(10,70)(30,40)
\Text(5,75)[r]{\mbox{$e^-$}}
\Vertex(30,40){2}
\Photon(30,40)(120,40){3}{8}
\Text(75,55)[c]{\mbox{$Z$}}
\Vertex(120,40){2}
\DashLine(120,40)(140,10){5}
\Text(145,5)[l]{\mbox{$h$}}
\Photon(132.5,57.5)(140,70){3}{1.5}
\DashLine(120,40)(127.5,52.5){5}
\GCirc(130,55){5}{0.3}                                                      
\Text(145,75)[l]{\mbox{$Z$}}
\Text(130,45)[l]{\mbox{$A,G$}}
\Text(75,0)[c]{\mbox{\bf g)}}
\end{picture}
&&
\begin{picture}(150,80)(0,0)
\ArrowLine(30,40)(10,10)
\Text(5,5)[r]{\mbox{$e^+$}}
\ArrowLine(10,70)(30,40)
\Text(5,75)[r]{\mbox{$e^-$}}
\Vertex(30,40){2}
\Photon(30,40)(120,40){3}{8}
\Text(75,55)[c]{\mbox{$Z$}}
\Vertex(120,40){2}
\DashLine(120,40)(128.5,27.5){5}
\DashLine(131.5,22.5)(140,10){5}
\GCirc(130,25){5}{0.3}                                                      
\Text(145,5)[l]{\mbox{$h$}}
\Text(130,37)[l]{\mbox{$h,H$}}
\Photon(120,40)(140,70){3}{3}
\Text(145,75)[l]{\mbox{$Z$}}
\Text(75,0)[c]{\mbox{\bf h)}}
\end{picture}
\\
\\
\begin{picture}(150,80)(0,0)
\ArrowLine(30,40)(10,10)
\Text(5,5)[r]{\mbox{$e^+$}}
\ArrowLine(10,70)(30,40)
\Text(5,75)[r]{\mbox{$e^-$}}
\Vertex(30,40){2}
\Photon(30,40)(120,40){3}{8}
\Text(75,55)[c]{\mbox{$Z$}}
\GCirc(120,40){10}{0.3}                                                      
\DashLine(128,30)(140,10){5}
\Text(145,5)[l]{\mbox{$h$}}
\Photon(127.5,47.5)(140,70){3}{3}
\Text(145,75)[l]{\mbox{$Z$}}
\Text(75,0)[c]{\mbox{\bf i)}}
\end{picture}
&&
\begin{picture}(150,80)(0,0)
\ArrowLine(67,36)(10,10)
\Text(5,5)[r]{\mbox{$e^+$}}
\ArrowLine(10,70)(67,46)
\Text(5,75)[r]{\mbox{$e^-$}}
\GCirc(75,40){10}{0.3}                                                      
\DashLine(86,35)(140,10){5}
\Text(145,5)[l]{\mbox{$h$}}
\Photon(83,46)(140,70){3}{3}
\Text(145,75)[l]{\mbox{$Z$}}
\Text(75,0)[c]{\mbox{\bf j)}}
\end{picture}
\\
\\
\begin{picture}(150,80)(0,0)
\ArrowLine(75,10)(10,10)
\Text(5,5)[r]{\mbox{$e^+$}}
\ArrowLine(10,70)(75,70)
\Text(5,75)[r]{\mbox{$e^-$}}
\GCirc(75,10){10}{0.3}                                                      
\ArrowLine(75,70)(75,20)
\Text(80,40)[l]{\mbox{$e$}}
\Vertex(75,70){2}
\DashLine(85,10)(140,10){5}
\Text(145,5)[l]{\mbox{$h$}}
\Photon(75,70)(140,70){3}{4}
\Text(145,75)[l]{\mbox{$Z$}}
\Text(75,-10)[c]{\mbox{\bf k)}}
\end{picture}
&&
\begin{picture}(150,80)(0,0)
\ArrowLine(75,10)(10,10)
\Text(5,5)[r]{\mbox{$e^+$}}
\ArrowLine(10,70)(75,70)
\Text(5,75)[r]{\mbox{$e^-$}}
\GCirc(75,70){10}{0.3}                                                      
\ArrowLine(75,60)(75,10)
\Text(80,40)[l]{\mbox{$e$}}
\Vertex(75,10){2}
\Photon(75,10)(140,10){3}{4}
\Text(145,5)[l]{\mbox{$Z$}}
\DashLine(85,70)(140,70){5}
\Text(145,75)[l]{\mbox{$h$}}
\Text(75,-10)[c]{\mbox{\bf l)}}
\end{picture}
\\
\\
\end{tabular}
\caption{Generic one-loop diagrams contributing to the $e^+e^-\to Zh$ 
cross section.} 
\label{fig:diagrams}
\end{center}
\end{figure}


For completeness, in \reffi{fig:diagrams} also contributions are
shown that are proportional to 
the electron mass or vanish completely after contraction with the
polarization vector of the $Z$~boson (e.g.\ the $A,G$--$Z$ mixing
contributions and the longitudinal 
parts of the $Z$ and $\ga$--$Z$ self-energies).
The different types of corrections can be summarized as follows:
\begin{itemize}
\item[(i)] Corrections to the $e$, $Z$, $\ga$ and $\ga$--$Z$
  self-energies on the internal and external lines and to the (initial
  state) $Ze^+e^-$ and $\ga e^+e^-$ vertices, b) -- g).
\item[(ii)] Corrections to the scalar and pseudoscalar propagators, h).
\item[(iii)] Corrections to the $ZZH_i$ ($ZAH_i$) vertex, i).
\item[(iv)] Box-diagram contributions and $t$-channel-exchange diagrams,
j) -- l).
\end{itemize}

\noindent The corrections (i)-(iv) have a 
          different relative impact:
\begin{itemize}
\item[-] Electroweak corrections of type (i) are typically of the order
  of a few percent (like in the Standard Model) and do not exhibit a
  strong dependence on any SUSY parameters.
\item[-] The main source of differences between the tree-level  
  and higher-order results are the corrections to the Higgs-boson
  self-energies (ii). They are responsible for changes in the physical
  masses $\Mh$ and $\MH$ and the effective mixing angle $\aeff$
  (via contributions to the renormalization constants, $\Zext$, for the
  external Higgs particles in the $S$-matrix elements, see
  \citere{eehZhA})
  predicted for given values of $\tb$ and
  $\MA$.  At the \onel\ level these propagator corrections
  constitute the only source for the
  large correction of \order{\alt}. At the \twol\ level they
  exclusively give rise to contributions of \order{\alt\als} and
  of \order{\alt^2}. In this sense the propagator corrections
  define a closed subset of diagrams, being responsible for a
  numerically large contribution.  
\item[-] Corrections to the final-state vertices (iii) are
  typically larger than those of type (i), but smaller than the
  Higgs-boson propagator corrections. At LEP2 energies they can reach at
  most 7--10\%~\cite{hprod2} for very low or very large values of $\tb$,
  when the Yukawa couplings of the top or bottom quarks become strong.
\item[-] Finally, the box-diagram contributions (iv) depend strongly 
  on the center-of-mass energy. They are of the order of 2--3\% at LEP2
  energies and may reach 20\% for $\sqrt{s}=500$ GeV~\cite{hprodf1l}.
\end{itemize}
It should be noted that initial-state QED corrections as well as
finite-width effects (allowing for off-shell decays of the Higgs and
the $Z$~boson) are not included
in our calculation. However, by incorporating our result into existing
codes, e.g.\ HZHA~\cite{hzha}, QED corrections and finite-width effects
can be taken into account.

\bigskip
In \citere{eehZhA} it has been shown analytically that the leading
Higgs boson propagator corrections can be incorporated by changing the
tree-level $\al$ in the Higgs--gauge boson couplings to $\aeff$, see
also \refse{subsec:hffaeff}. For instance, 
for $Zh$ production, the Born coupling $\tilde V_{ZZh} \sim \Samb$ 
is changed by the leading propagator corrections to 
$\sin(\aeff - \be)$. 
Analogous results hold for all Higgs vertices, including the $A\Hi$
vertices.  
While the $\aeff$ approximation, i.e.\ using an improved Born result
for the cross sections
where the tree-level angle $\alpha$ is replaced by $\aeff$,
incorporates the dominant one-loop and two-loop contributions,
it is obvious from the discussion above that this approximation
neglects many effects included in a full FD calculation.
These are, in particular, the
process-specific vertex and box corrections.


\subsubsection{Cross sections}
\label{subsec:crosssection}

In this subsection we briefly summarize the analytical formulae
for the cross 
sections for the on-shell production of the Higgs bosons $e^+e^-\to
Z\Hi$, $e^+e^-\to A\Hi$ including the corrections (i)-(iii). Box
diagrams (iv) give another, more complicated, set of formfactors that
make the expressions quite lengthy and are, hence, omitted here;
more details can be found in~\citere{hprodboxes}.  However, we include the
box-diagram contributions, as described in~\citere{hprodf1l}, in our
numerical programs~\cite{eehZhA} and in the figures shown in
this section.

The presented formalism for cross sections is general enough to
accommodate corrections of any order to 2- and 3-point vertex
functions.  Beyond the one-loop level, however, currently only \twol\
corrections to the scalar propagators have been
included~\cite{mhiggsletter,mhiggslong}.  Therefore, in the cross
section calculations we include all possible types of \onel\
corrections and the available \twol\ corrections to scalar
self-energies.  This is well justified because, as discussed above,
propagator corrections constitute a closed subset of the leading
\order{\alt\als} and \order{\alt^2} contributions. Therefore,
these \twol\ corrections are of particular relevance and interest.

\bigskip
The cross sections (in the center of mass system (CMS)) for both
processes (\ref{eetohZ}) and (\ref{eetohA}) have the form:
\BEA
{d\sigma_{Z(A)H_i}\over d\Omega}=
{\lambda \KL s,M^2_{Z(A)},M^2_{\Hi} \KR \over 64\pi^2
s^2\left|D_Z(s)\right|^2}  \KL {\cal A}_1+{\cal
A}_2\cos^2\theta_{\rm CMS} \KR ,
\EEA
where $\lambda$ is the standard phase space factor,
\BEA
\lambda \KL s,m^2_1,m^2_2 \KR =
\sqrt{s^2+m_1^4+m_2^4-2sm_1^2-2sm_2^2-2m^2_1m^2_2} ,
\label{def:lambda}
\EEA
and ${\cal A}_1,{\cal A}_2$ are defined by 
\BEA
\label{eq:cr_sum_mat}
{\cal A}_1+{\cal A}_2\cos^2\theta_{\rm CMS} ={1\over
4}\sum_{pol} \KL {\cal M}{\cal M}^{*} \KR ,
\EEA
with ${\cal M}_{ZS}$ and ${\cal M}_{PS}$ as given below.
In the following, $\theta_{\rm CMS}$ denotes the scattering angle
$\theta_{\rm CMS} = < \hspace{-.45em} )\,(e^-, H_i)$ in the CMS.
The momenta of the incoming electron and positron are
denoted as $k_1$ and $k_2$, respectively. The momentum of the outgoing
$h/H$ is labeled with $q$, whereas the outgoing $Z/A$ momentum is
denoted as $p$, see \reffi{fig:diagrams}a.  The matrix elements for the
Higgs-strahlung process and the associated Higgs production read 
%
\BEA
{\cal M}_{ZS}^i&=&e\overline v(k_2)\ga^{\nu} \KKL \tilde V^{\mu\nu
i}_{ZZS} \KL \hat c_V-\hat c_A\ga^5 \KR + {\tilde V}^{\mu\nu i}_{\ga
ZS} {D_Z(s)\over D_{\ga}(s)} - \tilde V^{(0)\mu\nu i}_{ZZS}
{\hat\Sigma^T_{\ga Z}(s)\over D_{\ga}(s)} \KKR u(k_1)\epsilon_{\mu}(p) \non\\
&& + {\rm ~box~corrections} , \\
&& \non \\
{\cal M}_{PS}^{i}&=&e\overline v(k_2)\ga_{\mu} \KKL \tilde V^{\mu
ij}_{ZPS} \KL \hat c_V-\hat c_A\ga^5 \KR + {\tilde V}^{\mu ij}_{\ga
PS}{D_Z(s)\over D_{\ga}(s)} - \tilde V^{(0)\mu ij}_{ZPS}
{\hat\Sigma^T_{\ga Z}(s)\over D_{\ga}(s)} \KKR u(k_1) \non\\
&& + {\rm ~box~corrections} .
\EEA
For the corresponding expressions for the box contributions see
\citere{hprodboxes}. 

In the above expressions $u(k_1)$ and $v(k_2)$ are spinors of the
incoming electron-positron pair, $\epsilon_{\mu}(p)$ is the
polarization vector of the outgoing $Z$. $\hat c_V, \hat c_A$ are the
renormalized vector and axial couplings of the $Z$ boson to an
electron-positron pair, at the \onel\ level $\hat c_A=-1/4s_Wc_W+ \hat
c_A^{(1)}$, $\hat c_V= \KL -1+4s^2_W \KR /4s_Wc_W+ \hat
c_V^{(1)}$, $\cw^2 \equiv 1 - \sw^2 \equiv \MW^2/\MZ^2$.
$\hat\Sigma^T_Z(s)$, $\hat\Sigma_{\ga}(s)$ and
$\hat\Sigma^T_{\ga Z}(s)$ denote the renormalized photon and transverse
$Z$~boson self-energies.  $D_Z(s)$ and $D_{\ga}(s)$ are the inverse
$Z$ and photon propagators defined as
\BEA
D_Z(s) = s - M_Z^2 + \hat\Sigma^T_Z(s)~, \nonumber\\
D_{\ga}(s) = s + \hSi_\ga(s)~.
\EEA
Finally, $\tilde V$ denotes the effective neutral Higgs--gauge-boson
vertices with the one-loop form factors.  The explicit expression for
those vertices and for the matrix elements for Higgs-strahlung and
associated Higgs production can be found in the Appendix of
\citere{eehZhA} and in \citere{hprod2}.


\subsection{Numerical results}
\label{sec:num}


In the following we present numerical examples for the dependence of
the neutral Higgs-boson couplings and cross sections on $\tb$, $\Mh$,
and the mixing in the scalar top sector.  
The results are obtained in the $\mhmax$ and the no-mixing
scenario~\cite{benchmark}, see the Appendix. Only for the scalar soft
SUSY-breaking parameters we have set the slepton mass parameters to
$M_{\tilde l} = 300 \gev$. The squark mass parameter is denoted as
$M_{\tilde q}$. 

Below we will also perform comparisons with
results obtained in the framework of the RG improved one-loop
EPA, where the
input parameters are understood as \msbar\ quantities.
To ensure consistency, in the latter case we have transformed
the on-shell SUSY input parameters into the corresponding
\msbar\ values as discussed in \citere{bse}.  
The results shown below for the higher-order corrected Higgs-boson 
masses and the mixing angle within the RG improved one-loop EPA have been 
obtained with the Fortran program {\em subhpole} 
(based on \citeres{mhiggsRG1,bse}).



\bigskip
Differences in the Higgs-production cross sections 
between our FD result (containing the complete one-loop result and the
dominant two-loop corrections) and the RG $\aeff$ approximation
have a two-fold origin: the different predictions for
the values of $\Mh$ and $\aeff$, and the additional contributions
contained in the FD result 
(i.e.\ the one-loop 3- and 4-point vertex functions.

\begin{figure}[htb!]
\begin{center}
\begin{tabular}{p{0.48\linewidth}p{0.48\linewidth}}
\mbox{
\epsfig{file=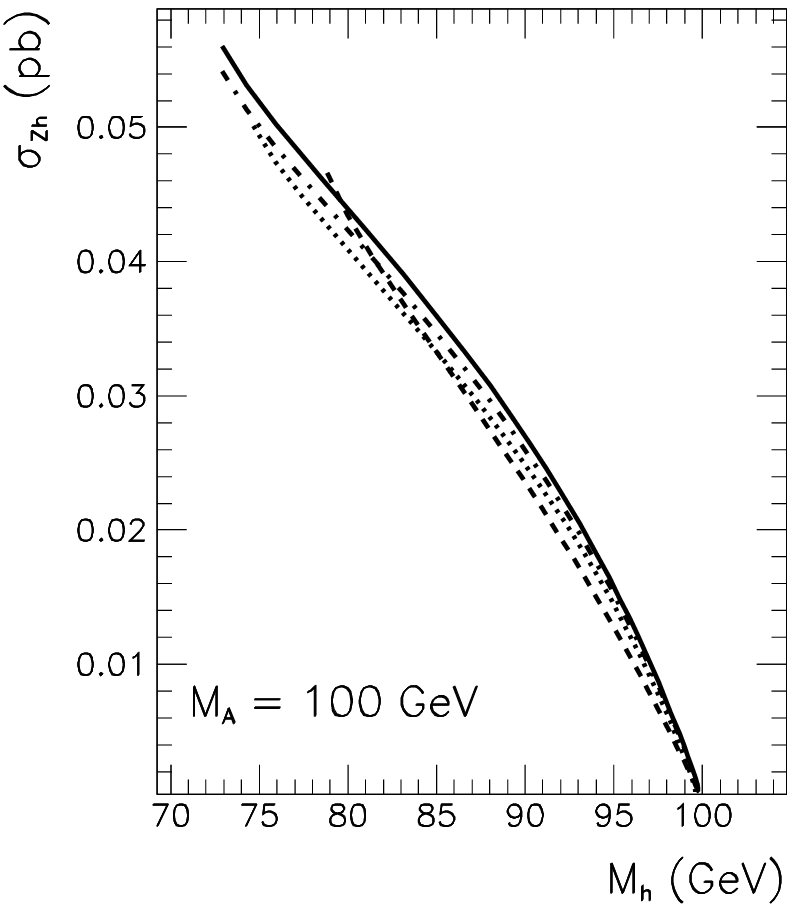,width=\linewidth,height=0.9\linewidth}}&
\mbox{
\epsfig{file=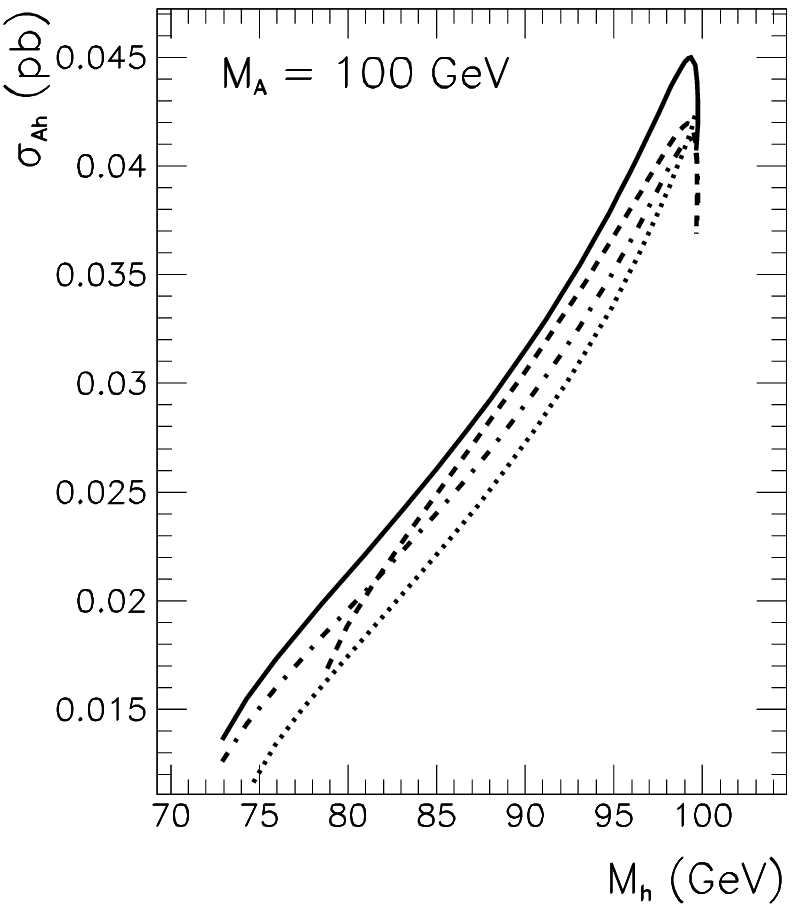,width=\linewidth,height=0.9\linewidth}}\\
\mbox{
\epsfig{file=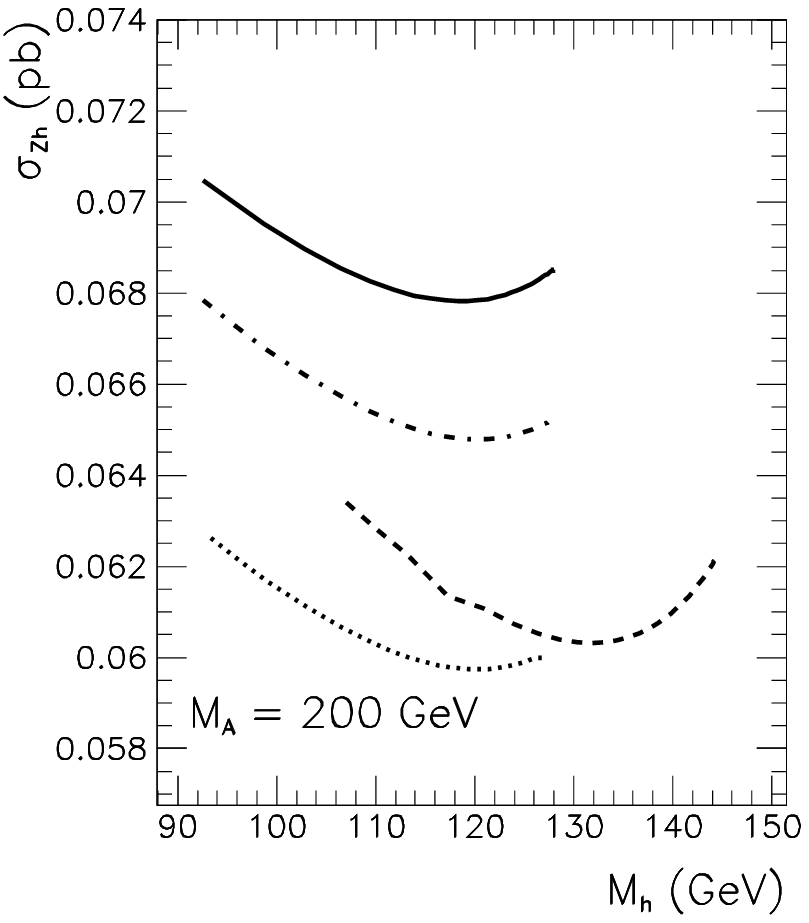,width=\linewidth,height=0.9\linewidth}}&
\mbox{
\epsfig{file=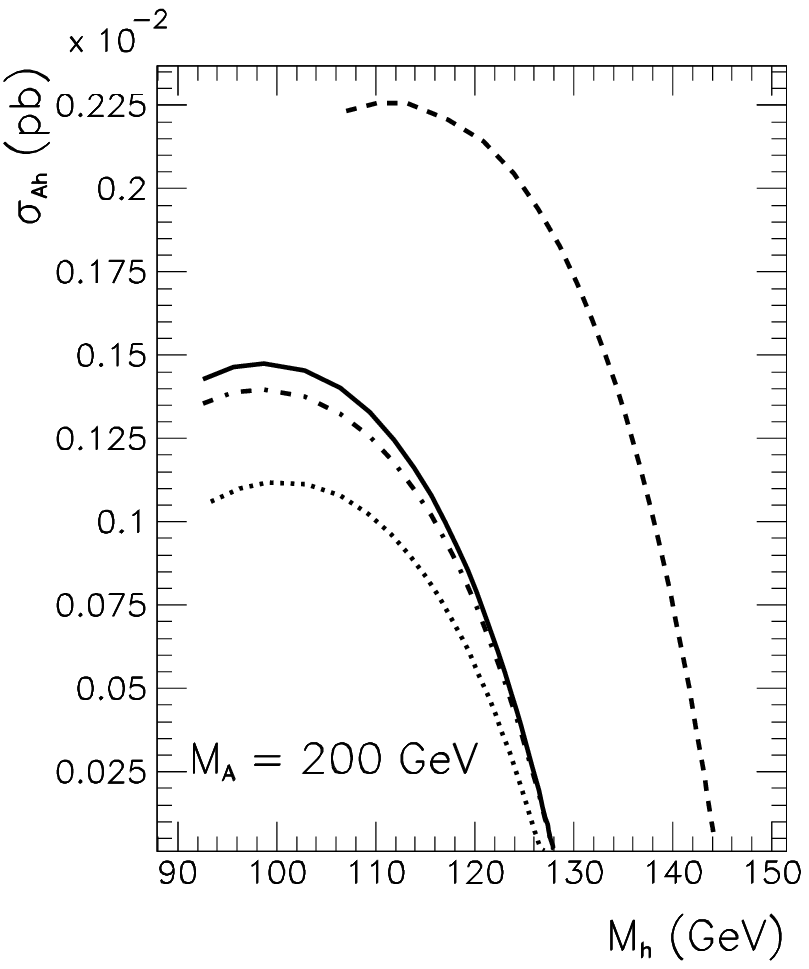,width=\linewidth,height=0.9\linewidth}}\\
\end{tabular}
\vskip -5mm 
\caption{$\sigma_{Zh}$ and $\sigma_{Ah}$ as a function of
  $\Mh$ at $\sqrt{s} = 500 \gev$, for two values of $\MA$, in the
  $\mhmax$ scenario. The solid (dot-dashed)
  line represents the \twol\ FD result including (excluding) box
  contributions, the dotted line shows the RG $\aeff$ approximation and the
  dashed line shows the \onel\ FD result.}
\label{fig:sigma_mh_s500}
\end{center}
\end{figure}

\begin{figure}[htb!]
\begin{center}
\begin{tabular}{p{0.48\linewidth}p{0.48\linewidth}}
\mbox{\epsfig{file=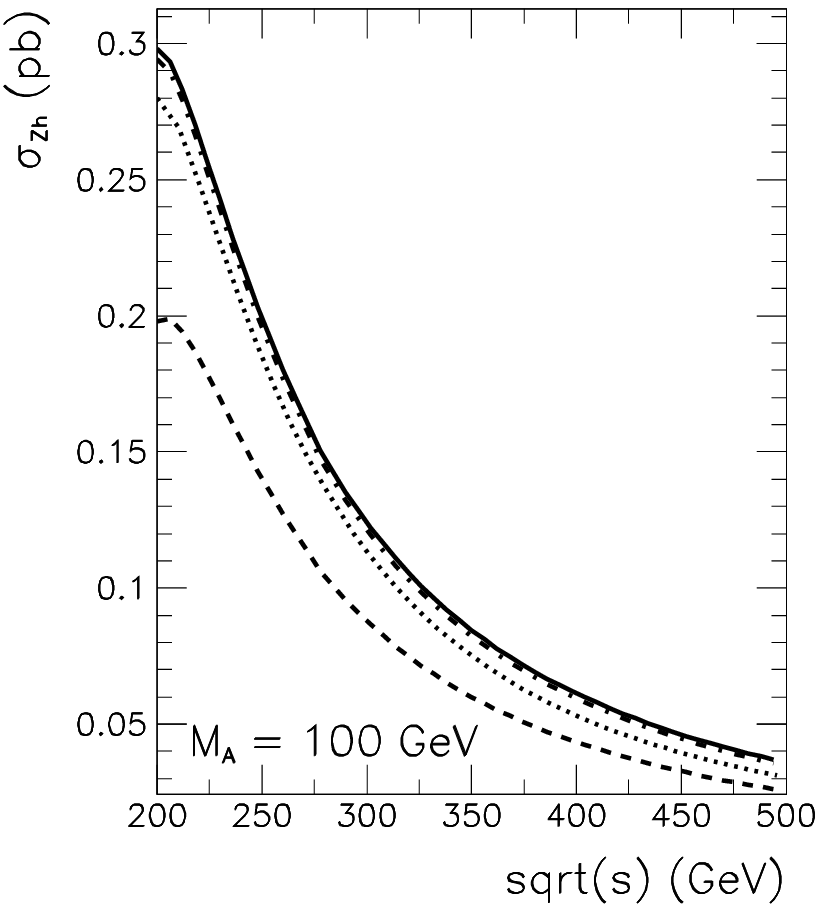,width=\linewidth,height=0.95\linewidth}}&
\mbox{\epsfig{file=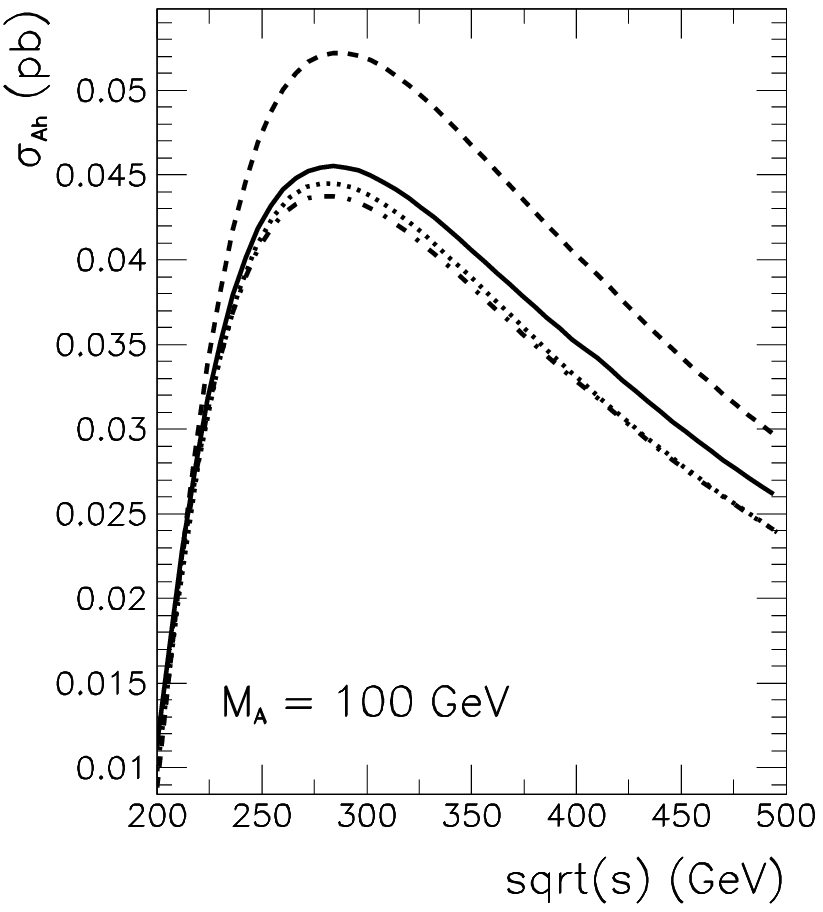,width=\linewidth,height=0.95\linewidth}}\\
\mbox{\epsfig{file=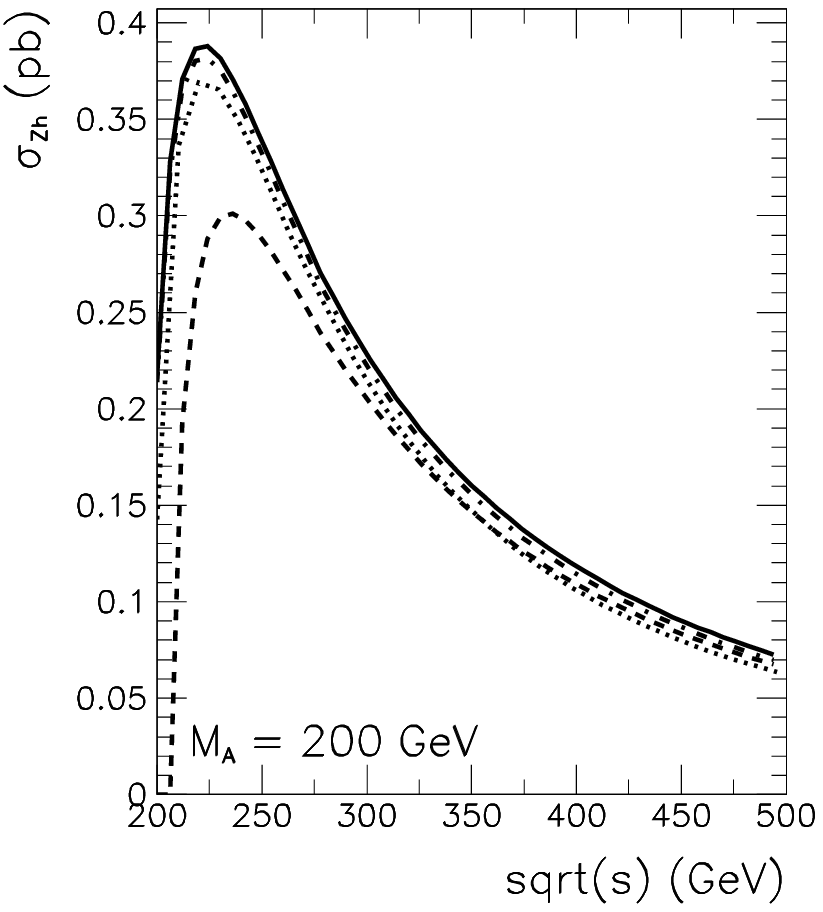,width=\linewidth,height=0.95\linewidth}}&
\mbox{\epsfig{file=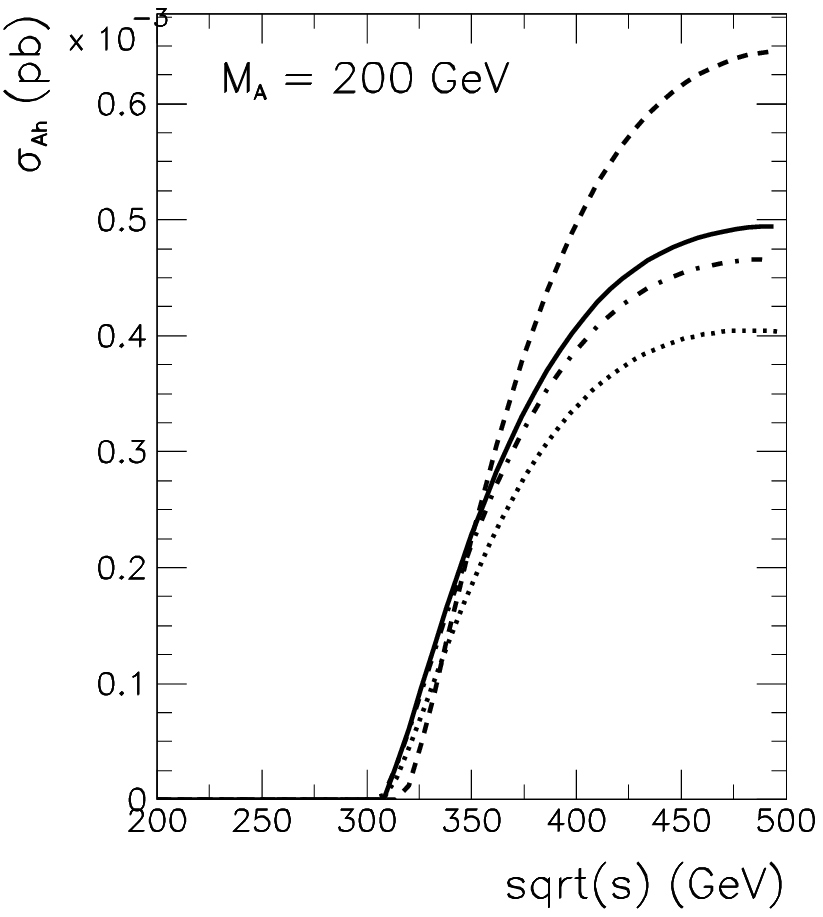,width=\linewidth,height=0.95\linewidth}}\\
\end{tabular}
\vskip -5mm 
\caption{$\sigma_{Zh}$ and $\sigma_{Ah}$ as a function of
  $\sqrt{s}$ for $\tb = 5$, in the no-mixing scenario.
  The solid (dot-dashed) line represents the
  \twol\ FD result including (excluding) box contributions, the dotted
  line shows the RG $\aeff$ approximation and the dashed line shows the
  \onel\ FD result.}
\label{fig:sigma_s_tb5}
\end{center}
\end{figure}

\begin{figure}[htb!]
\begin{center}
\begin{tabular}{p{0.48\linewidth}p{0.48\linewidth}}
\mbox{\epsfig{file=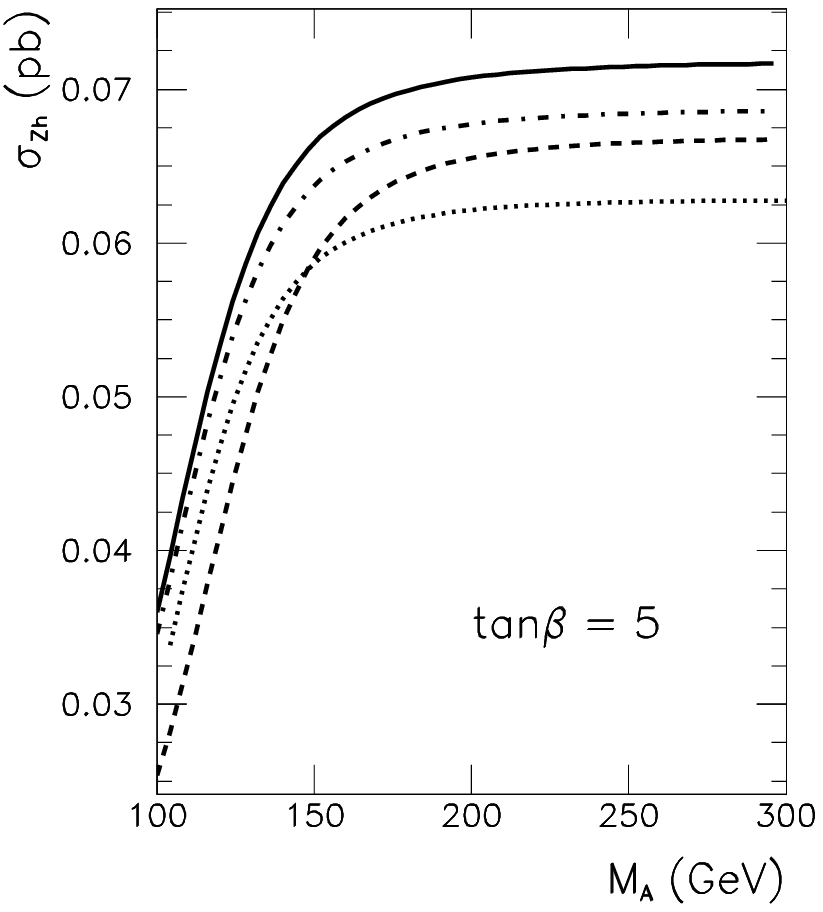,width=\linewidth,height=0.9\linewidth}}&
\mbox{\epsfig{file=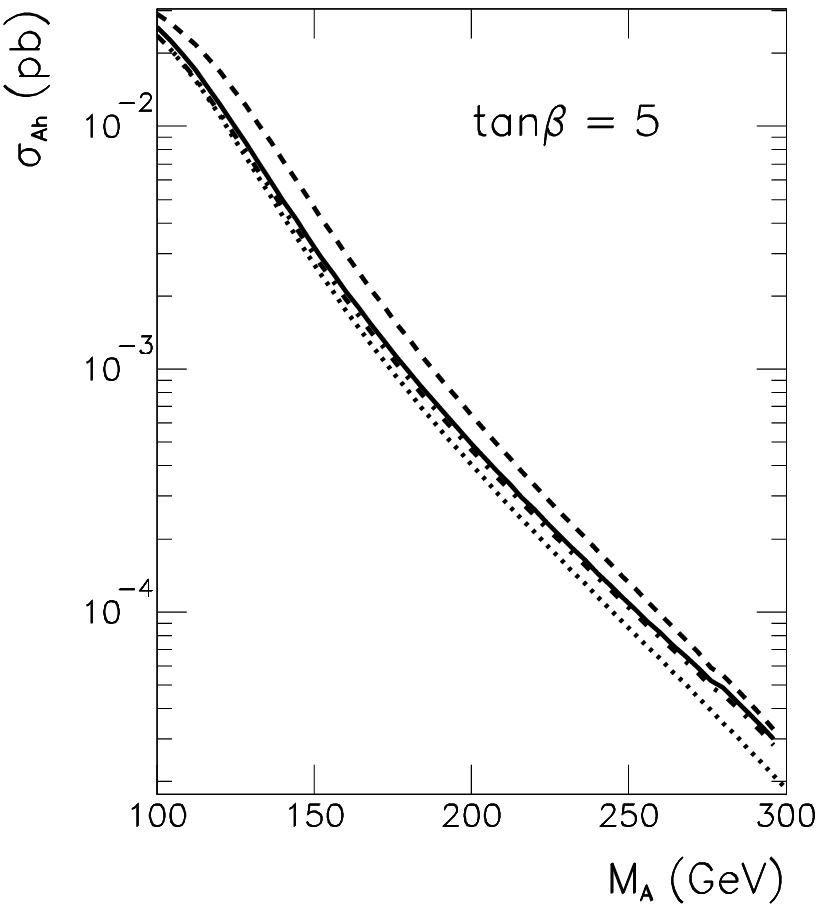,width=\linewidth,height=0.9\linewidth}}\\
\mbox{\epsfig{file=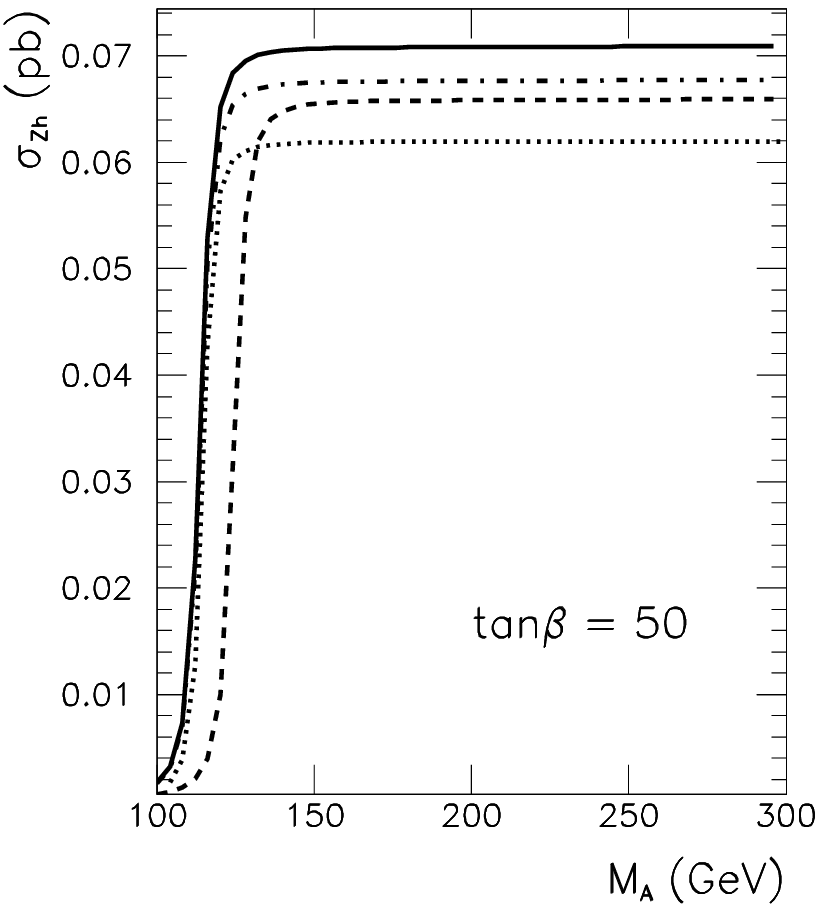,width=\linewidth,height=0.9\linewidth}}&
\mbox{\epsfig{file=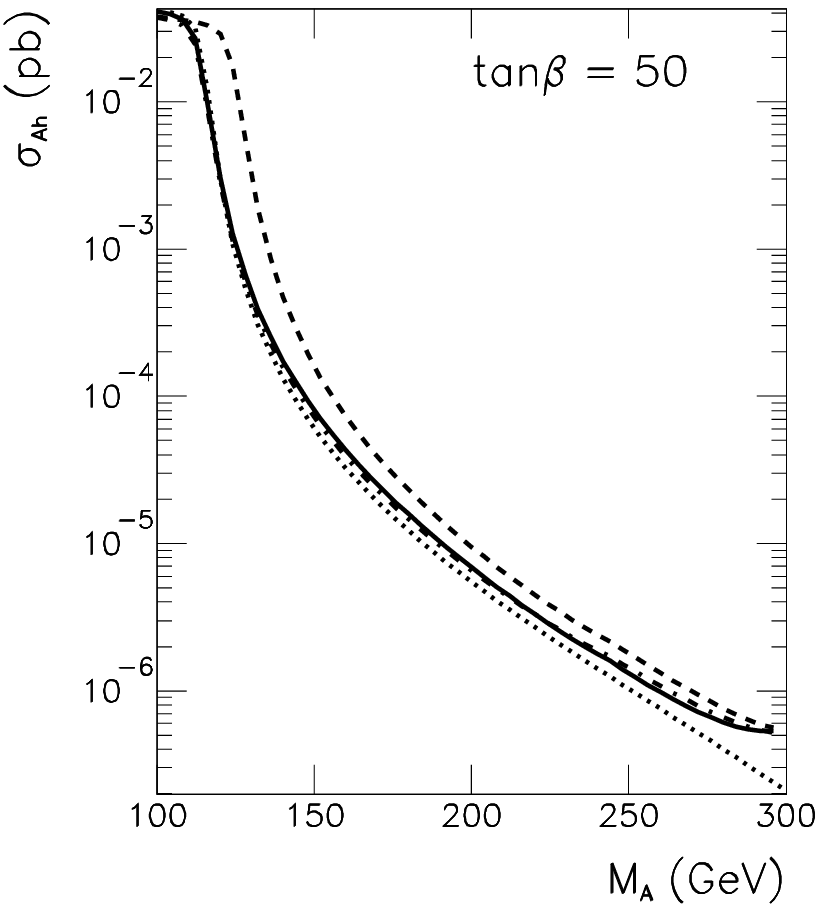,width=\linewidth,height=0.9\linewidth}}\\
\end{tabular}
\vskip -5mm 
\caption{$\sigma_{Zh}$ and $\sigma_{Ah}$ as a function of
  $\MA$ at $\sqrt{s} = 500 \gev$, shown for $\tb = 5$ and $\tb = 50$,
  in the no-mixing scenario.  The solid
  (dot-dashed) line represents the \twol\ FD result including
  (excluding) box contributions, the dotted line shows the RG $\aeff$
  approximation and the dashed line shows the \onel\ FD result.}
\label{fig:sigma_MA}
\end{center}
\end{figure}

In \reffi{fig:sigma_mh_s500} the cross sections for the
Higgs-strahlung process and the associated production are shown for a
typical Linear Collider energy, $\sqrt{s}=500$ GeV~\cite{teslatdr}, in
the $\mhmax$ scenario (in the no-mixing scenario similar results
have been obtained). We also show
the result for the \twol\ FD calculation where the box contributions
have not been included, in order to point out their relative
importance for high-energy collisions.  For $\sqrt{s} = 500 \gev$ the
differences between the FD result and the RG $\aeff$ approximation
can be large, 
an effect that is more pronounced for the higher value of
$\MA$.  For $\MA = 200 \gev$, typically they are of the order of
10--15\% for $\sigma_{Zh}$ and even up to 25\% for $\sigma_{Ah}$ (for
$\Mh \gsim 90 \gev$).
The difference between the \twol\ and \onel\ FD result can be
sizable. The \twol\ result for $\si_{Zh}$ is in general larger than
the \onel\ value, again increasing with $\MA$, where for $\MA =
200 \gev$ the difference can amount up to 15\%. $\si_{Ah}$, on the 
other hand, is
decreased at the \twol\ level for $\MA = 200 \gev$ and the difference
may be sizable.

The box contributions become more important for higher $\sqrt{s}$ and
change the total cross section by 5--10\%.  This result remains
unchanged even if sleptons are significantly heavier than
$M_{\tilde{l}}=300$ GeV used in our numerical analysis, as the
dominant contributions to box diagrams are given by $W$ and Higgs
boson exchanges~\cite{hprodboxes,hprodf1l}, which do not depend on
$M_{\tilde{l}}$.  Also, one should recall that box contributions
lead to an angular distribution of the final-state particles
different from the effective Born approximation and thus give much larger
corrections to the differential rather than to the total cross
section, at least for some range of the scattering angle.  Therefore,
the box diagrams have a significant effect at Linear Collider energies
and thus have to be included.  The same conclusions can be drawn for
$\sqrt{s} = 500 \gev$ in the no-mixing scenario, which we do not show
here. The differences between the FD result and the RG $\aeff$
approximation are only slightly smaller than in the $\mhmax$ case.

In \reffi{fig:sigma_s_tb5} the results for $\sigma_{Zh}$ and
$\sigma_{Ah}$ are shown as a function of $\sqrt{s}$ in the no-mixing
scenario for $\MA = 100, 200 \gev$.  Besides the obvious kinematical
drop-off of the cross sections, one can observe that the relative
differences between the FD two-loop result and the RG $\aeff$
approximation  grow with~$\sqrt{s}$. The differences
remain almost constant or even increase slowly in absolute terms,
whereas the full cross sections decrease.  $\sigma_{Ah}$ becomes very
small for large $\MA$, as can be seen in more detail in
\reffi{fig:sigma_MA}. There we show the dependence of
$\sigma_{Zh}$ and $\sigma_{Ah}$ on $\MA$ in the no-mixing scenario for
$\tb = 5$ and $\tb=50$.  For $\sigma_{Zh}$, the $A$ boson decouples
quickly; the dependence on $M_A$ becomes very weak for
$\MA \gsim 250 \gev$, when $\sigma_{Zh}$ is already practically
constant (compare e.g.~\citere{mhiggslong}).  In the same limit,
$\sigma_{Ah}$ goes quickly to zero due to suppression of the effective
$ZhA$ coupling, which is $\sim\cos(\aeff - \be)$; 
also the kinematical suppression
plays a role, but this becomes significant only for sufficiently large
$\MA$, $\MA>350 \gev$.  For large $\tb$ the decoupling of $M_A$ is
even more rapid. The differences between the FD two-loop result and
the RG $\aeff$ approximation
for the Higgs-strahlung cross section tend also to a constant, but
they increase with $\MA$ for the associated production. The latter can
be explained by the growing relative importance of 3- and 4-point
vertex function contributions compared to the strongly suppressed
Born-like diagrams. As can be seen from~\reffi{fig:sigma_MA}, for
$\tb=50$ and $\MA\geq 300 \gev$ the FD two-loop result is almost an order of
magnitude larger than the result of the RG $\aeff$ approximation, and 
starts to saturate. This can be
attributed to the fact that the (non-decoupling) vertex and box
contributions begin to dominate the cross section value. However, such
a situation occurs only for very small $\sigma_{Ah}$ values,
$\sigma_{Ah} \approx 10^{-3}$~fb, below the expected experimental LC
sensitivities.


\section{Corrections to the $WW$ fusion channel}

While the discovery of one light Higgs boson 
might well be compatible with the predictions both of the SM and the MSSM, the
discovery of one or more other heavy Higgs bosons would be a clear and
unambiguous signal for physics beyond the SM.

In the decoupling limit, i.e.\ for $\MA \gsim 200 \gev$, the heavy MSSM
Higgs bosons are nearly degenerate in mass, $\MA \sim \MH \sim \MHp$.  The
couplings of the neutral Higgs bosons to SM gauge bosons are
proportional to
\BEA
VVh &\sim& VHA \sim \Sba\,, \\
VVH &\sim& VhA \sim \Cba\,,
\qquad (V = Z, W^\pm)~.
\EEA
In the decoupling limit
one finds $\be - \al \to \pi/2$, i.e.\ $\Sba \to 1$, $\Cba \to 0$.

At the LC, the possible channels for neutral Higgs-boson production are
the production via $Z$-boson exchange, \refeq{eehZhA}, and the
$WW$~fusion channel, \refeq{eennH}. 
As a consequence of the coupling structure, in the decoupling limit
the heavy Higgs boson can only be produced in $(H,A)$~pairs. This
limits the LC reach to $\MH \lsim \sqrt{s}/2$. Higher-order
corrections to the $\WWH$ channel from loops of 
fermions and sfermions, however, involve potentially large
contributions from the top and bottom Yukawa couplings and could thus
significantly affect the decoupling behavior. 

In this section we review the one-loop corrections of fermions and sfermions
to the process \eennhH, i.e.\ to the production of a neutral
$\cp$-even Higgs boson in association with a neutrino pair, both via
the $WW$-fusion and the Higgs-strahlung 
mechanism. In the latter case the $Z$~boson is connected to a neutrino
pair, $e^+e^- \to Z \{h,H\} \to \bar\nu_l\nu_l \{h,H\}$, 
with $l = e, \mu, \tau$ (where the latter two neutrinos result in an
indistinguishable final state in the detector). 

While the well-known universal Higgs-boson propagator corrections
turned out not to significantly modify the decoupling behavior of the
heavy $\cp$-even Higgs boson, an analysis of the process-specific
contributions to the $WWH$ vertex has been obtained recently.
Taking into account all loop and counter-term contributions to
the process \eennhH\ from fermions and sfermions and including also
the effects of beam polarization in our analysis, we review in
this section the LC reach for the 
heavy $\cp$-even Higgs boson. We have obtained results for values of the MSSM
parameters according to the four benchmark scenarios defined in
\citere{benchmark}. While within these benchmark scenarios we find
that the loop corrections do not significantly enhance the LC reach
for heavy $\cp$-even Higgs boson production and in some cases even
slightly reduce the accessible parameter space, we have also
investigated MSSM parameter regions where the loop effects do in 
fact lead to a significant improvement of the LC reach. In
``favorable'' MSSM parameter regions an $e^+e^-$ LC running at
$\sqrt{s} = 1 \tev$ can be capable of producing a heavy $\cp$-even
Higgs boson with a mass up to $\MH \lsim 700 \gev$. 

Concerning the production of the light $\cp$-even Higgs boson, an
accurate prediction of the production cross section for precision
analyses will be necessary. Aiming for analyses at the percent
level~\cite{eennHexp} also requires a prediction of the production cross
section in this range of precision. Besides the already known universal
Higgs propagator corrections, in particular loops from fermions and
sfermions (especially from the third family) are expected to give
relevant contributions. We analyze our results for the parameters of the
four benchmark scenarios defined in \citere{benchmark} and study the
results as a function of different SUSY parameters. We discuss the relative
importance of the fermion- and the sfermion-loop contributions and
furthermore evaluate the fermion-loop correction within the SM for
comparison purposes (the complete \order{\al} corrections within the
SM can be found in \citere{eennHSMf1l}).


\subsection{Renormalization in the Higgs sector}

The renormalized Higgs-boson self-energies, see
\refeqs{renSEhh}-(\ref{renSEHH}), are as usual given by
\BEA
\hSi_{HH}(q^2)
&=& \Si_{HH}(q^2) + \de Z_{H} (q^2 - \mH^2) - \de \mH^2\,, \non \\
\hSi_{hH}(q^2)
&=& \Si_{hH}(q^2) + \edz \de Z_{Hh} (q^2 - \mH^2)
                  + \edz \de Z_{hH} (q^2 - \mh^2)
                  - \de m_{hH}^2\,, \non \\
\hSi_{hh}(q^2)
&=& \Si_{hh}(q^2) + \de Z_{h} (q^2 - \mh^2)
                  - \de \mh^2\,.
\label{eq:renSihiggs}
\EEA
The mass counter terms arise from the renormalization of the Higgs
potential, see~\citere{mhiggsrenorm}. They are evaluated in the on-shell
renormalization scheme.  The field-renormalization constants, see also
\refeq{eq:deltaZHiggsTB} can be
obtained in the \drbar\ scheme, %
%
leading to
\BEA
\de Z_{H} &=& -\KKL \re\Si_{HH}'(\mH^2) \KKR^{\rm div}, \non \\
\de Z_{h} &=& -\KKL \re\Si_{hh}'(\mh^2) \KKR^{\rm div} , \non \\
\de Z_{hH} &=& \frac{\Sa\Ca}{\CZa} (\de Z_{h} - \de Z_{H})\,, \non \\
\de Z_{Hh} &=& \de Z_{hH}\, ,
\label{eq:deltaZHiggs}
\EEA
i.e.\ only the divergent parts of the renormalization constants in
\refeqs{eq:deltaZHiggs} are taken into account.  As renormalization
scale we have chosen $\mu_{\drbarm} = \mt$.
The finite LSZ factors required in this renormalization scheme are
described in the next subsection. 


\subsection{The process \eenenehH}
\label{sec:WWhH}

\subsubsection{The tree-level process}
\label{subsec:tree}

\begin{figure}[ht!]
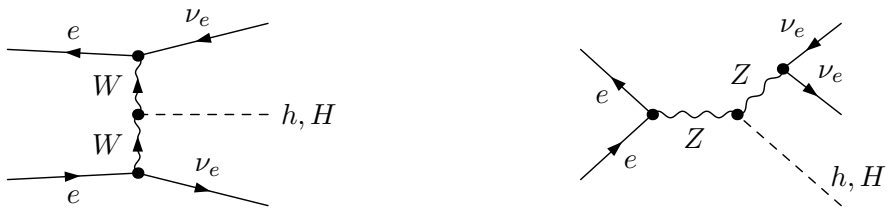

\begin{center}
\begin{small}
\unitlength=1.5bp%
\begin{feynartspicture}(216,72)(3,1)
\FADiagram{}
\FAProp(0.,15.)(10.,14.5)(0.,){/Straight}{-1}
\FALabel(5.0774,15.8181)[b]{$e$}
\FAProp(0.,5.)(10.,5.5)(0.,){/Straight}{1}
\FALabel(5.0774,4.18193)[t]{$e$}
\FAProp(20.,17.)(10.,14.5)(0.,){/Straight}{1}
\FALabel(14.6241,16.7737)[b]{$\nu_e$}
\FAProp(20.,10.)(10.,10.)(0.,){/ScalarDash}{0}
\FALabel(21.,8.82)[lb]{$h, H$}
\FAProp(20.,3.)(10.,5.5)(0.,){/Straight}{-1}
\FALabel(15.3759,5.27372)[b]{$\nu_e$}
\FAProp(10.,14.5)(10.,10.)(0.,){/Sine}{-1}
\FALabel(8.93,12.25)[r]{$W$}
\FAProp(10.,5.5)(10.,10.)(0.,){/Sine}{1}
\FALabel(8.93,7.75)[r]{$W$}
\FAVert(10.,14.5){0}
\FAVert(10.,5.5){0}
\FAVert(10.,10.){0}

\FADiagram{}

\FADiagram{}
\FAProp(0.,15.)(5.5,10.)(0.,){/Straight}{-1}
\FALabel(2.18736,11.8331)[tr]{$e$}
\FAProp(0.,5.)(5.5,10.)(0.,){/Straight}{1}
\FALabel(3.31264,6.83309)[tl]{$e$}
\FAProp(20.,17.)(15.5,13.5)(0.,){/Straight}{1}
\FALabel(17.2784,15.9935)[br]{$\nu_e$}
\FAProp(20.,10.)(15.5,13.5)(0.,){/Straight}{-1}
\FALabel(18.2216,12.4935)[bl]{$\nu_e$}
\FAProp(20.,3.)(12.,10.)(0.,){/ScalarDash}{0}
\FALabel(23.5,6.00165)[tr]{$h,H$}
\FAProp(5.5,10.)(12.,10.)(0.,){/Sine}{0}
\FALabel(8.75,8.93)[t]{$Z$}
\FAProp(15.5,13.5)(12.,10.)(0.,){/Sine}{0}
\FALabel(13.134,12.366)[br]{$Z$}
\FAVert(5.5,10.){0}
\FAVert(15.5,13.5){0}
\FAVert(12.,10.){0}
\end{feynartspicture}
\end{small}
\caption{%
The tree-level diagrams for the process \eennhH, consisting of the
$WW$-fusion contribution (left) and the Higgs-strahlung contribution
(right). 
}
\label{fig:tree}
\end{center}
\end{figure}

The tree-level process~\cite{eennHtree0,hprod}
consists of the two diagrams shown in
\reffi{fig:tree}. Besides the $WW$-fusion contribution (left diagram),
we also take into account the Higgs-strahlung contribution (right
diagram), where a virtual $Z$~boson is connected to two 
electron neutrinos.

An analytical expression for the tree-level cross section for a SM
Higgs boson can be found e.g.\ in \citere{eennHtree2}. 
For relatively low energies and moderate values of the SM Higgs-boson mass 
($\sqrt{s} \lsim 400 \gev$, $\MHSM \lsim 200 \gev$) the resonant
production via the Higgs-strahlung contribution dominates over the
$WW$-fusion contribution. 
At higher energies,
however, the $WW$-fusion contribution becomes dominant.
The cross section, containing both contributions, in the
high-energy limit takes the simple form~\cite{hprod}
\begin{equation}
\si(e^+ e^- \to \bar\nu_e\nu_e H_{\rm SM})\to
\frac{G_F^3 \MW^4}{4 \wz \, \pi^3} 
 \KKL \KL 1 + \frac{\MHSM^2}{s} \KR \log\KL\frac{s}{\MHSM^2}\KR
     - 2 \KL 1 - \frac{\MHSM^2}{s} \KR \KKR ,
\end{equation}
where the $t$-channel contribution from the $WW$-fusion diagram
gives rise to the logarithmic increase.

The coefficients for the couplings 
$WWh$ and $WWH$ are denoted by $\Ghn$ and $\GHn$ at the tree level,
respectively (and analogously for the $ZZh$ and $ZZH$ couplings): 
\BEA
\label{eq:WWhtree}
\Ghn &=& \frac{\ri\,e\,\MW}{\sw} \Sba\,, \\
\label{eq:WWHtree}
\GHn &=& \frac{\ri\,e\,\MW}{\sw} \Cba\,.
\EEA
The SM coupling $\Ga^{(0)}_{H_{\rm SM}}$ is obtained by dropping the
SUSY factors $\Sba$ or $\Cba$.  In the decoupling limit, $\MA\gsim 200
\gev$, $\be - \al \to \pi/2$, so that $\Sba \to 1$ and $\Cba \to 0$, i.e.\
the heavy neutral $\cp$-even Higgs boson decouples from the $W$~and
$Z$~bosons. 

We parametrize the Born matrix element by the Fermi constant, $\gf$,
i.e.\ we use the relation 
\begin{equation}
e = 2 \; \sw \; \MW \KKL \frac{\wz \, \gf}
                              {1 + \De r} \KKR^{1/2} ,
\label{gmueparam}
\end{equation}
where $\De r$ incorporates higher-order corrections, see
the next subsection.


\subsubsection{Higher-order corrections}
\label{subsec:higherorder}

In the description of our calculation below we will mainly concentrate 
on the $WW$-fusion contribution.  
The Higgs-strahlung contribution, which we describe in less 
detail, is taken into account in exactly the same way (with the only
exception that a finite $Z$~width has to be taken into account).

We evaluate the \onel\ \order{\al} contributions from loops involving
all fermions and sfermions.
Especially the corrections involving third-generation fermions and 
sfermions, i.e.\ 
$t, b, \tau, \nu_\tau$, and their corresponding superpartners, 
$\Stope, \Stopz$, $\Sbote, \Sbotz$, $\Staue, \Stauz$, $\Sneut$, are
expected to be sizable, 
since they contain potentially large Yukawa couplings, 
$y_t$, $y_b$, $y_\tau$, where the down-type couplings can be enhanced 
in the MSSM for large values of $\tb$. This class of diagrams
in particular contains contributions enhanced by $\mt^2/\MW^2$. 

\begin{figure}[ht!]
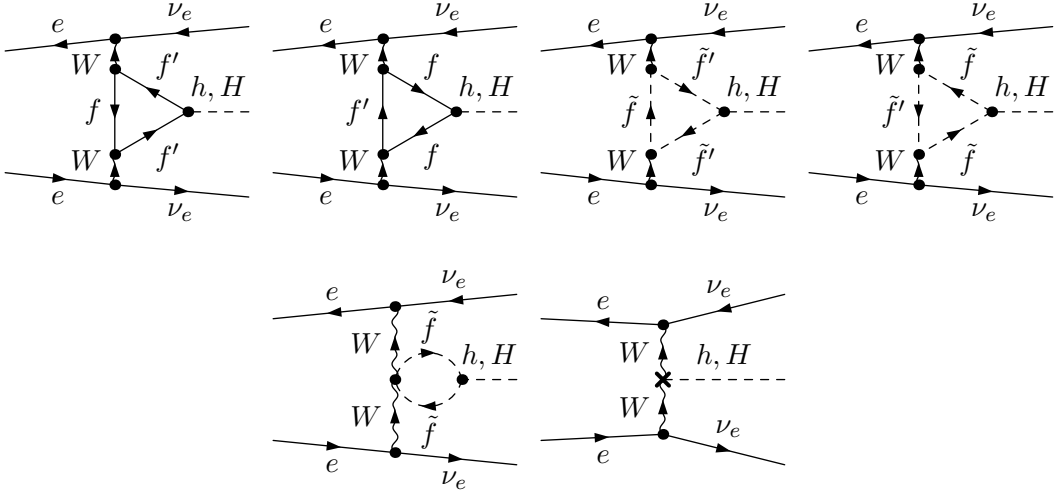

\vspace{2em}
\begin{center}
\begin{small}
\unitlength=1bp%
\begin{feynartspicture}(432,202)(4,2)
\FADiagram{}
\FAProp(0.,15.)(9.,16.)(0.,){/Straight}{-1}
\FALabel(4.32883,16.5605)[b]{$e$}
\FAProp(0.,5.)(9.,4.)(0.,){/Straight}{1}
\FALabel(4.32883,3.43948)[t]{$e$}
\FAProp(20.,17.)(9.,16.)(0.,){/Straight}{1}
\FALabel(14.3597,17.5636)[b]{$\nu_e$}
\FAProp(20.,10.)(15.,10.)(0.,){/ScalarDash}{0}
\FALabel(17.5,10.82)[b]{$h, H$}
\FAProp(20.,3.)(9.,4.)(0.,){/Straight}{-1}
\FALabel(14.3597,2.43637)[t]{$\nu_e$}
\FAProp(9.,16.)(9.,13.5)(0.,){/Sine}{-1}
\FALabel(7.93,14.75)[tr]{$W$}
\FAProp(9.,4.)(9.,6.5)(0.,){/Sine}{1}
\FALabel(7.93,5.25)[br]{$W$}
\FAProp(15.,10.)(9.,13.5)(0.,){/Straight}{1}
\FALabel(12.301,12.6089)[bl]{$f'$}
\FAProp(15.,10.)(9.,6.5)(0.,){/Straight}{-1}
\FALabel(12.301,7.39114)[tl]{$f'$}
\FAProp(9.,13.5)(9.,6.5)(0.,){/Straight}{1}
\FALabel(7.93,10.)[r]{$f$}
\FAVert(9.,16.){0}
\FAVert(9.,4.){0}
\FAVert(15.,10.){0}
\FAVert(9.,13.5){0}
\FAVert(9.,6.5){0}

\FADiagram{}
\FAProp(0.,15.)(9.,16.)(0.,){/Straight}{-1}
\FALabel(4.32883,16.5605)[b]{$e$}
\FAProp(0.,5.)(9.,4.)(0.,){/Straight}{1}
\FALabel(4.32883,3.43948)[t]{$e$}
\FAProp(20.,17.)(9.,16.)(0.,){/Straight}{1}
\FALabel(14.3597,17.5636)[b]{$\nu_e$}
\FAProp(20.,10.)(15.,10.)(0.,){/ScalarDash}{0}
\FALabel(17.5,10.82)[b]{$h, H$}
\FAProp(20.,3.)(9.,4.)(0.,){/Straight}{-1}
\FALabel(14.3597,2.43637)[t]{$\nu_e$}
\FAProp(9.,16.)(9.,13.5)(0.,){/Sine}{-1}
\FALabel(7.93,14.75)[tr]{$W$}
\FAProp(9.,4.)(9.,6.5)(0.,){/Sine}{1}
\FALabel(7.93,5.25)[br]{$W$}
\FAProp(15.,10.)(9.,13.5)(0.,){/Straight}{-1}
\FALabel(12.301,12.6089)[bl]{$f$}
\FAProp(15.,10.)(9.,6.5)(0.,){/Straight}{1}
\FALabel(12.301,7.39114)[tl]{$f$}
\FAProp(9.,13.5)(9.,6.5)(0.,){/Straight}{-1}
\FALabel(7.93,10.)[r]{$f'$}
\FAVert(9.,16.){0}
\FAVert(9.,4.){0}
\FAVert(15.,10.){0}
\FAVert(9.,13.5){0}
\FAVert(9.,6.5){0}

\FADiagram{}
\FAProp(0.,15.)(9.,16.)(0.,){/Straight}{-1}
\FALabel(4.32883,16.5605)[b]{$e$}
\FAProp(0.,5.)(9.,4.)(0.,){/Straight}{1}
\FALabel(4.32883,3.43948)[t]{$e$}
\FAProp(20.,17.)(9.,16.)(0.,){/Straight}{1}
\FALabel(14.3597,17.5636)[b]{$\nu_e$}
\FAProp(20.,10.)(15.,10.)(0.,){/ScalarDash}{0}
\FALabel(17.5,10.82)[b]{$h, H$}
\FAProp(20.,3.)(9.,4.)(0.,){/Straight}{-1}
\FALabel(14.3597,2.43637)[t]{$\nu_e$}
\FAProp(9.,16.)(9.,13.5)(0.,){/Sine}{-1}
\FALabel(7.93,14.75)[tr]{$W$}
\FAProp(9.,4.)(9.,6.5)(0.,){/Sine}{1}
\FALabel(7.93,5.25)[br]{$W$}
\FAProp(15.,10.)(9.,13.5)(0.,){/ScalarDash}{-1}
\FALabel(12.301,12.6089)[bl]{$\tilde f'$}
\FAProp(15.,10.)(9.,6.5)(0.,){/ScalarDash}{1}
\FALabel(12.301,7.39114)[tl]{$\tilde f'$}
\FAProp(9.,13.5)(9.,6.5)(0.,){/ScalarDash}{-1}
\FALabel(7.93,10.)[r]{$\tilde f$}
\FAVert(9.,16.){0}
\FAVert(9.,4.){0}
\FAVert(15.,10.){0}
\FAVert(9.,13.5){0}
\FAVert(9.,6.5){0}

\FADiagram{}
\FAProp(0.,15.)(9.,16.)(0.,){/Straight}{-1}
\FALabel(4.32883,16.5605)[b]{$e$}
\FAProp(0.,5.)(9.,4.)(0.,){/Straight}{1}
\FALabel(4.32883,3.43948)[t]{$e$}
\FAProp(20.,17.)(9.,16.)(0.,){/Straight}{1}
\FALabel(14.3597,17.5636)[b]{$\nu_e$}
\FAProp(20.,10.)(15.,10.)(0.,){/ScalarDash}{0}
\FALabel(17.5,10.82)[b]{$h, H$}
\FAProp(20.,3.)(9.,4.)(0.,){/Straight}{-1}
\FALabel(14.3597,2.43637)[t]{$\nu_e$}
\FAProp(9.,16.)(9.,13.5)(0.,){/Sine}{-1}
\FALabel(7.93,14.75)[tr]{$W$}
\FAProp(9.,4.)(9.,6.5)(0.,){/Sine}{1}
\FALabel(7.93,5.25)[br]{$W$}
\FAProp(15.,10.)(9.,13.5)(0.,){/ScalarDash}{1}
\FALabel(12.301,12.6089)[bl]{$\tilde f$}
\FAProp(15.,10.)(9.,6.5)(0.,){/ScalarDash}{-1}
\FALabel(12.301,7.39114)[tl]{$\tilde f$}
\FAProp(9.,13.5)(9.,6.5)(0.,){/ScalarDash}{1}
\FALabel(7.93,10.)[r]{$\tilde f'$}
\FAVert(9.,16.){0}
\FAVert(9.,4.){0}
\FAVert(15.,10.){0}
\FAVert(9.,13.5){0}
\FAVert(9.,6.5){0}

\FADiagram{}

\FADiagram{}
\FAProp(0.,15.)(10.,16.)(0.,){/Straight}{-1}
\FALabel(4.84577,16.5623)[b]{$e$}
\FAProp(0.,5.)(10.,4.)(0.,){/Straight}{1}
\FALabel(4.84577,3.43769)[t]{$e$}
\FAProp(20.,17.)(10.,16.)(0.,){/Straight}{1}
\FALabel(14.8458,17.5623)[b]{$\nu_e$}
\FAProp(20.,10.)(15.5,10.)(0.,){/ScalarDash}{0}
\FALabel(17.75,10.82)[b]{$h, H$}
\FAProp(20.,3.)(10.,4.)(0.,){/Straight}{-1}
\FALabel(14.8458,2.43769)[t]{$\nu_e$}
\FAProp(10.,16.)(10.,10.)(0.,){/Sine}{-1}
\FALabel(8.93,13.)[r]{$W$}
\FAProp(10.,4.)(10.,10.)(0.,){/Sine}{1}
\FALabel(8.93,7.)[r]{$W$}
\FAProp(15.5,10.)(10.,10.)(0.8,){/ScalarDash}{-1}
\FALabel(12.75,13)[b]{$\tilde f$}
\FAProp(15.5,10.)(10.,10.)(-0.8,){/ScalarDash}{1}
\FALabel(12.75,7)[t]{$\tilde f$}
\FAVert(10.,16.){0}
\FAVert(10.,4.){0}
\FAVert(15.5,10.){0}
\FAVert(10.,10.){0}

\FADiagram{}
\FAProp(0.,15.)(10.,14.5)(0.,){/Straight}{-1}
\FALabel(5.0774,15.8181)[b]{$e$}
\FAProp(0.,5.)(10.,5.5)(0.,){/Straight}{1}
\FALabel(5.0774,4.18193)[t]{$e$}
\FAProp(20.,17.)(10.,14.5)(0.,){/Straight}{1}
\FALabel(14.6241,16.7737)[b]{$\nu_e$}
\FAProp(20.,10.)(10.,10.)(0.,){/ScalarDash}{0}
\FALabel(15.,10.82)[b]{$h, H$}
\FAProp(20.,3.)(10.,5.5)(0.,){/Straight}{-1}
\FALabel(15.3759,5.27372)[b]{$\nu_e$}
\FAProp(10.,14.5)(10.,10.)(0.,){/Sine}{-1}
\FALabel(8.93,12.25)[r]{$W$}
\FAProp(10.,5.5)(10.,10.)(0.,){/Sine}{1}
\FALabel(8.93,7.75)[r]{$W$}
\FAVert(10.,14.5){0}
\FAVert(10.,5.5){0}
\FAVert(10.,10.){1}
\end{feynartspicture}
\end{small}
\vspace{1em}
\caption{
Corrections to the $WW\{h,H\}$ vertex and the corresponding counter-term
diagram.  The label $\xtilde f$ denotes all (s)fermions, except in the
presence of a $\xtilde f\,'$, in which case the former denotes only the
isospin-up and the latter the isospin-down members of the (s)fermion
doublets.
}
\label{fig:WWhHvert}
\end{center}
\end{figure}

The contributions involve corrections to the $WW\{h,H\}$ vertex and the
corresponding counter-term diagram, shown in \reffi{fig:WWhHvert},
corrections to the $W$-boson propagators and the corresponding counter
terms, shown in \reffi{fig:WWhHself}, and the counter-term contributions
to the $e\nu_e W$ vertex as shown in \reffi{fig:enuWCT}. Furthermore,
Higgs propagator corrections enter via the wave-function normalization of
the external Higgs boson, see below. There are
also $W$-boson propagator corrections inducing a transition from the
$W^\pm$ to either $G^\pm$ or $H^\pm$.  These corrections affect only the
longitudinal part of the $W$~boson, however, and are thus $\propto
m_e/\MW$ and have been neglected.

\begin{figure}[ht!]
\vspace{2em}
\begin{center}
\begin{small}
\unitlength=1bp%
\begin{feynartspicture}(432,202)(4,2)
\FADiagram{}
\FAProp(0.,15.)(10.,16.5)(0.,){/Straight}{-1}
\FALabel(4.77007,16.8029)[b]{$e$}
\FAProp(0.,5.)(10.,3.5)(0.,){/Straight}{1}
\FALabel(4.77007,3.19715)[t]{$e$}
\FAProp(20.,17.)(10.,16.5)(0.,){/Straight}{1}
\FALabel(14.9226,17.8181)[b]{$\nu_e$}
\FAProp(20.,10.)(10.,13.5)(0.,){/ScalarDash}{0}
\FALabel(15.3153,12.17)[lb]{$h, H$}
\FAProp(20.,3.)(10.,3.5)(0.,){/Straight}{-1}
\FALabel(14.9226,2.18193)[t]{$\nu_e$}
\FAProp(10.,16.5)(10.,13.5)(0.,){/Sine}{-1}
\FALabel(11.07,15.)[l]{$W$}
\FAProp(10.,3.5)(10.,6.)(0.,){/Sine}{1}
\FALabel(11.07,4.75)[l]{$W$}
\FAProp(10.,13.5)(10.,11.)(0.,){/Sine}{-1}
\FALabel(8.93,12.25)[r]{$W$}
\FAProp(10.,6.)(10.,11.)(0.8,){/Straight}{-1}
\FALabel(13.07,8.5)[l]{$f$}
\FAProp(10.,6.)(10.,11.)(-0.8,){/Straight}{1}
\FALabel(6.93,8.5)[r]{$f'$}
\FAVert(10.,16.5){0}
\FAVert(10.,3.5){0}
\FAVert(10.,13.5){0}
\FAVert(10.,6.){0}
\FAVert(10.,11.){0}

\FADiagram{}
\FAProp(0.,15.)(10.,16.5)(0.,){/Straight}{-1}
\FALabel(4.77007,16.8029)[b]{$e$}
\FAProp(0.,5.)(10.,3.5)(0.,){/Straight}{1}
\FALabel(4.77007,3.19715)[t]{$e$}
\FAProp(20.,17.)(10.,16.5)(0.,){/Straight}{1}
\FALabel(14.9226,17.8181)[b]{$\nu_e$}
\FAProp(20.,10.)(10.,13.5)(0.,){/ScalarDash}{0}
\FALabel(15.3153,12.17)[lb]{$h, H$}
\FAProp(20.,3.)(10.,3.5)(0.,){/Straight}{-1}
\FALabel(14.9226,2.18193)[t]{$\nu_e$}
\FAProp(10.,16.5)(10.,13.5)(0.,){/Sine}{-1}
\FALabel(11.07,15.)[l]{$W$}
\FAProp(10.,3.5)(10.,6.)(0.,){/Sine}{1}
\FALabel(11.07,4.75)[l]{$W$}
\FAProp(10.,13.5)(10.,11.)(0.,){/Sine}{-1}
\FALabel(8.93,12.25)[r]{$W$}
\FAProp(10.,6.)(10.,11.)(0.8,){/ScalarDash}{-1}
\FALabel(13.07,8.5)[l]{$\tilde f$}
\FAProp(10.,6.)(10.,11.)(-0.8,){/ScalarDash}{1}
\FALabel(6.93,8.5)[r]{$\tilde f'$}
\FAVert(10.,16.5){0}
\FAVert(10.,3.5){0}
\FAVert(10.,13.5){0}
\FAVert(10.,6.){0}
\FAVert(10.,11.){0}

\FADiagram{}
\FAProp(0.,15.)(10.,16.)(0.,){/Straight}{-1}
\FALabel(4.84577,16.5623)[b]{$e$}
\FAProp(0.,5.)(10.,4.)(0.,){/Straight}{1}
\FALabel(4.84577,3.43769)[t]{$e$}
\FAProp(20.,17.)(10.,16.)(0.,){/Straight}{1}
\FALabel(14.8458,17.5623)[b]{$\nu_e$}
\FAProp(20.,10.)(10.,12.5)(0.,){/ScalarDash}{0}
\FALabel(15.3153,12.0312)[lb]{$h, H$}
\FAProp(20.,3.)(10.,4.)(0.,){/Straight}{-1}
\FALabel(14.8458,2.43769)[t]{$\nu_e$}
\FAProp(10.,16.)(10.,12.5)(0.,){/Sine}{-1}
\FALabel(8.93,14.25)[r]{$W$}
\FAProp(10.,4.)(10.,8.5)(0.,){/Sine}{1}
\FALabel(11.07,6.25)[l]{$W$}
\FAProp(10.,12.5)(10.,8.5)(0.,){/Sine}{-1}
\FALabel(11.12,10.15)[l]{$W$}
\FAProp(10.,8.5)(10.,8.5)(5.5,8.5){/ScalarDash}{-1}
\FALabel(4.43,8.5)[r]{$\tilde f$}
\FAVert(10.,16.){0}
\FAVert(10.,4.){0}
\FAVert(10.,12.5){0}
\FAVert(10.,8.5){0}

\FADiagram{}
\FAProp(0.,15.)(10.,15.5)(0.,){/Straight}{-1}
\FALabel(4.9226,16.3181)[b]{$e$}
\FAProp(0.,5.)(10.,4.5)(0.,){/Straight}{1}
\FALabel(4.9226,3.68193)[t]{$e$}
\FAProp(20.,17.)(10.,15.5)(0.,){/Straight}{1}
\FALabel(14.7701,17.3029)[b]{$\nu_e$}
\FAProp(20.,10.)(10.,11.5)(0.,){/ScalarDash}{0}
\FALabel(15.3153,11.5556)[lb]{$h, H$}
\FAProp(20.,3.)(10.,4.5)(0.,){/Straight}{-1}
\FALabel(15.2299,4.80285)[b]{$\nu_e$}
\FAProp(10.,8.)(10.,4.5)(0.,){/Sine}{-1}
\FALabel(8.93,6.25)[r]{$W$}
\FAProp(10.,8.)(10.,11.5)(0.,){/Sine}{1}
\FALabel(8.93,9.75)[r]{$W$}
\FAProp(10.,15.5)(10.,11.5)(0.,){/Sine}{-1}
\FALabel(8.93,13.5)[r]{$W$}
\FAVert(10.,15.5){0}
\FAVert(10.,4.5){0}
\FAVert(10.,11.5){0}
\FAVert(10.,8.){1}

\FADiagram{}
\FAProp(0.,15.)(10.,16.5)(0.,){/Straight}{-1}
\FALabel(4.77007,16.8029)[b]{$e$}
\FAProp(0.,5.)(10.,3.5)(0.,){/Straight}{1}
\FALabel(4.77007,3.19715)[t]{$e$}
\FAProp(20.,17.)(10.,16.5)(0.,){/Straight}{1}
\FALabel(14.9226,17.8181)[b]{$\nu_e$}
\FAProp(20.,10.)(10.,6.5)(0.,){/ScalarDash}{0}
\FALabel(15.3153,7.68)[lt]{$h, H$}
\FAProp(20.,3.)(10.,3.5)(0.,){/Straight}{-1}
\FALabel(14.9226,2.18193)[t]{$\nu_e$}
\FAProp(10.,16.5)(10.,14.)(0.,){/Sine}{-1}
\FALabel(11.07,15.25)[l]{$W$}
\FAProp(10.,3.5)(10.,6.5)(0.,){/Sine}{1}
\FALabel(11.07,5.)[l]{$W$}
\FAProp(10.,6.5)(10.,9.)(0.,){/Sine}{1}
\FALabel(8.93,7.75)[r]{$W$}
\FAProp(10.,14.)(10.,9.)(0.8,){/Straight}{1}
\FALabel(6.93,11.5)[r]{$f$}
\FAProp(10.,14.)(10.,9.)(-0.8,){/Straight}{-1}
\FALabel(13.07,11.5)[l]{$f'$}
\FAVert(10.,16.5){0}
\FAVert(10.,3.5){0}
\FAVert(10.,6.5){0}
\FAVert(10.,14.){0}
\FAVert(10.,9.){0}

\FADiagram{}
\FAProp(0.,15.)(10.,16.5)(0.,){/Straight}{-1}
\FALabel(4.77007,16.8029)[b]{$e$}
\FAProp(0.,5.)(10.,3.5)(0.,){/Straight}{1}
\FALabel(4.77007,3.19715)[t]{$e$}
\FAProp(20.,17.)(10.,16.5)(0.,){/Straight}{1}
\FALabel(14.9226,17.8181)[b]{$\nu_e$}
\FAProp(20.,10.)(10.,6.5)(0.,){/ScalarDash}{0}
\FALabel(15.3153,7.68)[lt]{$h, H$}
\FAProp(20.,3.)(10.,3.5)(0.,){/Straight}{-1}
\FALabel(14.9226,2.18193)[t]{$\nu_e$}
\FAProp(10.,16.5)(10.,14.)(0.,){/Sine}{-1}
\FALabel(11.07,15.25)[l]{$W$}
\FAProp(10.,3.5)(10.,6.5)(0.,){/Sine}{1}
\FALabel(11.07,5.)[l]{$W$}
\FAProp(10.,6.5)(10.,9.)(0.,){/Sine}{1}
\FALabel(8.93,7.75)[r]{$W$}
\FAProp(10.,14.)(10.,9.)(0.8,){/ScalarDash}{1}
\FALabel(6.93,11.5)[r]{$\tilde f$}
\FAProp(10.,14.)(10.,9.)(-0.8,){/ScalarDash}{-1}
\FALabel(13.07,11.5)[l]{$\tilde f'$}
\FAVert(10.,16.5){0}
\FAVert(10.,3.5){0}
\FAVert(10.,6.5){0}
\FAVert(10.,14.){0}
\FAVert(10.,9.){0}

\FADiagram{}
\FAProp(0.,15.)(10.,16.)(0.,){/Straight}{-1}
\FALabel(4.84577,16.5623)[b]{$e$}
\FAProp(0.,5.)(10.,4.)(0.,){/Straight}{1}
\FALabel(4.84577,3.43769)[t]{$e$}
\FAProp(20.,17.)(10.,16.)(0.,){/Straight}{1}
\FALabel(14.8458,17.5623)[b]{$\nu_e$}
\FAProp(20.,10.)(10.,8.)(0.,){/ScalarDash}{0}
\FALabel(15.3153,8.20525)[lt]{$h, H$}
\FAProp(20.,3.)(10.,4.)(0.,){/Straight}{-1}
\FALabel(14.8458,2.43769)[t]{$\nu_e$}
\FAProp(10.,16.)(10.,11.5)(0.,){/Sine}{-1}
\FALabel(11.07,13.75)[l]{$W$}
\FAProp(10.,4.)(10.,8.)(0.,){/Sine}{1}
\FALabel(8.93,6.)[r]{$W$}
\FAProp(10.,8.)(10.,11.5)(0.,){/Sine}{1}
\FALabel(11.12,10.1)[l]{$W$}
\FAProp(10.,11.5)(10.,11.5)(5.5,11.5){/ScalarDash}{-1}
\FALabel(4.43,11.5)[r]{$\tilde f$}
\FAVert(10.,16.){0}
\FAVert(10.,4.){0}
\FAVert(10.,8.){0}
\FAVert(10.,11.5){0}

\FADiagram{}
\FAProp(0.,15.)(10.,15.5)(0.,){/Straight}{-1}
\FALabel(5.0774,14.1819)[t]{$e$}
\FAProp(0.,5.)(10.,4.5)(0.,){/Straight}{1}
\FALabel(4.9226,3.68193)[t]{$e$}
\FAProp(20.,17.)(10.,15.5)(0.,){/Straight}{1}
\FALabel(14.7701,17.3029)[b]{$\nu_e$}
\FAProp(20.,10.)(10.,8.5)(0.,){/ScalarDash}{0}
\FALabel(15.3153,10.556)[lb]{$h, H$}
\FAProp(20.,3.)(10.,4.5)(0.,){/Straight}{-1}
\FALabel(14.7701,2.69715)[t]{$\nu_e$}
\FAProp(10.,12.)(10.,15.5)(0.,){/Sine}{1}
\FALabel(11.07,13.75)[l]{$W$}
\FAProp(10.,12.)(10.,8.5)(0.,){/Sine}{-1}
\FALabel(8.93,10.25)[r]{$W$}
\FAProp(10.,4.5)(10.,8.5)(0.,){/Sine}{1}
\FALabel(11.07,6.5)[l]{$W$}
\FAVert(10.,15.5){0}
\FAVert(10.,4.5){0}
\FAVert(10.,8.5){0}
\FAVert(10.,12.){1}
\end{feynartspicture}
\end{small}
\vspace{1em}
\caption{
Corrections to the $W$-boson propagator and the corresponding counter-term
diagrams.  The label $\xtilde f$ denotes all (s)fermions, except
in the presence of a $\xtilde f\,'$, in which case the former denotes only
the isospin-up and the latter the isospin-down members of the
(s)fermion doublets.
}
\label{fig:WWhHself}
\end{center}
\end{figure}


\begin{figure}[ht!]
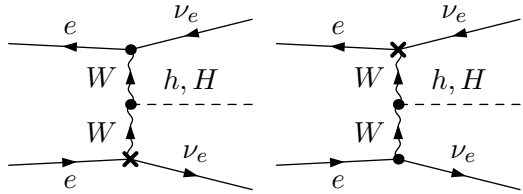

\begin{center}
\begin{small}
\unitlength=1bp%
\begin{feynartspicture}(216,101)(2,1)
\FADiagram{}
\FAProp(0.,15.)(10.,14.5)(0.,){/Straight}{-1}
\FALabel(5.0774,15.8181)[b]{$e$}
\FAProp(0.,5.)(10.,5.5)(0.,){/Straight}{1}
\FALabel(5.0774,4.18193)[t]{$e$}
\FAProp(20.,17.)(10.,14.5)(0.,){/Straight}{1}
\FALabel(14.6241,16.7737)[b]{$\nu_e$}
\FAProp(20.,10.)(10.,10.)(0.,){/ScalarDash}{0}
\FALabel(15.,10.82)[b]{$h, H$}
\FAProp(20.,3.)(10.,5.5)(0.,){/Straight}{-1}
\FALabel(15.3759,5.27372)[b]{$\nu_e$}
\FAProp(10.,14.5)(10.,10.)(0.,){/Sine}{-1}
\FALabel(8.93,12.25)[r]{$W$}
\FAProp(10.,5.5)(10.,10.)(0.,){/Sine}{1}
\FALabel(8.93,7.75)[r]{$W$}
\FAVert(10.,14.5){0}
\FAVert(10.,10.){0}
\FAVert(10.,5.5){1}

\FADiagram{}
\FAProp(0.,15.)(10.,14.5)(0.,){/Straight}{-1}
\FALabel(5.0774,15.8181)[b]{$e$}
\FAProp(0.,5.)(10.,5.5)(0.,){/Straight}{1}
\FALabel(5.0774,4.18193)[t]{$e$}
\FAProp(20.,17.)(10.,14.5)(0.,){/Straight}{1}
\FALabel(14.6241,16.7737)[b]{$\nu_e$}
\FAProp(20.,10.)(10.,10.)(0.,){/ScalarDash}{0}
\FALabel(15.,10.82)[b]{$h, H$}
\FAProp(20.,3.)(10.,5.5)(0.,){/Straight}{-1}
\FALabel(15.3759,5.27372)[b]{$\nu_e$}
\FAProp(10.,14.5)(10.,10.)(0.,){/Sine}{-1}
\FALabel(8.93,12.25)[r]{$W$}
\FAProp(10.,5.5)(10.,10.)(0.,){/Sine}{1}
\FALabel(8.93,7.75)[r]{$W$}
\FAVert(10.,5.5){0}
\FAVert(10.,10.){0}
\FAVert(10.,14.5){1}
\end{feynartspicture}
\end{small}
\caption{
Counter-term contributions entering via the $e\,\nu_e\,W$ vertex.
}
\label{fig:enuWCT}
\end{center}
\end{figure}


While the renormalization in the counter terms depicted in
\reffis{fig:WWhHself} and \ref{fig:enuWCT} is as in the SM (see e.g.\
\citere{ansgarhabil}), the
$WW\{h,H\}$ vertices are renormalized as follows,
\BEA
\label{eq:WWhct}
WWh:\,
\Ga_{WWh}^{(0), {\rm CT}} = \Ghn && 
  \Biggl[ 1 + \de \tilde Z_e 
            + \edz \frac{\de \MW^2}{\MW^2}
            + \de Z_W 
            + \frac{\de \sw}{\sw} \\
&&{}
            + \sinb\Cb \frac{\Cba}{\Sba} \de\tb
            + \edz \de Z_h
            + \edz \frac{\GHn}{\Ghn} \de Z_{Hh}
  \Biggr] \notag \\
\label{eq:WWHct}
WWH:\,
\Ga_{WWH}^{(0), {\rm CT}} = \GHn && 
  \Biggl[ 1 + \de \tilde Z_e 
            + \edz \frac{\de \MW^2}{\MW^2}
            + \de Z_W 
            + \frac{\de \sw}{\sw} \\
&&{}
            - \sinb\Cb \frac{\Sba}{\Cba} \de \tb
            + \edz \de Z_H
            + \edz \frac{\Ghn}{\GHn} \de Z_{hH}
  \Biggr] \notag
\EEA
Analogous expressions are obtained for $\Ga_{ZZ\Phi}^{(0),{\rm CT}}$ 
($\Phi = h,H$). In the above expressions
$\de \tilde Z_e$ incorporates the charge renormalization and the
$\De r$ contribution arising from \refeq{gmueparam}, 
\BE
\de \tilde Z_e = \de Z_e - \edz \De r, \quad
\de Z_e = \edz\Pi^\ga(0) - \frac{\sw}{\cw} 
                           \frac{\Si_{\ga Z}^T(0)}{\MZ^2}\,,
\end{equation}
where $\Si^T$ denotes the transverse part of a self-energy.
$\de\MW^2$ is
the $W$-mass counter term, $\de Z_W$ is the corresponding
field-renormalization constant, and $\de\sw$ denotes the renormalization 
constant for
the weak mixing angle. The field-renormalization constants, $\de
Z_H, \de Z_h$, and $\de Z_{Hh} = \de Z_{hH}$ are given in
\refeq{eq:deltaZHiggs}.  The counter term for $\tb$ (with $\tb \to \tb
(1 + \de \tb)$) is derived in the \drbar\ renormalization
scheme~\cite{mhiggsrenorm} (using DRED). The parameter $\tb$ in our result thus
corresponds to the \drbar\ parameter, taken at the scale 
$\mu_{\drbarm} = \mt$.
We list here all contributing
counter terms except for the Higgs field renormalization which has already
been given in \refeq{eq:deltaZHiggs}:
\BEA
\de\MW^2 &=& \re\Si_W^T(\MW^2)\,, \\[.5em]
\de\MZ^2 &=& \re\Si_Z^T(\MZ^2)\,, \\[.5em]
\frac{\de\sw}{\sw} &=& \edz \frac{\cw^2}{\sw^2}
  \KL \frac{\de\MZ^2}{\MZ^2} - \frac{\de\MW^2}{\MW^2} \KR, \\
\de\tb &=& \de\tb^{\drbarm} = - \ed{2\CZa} 
  \KKL \re\Si_{hh}'(\mh^2) - \re\Si_{HH}'(\mH^2) \KKR^{\rm div}.
\label{eq:rcs}
\EEA
For $\de \tilde Z_e$ we find, taking into account only contributions
from fermion and sfermion loops,
\BE
\de \tilde Z_e = \edz \KKKL \frac{\cw^2}{\sw^2} 
                \KL \frac{\de\MZ^2}{\MZ^2} - \frac{\de\MW^2}{\MW^2} \KR
                - \KKL \frac{\Si_W^T(0) - \de\MW^2}{\MW^2} \KKR 
                \KKKR \,.
\end{equation}
The gauge-boson field-renormalization constants,
$\de Z_W, \de Z_Z, \de Z_{\ga Z}$, drop out in the result for the
complete $S$-matrix element. 

In order to ensure the correct on-shell properties of the outgoing
Higgs boson, which are necessary for the correct renormalization of 
the $S$-matrix element, furthermore finite wave-function normalizations 
have to be incorporated, see below.
For the various checks that have been performed to ensure the
reliability of our result, see \citere{eennH}.

The individual contributions from fermions and sfermions constitute two
subsets of the full result which are individually UV-finite. This is in
contrast to the evaluation of renormalized Higgs boson self-energies
(see e.g.\ \citeres{mhiggsf1lB,mhiggsf1lC}), where a UV-finite result
is obtained only after adding the fermion- and sfermion-loop contributions.


\subsubsection{The Higgs-boson propagator corrections and the effective Born
  approximation}
\label{subsec:higgsprop}

For the correct normalization of the $S$-matrix element, finite
Higgs-boson propagator corrections have to be included such that the
residues of the outgoing Higgs bosons are set to unity and no mixing
between $h$ and $H$ occurs on the mass shell of the two particles. 
The corrections affecting the Higgs-boson propagators and the Higgs-boson
masses are numerically very important. Therefore we go beyond the
\onel\ fermion/sfermion contribution used for the evaluation of the
genuine \onel\ diagrams and include Higgs-boson corrections also from
other sectors of the model~\cite{mhiggsf1lB,mhiggsf1lC} as well as the
dominant \twol\
contributions~\cite{mhiggslong,mhiggsAEC} as 
incorporated in the program \fh~\cite{feynhiggs}.

For the $WW\{h,H\}$ vertex, these contributions can be
included as follows, yielding the correct normalization of the
$S$ matrix at \onel\ order%
\footnote{
Note that our notation is slightly different from
\citeres{eehZhA,hff}. 
}%
:
\BEA
\label{eq:ewdecayamplitude}
WWh &:& \quad \wrZh \KL \Ghn + \edz \rZHh\; \GHn \KR, \\
WWH &:& \quad \wrZH \KL \GHn + \edz \rZhH\; \Ghn \KR\,.
\EEA
This gives rise to the following terms: 
\BEA
WWh:\quad
\Ga_{WWh}^{{\rm WF}} &=& \Ghn \;
  \Biggl[   \KL \sqrt{\rZh} - 1 \KR 
                 + \edz \frac{\GHn}{\Ghn} \sqrt{\rZh} \; \rZHh 
  \Biggr]\,,
\label{eq:WWhct2} \\
WWH:\quad
\Ga_{WWH}^{{\rm WF}} &=& \GHn \;
  \Biggl[ \KL \sqrt{\rZH} - 1 \KR 
                 + \edz \frac{\Ghn}{\GHn} \sqrt{\rZH} \; \rZhH 
  \Biggr]\,.
\label{eq:WWHct2}
\EEA

\noindent
Analogous expressions are obtained for $\Ga_{ZZ\Phi}^{\rm WF}$ 
($\Phi = h,H$).
In the above expressions, the finite Higgs-mixing contributions enter,
\BEA
\label{eq:ZhH}
\rZHh &=& -2 \; \frac{\re\hSihH(\Mh^2)}{\Mh^2 - \mH^2 + \re\hSiH(\Mh^2)}\,, \\
\label{eq:ZHh}
\rZhH &=& -2 \; \frac{\re\hSihH(\MH^2)}{\MH^2 - \mh^2 + \re\hSih(\MH^2)}\,,
\EEA
involving the renormalized self-energies $\hSi(q^2)$, see
\refeq{eq:renSihiggs}, which contain corrections up to the \twol\ level.
The wave-function normalization factors $\rZh, \rZH$ are
related to the finite residue of the Higgs-boson propagators:
\BEA
\label{eq:zlh}
\rZh &=& \ed{1 + \re\hSiph(q^2) - 
        \KL \frac{\KL \re\hSihH(q^2) \KR^2}
            {q^2 - \mH^2 + \re\hSiH(q^2)} \KR '}~_{\Bigr| q^2 = \Mh^2}\,, 
                                                                 \\[.5em]
\label{eq:zhh}
\rZH &=& \ed{1 + \re\hSipH(q^2) -
        \KL \frac{\KL \re\hSihH(q^2) \KR^2}
            {q^2 - \mh^2 + \re\hSih(q^2)} \KR '}~_{\Bigr| q^2 = \MH^2}\,.
\EEA
$'$ denotes the derivative with respect to the momentum squared.

If in \refeqs{eq:ZhH}--(\ref{eq:zhh}) the renormalized self-energies
were evaluated at $q^2 = 0$, the above wave-function correction would
reduce to the $\aeff$~approximation~\cite{hff,eehZhA}. In this
approximation, however, the outgoing Higgs boson does not have the
correct on-shell properties. 

In order to analyze the effect of those corrections that go beyond the
universal Higgs propagator corrections, we 
include the Higgs propagator corrections according to
\refeqs{eq:WWhct2}--(\ref{eq:zhh}) into our Born matrix element, see
the next subsection.
Concerning our numerical analysis, see \refse{sec:numeval_eennH}, we either
use this Born cross section
(thus the difference between our tree-level
and the one-loop cross sections indicates the effect of the new genuine
loop corrections), or we use the $\aeff$~approximation (so that the
difference between the tree-level and the \onel\ cross section
directly shows the effect of our new calculation compared to the
previously used results).


\subsubsection{The higher-order production cross section}
\label{subsec:HOXS}

The amplitude for the process \eenenehH\ is denoted as
\BE
\cM^{(i)}_{\Phi,e}, \quad (\Phi = h,H;\ i = 0,1)\,,
\end{equation}
where $i = 0$ denotes the lowest-order contribution and $i = 1$ the
\onel\ correction.

The tree-level amplitude involves the $WW$-fusion channel (left diagram
of \reffi{fig:tree}) and the Higgs-strahlung process (right diagram of
\reffi{fig:tree}) where the virtual $Z$~boson is connected to two
electron neutrinos. As explained above, we include the Higgs
propagator corrections into our lowest-order matrix element. We use
\BE
\cM^{(0)}_{\Phi,e} = \cM^{\rm tree}_{\Phi,e} + \cM^{\rm WF}_{\Phi,e}\,,
\end{equation}
where $\cM^{\rm tree}_{\Phi,e}$ is the contribution of the two tree-level
diagrams, parametrized with $\al = \al_{\rm tree}$, \refeq{alphaborn},
and $\cM^{\rm WF}_{\Phi,e}$ denotes the wave-function normalization
contributions given in \refeqs{eq:WWhct2}--(\ref{eq:zhh}) (and
analogously for the $ZZ\{h,H\}$ vertices).

At \onel\ order ($i = 1$), the diagrams shown in
\reffis{fig:WWhHvert}--\ref{fig:enuWCT} contribute (and corresponding
diagrams for the Higgs-strahlung process), involving fermion and sfermion
loops.  The counter-term contributions given in \refeqs{eq:WWhct},
(\ref{eq:WWHct}) enter via the $WW\{h,H\}$ vertices (and analogously for
the $ZZ\{h,H\}$ vertices), while the other counter-term contributions
have the same form as in the SM.

In order to evaluate the cross section that is actually observed in the
detector, $e^+e^- \to (h, H \; + \; {\rm missing~energy})$,
we furthermore take into account the 
amplitude of the 
Higgs-strahlung process where the $Z$~boson is connected to 
$\nu_f \bar \nu_f$ ($f = \mu, \tau$),
\BE
\label{eq:treeWF}
\cM^{(i)}_{\Phi,f} = \cM^{\rm tree}_{\Phi,f} + 
                     \cM^{\rm WF}_{\Phi,f}, 
\quad (\Phi = h,H;\ i = 0, 1;\ f = \mu,\tau)\,.
\end{equation}
Of course there is no interference between the $\cM^{(i)}_{\Phi,f}$
for different flavors.

For all flavors, on the other hand, the $Z$-boson propagator connected to the
two outgoing neutrinos can become resonant when integrating over the
full phase-space, and therefore a width has to be included in
that propagator. We have incorporated this by using the running width
in the $Z$-boson propagators, $\Ga_Z(s) = (s/\MZ^2) \, \Ga_Z^{\rm exp}$,
where $\Ga_Z^{\rm exp} = 2.4952 \gev$, and dropping the imaginary
parts of the light-fermion contributions to the $Z$-boson
self-energies.  

The cross-section formulas for $h$~production thus become
\BEA
\label{sih0}
\si^0_h &\propto& \sum_{f=e,\mu,\tau} |\cM^{(0)}_{h,f}|^2 \, , \\
\si^1_h &\propto& \sum_{f=e,\mu,\tau} \KL |\cM^{(0)}_{h,f}|^2 
               + 2\, \re \bigl[ (\cM^{(0)}_{h,f})^*\, \cM^{(1)}_{h,f} \bigr]
               \KR \, .
\label{sih1}
\EEA
The formulas for $H$~production are analogous, except that we have also
included the square of the one-loop amplitude.  This is because the
decoupling behavior of the $WWH$~coupling can make the tree-level cross
section 
very small so that the square of the one-loop amplitude
becomes of comparable size:
\BEA
\label{siH0}
\si^0_H &\propto& \sum_{f=e,\mu,\tau} |\cM^{(0)}_{H,f}|^2 \, , \\
\si^1_H &\propto& \sum_{f=e,\mu,\tau} \KL |\cM^{(0)}_{H,f} \, 
               + \, \cM^{(1)}_{H,f}|^2 \KR \, .
\label{siH1}
\EEA
In this way, 
at \order{\al^2} only contributions 
$\sim (\cM^{(0)}_{H,f})^*\, \cM^{(2)}_{H,f}$ are neglected, 
which are expected to be very small, if $\cM^{(0)}_{H,f}$ is
suppressed.


\subsubsection{Parameters for the numerical evaluation}
\label{sec:numeval_eennH}

For the numerical evaluation we followed the procedure outlined in
\citere{fa-fc-lt}: The Feynman diagrams for the contributions mentioned
above were 
generated using the \FA~\cite{feynarts} package.  The only necessary
addition was the implementation of 
the counter terms for the $VV\{h,H\}$ ($V = W, Z$) vertices,
\refeqs{eq:WWhct}--(\ref{eq:rcs}), into the existing MSSM model
file~\cite{fa-mssm}.  
The resulting amplitudes were algebraically simplified using \FC\
\cite{formcalc} 
and then automatically converted to a Fortran program.  The \LT\ package
\cite{formcalc,ff} was used to evaluate the one-loop scalar and
tensor integrals. 
The numerical results presented in the following subsections were
obtained with 
this Fortran program.%
\footnote{The code is available at {\tt www.hep-processes.de} .}

While we have obtained results both for the total cross sections and
differential distributions, in the numerical examples below we will
focus on total cross sections only.
For the various cross checks of our results with (existing)
literature, see \citere{eennH} and the second and third reference in
\cite{eennHSMf1l}.

For our numerical
evaluation we have chosen for simplicity a common soft SUSY-breaking 
parameter in the
diagonal entries of the sfermion mass matrices, $\msusy$, and the
same trilinear couplings for all generations. 
Our analytical result, however, holds for general values of the
parameters in the sfermion sector. The SM parameter are chosen as in
the Appendix. 

The further SUSY parameters entering our result via the Higgs boson
propagator corrections are 
the SU(2) gaugino mass parameter, $M_2$, (the U(1)
gaugino mass parameter is obtained via the GUT relation, 
$M_1 = (5/3)\, (\sw^2/\cw^2)\, M_2$), and the gluino mass, $\mgl$.

For our numerical analyses we assume all 
soft SUSY-breaking parameters to be real.
Our analytical result, however,
holds also for complex parameters entering the loop corrections to
\eennhH.


\subsection{SM Higgs-boson production}
\label{subsec:SMprod}

For comparison purposes, we start our analysis with the fermion-loop
corrections to the process $e^+e^- \to \bar \nu \nu \HSM$
in the SM. For the full one-loop calculation see \citere{eennHSMf1l}. 

\begin{figure}[ht!]
\begin{center}
\includegraphics[width=13cm]{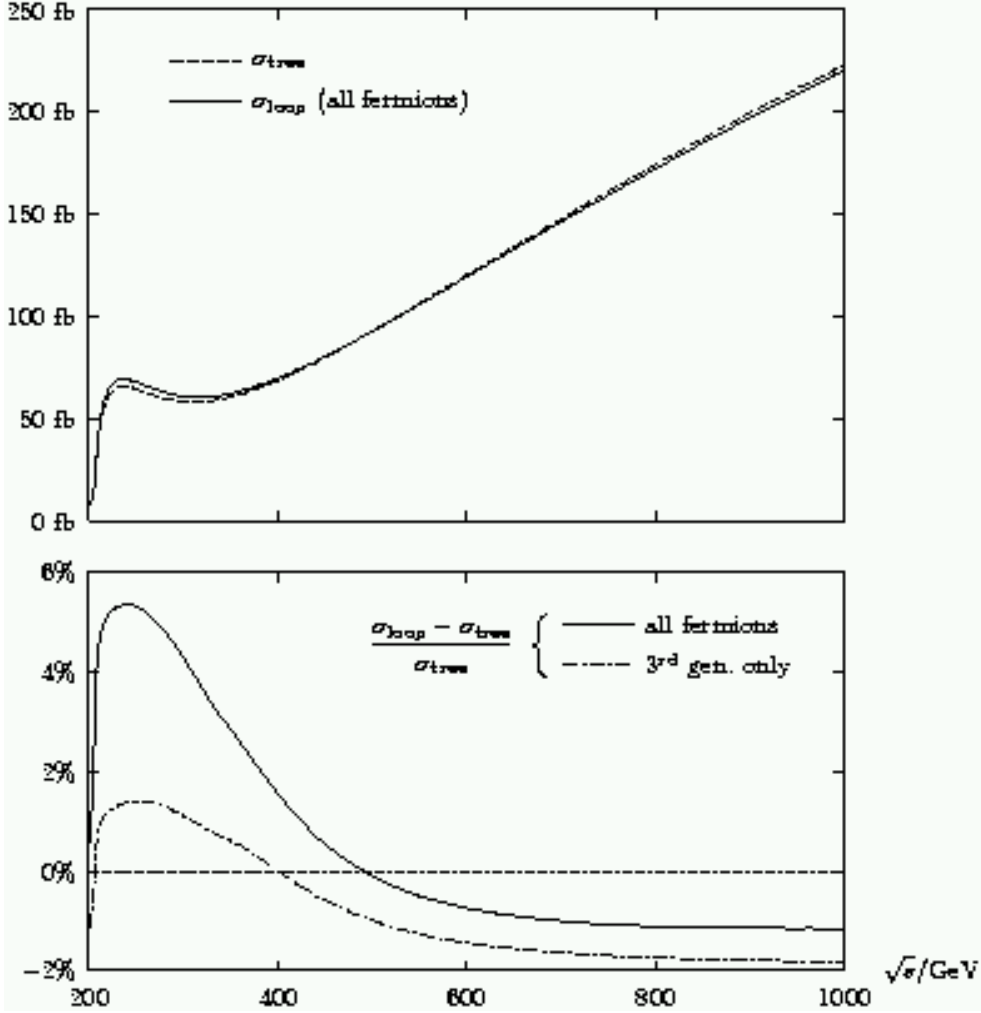}
\caption{
The tree-level and the one-loop cross sections for 
$e^+e^- \to \bar \nu \nu \HSM$ in the SM,
$\si_{\HSM}^0$ and $\si_{\HSM}^1$, are shown as a function of
$\sqrt{s}$ for  $\MHSM = 115 \gev$. The upper plot shows the absolute
values, the lower plot shows the relative corrections for all fermions
and for the third-generation fermions only.
}
\label{fig:HSM_sqrts}
\end{center}
\vspace{-1em}
\end{figure}

\reffi{fig:HSM_sqrts} shows the tree-level and \onel-corrected
production cross section for a SM Higgs-boson mass of 
$\MHSM = 115 \gev$. The absolute values are shown in the upper
plot. The sharp rise in the cross section for $\sqrt{s} \gsim 200$~GeV
is due to the threshold for on-shell production of the $Z$~boson in the
Higgs-strahlung contribution, see the right diagram of \reffi{fig:tree}.
Above the threshold the $1/s$ behavior of the Higgs-strahlung
contribution competes with the logarithmically rising $t$-channel
contribution from $WW$ fusion.

The lower plot shows the relative correction coming from all fermions,
as well as the correction from 
the third-generation fermions only. The correction from all
fermions ranges from about
$+5\%$ at low $\sqrt{s}$ to $-1.2\%$ at high $\sqrt{s}$.
Restricting to the contribution of third-generation fermions only,
we obtain corrections in the range
from $+1.3\%$ to $-1.8\%$. 
These corrections (both from the third family only as well as from the 
first two generations) are at the level of the expected sensitivity for the
$WW$-fusion channel at the LC%
\footnote{
The full \order{\al} corrections in the SM can be even larger due to
the QED/bremsstrahlung contributions~\cite{eennHSMf1l}. 
}%
. For a LC running in its high-energy mode
with $\sqrt{s} \approx 800$~GeV, in particular, a
measurement of the total cross section with an accuracy of better than
2\% seems to be feasible~\cite{eennHexp}.

\begin{figure}[ht!]
\begin{center}
\includegraphics[width=13.5cm]{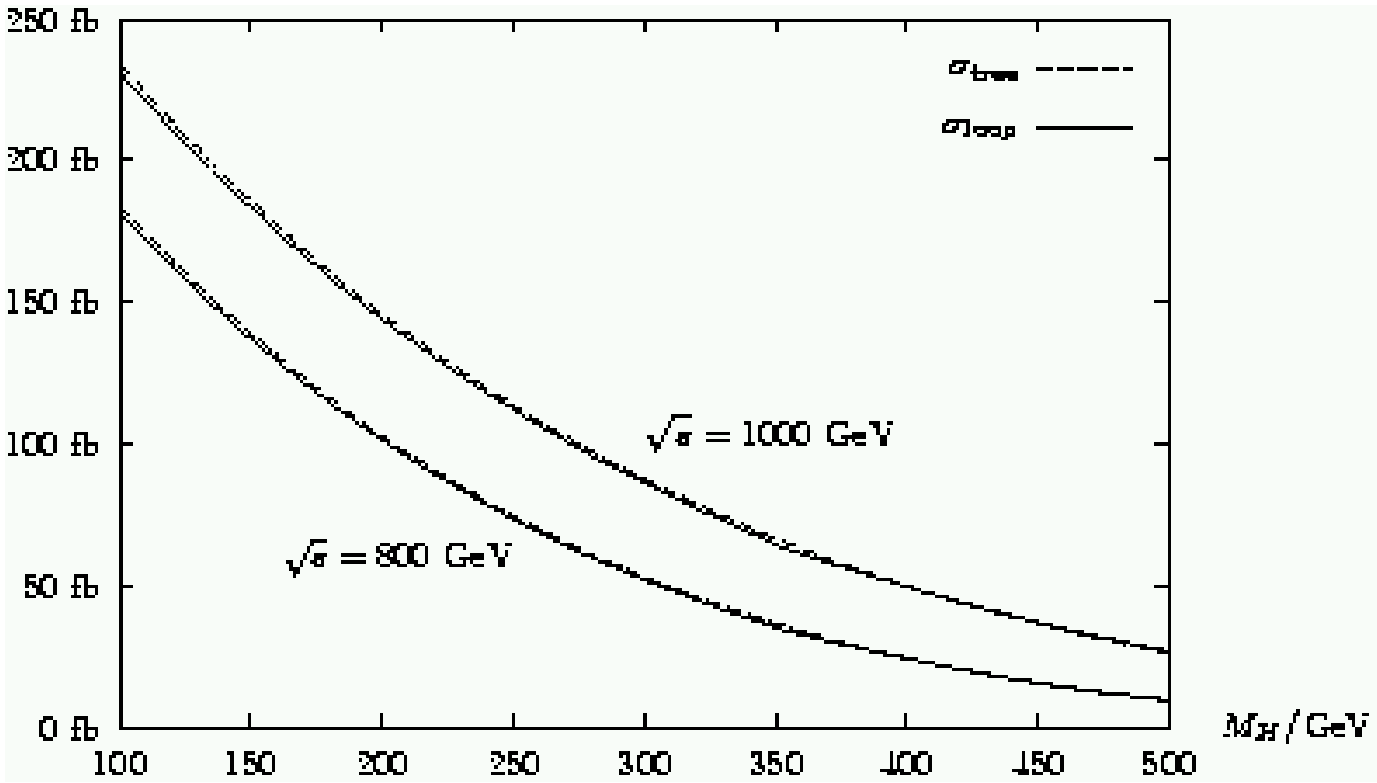}
\caption{
The tree-level and the one-loop cross sections for 
$e^+e^- \to \bar \nu \nu \HSM$ in the SM,
$\si_{\HSM}^0$ and $\si_{\HSM}^1$, are shown as a function of
$\MHSM$ for $\sqrt{s} = 800, 1000 \gev$. 
}
\label{fig:HSM_MH}
\end{center}
\end{figure}

In \reffi{fig:HSM_MH} the SM production cross section is shown
as a function of $\MHSM$ for $\sqrt{s} = 800, 1000 \gev$. A SM Higgs
boson possesses a relatively large production cross section,
\order{10\fb}, depending on the available energy, even for 
$\MHSM \gsim 500 \gev$. Thus it should easily be detectable at a 
high-luminosity LC.


\subsection{Light $\cp$-even Higgs-boson production}
\label{subsec:hprod}

Since the mass of the lightest $\cp$-even Higgs boson in the MSSM is
bounded from above by $\Mh \lsim 140 \gev$~\cite{mhiggslong,mhiggsAEC},
its detection at the LC is guaranteed~\cite{lc}.  In order to exploit
the precision measurements possible at the LC, a precise prediction at
the percent level of its
production cross section (and its decay rates) is necessary.

In the following we analyze the $h$~production cross
section. To begin with, we focus on the four benchmark
scenarios given in the Appendix \cite{benchmark} 
(proposed for MSSM Higgs-boson searches at hadron colliders and
beyond). $\MA$ and $\tb$ are kept as free parameters.

\begin{figure}[ht!]
\begin{center}
\vspace{-1em}
\includegraphics[width=16.7cm]{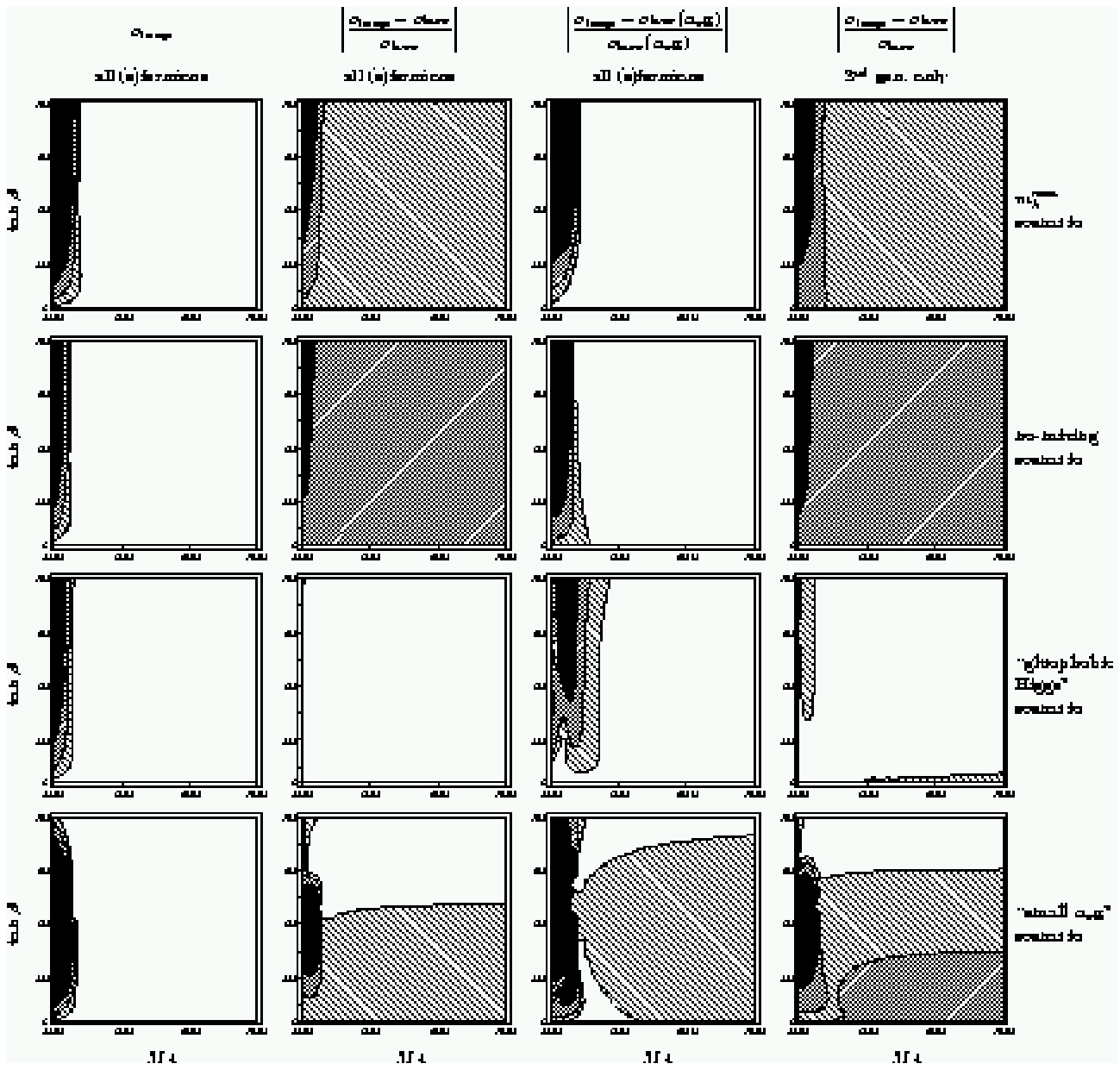}
\caption{
Results for the cross section for \eennh\ at one-loop order,
$\si^1_h$ (left column), the relative corrections including all three
fermion and sfermion generations (two middle columns), and the relative
corrections including only the third generation (right column) are shown
in the $\MA$--$\tb$ plane for four benchmark scenarios at
$\sqrt{s} = 1 \tev$. 
The universal Higgs propagator corrections have
been absorbed into the tree-level cross section and thus do not appear
in the relative corrections. In the second and fourth column 
the Higgs propagator corrections are implemented according 
to \refeqs{eq:treeWF}, (\ref{sih0}), while in
the third column the $\aeff$ approximation is used.
In the results for
$\si^1_h$ (left column) the black region corresponds to 
$\si^1_h < 50 \fb$, the dark-shaded region to
$50 \fb \le \si^1_h \le 100 \fb$, the light-shaded region to
$100 \fb \le \si^1_h \le 150 \fb$, and the white region to 
$\si^1_h \ge 150 \fb$. 
For the relative corrections (second to fourth column) the black region
corresponds to  $R^1_h > 5\%$, the dark-shaded region to
$2\% \le R^1_h \le 5\%$, the light-shaded region to
$1\% \le R^1_h \le 2\%$, and the white region corresponds to 
$R^1_h \le 1\%$ (see text). 
}
\label{fig:h_benchmark}
\vspace{-2em}
\end{center}
\end{figure}

Figure (\ref{fig:h_benchmark}) shows in the four benchmark scenarios the
$h$~production cross section, $\si^1_h$, \refeq{sih1}, as well as the
relative size of the loop 
corrections,
\BE
R^1_h = \Bigg| \frac{\si^1_h - \si^0_h}{\si^0_h} \Bigg|
\label{relcor}
\end{equation}
(which is nearly always negative), 
including the contributions from all generations as well as from the
third generation only. 
The results are shown in the
$\MA$--$\tb$ plane for $100 \gev \le \MA \le 400 \gev$ and 
$2 \le \tb \le 50$. For larger $\MA$ values the behavior of the
lightest $\cp$-even MSSM Higgs boson is very SM-like, i.e.\ the
results hardly vary with $\MA$ any more. 

For very low values of $\MA$, $\MA < 150 \gev$, the cross section is
relatively small. This is due to the fact that the $WWh$~coupling at
tree level, being $\sim \Sba$, can become very small. In this region
of parameter space, however, the heavy $\cp$-even Higgs boson is still 
very light
and couples to the gauge bosons with approximately SM strength, the 
tree-level coupling of
$WWH$ being $\sim \Cba \approx 1$.

For the interpretation of the two middle and the right column of
\reffi{fig:h_benchmark} it is important to 
keep in mind that we have absorbed the universal Higgs propagator
corrections, which are numerically very important, into our tree-level
cross section. Thus, the relative corrections, $R^1_h$, shown in
\reffi{fig:h_benchmark}, display the effects of the other genuine
one-loop corrections only. We first compare the corrections in the 
two cases where the Higgs propagator corrections are implemented
according to \refeqs{eq:treeWF}, (\ref{sih0}), which ensures the correct
on-shell properties of the outgoing Higgs boson (second column), and
where an $\aeff$ approximation is used (third column), which is often
done in the literature. In the $\aeff$ approximation, the leading
contribution of the process-independent corrections entering via the
Higgs-boson propagators is included by replacing
the tree-level coupling of $\{h,H\}WW = \{\sin,\cos\}(\be - \al)$
by $\{\sin,\cos\}(\be - \aeff)$.
The difference between the full on-shell prescription
and the $\aeff$ approximation turns out to be sizable. 
It amounts to
several percent even for relatively large values of $\MA$. As a
consequence, including the Higgs propagator corrections in an $\aeff$
approximation will not be sufficient in view of prospective precision
measurements of the \eennh\ cross section.

The results in the second column of \reffi{fig:h_benchmark} show that
the size of the corrections from fermion and sfermion loops is somewhat
different in the four scenarios. While corrections of more than 5\%
only occur for $\MA \lsim 130 \gev$, we obtain corrections of 2--5\% in
the whole parameter space of the no-mixing scenario. 
Corrections of 1--2\% can be found in large  
parts of the parameter space of the $\mhmax$ and the small-$\aeff$ scenario.
The situation in the four benchmark scenarios, which have been chosen to
represent different aspects of MSSM phenomenology, shows that the
corrections investigated here are typically of the order of about 1--5\%.
A measurement of the \eennh\ cross section at the percent level will
thus be sensitive to this kind of corrections. 

In the right column of \reffi{fig:h_benchmark} we show $R^1_h$ derived
including the contributions from the third family of fermions and
sfermions only. Thus the differences between the second and the fourth
column reflect the relevance of the loop corrections coming from the
first two families. While in the $\mhmax$~and the no-mixing scenario
differences can mostly be found for small $\MA$, $\MA \lsim 150 \gev$,
in the other two scenarios the effect of the first two families can be
relevant also for larger $\MA$. 
Within the gluophobic-Higgs scenario, the first two families play a
role for small $\tb$ and large $\MA$. In the small-$\aeff$ scenario
differences can be found for larger $\tb$ over the whole $\MA$ range.
In the latter two scenarios, the corrections coming from the first and
second family lead to a partial compensation of the corrections from the
third family. The light fermion generations can give rise to a
contribution of $\sim 1\%$, which is non-negligible for
cross section measurements at the percent level.

The Higgs propagator corrections, which we have absorbed into our 
tree-level cross section, mainly affect the numerical value of $\Mh$,
which enters the final-state kinematics, while the numerical effect of
the corrections to the 
$WWh$ coupling is less important. The comparison between the
prediction for \eennh\ in the MSSM and the corresponding process in the
SM for the same value of the Higgs boson mass (which is not shown here)
yields deviations of more than 5\% for $\MA \lsim 200 \gev$, which to a
large extent are due to the suppressed $WWh$~coupling in the MSSM case. 
Deviations of more than 1\% are found in all scenarios up to rather 
large values of $\MA$.

\begin{figure}[ht!]
\vspace{2em}
\begin{center}
\includegraphics[width=15.5cm]{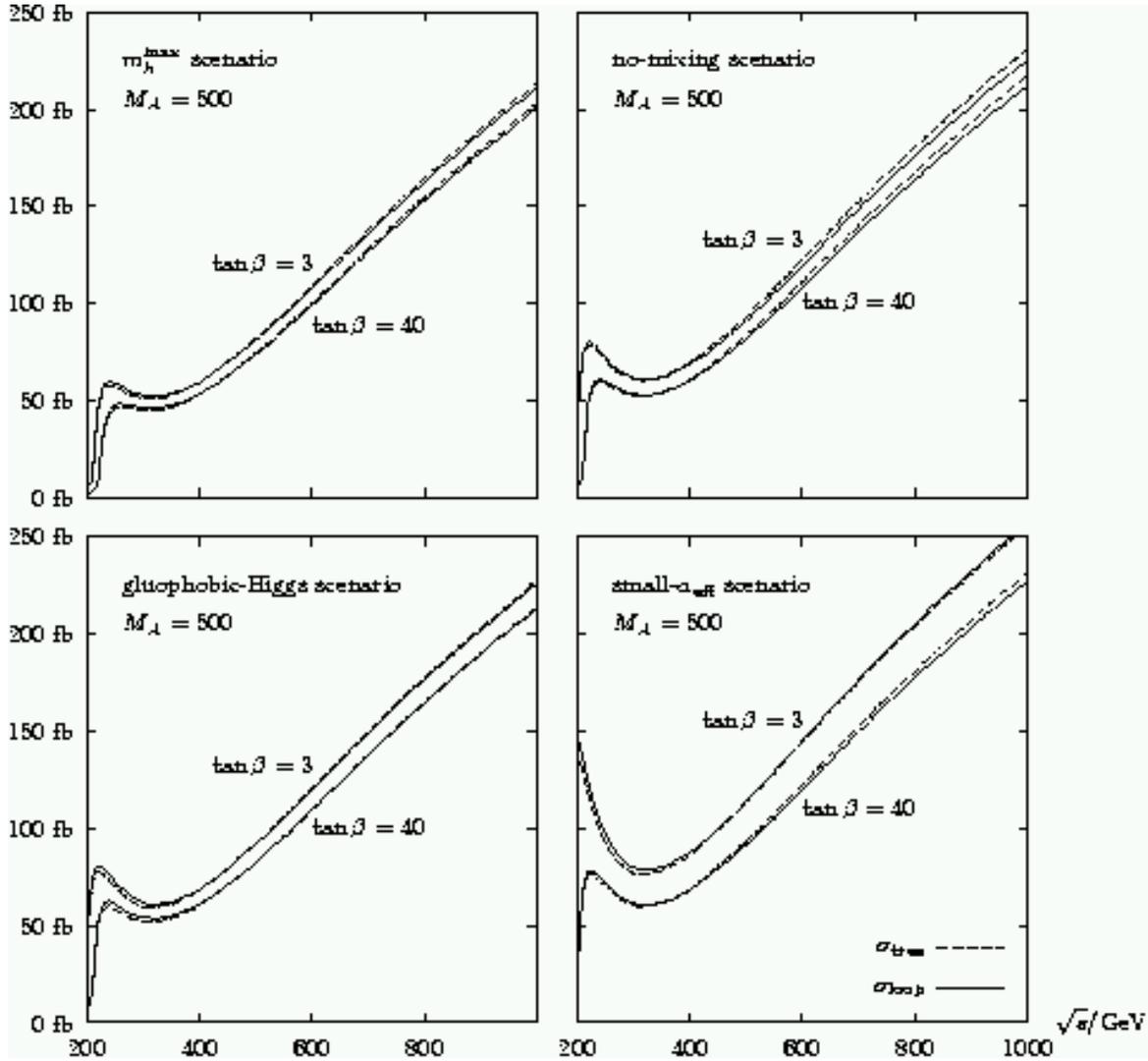}
\caption{
The tree-level and the one-loop cross sections for 
\eennh,
$\si_h^0$ and $\si_h^1$, are shown as a function of $\sqrt{s}$ in the
four benchmark scenarios for $\MA = 500 \gev$ and
$\tb = 3, 40$. 
}
\label{fig:sih_sqrts}
\end{center}
\end{figure}

\reffi{fig:sih_sqrts} shows our results for \eennh\ in the four 
benchmark scenarios as a function of $\sqrt{s}$ for 
$\MA = 500 \gev$ and $\tb = 3, 40$. Note that the
difference in the cross sections for the four benchmark scenarios for
given $\MA$ and $\tb$ is entirely due to SUSY loop corrections
(which, as explained above, affect in particular the value of $\Mh$).

\begin{figure}[ht!]
\begin{center}
\includegraphics[width=13.0cm]{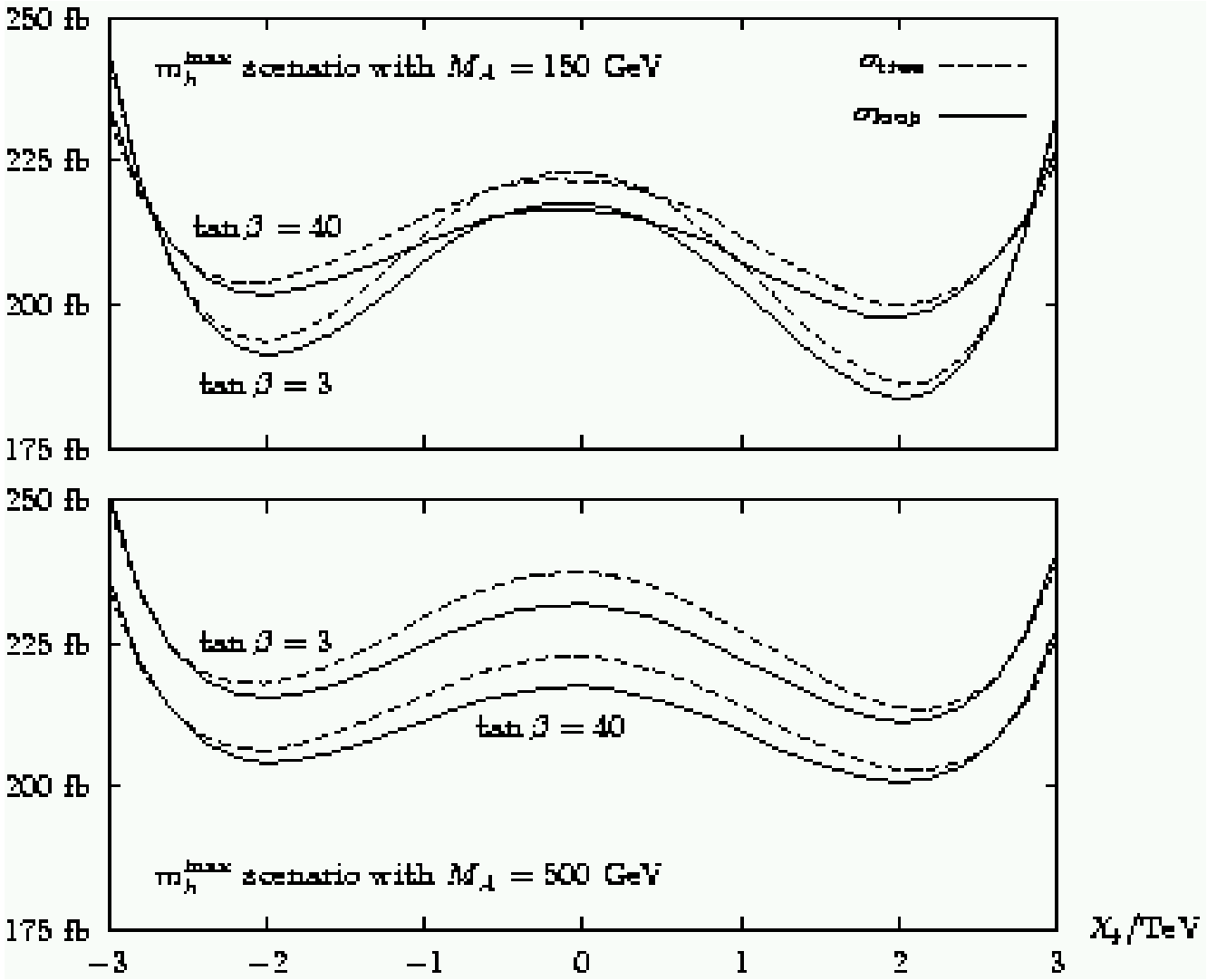}
\caption{
The tree-level and the one-loop cross sections for 
\eennh,
$\si_h^0$ and $\si_h^1$, are shown as a function of $\Xt$.
In the upper (lower) plot $\MA$ has been set to $\MA = 150$ $(500) \gev$, 
$\tb$ is fixed at $\tb = 3, 40$. The other parameters are as given 
in \refeq{mhmax}.
}
\label{fig:sih_Xt}
\end{center}
\end{figure}

The numerically important effects of the Higgs propagator corrections
become apparent in particular 
from \reffi{fig:sih_Xt}, where the tree-level 
and the one-loop
cross sections are shown as a function of $\Xt$, i.e.\ the mixing in the
scalar top sector. The plots are given for the four combinations of 
$\MA = 150, 500 \gev$ and $\tb = 3, 40$, and the other parameters 
(besides $\Xt$) 
are chosen as in the 
$\mhmax$ scenario (see the Appendix). The variation of the tree-level
cross sections 
indicates the effect of the Higgs propagator corrections affecting both
the value of $\Mh$ and the 
Higgs coupling to gauge bosons. These
corrections can change the cross section by up to $\sim 25\%$,
while the other loop corrections typically stay below 2.5\%.

\begin{figure}[ht!]
\begin{center}
\includegraphics[width=15.8cm]{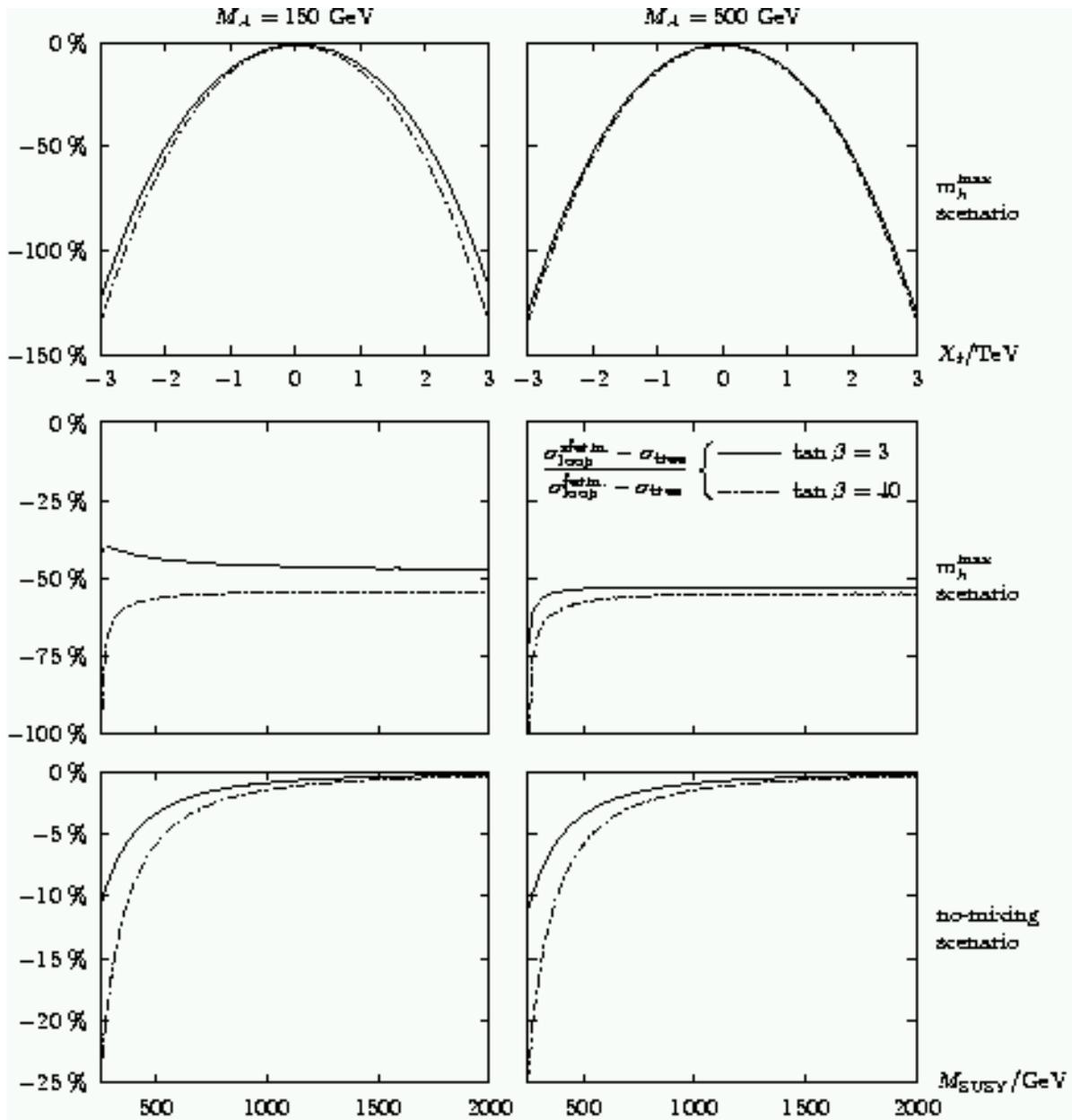}
\vspace{1em}
\caption{
The relative size of the purely sfermionic loop corrections to $\si^1_h$
as compared to the purely fermionic \onel\ corrections is shown as a 
function of $\Xt$ (upper) and $\msusy$ (middle and lower row) for all
combinations  
of $\MA = 150, 500 \gev$ and $\tb = 3, 40$. The other parameters are
chosen as in the $\mhmax$~(upper and middle) or as in the no-mixing
scenario (lower row). 
}
\label{fig:sih_sf_effects}
\end{center}
\vspace{3em}
\end{figure}

Finally, in \reffi{fig:sih_sf_effects} we analyze the relative
importance of the purely sfermionic loop corrections (corresponding to
the Feynman diagrams with sfermion loops in 
\reffis{fig:WWhHvert}--\ref{fig:enuWCT}; as before, the Higgs propagator
corrections absorbed into the tree-level result contain both fermion- and
sfermion-loop contributions). These corrections
constitute, as explained earlier, a UV-finite and gauge-invariant
subset of the loop contributions. The relative size of the sfermion
corrections as compared to the purely fermionic \onel\ corrections 
is shown in \reffi{fig:sih_sf_effects}. The upper row shows the
relative size as a function of $\Xt$ in the 
$\mhmax$ scenario for all combinations of $\MA = 150, 500 \gev$ and
$\tb = 3, 40$. While for $\Xt \approx 0$, i.e.\ for small splitting in
the scalar top sector, the sfermionic corrections are small, their
contribution becomes more important for increasing $|\Xt|$.
They have the opposite sign of
the purely fermionic corrections and thus partially compensate
their effects. For
$|\Xt/\msusy| \approx 2$ the sfermionic corrections are about 
half as large
as the fermion corrections. For very large $|\Xt|$ (which also lowers
$\Mh$ substantially) they can become even bigger than the fermionic ones.

In the lower part of \reffi{fig:sih_sf_effects} (middle and lower row)
we analyze the relative size of the purely sfermionic corrections in
the $\mhmax$~(middle) and the no-mixing scenario (lower row) as a
function of $\msusy$.  
In the no-mixing scenario, for increasing $\msusy$ the relative size
of the sfermion corrections 
becomes smaller, as can be expected in the decoupling
limit~\cite{decoupling1l,decoupling2l}.
In the $\mhmax$ scenario, however, the situation is different. Here
$\Xt \approx \At$ is fixed to $\At \approx \Xt = 2 \, \msusy$. In the
$h\Stope\Stopz$ vertex, being $\sim \At\Ca + \mu\Sa$, the coupling is
proportional to the SUSY mass scale. This results in a term 
$\sim \At/\msusy$ in the \onel\ corrected $WWh$ vertex, which for
large $\msusy$ goes to a constant and can be of the order of the
purely fermionic correction. The cross section then behaves as 
$\sim \Xt^2/\msusy^2$ as can be seen in the upper row of
\reffi{fig:sih_sf_effects}. 

\subsection{Heavy $\cp$-even Higgs-boson production}
\label{subsec:Hprod}

We now investigate the effects of loop corrections on the cross section
for heavy $\cp$-even Higgs-boson production in the MSSM. As explained
above, these corrections are of particular interest in the decoupling
region, i.e.\ for large values of $\MA$.
If $\MA \lsim \sqrt{s}/2$, the heavy Higgs bosons can be pair-produced at
the LC via $e^+e^- \to Z^* \to HA$. Beyond this kinematical limit,
$H$~production is in principle possible via the $WW$-fusion and the
Higgs-strahlung channels. This production mechanism is heavily
suppressed at tree-level, however, owing to the decoupling property of
the $H$~coupling to gauge bosons. If loop-induced contributions turn out
to be sizable in the mass range $\MH > \sqrt{s}/2$, an enhanced reach of
the LC for $H$~production could result.

\begin{figure}[ht!]
\vspace{3em}
\mbox{} \hspace{5em}
\includegraphics[width=14.5cm]{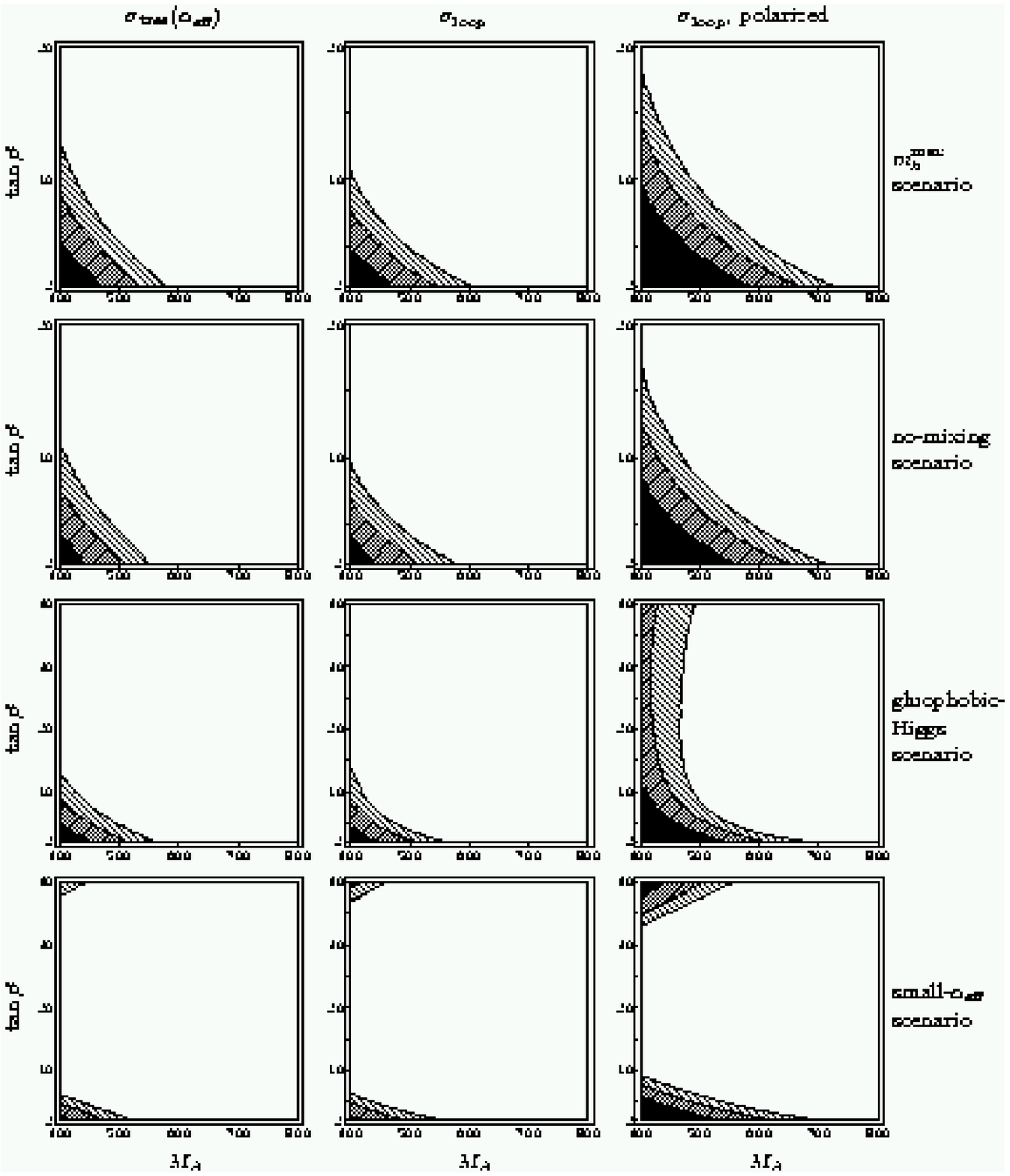}
\vspace{1em}
\caption{
The tree-level cross section for \eennH, $\si^0_H(\aeff)$, 
using an $\aeff$~approximation (left), and the one-loop cross 
sections, $\si^1_H$, without (middle), and with beam polarization
(right column), are shown in the $\MA$--$\tb$ plane for four benchmark
scenarios for $\sqrt{s} = 1 \tev$. The different shadings correspond to:
white: $\si \le 0.01 \fb$, light shaded: $0.01 \fb \le \si \le 0.02 \fb$, 
dark shaded: $0.02 \fb \le \si \le 0.05 \fb$, black: $\si \ge 0.05 \fb$.
}
\label{fig:H_benchmark}
\end{figure}

In \reffi{fig:H_benchmark} we first compare the tree-level cross section
evaluated in the $\aeff$~approximation, $\si_H^0(\aeff)$ (left column),
and the one-loop cross section according to \refeq{siH1}, $\si_H^1$
(middle column), in the four benchmark scenarios. 
For phenomenological analyses of MSSM Higgs-boson production in this 
channel the cross
section has so far mostly been evaluated using an $\aeff$~Born
approximation (see also \refse{subsec:hprod}).

We concentrate on the case of $\sqrt{s} = 1 \tev$. Since we are
interested in $\MH \approx \MA > \sqrt{s}/2 = 500 \gev$, we focus on
the region $400 \gev \le \MA \le 900 \gev$, and scan over the whole $\tb$
region. For a LC like TESLA, the anticipated integrated luminosity is
of \order{2 \iabm}. For this luminosity, 
a production cross section of about $\si_H = 0.01 \fb$ constitutes a
lower limit for the observation of the heavy $\cp$-even Higgs boson.
(For the scenarios discussed below, the dominant decay channel of $H$
is the decay into $t \bar t$ or $b \bar b$, depending on the value of
$\tb$, and also sizable branching ratios into SUSY particles are 
possible in some regions of parameter space; a more detailed simulation of 
this process should of course take into account the impact of the decay
characteristics on the lower limit of observability, while in this work
we use the approximation of a universal limit.)
The area with $\si_H \le 0.01 \fb$ is shown in white in
\reffi{fig:H_benchmark}.  
In the four benchmark scenarios shown in \reffi{fig:H_benchmark}, the
inclusion of the loop corrections that go beyond the $\aeff$~Born
approximation turns out to have only a moderate effect on the
area in the $\MA$--$\tb$ plane in which $H$~production could be
observable%
\footnote{
The neglected higher-order corrections (e.g.\ from bosonic loops and
initial state radiation) can in principle be sizable, see e.g.\ Fig.~3
in the second reference in \cite{eennHSMf1l}. However, they constitute
corrections to 
the Born matrix element, i.e.\ they contain the $WWH$ coupling and
thus show the decoupling for the $\MA$~values considered
here. Therefore the neglected corrections would not substantially
change the contours shown in this section.
}%
. While for the $\mhmax$ and the no-mixing scenario the area
is slightly decreased to smaller $\tb$ values and somewhat enlarged to
higher $\MA$ values, the area is slightly decreased in $\tb$ in the
gluophobic-Higgs scenario (and stays approximately the same in $\MA$),
while the area is slightly enlarged both in $\MA$ and $\tb$ in 
the small-$\aeff$ scenario. 
For the four benchmark scenarios, an
observation with $\MH > 500 \gev$ is only possible for low $\tb$, 
$\tb \lsim 5$, where the LC reach in $\MH$ can be extended by up to 
$100 \gev$. It should be noted at this point that while the area of
observability is modified only slightly in the plots as a consequence of 
including the loop corrections, the relative changes between the
tree-level and the one-loop values of the cross sections can be very large,
owing to the suppressed $WWH$ coupling in the tree-level cross section.

The prospects for observing a heavy Higgs boson beyond the kinematical
limit of the $HA$ pair production channel become more favorable, 
however, if polarized beams are
used. The cross section becomes enhanced for left-handedly polarized
electrons and right-handedly polarized positrons. While a 100\%
polarization results in a cross section that is enhanced roughly by a
factor of 4, more realistic values of 80\% polarization for
electrons and 60\% polarization for positrons~\cite{polarization}
would yield roughly an enhancement by a factor of 3. 
The right column of \reffi{fig:H_benchmark} shows the four benchmark
scenarios with 100\% polarization of both beams. The area in the 
$\MA$--$\tb$ plane in which observation of the $H$~boson might become
possible is strongly increased in this case. In the
$\mhmax$ and the no-mixing scenario, $H$ observation could be possible
for small $\tb$ up to $\MA \lsim 700 \gev$. In the gluophobic-Higgs
and the small-$\aeff$ scenario this effect is somewhat smaller. In the
latter scenario the discovery of a heavy Higgs boson in the parameter
region $\MA > 500 \gev$ will become possible also for large $\tb$, $\tb
\gsim 35$.

While without the inclusion of beam polarization we do not find a
significant enhancement of the LC discovery reach as a consequence of 
the loop corrections in the four benchmark scenarios analyzed in
\reffi{fig:H_benchmark}, this behavior changes in other
regions of the MSSM parameter space. As a particular example, we
investigate the $\MA$--$\tb$ plane in the ``$\siHenh$'' (``enhanced cross
section'') scenario, which is defined by
\BE 
\msusy = 350 \gev, \quad \mu = 1000 \gev, 
\label{xsmax}
\end{equation}
with the other parameters as in the $\mhmax$ scenario, \refeq{mhmax}.
The $\siHenh$ scenario is thus characterized by a relatively small
value of $\msusy$ 
and a relatively large value of~$\mu$. Large $\tb$ values, $\tb \gsim 30$, 
can result in low and experimentally ruled-out $\Sbot$~masses and are
therefore omitted.

\begin{figure}[htb!]
\begin{center}
\includegraphics{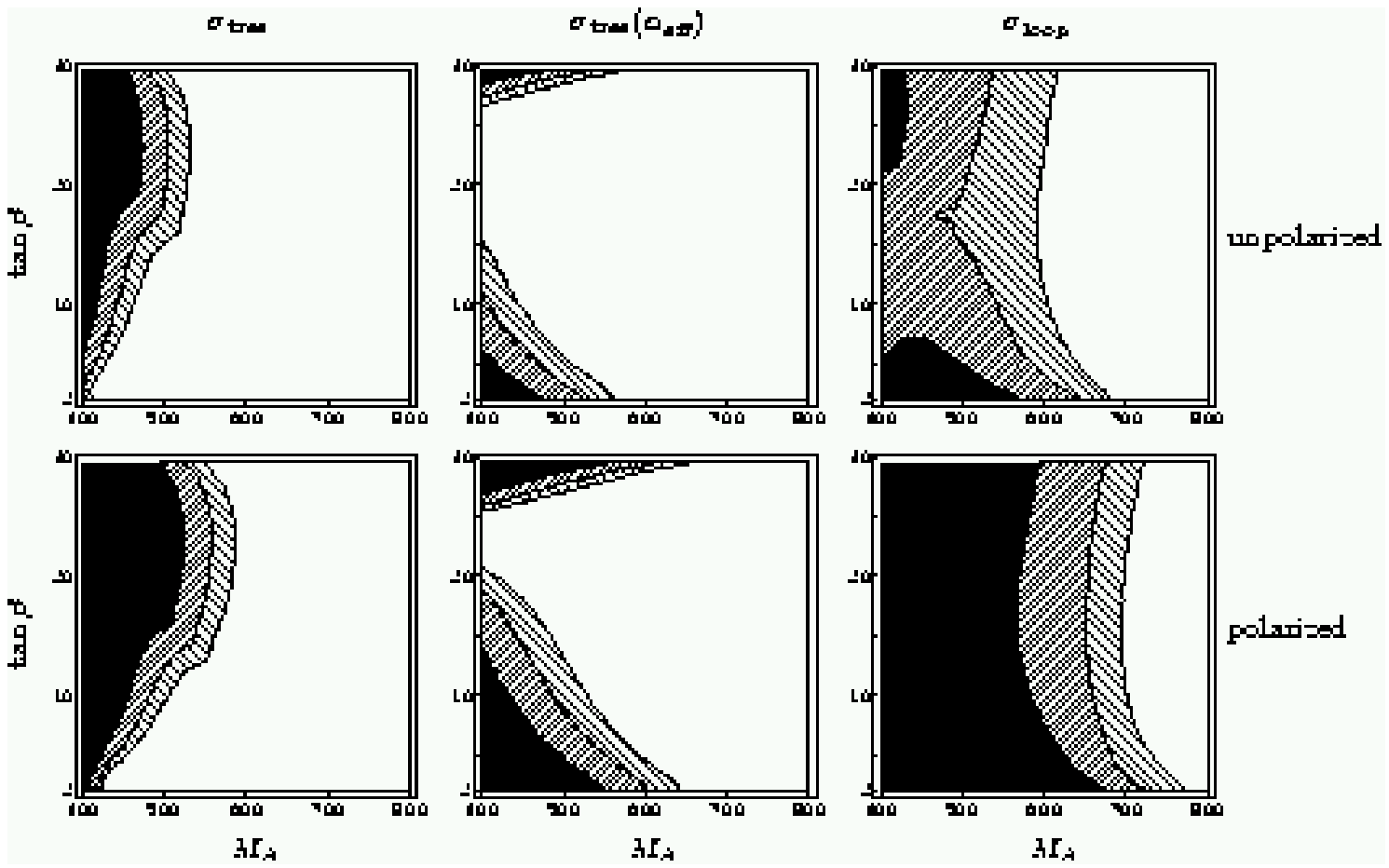}
\caption{
The cross sections for \eennH\
are shown in the $\MA$--$\tb$ plane for the $\siHenh$ scenario, 
\refeq{xsmax}. The tree-level cross section (left) including the
finite wave-function corrections is compared to the
$\aeff$~approximation (middle) and the \onel\ corrected cross section
(right column). 
The upper (lower) row shows the production cross section for
unpolarized (100\% polarized) electron and positron beams. 
The color coding is as in \reffi{fig:H_benchmark}.
}
\label{fig:xsmax}
\end{center}
\end{figure}

\begin{figure}[ht!]
\vspace{-2em}
\begin{center}
\includegraphics[width=8.5cm]{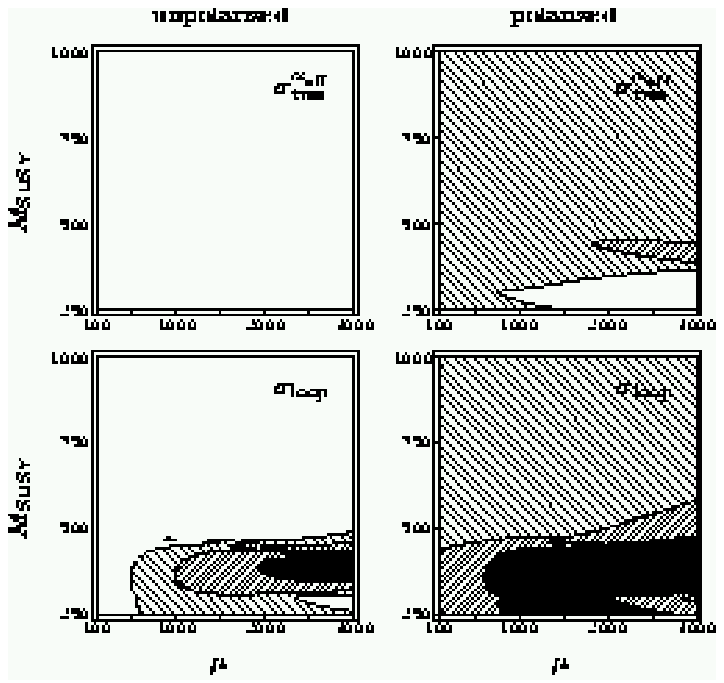}
\caption{
The tree-level cross section for \eennH, $\si^0_H(\aeff)$, 
using an $\aeff$~approximation (upper), and the one-loop cross 
section, $\si^1_H$ (lower row), are
shown in the $\mu$--$\msusy$ plane for the parameters of
\refeq{mhmax} and $\MA = 600 \gev$ and $\tb = 4$. The left column shows
the unpolarized case, while in the right column the effects of beam
polarization are included. The color coding
is as in \reffi{fig:H_benchmark}.
}
\label{fig:mumsusy}
\end{center}
\end{figure}

\begin{figure}[htb!]
\vspace{-1em}
\begin{center}
\includegraphics[width=8.5cm]{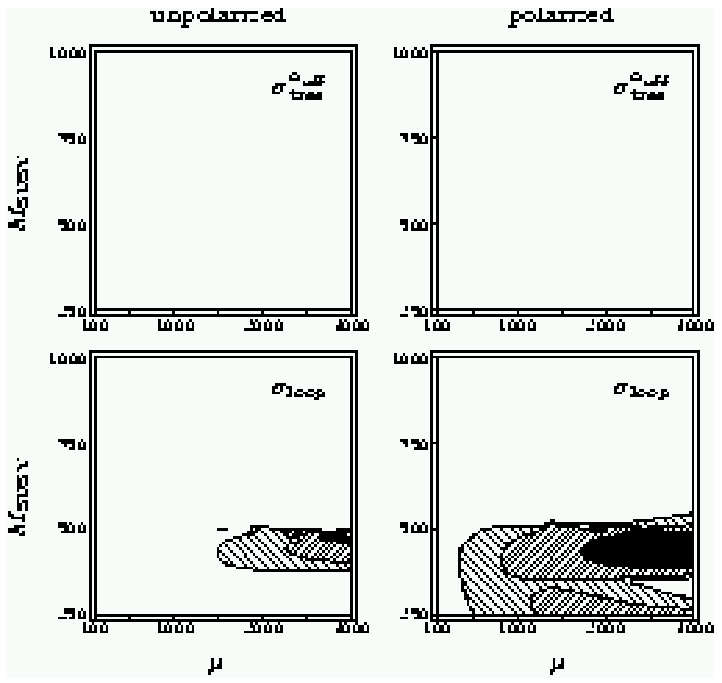}
\caption{
The tree-level cross section for \eennH, $\si^0_H(\aeff)$, 
using an $\aeff$~approximation (upper), and the one-loop cross 
section, $\si^1_H$ (lower row), is 
shown in the $\mu$--$\msusy$ plane for the parameters of
\refeq{mhmax} and $\MA = 700 \gev$ and $\tb = 4$. The left column shows
the unpolarized case, while in the right column the effects of beam
polarization are included. The color coding
is as in \reffi{fig:H_benchmark}.
}
\label{fig:mumsusy_MA700}
\end{center}
\vspace{-1em}
\end{figure}

In the upper row of \reffi{fig:xsmax} we compare $\si_H^0$,
$\si_H^0(\aeff)$, and $\si_H^1$ for the parameters of the
$\siHenh$ scenario, \refeq{xsmax}, in the unpolarized case. 
The figure shows that both the inclusion of the finite Higgs propagator
wave-function corrections, see \refeqs{eq:WWhct2}-(\ref{eq:WWHct2}),
as compared to the $\aeff$ approximation  
(left vs.\ middle column) and of the genuine one-loop corrections
(right vs.\ left column) is very important in this scenario. 
According to the $\aeff$ Born approximation, the parameter area in which 
observation of $H$ is possible would not be significantly larger than in the
benchmark scenarios discussed in \reffi{fig:H_benchmark}. For $\MA \gsim
500 \gev$, observation of the $H$ boson is only possible for either rather
small, $\tb \lsim 5$, or rather large, $\tb \gsim 28$, values of $\tb$.
Inclusion of the finite Higgs propagator wave-function corrections,
which ensure the correct on-shell properties of the outgoing Higgs
boson, changes the situation considerably. While for small and large
values of $\tb$ the additional corrections suppress the \eennH\ cross section,
restricting the $H$ observability to values of $\MA$ below 500~GeV,
observation of the heavy $\cp$-even Higgs boson of the MSSM becomes
possible for  
$\MA\lsim 550 \gev$ for a significant range of intermediate values of $\tb$. 
Including also the genuine one-loop
corrections (right column) leads to a further drastic enhancement of the
parameter space in which the $H$ boson could be observed. The
observation of the $H$ boson will be possible in this scenario for all
values of $\tb$ if $\MA \lsim 600 \gev$, i.e.\ the discovery reach of
the LC in this case is enhanced by about 100~GeV compared to the 
$HA$ pair production channel.

The prospects in this scenario for observation of the heavy $\cp$-even 
Higgs boson of the MSSM become even more favorable if
polarized beams are used.
The lower row of \reffi{fig:xsmax} shows the situation with 100\%
polarization of both beams for the 
$\siHenh$ scenario, \refeq{xsmax}. While in this case the
tree-level result (both for the case including the finite wave-function 
corrections and for the $\aeff$~approximation) gives rise to
observable rates for $\MA \lsim 600 \gev$ for a certain range of $\tb$
values only, the further genuine loop
corrections enhance the cross section significantly. In this situation
the observation of the heavy $\cp$-even Higgs boson might be possible for
values of $\MA$ up to about 700--750~GeV for all $\tb$ values,
corresponding to an enhancement of the LC reach by more than $200 \gev$.
Cross-section values in excess of $0.05 \fb$ are obtained in this example
for all $\tb$ values for $\MA \lsim 600 \gev$.

In order to investigate whether this result is a consequence of a very
special choice of SUSY parameters or a more general feature, in
\reffi{fig:mumsusy} we choose the parameters of the $\mhmax$ scenario for
a fixed combination of $\MA$ and $\tb$, $\MA = 600 \gev$, $\tb = 4$, but
scan over $\mu$ and $\msusy$. 
The choice $\MH \approx \MA = 600 \gev$
implies that the $HA$ production channel is clearly beyond the reach of a
1~TeV LC. The upper row of \reffi{fig:mumsusy} shows that according
to the tree-level cross section (using the $\aeff$ approximation) an
observable rate for a heavy $\cp$-even Higgs boson with $\MA = 600 \gev$
cannot be found for any of the scanned values of $\mu$ and $\msusy$.
Inclusion of the further loop corrections changes this
situation significantly and gives rise to observable rates
in this example for nearly all $\msusy \lsim 500 \gev$ if 
$\mu \gsim 500 \gev$. 
The visible ``structure'' at $\msusy \approx 500 \gev$ is the result
of several competing effects that affect the finite Higgs
wave-function corrections. 

The same analysis, but with 100\% polarization of both beams, 
is shown in the right column
of \reffi{fig:mumsusy}. The tree-level $\aeff$~approximation
results in an observable rate in nearly the whole plane apart
from the area with $\mu \gsim 1000 \gev$ and $\msusy \lsim 350 \gev$.
Adding the loop corrections again improves the
situation. No unobservable holes remain in the $\mu$--$\msusy$ plane,
i.e.\ the heavy $\cp$-even Higgs boson with $\MH \approx 600 \gev$
should be visible at a 1~TeV LC with (idealized) beam polarization
in this scenario. The
production cross section is larger than 
$0.02 \fb$ for all $\mu$ and $\msusy \lsim 500 \gev$ and mostly even
larger than $0.05 \fb$ for $\mu \gsim 500 \gev$.

The same analysis, but for an even larger Higgs boson mass scale, 
$\MA = 700 \gev$ is shown in \reffi{fig:mumsusy_MA700}. For the
$\aeff$~approximation the production cross section is lower than 
$0.01 \fb$ in the whole $\mu$--$\msusy$-plane. However, for the loop
corrected cross section, especially including polarization, the heavy
$\cp$-even 
Higgs boson could be observable for all $\mu$~values for 
$\msusy \lsim 500 \gev$.

\begin{figure}[ht!]
\vspace{-3em}
\mbox{} \hspace{3em}
\includegraphics[width=14.5cm]{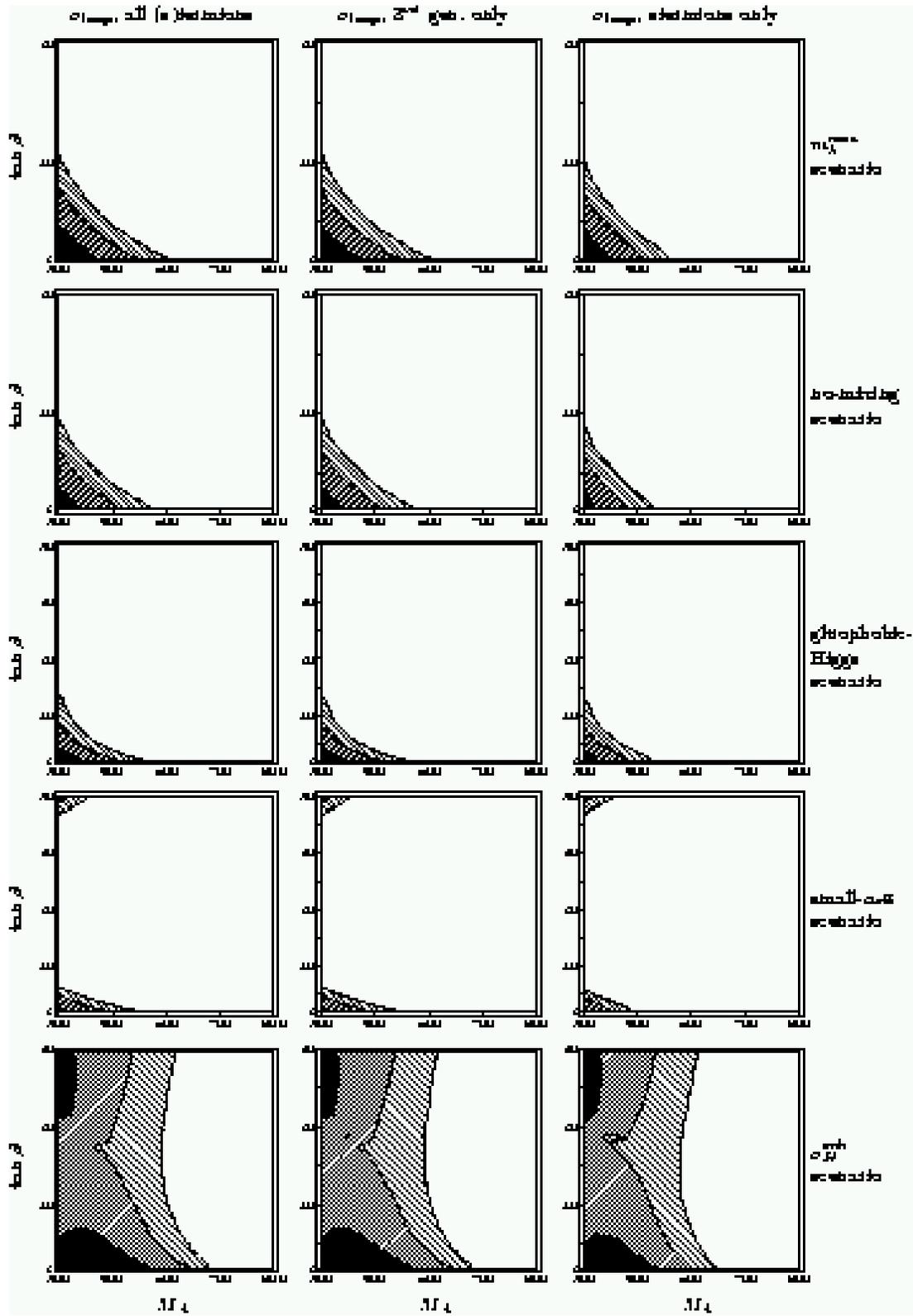}
\caption{
$\si_H^1$ is shown in the $\MA$--$\tb$ plane for the four benchmark scenarios
defined in \refeqs{mhmax}--(\ref{smallaeff})
and the $\siHenh$ scenario for $\sqrt{s} = 1 \tev$ in the unpolarized
case. The \onel\ result containing the corrections from all fermion and
sfermion loops (left) is compared with the result from only
third-generation fermions and sfermions (middle) and the purely
sfermionic corrections from all families (right column). 
The color coding is as in \reffi{fig:H_benchmark}.
}
\label{fig:H_benchmark_contrib}
\vspace{-2em}
\end{figure}

In \reffi{fig:H_benchmark_contrib} we compare for all five scenarios
analyzed in this section the results for the cross section when different
parts of the generic one-loop corrections are taken into account. In all
cases the full result for the Higgs propagator corrections being
absorbed in the lowest-order cross section are employed.
The left column shows the result containing the
corrections from all fermion and sfermion loops, $\si_H^1$, which is
repeated from previous plots for comparison purposes.
The middle column shows the cross-section prediction based on 
taking into account only the corrections
from the third-generation fermions and sfermions,
which have turned out to be the leading
corrections for the light Higgs-boson production. The result shown in the
right column have been obtained by including only the purely sfermionic
contributions from all generations, i.e.\ the fermion-loop corrections
are omitted in this case. As expected from \reffi{fig:H_benchmark},
in the four scenarios defined in the Appendix, 
where the loop corrections turned out to modify the parameter regions
in which $H$ observation becomes possible only slightly, omitting the 
contributions of the first two generations of fermions and sfermions 
and of the fermion loops
of all three generations does not lead to a qualitative change in the
$H$ discovery reach. In the $\siHenh$ scenario, on the other hand, the 
genuine one-loop corrections had a considerable impact on the area in
the $\MA$--$\tb$ plane in which $H$ observation is possible, see 
\reffi{fig:xsmax}. The result displayed in the middle column for this
scenario shows that the bulk of the corrections comes from the third
generation of fermions and sfermions, i.e.\ omission of the first two
generations does not lead to significant effects in the $\MA$--$\tb$ plane.
The result in the right column for this scenario shows furthermore that
the omission of all fermion-loop corrections leads only to very moderate
changes of the parameter regions where $H$ observation is possible. As a
consequence, the by far dominant corrections in this scenario can be
identified as the ones from the sfermions of the third generation.
This is contrary to the $h$~production, where we found that the
fermionic corrections are mostly larger than the sfermionic ones.


\section{Single charged Higgs boson production}

In this section the dominant contributions to the process
\eeenH~\cite{eeenH} are briefly described.
Since there is no $\{\ga,Z\}W^\pm H^\mp$ coupling, the single
charged-Higgs production starts at the one-loop level. As for
$e^+e^- \to \nu \bar\nu \{h,H\}$, we take into
account the leading corrections arising from the full set of SM fermion and
sfermion loops. Bremsstrahlung diagrams are suppressed by an additional
\order{\al} as compared to the leading contribution and have thus been
omitted. 


\subsection{Calculation}

The loop corrections that enter the process \eeenH\ at the \onel\ level
are generically depicted in \reffis{fig:FD_eeenHself},
\ref{fig:FD_eeenH}. 
The contributions 
involve all corrections from fermion and sfermion loops (which give
contributions only to self-energies and vertices). 
Contributions $\propto m_e/\MW$ were neglected.

\begin{figure}[ht]
\begin{center}
\unitlength=1.4bp%
\begin{feynartspicture}(324,200)(3, 2)

\FADiagram{}
\FAProp(0.,15.)(3.,9.)(0.,){/Straight}{-1}
\FALabel(2.40636,12.2132)[bl]{$e$}
\FAProp(0.,5.)(3.,9.)(0.,){/Straight}{1}
\FALabel(0.74,7.45)[br]{$e$}
\FAProp(20.,17.)(13.5,16.5)(0.,){/Straight}{1}
\FALabel(16.6311,17.8154)[b]{$e$}
\FAProp(20.,10.)(6.,7.)(0.,){/ScalarDash}{-1}
\FALabel(10.9524,6.83555)[t]{$H$}
\FAProp(20.,3.)(13.5,16.5)(0.,){/Straight}{-1}
\FALabel(16.6453,12.5785)[bl]{$\nu_e$}
\FAProp(3.,9.)(6.,7.)(0.,){/Sine}{0}
\FALabel(4.7,7.19032)[tr]{$\gamma,Z$}
\FAProp(13.5,16.5)(11.45,14.1)(0.,){/Sine}{1}
\FALabel(11.7764,15.8267)[br]{$W$}
\FAProp(6.,7.)(8.4,10.)(0.,){/ScalarDash}{-1}
\FALabel(6.46965,8.98828)[br]{$H$}
\FAProp(11.45,14.1)(8.4,10.)(-0.8,){/Straight}{-1}
\FALabel(12.3286,11.3849)[tl]{$f'$}
\FAProp(11.45,14.1)(8.4,10.)(0.8,){/Straight}{1}
\FALabel(7.52137,13.7151)[br]{$f$}
\FAVert(3.,9.){0}
\FAVert(13.5,16.5){0}
\FAVert(6.,7.){0}
\FAVert(11.45,14.1){0}
\FAVert(8.4,10.){0}

\FADiagram{}
\FAProp(0.,15.)(3.,9.)(0.,){/Straight}{-1}
\FALabel(2.40636,12.2132)[bl]{$e$}
\FAProp(0.,5.)(3.,9.)(0.,){/Straight}{1}
\FALabel(0.74,7.45)[br]{$e$}
\FAProp(20.,17.)(13.5,16.5)(0.,){/Straight}{1}
\FALabel(16.6311,17.8154)[b]{$e$}
\FAProp(20.,10.)(6.,7.)(0.,){/ScalarDash}{-1}
\FALabel(10.9524,6.83555)[t]{$H$}
\FAProp(20.,3.)(13.5,16.5)(0.,){/Straight}{-1}
\FALabel(16.6453,12.5785)[bl]{$\nu_e$}
\FAProp(3.,9.)(6.,7.)(0.,){/Sine}{0}
\FALabel(4.7,7.19032)[tr]{$\gamma,Z$}
\FAProp(13.5,16.5)(11.45,14.1)(0.,){/Sine}{1}
\FALabel(11.7764,15.8267)[br]{$W$}
\FAProp(6.,7.)(8.4,10.)(0.,){/ScalarDash}{-1}
\FALabel(6.46965,8.98828)[br]{$H$}
\FAProp(11.45,14.1)(8.4,10.)(-0.8,){/ScalarDash}{-1}
\FALabel(12.3286,11.3849)[tl]{$\tilde f'$}
\FAProp(11.45,14.1)(8.4,10.)(0.8,){/ScalarDash}{1}
\FALabel(7.52137,13.7151)[br]{$\tilde f$}
\FAVert(3.,9.){0}
\FAVert(13.5,16.5){0}
\FAVert(6.,7.){0}
\FAVert(11.45,14.1){0}
\FAVert(8.4,10.){0}

\FADiagram{}
\FAProp(0.,15.)(3.,10.)(0.,){/Straight}{-1}
\FALabel(0.650886,12.1825)[tr]{$e$}
\FAProp(0.,5.)(3.,10.)(0.,){/Straight}{1}
\FALabel(2.34911,7.18253)[tl]{$e$}
\FAProp(20.,17.)(15.5,5.5)(0.,){/Straight}{1}
\FALabel(19.2028,12.2089)[l]{$e$}
\FAProp(20.,10.)(8.,12.)(0.,){/ScalarDash}{-1}
\FALabel(14.2548,12.0489)[b]{$H$}
\FAProp(20.,3.)(15.5,5.5)(0.,){/Straight}{-1}
\FALabel(17.4773,3.37506)[tr]{$\nu_e$}
\FAProp(12.,8.5)(15.5,5.5)(0.,){/Sine}{-1}
\FALabel(13.2213,6.30315)[tr]{$W$}
\FAProp(12.,8.5)(8.,12.)(0.,){/ScalarDash}{1}
\FALabel(9.45932,9.56351)[tr]{$H$}
\FAProp(3.,10.)(8.,12.)(0.,){/Sine}{0}
\FALabel(4.92434,11.9591)[b]{$\gamma,Z$}
\FAVert(3.,10.){0}
\FAVert(15.5,5.5){0}
\FAVert(8.,12.){0}
\FAVert(12.,8.5){1}

\FADiagram{}
\FAProp(0.,15.)(10.,16.5)(0.,){/Straight}{-1}
\FALabel(4.77007,16.8029)[b]{$e$}
\FAProp(0.,5.)(10.,3.5)(0.,){/Straight}{1}
\FALabel(4.77007,3.19715)[t]{$e$}
\FAProp(20.,17.)(10.,16.5)(0.,){/Straight}{1}
\FALabel(14.9226,17.8181)[b]{$e$}
\FAProp(20.,10.)(10.,13.5)(0.,){/ScalarDash}{-1}
\FALabel(16.6326,12.406)[b]{$H$}
\FAProp(20.,3.)(10.,3.5)(0.,){/Straight}{-1}
\FALabel(14.9226,2.18193)[t]{$\nu_e$}
\FAProp(10.,16.5)(10.,13.5)(0.,){/Sine}{0}
\FALabel(11.07,15.)[l]{$\gamma,Z$}
\FAProp(10.,3.5)(10.,6.)(0.,){/Sine}{1}
\FALabel(11.07,4.75)[l]{$W$}
\FAProp(10.,13.5)(10.,11.)(0.,){/ScalarDash}{-1}
\FALabel(8.93,12.25)[r]{$H$}
\FAProp(10.,6.)(10.,11.)(0.8,){/Straight}{-1}
\FALabel(13.07,8.5)[l]{$f'$}
\FAProp(10.,6.)(10.,11.)(-0.8,){/Straight}{1}
\FALabel(6.93,8.5)[r]{$f$}
\FAVert(10.,16.5){0}
\FAVert(10.,3.5){0}
\FAVert(10.,13.5){0}
\FAVert(10.,6.){0}
\FAVert(10.,11.){0}

\FADiagram{}
\FAProp(0.,15.)(10.,16.5)(0.,){/Straight}{-1}
\FALabel(4.77007,16.8029)[b]{$e$}
\FAProp(0.,5.)(10.,3.5)(0.,){/Straight}{1}
\FALabel(4.77007,3.19715)[t]{$e$}
\FAProp(20.,17.)(10.,16.5)(0.,){/Straight}{1}
\FALabel(14.9226,17.8181)[b]{$e$}
\FAProp(20.,10.)(10.,13.5)(0.,){/ScalarDash}{-1}
\FALabel(16.6326,12.406)[b]{$H$}
\FAProp(20.,3.)(10.,3.5)(0.,){/Straight}{-1}
\FALabel(14.9226,2.18193)[t]{$\nu_e$}
\FAProp(10.,16.5)(10.,13.5)(0.,){/Sine}{0}
\FALabel(11.07,15.)[l]{$\gamma,Z$}
\FAProp(10.,3.5)(10.,6.)(0.,){/Sine}{1}
\FALabel(11.07,4.75)[l]{$W$}
\FAProp(10.,13.5)(10.,11.)(0.,){/ScalarDash}{-1}
\FALabel(8.93,12.25)[r]{$H$}
\FAProp(10.,6.)(10.,11.)(0.8,){/ScalarDash}{-1}
\FALabel(13.07,8.5)[l]{$\tilde f'$}
\FAProp(10.,6.)(10.,11.)(-0.8,){/ScalarDash}{1}
\FALabel(6.93,8.5)[r]{$\tilde f$}
\FAVert(10.,16.5){0}
\FAVert(10.,3.5){0}
\FAVert(10.,13.5){0}
\FAVert(10.,6.){0}
\FAVert(10.,11.){0}

\FADiagram{}
\FAProp(0.,15.)(10.,16.)(0.,){/Straight}{-1}
\FALabel(4.84577,16.5623)[b]{$e$}
\FAProp(0.,5.)(10.,4.)(0.,){/Straight}{1}
\FALabel(4.84577,3.43769)[t]{$e$}
\FAProp(20.,17.)(10.,16.)(0.,){/Straight}{1}
\FALabel(14.8458,17.5623)[b]{$e$}
\FAProp(20.,10.)(10.,12.)(0.,){/ScalarDash}{-1}
\FALabel(15.304,12.0399)[b]{$H$}
\FAProp(20.,3.)(10.,4.)(0.,){/Straight}{-1}
\FALabel(15.1542,4.56231)[b]{$\nu_e$}
\FAProp(10.,8.)(10.,4.)(0.,){/Sine}{-1}
\FALabel(8.93,6.)[r]{$W$}
\FAProp(10.,8.)(10.,12.)(0.,){/ScalarDash}{1}
\FALabel(8.93,10.)[r]{$H$}
\FAProp(10.,16.)(10.,12.)(0.,){/Sine}{0}
\FALabel(8.93,14.)[r]{$\gamma,Z$}
\FAVert(10.,16.){0}
\FAVert(10.,4.){0}
\FAVert(10.,12.){0}
\FAVert(10.,8.){1}

\end{feynartspicture}
\caption{Generic self-energy diagrams for the process \eeenH.
\label{fig:FD_eeenHself}}
\end{center}
\end{figure}


\begin{figure}[htb!]
\vspace{3em}
\begin{center}
\unitlength=1.4bp%
\begin{feynartspicture}(324,160)(4,2)
\FADiagram{}
\FAProp(0.,15.)(9.,16.)(0.,){/Straight}{-1}
\FALabel(4.32883,16.5605)[b]{$e$}
\FAProp(0.,5.)(9.,4.)(0.,){/Straight}{1}
\FALabel(4.32883,3.43948)[t]{$e$}
\FAProp(20.,17.)(9.,16.)(0.,){/Straight}{1}
\FALabel(14.3597,17.5636)[b]{$e$}
\FAProp(20.,10.)(15.,10.)(0.,){/ScalarDash}{-1}
\FALabel(17.5,11.07)[b]{$H$}
\FAProp(20.,3.)(9.,4.)(0.,){/Straight}{-1}
\FALabel(14.3597,2.43637)[t]{$\nu_e$}
\FAProp(9.,16.)(9.,13.5)(0.,){/Sine}{0}
\FALabel(7.93,14)[r]{$\gamma,Z$}
\FAProp(9.,4.)(9.,6.5)(0.,){/Sine}{1}
\FALabel(7.93,5.5)[r]{$W$}
\FAProp(15.,10.)(9.,13.5)(0.,){/Straight}{-1}
\FALabel(12.301,12.6089)[bl]{$f$}
\FAProp(15.,10.)(9.,6.5)(0.,){/Straight}{1}
\FALabel(12.301,7.39114)[tl]{$f'$}
\FAProp(9.,13.5)(9.,6.5)(0.,){/Straight}{-1}
\FALabel(7.93,10.)[r]{$f$}
\FAVert(9.,16.){0}
\FAVert(9.,4.){0}
\FAVert(15.,10.){0}
\FAVert(9.,13.5){0}
\FAVert(9.,6.5){0}

\FADiagram{}
\FAProp(0.,15.)(9.,16.)(0.,){/Straight}{-1}
\FALabel(4.32883,16.5605)[b]{$e$}
\FAProp(0.,5.)(9.,4.)(0.,){/Straight}{1}
\FALabel(4.32883,3.43948)[t]{$e$}
\FAProp(20.,17.)(9.,16.)(0.,){/Straight}{1}
\FALabel(14.3597,17.5636)[b]{$e$}
\FAProp(20.,10.)(15.,10.)(0.,){/ScalarDash}{-1}
\FALabel(17.5,11.07)[b]{$H$}
\FAProp(20.,3.)(9.,4.)(0.,){/Straight}{-1}
\FALabel(14.3597,2.43637)[t]{$\nu_e$}
\FAProp(9.,16.)(9.,13.5)(0.,){/Sine}{0}
\FALabel(7.93,14)[r]{$\gamma,Z$}
\FAProp(9.,4.)(9.,6.5)(0.,){/Sine}{1}
\FALabel(7.93,5.5)[r]{$W$}
\FAProp(15.,10.)(9.,13.5)(0.,){/ScalarDash}{-1}
\FALabel(12.301,12.6089)[bl]{$\tilde f$}
\FAProp(15.,10.)(9.,6.5)(0.,){/ScalarDash}{1}
\FALabel(12.301,7.39114)[tl]{$\tilde f'$}
\FAProp(9.,13.5)(9.,6.5)(0.,){/ScalarDash}{-1}
\FALabel(7.93,10.)[r]{$\tilde f$}
\FAVert(9.,16.){0}
\FAVert(9.,4.){0}
\FAVert(15.,10.){0}
\FAVert(9.,13.5){0}
\FAVert(9.,6.5){0}

\FADiagram{}
\FAProp(0.,15.)(10.,16.)(0.,){/Straight}{-1}
\FALabel(4.84577,16.5623)[b]{$e$}
\FAProp(0.,5.)(10.,4.)(0.,){/Straight}{1}
\FALabel(4.84577,3.43769)[t]{$e$}
\FAProp(20.,17.)(10.,16.)(0.,){/Straight}{1}
\FALabel(14.8458,17.5623)[b]{$e$}
\FAProp(20.,10.)(15.5,10.)(0.,){/ScalarDash}{-1}
\FALabel(17.75,11.07)[b]{$H$}
\FAProp(20.,3.)(10.,4.)(0.,){/Straight}{-1}
\FALabel(14.8458,2.43769)[t]{$\nu_e$}
\FAProp(10.,16.)(10.,10.)(0.,){/Sine}{0}
\FALabel(8.93,13.)[r]{$\gamma,Z$}
\FAProp(10.,4.)(10.,10.)(0.,){/Sine}{1}
\FALabel(8.93,7.)[r]{$W$}
\FAProp(15.5,10.)(10.,10.)(0.8,){/ScalarDash}{1}
\FALabel(12.75,13.27)[b]{$\tilde f'$}
\FAProp(15.5,10.)(10.,10.)(-0.8,){/ScalarDash}{-1}
\FALabel(12.75,7.2)[t]{$\tilde f$}
\FAVert(10.,16.){0}
\FAVert(10.,4.){0}
\FAVert(15.5,10.){0}
\FAVert(10.,10.){0}

\FADiagram{}
\FAProp(0.,15.)(10.,14.5)(0.,){/Straight}{-1}
\FALabel(5.0774,15.8181)[b]{$e$}
\FAProp(0.,5.)(10.,5.5)(0.,){/Straight}{1}
\FALabel(5.0774,4.18193)[t]{$e$}
\FAProp(20.,17.)(10.,14.5)(0.,){/Straight}{1}
\FALabel(14.6241,16.7737)[b]{$e$}
\FAProp(20.,10.)(10.,10.)(0.,){/ScalarDash}{-1}
\FALabel(15.,11.07)[b]{$H$}
\FAProp(20.,3.)(10.,5.5)(0.,){/Straight}{-1}
\FALabel(15.3759,5.27372)[b]{$\nu_e$}
\FAProp(10.,14.5)(10.,10.)(0.,){/Sine}{0}
\FALabel(8.93,12.25)[r]{$\gamma,Z$}
\FAProp(10.,5.5)(10.,10.)(0.,){/Sine}{1}
\FALabel(8.93,7.75)[r]{$W$}
\FAVert(10.,14.5){0}
\FAVert(10.,5.5){0}
\FAVert(10.,10.){1}

\FADiagram{}
\FAProp(0.,15.)(2.5,10.)(0.,){/Straight}{-1}
\FALabel(0.343638,12.2868)[tr]{$e$}
\FAProp(0.,5.)(2.5,10.)(0.,){/Straight}{1}
\FALabel(2.15636,7.28682)[tl]{$e$}
\FAProp(20.,17.)(16.,15.)(0.,){/Straight}{1}
\FALabel(17.7868,16.9064)[br]{$e$}
\FAProp(20.,10.)(12.,6.5)(0.,){/ScalarDash}{-1}
\FALabel(14.6702,6.55096)[tl]{$H$}
\FAProp(20.,3.)(16.,15.)(0.,){/Straight}{-1}
\FALabel(18.1072,12.6291)[l]{$\nu_e$}
\FAProp(2.5,10.)(6.,10.)(0.,){/Sine}{0}
\FALabel(4.25,11.07)[b]{$\gamma,Z$}
\FAProp(16.,15.)(12.,13.5)(0.,){/Sine}{1}
\FALabel(13.4558,15.2213)[b]{$W$}
\FAProp(12.,6.5)(6.,10.)(0.,){/Straight}{-1}
\FALabel(8.699,7.39114)[tr]{$f$}
\FAProp(12.,6.5)(12.,13.5)(0.,){/Straight}{1}
\FALabel(13.07,10.)[l]{$f'$}
\FAProp(6.,10.)(12.,13.5)(0.,){/Straight}{-1}
\FALabel(8.699,12.6089)[br]{$f$}
\FAVert(2.5,10.){0}
\FAVert(16.,15.){0}
\FAVert(12.,6.5){0}
\FAVert(6.,10.){0}
\FAVert(12.,13.5){0}

\FADiagram{}
\FAProp(0.,15.)(2.5,10.)(0.,){/Straight}{-1}
\FALabel(0.343638,12.2868)[tr]{$e$}
\FAProp(0.,5.)(2.5,10.)(0.,){/Straight}{1}
\FALabel(2.15636,7.28682)[tl]{$e$}
\FAProp(20.,17.)(16.,15.)(0.,){/Straight}{1}
\FALabel(17.7868,16.9064)[br]{$e$}
\FAProp(20.,10.)(12.,6.5)(0.,){/ScalarDash}{-1}
\FALabel(14.6702,6.55096)[tl]{$H$}
\FAProp(20.,3.)(16.,15.)(0.,){/Straight}{-1}
\FALabel(18.1072,12.6291)[l]{$\nu_e$}
\FAProp(2.5,10.)(6.,10.)(0.,){/Sine}{0}
\FALabel(4.25,11.07)[b]{$\gamma,Z$}
\FAProp(16.,15.)(12.,13.5)(0.,){/Sine}{1}
\FALabel(13.4558,15.2213)[b]{$W$}
\FAProp(12.,6.5)(6.,10.)(0.,){/ScalarDash}{-1}
\FALabel(8.699,7.39114)[tr]{$\tilde f$}
\FAProp(12.,6.5)(12.,13.5)(0.,){/ScalarDash}{1}
\FALabel(13.07,10.)[l]{$\tilde f'$}
\FAProp(6.,10.)(12.,13.5)(0.,){/ScalarDash}{-1}
\FALabel(8.699,12.6089)[br]{$\tilde f$}
\FAVert(2.5,10.){0}
\FAVert(16.,15.){0}
\FAVert(12.,6.5){0}
\FAVert(6.,10.){0}
\FAVert(12.,13.5){0}

\FADiagram{}
\FAProp(0.,15.)(3.5,10.)(0.,){/Straight}{-1}
\FALabel(0.960191,12.0911)[tr]{$e$}
\FAProp(0.,5.)(3.5,10.)(0.,){/Straight}{1}
\FALabel(2.53981,7.09113)[tl]{$e$}
\FAProp(20.,17.)(15.5,16.)(0.,){/Straight}{1}
\FALabel(17.4138,17.5331)[b]{$e$}
\FAProp(20.,10.)(11.45,7.9)(0.,){/ScalarDash}{-1}
\FALabel(14.3441,7.31976)[t]{$H$}
\FAProp(20.,3.)(15.5,16.)(0.,){/Straight}{-1}
\FALabel(17.7562,12.5818)[l]{$\nu_e$}
\FAProp(3.5,10.)(8.,12.5)(0.,){/Sine}{0}
\FALabel(6.47725,12.8)[br]{$\gamma,Z$}
\FAProp(15.5,16.)(8.,12.5)(0.,){/Sine}{1}
\FALabel(11.5745,15.1746)[br]{$W$}
\FAProp(11.45,7.9)(8.,12.5)(0.8,){/ScalarDash}{1}
\FALabel(12.325,12.03)[bl]{$\tilde f'$}
\FAProp(11.45,7.9)(8.,12.5)(-0.8,){/ScalarDash}{-1}
\FALabel(7.125,8.37)[tr]{$\tilde f$}
\FAVert(3.5,10.){0}
\FAVert(15.5,16.){0}
\FAVert(11.45,7.9){0}
\FAVert(8.,12.5){0}

\FADiagram{}
\FAProp(0.,15.)(4.5,10.)(0.,){/Straight}{-1}
\FALabel(1.57789,11.9431)[tr]{$e$}
\FAProp(0.,5.)(4.5,10.)(0.,){/Straight}{1}
\FALabel(2.92211,6.9431)[tl]{$e$}
\FAProp(20.,17.)(13.,14.5)(0.,){/Straight}{1}
\FALabel(15.9787,16.7297)[b]{$e$}
\FAProp(20.,10.)(10.85,8.4)(0.,){/ScalarDash}{-1}
\FALabel(18.4569,10.7663)[b]{$H$}
\FAProp(20.,3.)(13.,14.5)(0.,){/Straight}{-1}
\FALabel(17.7665,4.80001)[tr]{$\nu_e$}
\FAProp(4.5,10.)(10.85,8.4)(0.,){/Sine}{0}
\FALabel(7.29629,8.17698)[t]{$\gamma,Z$}
\FAProp(13.,14.5)(10.85,8.4)(0.,){/Sine}{1}
\FALabel(10.9431,11.9652)[r]{$W$}
\FAVert(4.5,10.){0}
\FAVert(13.,14.5){0}
\FAVert(10.85,8.4){1}

\end{feynartspicture}
\caption{Generic vertex diagrams for the process \eeenH.
\label{fig:FD_eeenH}}
\end{center}
\end{figure}


Furthermore, counterterm contributions are needed for the $W^\pm H^\mp$
self-energy corrections, see \citere{eeWH2}.  
In order to generate the counterterms
it is sufficient to introduce the field renormalization for the 
$H^\pm - W^\pm$ mixing, $\de Z_{HW}$. 
This yields the Feynman rules:
\begin{align}
\label{ct-wh}
\Gamma_{\text{CT}}[H^\mp W^\pm(k^\mu)] & = 
        i \frac{k^\mu}{\MW} \MW^2\, \delta Z_{HW}\;,\\
\Gamma_{\text{CT}}[\gamma_\mu W^\pm_\nu H^\mp] & = 
        - i e \MW g_{\mu\nu}\,\delta Z_{HW}\;,\\
\label{ct-zwh}
\Gamma_{\text{CT}}[Z_\mu W^\pm_\nu H^\mp] & =
          i e \MW \frac{s_w}{c_w} g_{\mu\nu}\, \delta Z_{HW}
\end{align}
In the on-shell scheme $\de Z_{HW}$ is given by
\begin{align}
\de Z_{HW} & = \ed{\MW^2}\,\re \Si_{HW}(\MHp^2)~.
\end{align}


\subsection{Results}

The results for $H^+$ and $H^-$ production are the same if CP is not
violated (which we assume in this section).  In \reffi{fig:eeenHresults}
we show the typical size of the production cross section for \eeenHm\
for unpolarized external particles.  The parameters are chosen according
to the four benchmark scenarios described in the Appendix, with
$\MA = 250 \gev$ and $\tb = 2$ and~$10$ (with $\MHp \approx 262 \gev$). 
Concerning the discovery of the charged Higgs boson, the cross section 
has to be doubled, due to the production of both, $H^+$ and $H^-$.
%

%
\setlength{\unitlength}{1bp}
\begin{wrapfigure}[4]{r}{80bp}
\vspace*{-6ex}
\begin{feynartspicture}(80,80)(1,1)   
\FADiagram{}
\FAProp(0.,15.)(5.5,10.)(0.,){/Straight}{-1}
\FALabel(2.18736,11.8331)[tr]{$e$}
\FAProp(0.,5.)(5.5,10.)(0.,){/Straight}{1}
\FALabel(3.31264,6.83309)[tl]{$e$}
\FAProp(20.,17.)(15.5,13.5)(0.,){/Straight}{1}
\FALabel(17.2784,15.9935)[br]{$e$}
\FAProp(20.,10.)(15.5,13.5)(0.,){/Straight}{-1}
\FALabel(18.2216,12.4935)[bl]{$\nu_e$}
\FAProp(20.,3.)(12.,10.)(0.,){/ScalarDash}{-1}
\FALabel(15.4593,5.81351)[tr]{$H$}
\FAProp(5.5,10.)(12.,10.)(0.,){/Sine}{0}
\FALabel(8.25,8.93)[t]{$\gamma,Z$}
\FAProp(15.5,13.5)(12.,10.)(0.,){/Sine}{0}
\FALabel(13.134,12.366)[br]{$W$}
\FAVert(5.5,10.){0}
\FAVert(15.5,13.5){0}
\FAVert(12.,10.){-1}
\vspace*{-5ex}
\end{feynartspicture}
\end{wrapfigure}
In \reffi{fig:eeenHresults} the cross section for \eeenHm\ is shown as a
function of $\sqrt{s}$.  
The rise of the cross section at $\sqrt{s} \approx \MHp + \MW$ is due
to the $W$ propagator in the type of diagram on the right becoming 
resonant.  The resonance was treated with a fixed $W$~width. 
No cut on the electron emission angle had to be put, since the result
is finite in the limit of forward electron scattering.

%
\begin{figure}[ht]
\centerline{\includegraphics{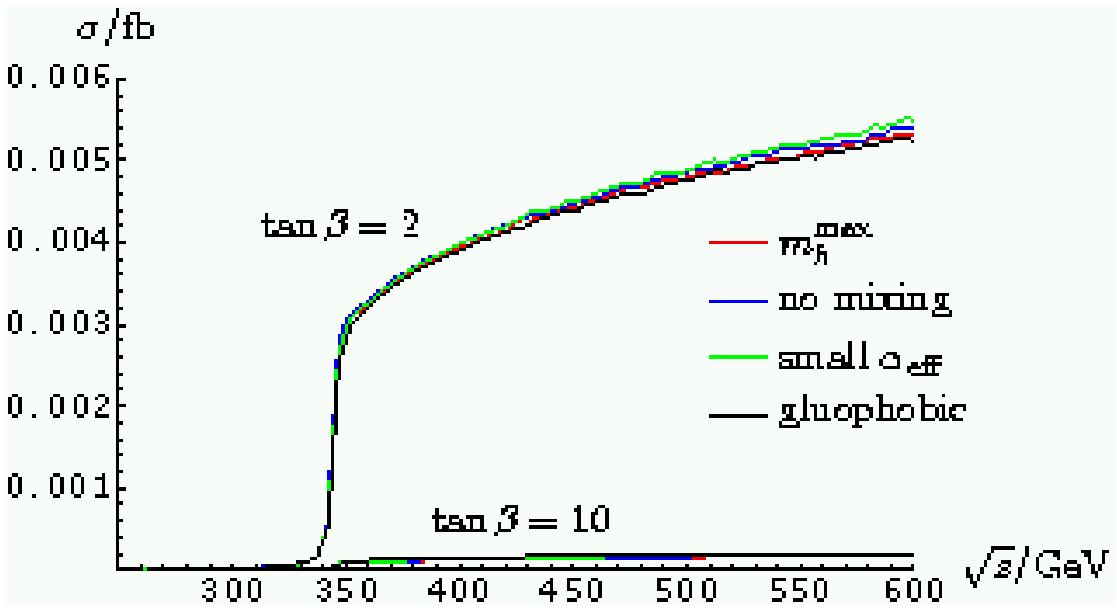}}
\caption{The \eeenHm\ production cross section as a function of $\sqrt{s}$.
\label{fig:eeenHresults}}
\end{figure}
%

The variation within the four benchmark scenarios is small.  For $\tb =
10$ the charged-Higgs production cross section stays at a negligible
level.  Even for $\tb = 2$ it stays below $0.01 \fb$ for $2\,\MHp
\approx \sqrt{s} \lesssim 500 \gev$.  Using polarized $e^+$ and $e^-$
beams, the cross section could be enhanced by about a factor of 2.  In
summary, however, the single charged-Higgs production, \eeenH, could
increase the potential of a LC for the detection of the heavy MSSM
Higgs-boson spectrum only for parameters beyond the typical benchmark
scenarios.



\newpage

\chapter{Higgs boson decays to SM fermions}

After the detection of a scalar particle it is mandatory as a next step
to measure its couplings to gauge bosons and fermions and also its
self-couplings very accurately, in order to establish the Higgs
mechanism and the Yukawa interactions experimentally. The determination
of the trilinear Higgs-boson self-couplings 
might be possible at a future linear \epem\ collider with high
luminosity~\cite{hhh}.

In this section we concentrate on the Feynman-diagrammatic corrections
(i.e.\ beyond the effective coupling approximation) to the coupling of
the lightest MSSM Higgs boson to SM fermions. 
Since the $b$-, the $c$-quark and the $\tau$-lepton are the heaviest
particles for which the decay $\hff$ is kinematically allowed, it is of
particular interest to calculate the corresponding decay rates and
branching ratios with high precision
\cite{mhiggsf1lC,gammagluon1,gammagluon2,gammagluon3,hprod2,gluongluino1,hffhigherorder,hff}.
We analyze these decay rates and branching ratios, taking into account
the Higgs-boson propagator corrections, where at the
\onel\ level the full momentum dependence is kept~\cite{hff}. 
We also take into
account the \onel\ vertex corrections resulting from gluon, gluino, 
and photon exchange together with real gluon and photon emission as
given in \citere{mhiggsf1lC}. 
Only the purely weak $\oa$ vertex corrections have been neglected (see
e.g.\ \citere{hfff1l} for a recent evaluation).
We numerically investigate the effect of the \twol\ propagator
contributions and the \onel\ gluino-exchange vertex correction. 
The latter is supplemented with a resummation of the leading
contributions of \order{(\als\tb)^n}, see
\refse{subsec:recentHO}. Some  
phenomenological consequences are investigated.

We show analytically that the Higgs-boson propagator correction with
neglected momentum dependence can be absorbed into the tree-level
coupling using the effective mixing angle from the neutral $\cp$-even
Higgs boson sector~\cite{hff}. The result in this approximation
is then compared with the full
result. 


\section{Higher-order corrections to $\hff$}
\label{sec:hff}

\subsection{Calculation of the decay amplitude}
\label{subsec:hffamp}

Our main emphasis in this section is on the fermionic decays of the
light Higgs boson, but for completeness we list the expressions for both
$h$ and $H$.
The amplitudes for the decays $h,H \to f\bar{f}$
can be written as follows:

\BEA
\label{ewdecayamplitude}
A(\hff) &=& \sqrt{\rZh} \KL \Gh + \rZhH\; \GH \KR~, \\
A(\Hff) &=& \sqrt{\rZH} \KL \GH + \rZHh\; \Gh \KR~, 
\EEA
with $\rZh, \rZH, \rZhH$ and $\rZHh$ given in \refeqs{eq:ZhH}-(\ref{eq:zhh}).


\subsection{The $\aeff$ approximation for decay amplitudes} 
\label{subsec:hffaeff}


The dominant contributions for the Higgs boson self-energies can be
obtained by setting $q^2=0$, see \refse{subsec:aeff}:
\BE
\hSi(q^2) \to \hSi(0) \equiv \hSi~.
\label{hffzeroexternalmomentum}
\end{equation}
With the approximation~(\ref{hffzeroexternalmomentum}) (see also
\refeqs{tandeltaalpha}, (\ref{tandeltaalpha2})) one deduces
\BEA
\label{ZhHTanDeltaalpha}
\rZhH &=& - \frac{\hSihH}{\Mh^2 - \mH^2 + \hSiH} = - \TDea, \\
\label{ZHhTanDeltaalpha}
\rZHh &=& - \frac{\hSihH}{\MH^2 - \mh^2 + \hSih} = + \TDea,
\EEA
and $\Zh$ can be expressed as 
\BEA
\rZh &=&
        \ed{1 + \KL \frac{\hSihH}{\Mh^2 - \mH^2 + \hSiH} \KR^2} \non \\
    &=& \ed{1 + \TQDea} = \CQDea~.
\EEA
Analogously one obtains 
\BE
\rZH = \CQDea~.  
\end{equation}

\bigskip \noindent
At tree level, the vertex functions can be written as
\BEA
\renewcommand{\arraystretch}{1.5}
\left. \begin{array}{l}
\Gh = \frac{i e \mf \Sa}{2\sw\MW\Cb} = \cfpfp \; \Sa  \\
\GH = \frac{-i e \mf \Ca}{2\sw\MW\Cb} = - \cfpfp \; \Ca 
\end{array} \KKKR \mbox{for d-type fermions}, \\
\renewcommand{\arraystretch}{1.5}
\left. \begin{array}{l}
\Gh = \frac{-i e \mf \Ca}{2\sw\MW\Sbe} = \cff \; \Ca  \\
\GH = \frac{-i e \mf \Sa}{2\sw\MW\Sbe} = \cff \; \Sa 
\end{array} \KKKR \mbox{for u-type fermions}~.
\renewcommand{\arraystretch}{1}
\EEA
Incorporating them into the decay amplitude yields the corresponding
$\aeff$~approximation: 
\BEA
A_{\rm eff}(\hff) &=& \wZh \KL \Gh + \ZhH\; \GH \KR \non \\
 &=& \cfpfp \CDea \KL \Sa - \TDea \KL - \Ca \KR \KR \non \\
 &=& \cfpfp \sin(\al + \De\al) \non \\
\label{ampeffhbb}
 &\equiv& \cfpfp \Saeff~~~\mbox{(for d-type fermions)}, \\
 A_{\rm eff}(\hff) &\equiv& \cff \Caeff~~~\mbox{(for u-type fermions)}, \\
 A_{\rm eff}(\Hff) &\equiv& -\cfpfp \Caeff~~~\mbox{(for d-type fermions)}, \\
 A_{\rm eff}(\Hff) &\equiv& \cff \Saeff~~~\mbox{(for u-type fermions)}~.
\EEA


\subsection{Evaluation of the Higgs-boson propagator corrections}
\label{subsec:evaloaas}

The Higgs boson self-energies employed in the field renormalization
constants, see \refeqs{eq:ZhH}--(\ref{eq:zhh}), have been evaluated in the
Feynman-diagrammatic approach according to 
\BE
\hSi_s(q^2) = \hSie_s(q^2) + \hSiz_s(0)~,~~s = h, H, hH~,
\label{hsefull}
\end{equation}
where the momentum dependence has been neglected only at the \twol\ level,
while the full momentum dependence is kept in the \onel\
contributions, see \refse{subsec:recentHO}.

For the numerical evaluation in a first step of approximation
for the calculation of the decay width $\Ga(\hff)$ 
the momentum dependence is
neglected everywhere in the Higgs boson self-energies, the $q^2=0$
approximation (see \refeq{hffzeroexternalmomentum}):
\BE
\hSi_s(q^2) \to \hSie_s(0) + \hSiz_s(0)~,~~s = h, H, hH~.
\label{hseq2zero}
\end{equation}
This corresponds to the $\aeff$-approximation, as described in
\refse{subsec:hffaeff}.

In a second step of approximation 
we approximate the Higgs boson self-energies by
the compact analytical formulas given in \citere{mhiggslle}:
\BE
\hSi_s(q^2) = \hSi_s^{(1)\rm\, approx}(0) 
            + \hSi_s^{(2)\rm\, approx}(0)~,~~s = h, H, hH ~.
\label{hselle}
\end{equation}
Here the full result of the self-energy corrections is approximated by
an expansion in terms of $\mt/\msusy$ and $\Xt/\msusy$, 
yielding relatively short expressions which allow a very fast
numerical evaluation. In the following, this approximation
is labeled by $\aeffapprox$.


\subsection{Decay width of the lightest Higgs boson}
\label{subsec:hdecaywidth}

At the tree level, the decay width for $\hff$ is given by
\BE
\label{decaywidthtree}
\Gz(\hff) = N_C \frac{\mh}{8\,\pi}
           \KL 1 - \frac{4\,\mf^2}{\mh^2} \KR^\frac{3}{2}~|\Gh|^2~.
\end{equation}
The electroweak propagator corrections 
are incorporated by using the higher-order decay
amplitude~(\ref{ewdecayamplitude})
\BE
\label{decaywidthoneloopew}
\Ge \equiv \Ge(\hff) =  N_C \frac{\Mh}{8\,\pi}
            \KL 1 - \frac{4\,\mf^2}{\Mh^2} \KR^\frac{3}{2}~|A(\hff)|^2~.
\end{equation}
The $\aeff$-approximation is given by
\BE
\Geeff \equiv \Geeff(\hff) = N_C \frac{\Mh}{8\,\pi}
       \KL 1 - \frac{4\,\mf^2}{\Mh^2} \KR^\frac{3}{2}~|A_{\rm eff}(\hff)|^2~.
\label{gameff}
\end{equation}
In this section we consider only those electroweak higher-order contributions
that enter via the Higgs boson self-energies. 
These corrections contain the Yukawa contributions of \order{\gf\mt^4/\MW^2},
which are the dominant electroweak \onel\ corrections to the Higgs-boson
decay width, and the corresponding dominant \twol\ corrections, see
\refse{subsec:evaloaas}. The pure weak $\oa$
vertex corrections are neglected (they have been calculated in
\citeres{mhiggsf1lC,hfff1l}. They were found to be 
at the level of only a few \% for most parts of the MSSM parameter
space, see also \citere{hff}).


\subsubsection{QED corrections}
\label{subsec:decaywidthgamma}

Here we follow the results given in
\citeres{gammagluon1,gammagluon2,gammagluon3,mhiggsf1lC}.
The IR-divergent virtual photon contribution is taken into account
in combination with real-photon bremsstrahlung yielding the QED 
corrections. The contribution to the decay width induced by
$\ga$-exchange and final-state photon radiation can be cast into the
very compact formula
\BE
\De\Gga^\phi = \Ge^\phi\cdot\deqed^\phi, ~\phi = h, H~,
\end{equation}
where for $\mf^2 \ll \Mphi^2$ the factor $\deqed^\phi$ has the simple form
\BE
\deqed^\phi = \frac{\al}{\pi}Q_f^2 \KKL - 3 \log \KL \frac{\Mphi}{\mf} \KR
                                   + \frac{9}{4} \KKR~.
\end{equation}


\subsubsection{QCD corrections: gluino contributions}
\label{subsubsec:decaywidthgluino}

In this section we will first focus on the decay $\phibb, \phi = h,H$
only, since for this decay channel the gluino contributions are
especially relevant. 
Concerning the evaluation of the corresponding gluino-exchange Feynman
diagrams for $\phibb$, we 
follow the calculation given in \citere{mhiggsf1lC} (see also
\citere{hff}), similar results can also be found in
\citere{gluongluino1}. 
The additional contributions to the decay amplitude induced by
gluino-exchange are incorporated, following the effective Lagrangian
\refeq{effL}: 
\BEA
\label{decayampgluino}
A^h_{\gl} &=& \ed{1 + \dmb} \KKKL
1 + \ed{\Gh + \ZhH \GH}
 \re \KKL \Ggl^{h} + \ZhH \Ggl^{H} \KKR \KKKR , \\
 && \non \\
\label{decayampHgluino}
A^H_{\gl} &=& \ed{1 + \dmb} \KKKL
1 + \ed{\GH + \ZHh \Gh}
 \re \KKL \Ggl^{H} + \ZHh \Ggl^{h} \KKR \KKKR 
\EEA
for real $\ZhH, \ZHh$ (i.e.\ neglecting the imaginary part in
\refeqs{decayampgluino}, (\ref{decayampHgluino}));
$\Ggl^h$ and $\Ggl^H$ are given by
\BEA
\Ggl^h &=& \Gh \KKL \De T_{\gl}^h~_{\Bigr| q^2 = \Mh^2}
         + \Si^b_{S,\gl}(\mb^2) + \dmb
         - 2 \mb^2 \KL \Si_{S,\gl}^{b\prime}(\mb^2)
                     + \Si_{V,\gl}^{b\prime}(\mb^2) \KR \KKR , \\
 && \non \\
\Ggl^H &=& \GH \KKL \De T_{\gl}^H~_{\Bigr| q^2 = \Mh^2}
         + \Si^b_{S,\gl}(\mb^2) + \dmb
         - 2 \mb^2 \KL \Si_{S,\gl}^{b\prime}(\mb^2)
                     + \Si_{V,\gl}^{b\prime}(\mb^2) \KR\KKR~. \\
 && \non
\EEA
$\De T_{\gl}^{h,H}$ denote the gluino vertex-corrections, whereas 
$\Si^b$ represents the gluino contribution to the bottom self-energy
corrections. Explicit expressions for these terms can be found
in~\citere{mhiggsf1lC}.

For large values of $\tb$ in combination with large values of  $|\mu|$,
the gluino-exchange corrections to 
$A(\phibb)$ can become very large.
They are resummed to all orders of $(\als\tb)^n$ via the inclusion of
$\dmb$ into $A^{h,H}_{\gl}$ (\refeqs{decayampgluino},
(\ref{decayampHgluino})). The results for $A^{h,H}_{\gl}$ constitutes
the currently best available  
evaluation of the gluino-exchange corrections to $\phibb$. It contains
the resummation of potentially large corrections $\sim(\als \tb)^n$ to
all orders as well as the full evaluation of the $\gl-\Sbot$~vertex
corrections including momentum effects.

\bigskip
In the case of $\phiff, \; f \neq b$ the expressions are similar, but
do not contain the resummed part:
\BEA
\label{decayampresummed}
A^h_{\gl} &=& 
1 + \ed{\Gh + \ZhH \GH}
 \re \KKL \Ggl^{h} + \ZhH \Ggl^{H} \KKR , \\
 && \non \\
A^H_{\gl} &=& 
1 + \ed{\GH + \ZHh \Gh}
 \re \KKL \Ggl^{H} + \ZHh \Ggl^{h} \KKR , \\
 && \non \\
\Ggl^h &=& \Gh \KKL \De T_{\gl}^h~_{\Bigr| q^2 = \Mh^2}
         + \Si^b_{S,\gl}(\mb^2) 
         - 2 \mb^2 \KL \Si_{S,\gl}^{b\prime}(\mb^2)
                     + \Si_{V,\gl}^{b\prime}(\mb^2) \KR \KKR , \\
 && \non \\
\Ggl^H &=& \GH \KKL \De T_{\gl}^H~_{\Bigr| q^2 = \Mh^2}
         + \Si^b_{S,\gl}(\mb^2) 
         - 2 \mb^2 \KL \Si_{S,\gl}^{b\prime}(\mb^2)
                     + \Si_{V,\gl}^{b\prime}(\mb^2) \KR\KKR~. \\
 && \non
\EEA


\subsubsection{QCD corrections: gluon contributions}
\label{subsec:decayampgluon}

The corresponding results have been obtained in 
\citeres{gammagluon1,gammagluon2,gammagluon3,hprod2,gluongluino1,mhiggsf1lC,hffhigherorder}.
The additional contribution to the decay width induced by
gluon exchange and final-state gluon radiation can be incorporated
into~(\ref{decayampgluino}) by 
\BE
\label{decayampgluon}
A^{\phi}_{\gl g} = A^{\phi}_{\gl} \cdot \frac{\mq(\Mphi^2)}{\mq}
       \KKL 1 + \frac{\als(\Mphi^2)}{2\,\pi} 
  \KL \cf \frac{9}{4} + \frac{8}{3} \KR \KKR~.
\end{equation}
The running quark mass $\mq(\Mphi^2)$ is calculated via
\BEA
\mq(q^2) &=& \mq \frac{c(q^2)}{c(\mq^2)}~,\\
c(q^2) &=& \KL \frac{\ben\als(q^2)}{2\,\pi}\KR^{-\gan/2\ben}
\Bigg[ 1 + \frac{\KL \bee\gan - \ben\gae \KR}{\ben^2}
           \frac{\als(q^2)}{8\,\pi} \Bigg]~,
\label{cfkt}
\EEA
where $\mq$ is the pole mass and $\mq(\mq^2) = \mq$. The coefficients in
\refeq{cfkt} are:
\BEA
\ben &=& \frac{33 - 2\nf}{3}~, \non \\
\bee &=& 102 - \frac{38}{3}\nf~, \non \\
\gan &=& -8~, \non \\
\gae &=& -\frac{404}{3} + \frac{40}{9}\nf~, 
\EEA
where 
$\nf = 5$ for 
$f = c, b$, which are considered here.
The strong coupling constant $\als$ is given up to two loops by:
\BE
\als(q^2) = \frac{4\,\pi}{\ben L_q}
  \KKL 1 - \frac{\bee}{\ben^2}\frac{\log L_q}{L_q} 
         + \frac{\bee^2}{\ben^4}\frac{\log^2 L_q}{L_q^2}
         - \frac{\bee^2}{\ben^4}\frac{\log L_q}{L_q^2}
\KKR~,
\end{equation}
where $L_q = \log(q^2/\Laqcd^2)$. (For the numerical evaluation 
$\Laqcd = 220 \mev$ has been used.)
Numerically, more than 80\% of the gluon-exchange contribution is
absorbed into the running quark mass.


\subsubsection{Decay width and branching ratio}

The result for the decay width of $\hff$, including strong
corrections, is given by 
\BEA
\Ga^h_{\gl g} &=& N_C \frac{\Mh}{8\,\pi}
            \KL 1 - \frac{4\,\mb^2}{\Mh^2} 
             \KR^\frac{3}{2}~|A^h_{\gl g}(\hbb)|^2~, \\
\Ga^H_{\gl g} &=& N_C \frac{\MH}{8\,\pi}
            \KL 1 - \frac{4\,\mb^2}{\MH^2} 
             \KR^\frac{3}{2}~|A^H_{\gl g}(\hbb)|^2~.
\EEA
Including also the QED corrections, the full decay width
is given by
\BEA
\label{decaywidthfull}
\Ga(\hbb) &=& \Ga^h_{\gl g} + \De\Gga^h~, \\
\Ga(\Hbb) &=& \Ga^H_{\gl g} + \De\Gga^H~.
\EEA

Summing over $f = b, c, \tau$ and adding $\Ga(h \to gg)$ 
(which can be numerically relevant~\cite{hgg}),
results in an approximation for the total decay width
\BE
\Gtot = \sum_{f=b,c,\tau} \Ga(\hff) + \Ga(h \to gg)~.
\end{equation}
We do not take into account the decay $h \to AA$ 
(see e.g.~\citere{haa} for a detailed study).
Although it is
dominant whenever it is kinematically allowed, 
it plays a role only for very small values of $\tb$ ($\tb \lsim 1.5$)
which will not be considered here because of the limits obtained at
LEP2~\cite{tbexcl}. We also assume that all other 
SUSY particles are too heavy to allow further decay channels.
In addition, we neglect the decay $h \to WW^*$ which can become
substantial for $\Mh \gsim 120 \gev$. The quantitative change in
our results due to this approximation is small, the qualitative change
is negligible.

The fermionic branching ratio is defined by
\BE
\rf \equiv \br(\hff) = \frac{\Ga(\hff)}{\Gtot}~.
\end{equation}
The results of this section, including {\em all} decay modes, is
incorporated into the Fortran code \fhto. In a comparison with the
code {\em Hdecay}~\cite{hdecay} we found good agreement.





\section{Phenomenological implications}
\label{sec:hffphenoimp}

Concerning the numerical evaluation of the Higgs-boson propagator
corrections, we follow \refse{subsec:evaloaas}. For
$\tb$ we mostly concentrate on two representative values,
a relatively low value, $\Tb = 3$, and a high value, $\Tb = 40$.
For sake of comparison we also consider 
an intermediate value of $\Tb = 20$ in some cases.
If not indicated differently, the other MSSM parameters
are chosen as follows:
$\mu = -100 \gev$, $M_2 = \msusy \equiv \msq$, 
$\mgl = 500 \gev$, $\Ab = \At$. With ``no mixing'' we denote the case
$\Xt = 0$, whereas ``maximal mixing'' denotes $\Xt = 2 \msusy$. 
The other SM parameters are given in the Appendix.
The mass $\MA$
of the $\cp$-odd Higgs boson is treated as an input
parameter and is varied in the interval $50 \gev \le \MA \le 500 \gev$. 
In some cases, also $\tb$ is varied in the interval $2 \le \tb \le 50$.

The corresponding values for $\Mh$ follow from 
\refeq{higgsmassmatrixnondiag}. 
$\Mh$, derived in this way, subsequently enters
the numerical evaluation of the formulas presented in
section~\ref{sec:hff}. Thus the variation of
$\Mh$ in the plots stems from the variation of $\MA$ (or $\tb$) in the
above given range.


\subsection{Effects of the \twol\ Higgs-propagator corrections}
\label{subsec:twoloopeffect}

\begin{figure}[htb!]
\vspace{1em}
\begin{center}
\mbox{
\psfig{figure=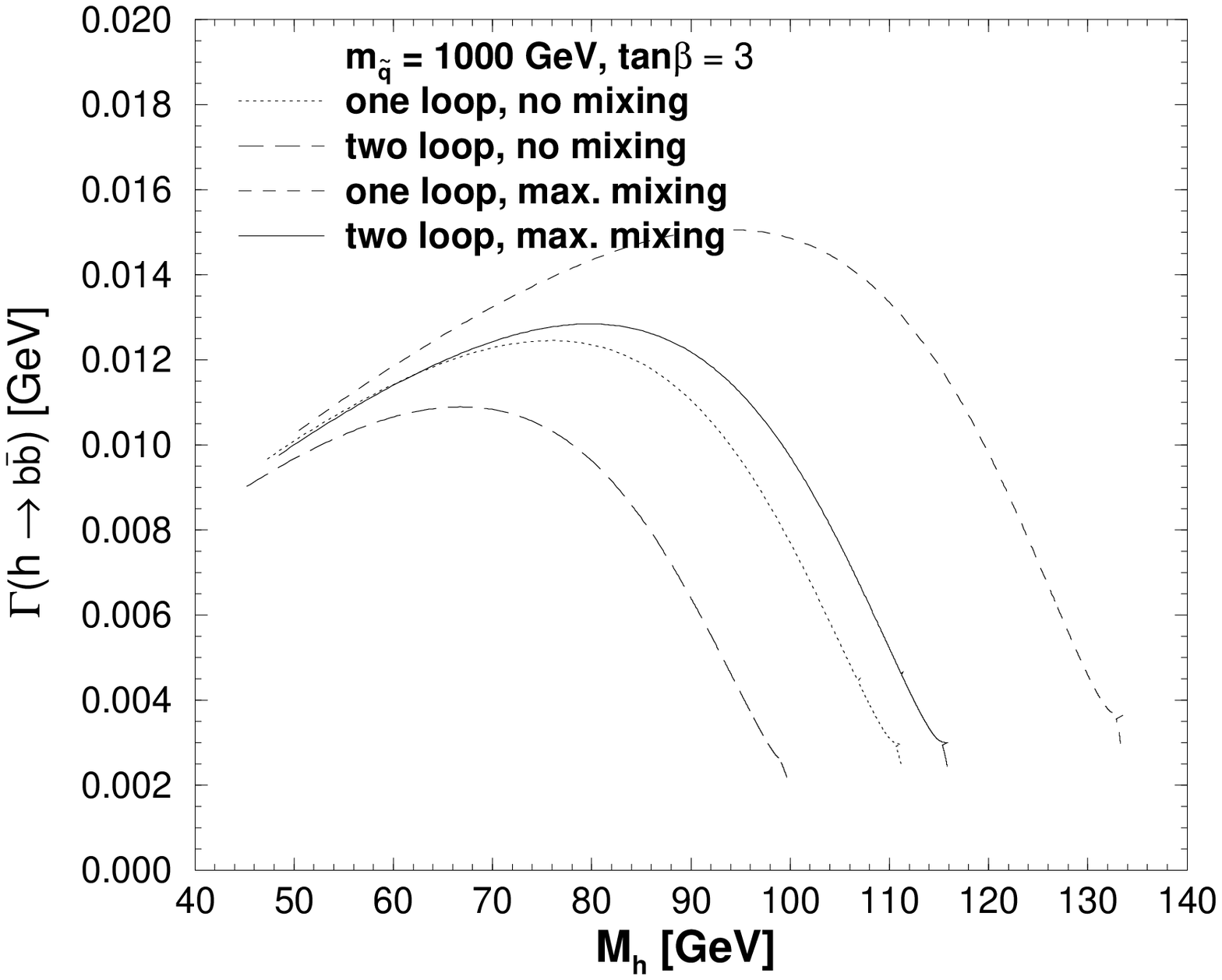,width=7cm,height=7cm}}
\hspace{1.5em}
\mbox{
\psfig{figure=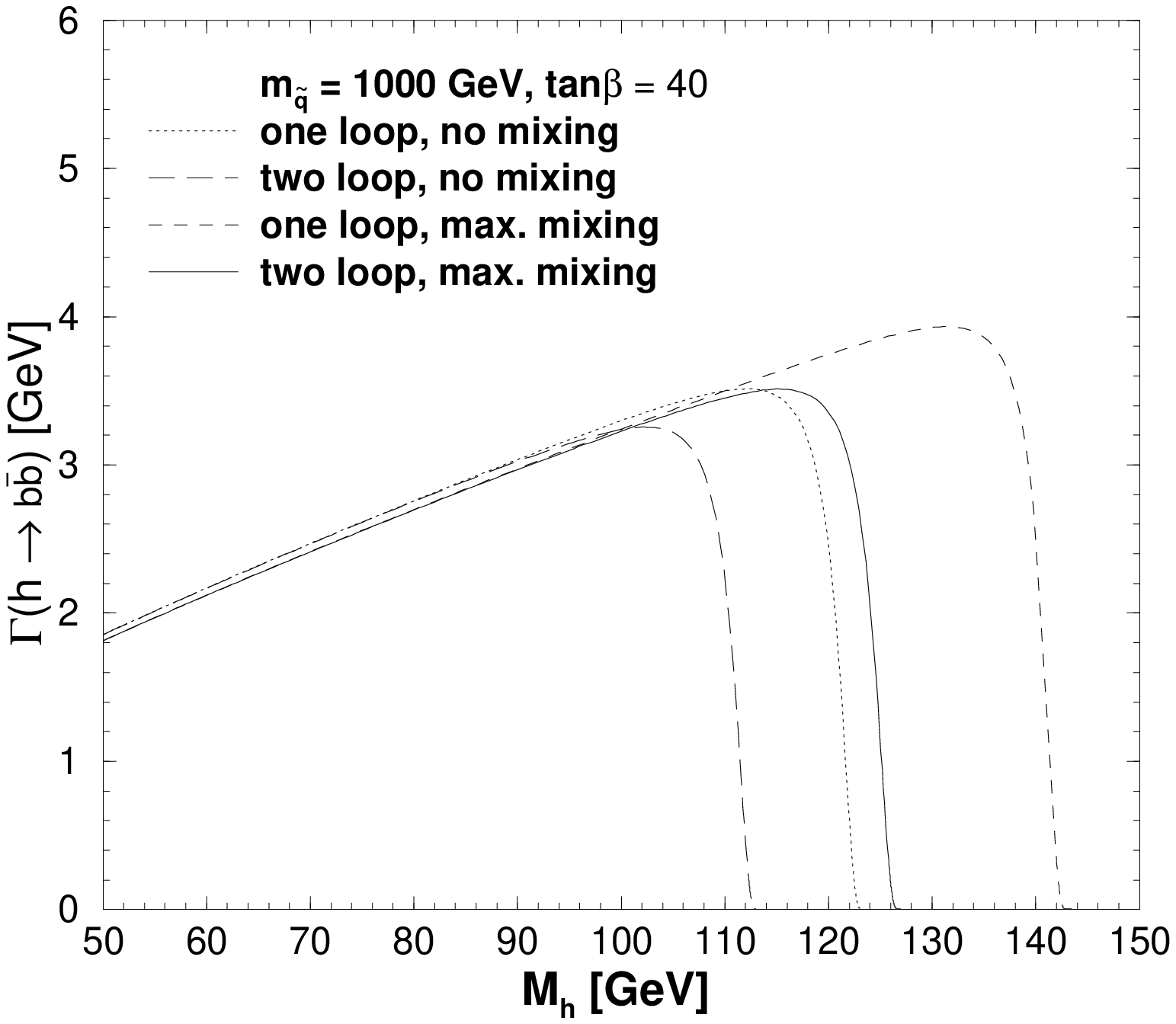,width=7cm,height=7cm}}
\end{center}
\caption[]{
$\Ga(\hbb)$ is shown as a function of $\Mh$. The Higgs-propagator
corrections have 
been evaluated at the one- and at the \twol\ level. The QED, 
gluon  and gluino contributions are included. The other parameters are
$\mu = -100 \gev$, $M_2 = \msusy$, $\mgl = 500 \gev$, $\Ab = \At$, 
$\Tb = 3, 40$. The result is given in the no-mixing and maximal-mixing
scenario.
}
\label{fig:Ghbb}
\end{figure}

We first focus on the effects of the \twol\ Higgs-boson propagator
corrections in comparison with the one-loop case. They have
been evaluated at the one- and at the \twol\ level as described in
\refse{subsec:evaloaas}. 
Figure~\ref{fig:Ghbb} shows the 
results for $\Ga(\hbb)$ for a common scalar quark mass 
$\msq \equiv \msusy = 1000 \gev$ and $\Tb = 3$ and $\Tb = 40$ in the
no-mixing and the maximal-mixing scenario. 
The QED and the QCD
gluon and gluino vertex contributions%
\footnote{
The gluino vertex corrections are included without the resummation
formula, since either the small value of $|\mu|$ or of $\tb$ results
in very small values of  
$\De \mb$. In \refse{subsec:hbbzero}, however, also the full
correction including resummation is taken into account.
}%
~are also included.

In the small $\tb$ scenario, larger values for $\Ga(\hbb)$ are obtained
for maximal mixing. The \twol\ corrections strongly reduce the decay width.
In the large $\tb$ scenario 
the variation is mainly a kinematical effect from the different values
of $\Mh$ at the one- and \twol\ level.
The absolute values obtained for $\Ga(\hbb)$ are three
orders of magnitude higher in the $\Tb = 40$ scenario, which is due
to the fact that $\Ga(\hbb) \sim 1/\CQb$. 

\begin{figure}[htb!]
\vspace{1em}
\begin{center}
\mbox{
\psfig{figure=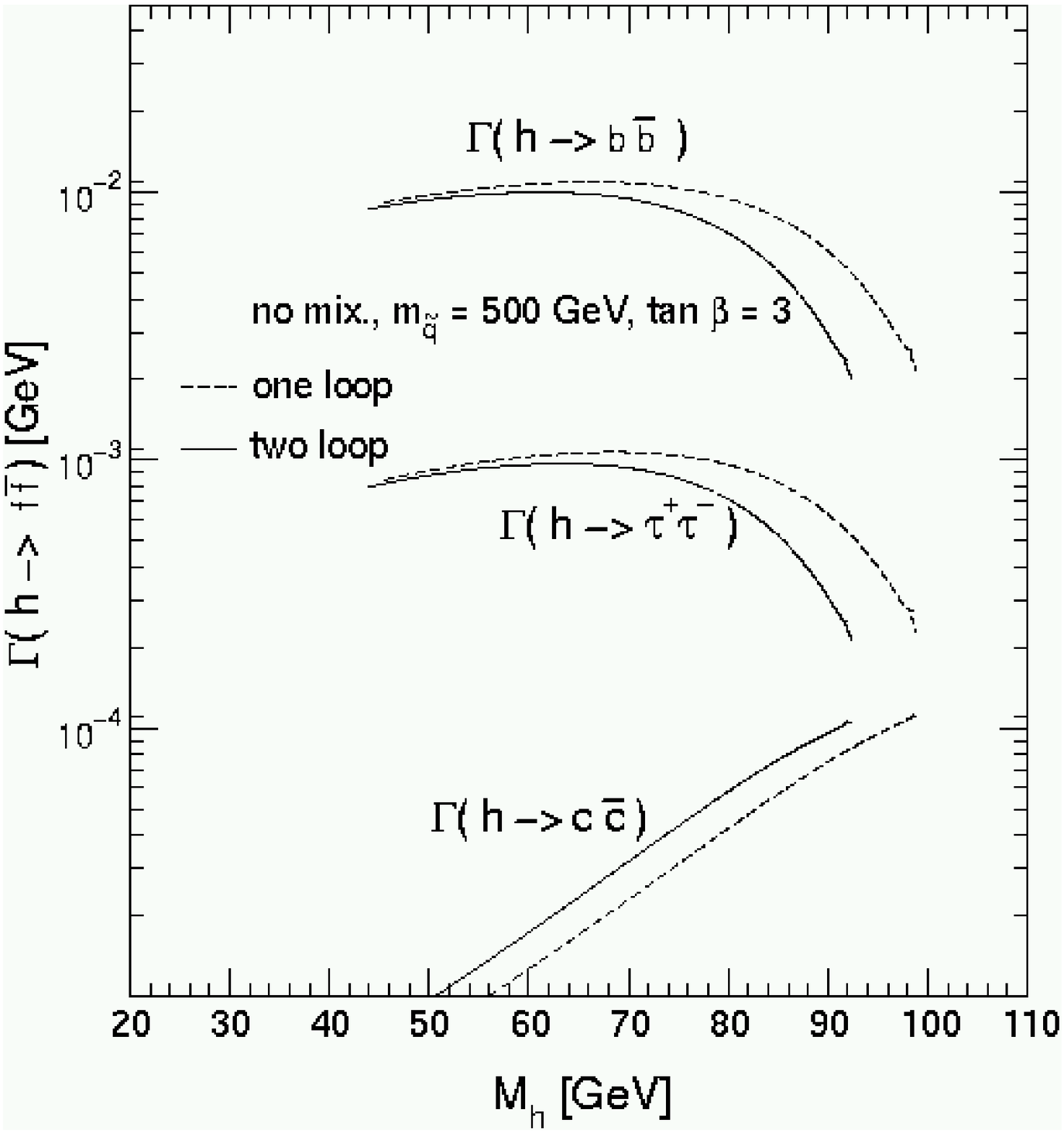,width=7cm,height=7cm}}
\hspace{1.5em}
\mbox{
\psfig{figure=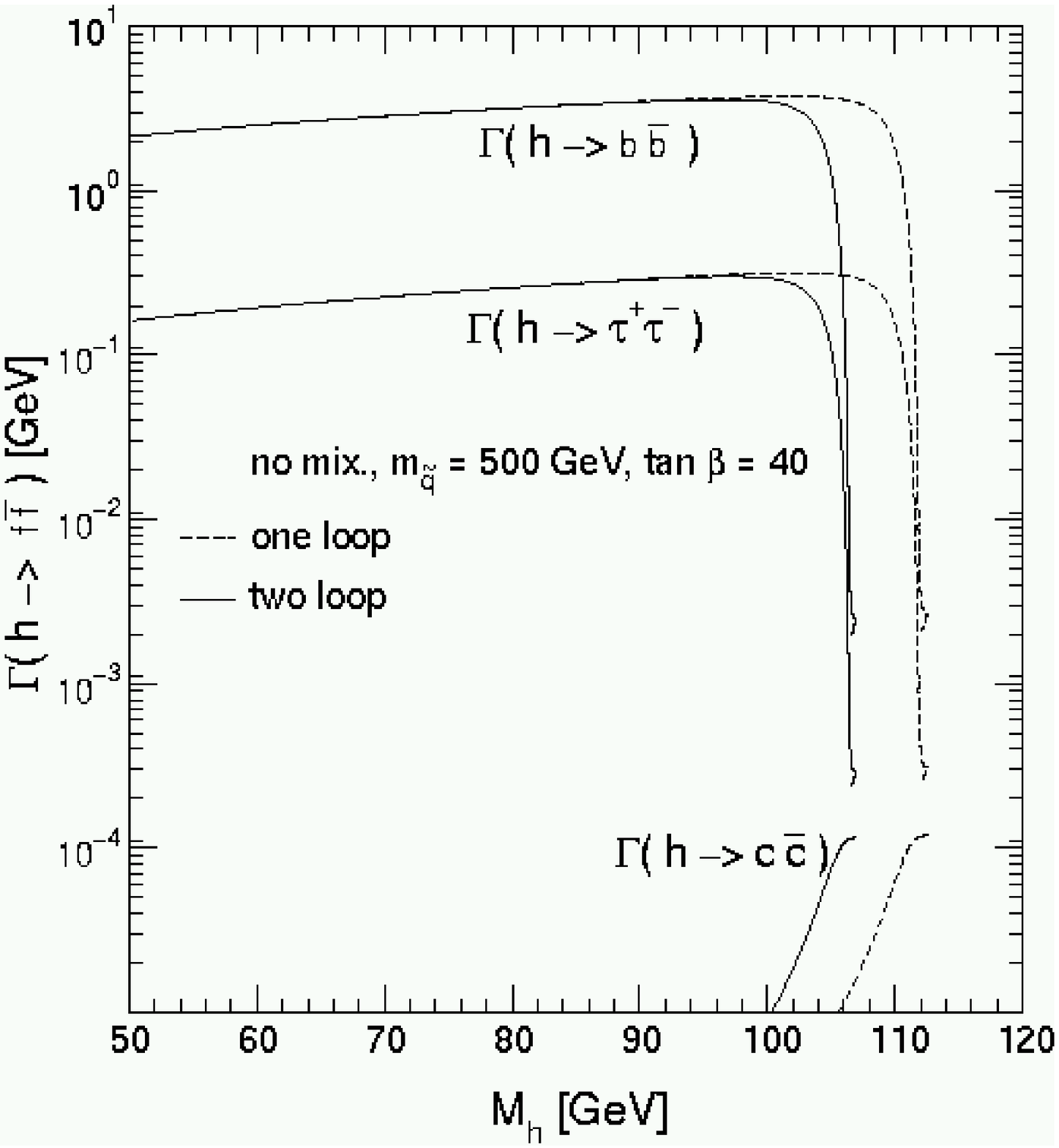,width=7cm,height=7cm}}
\end{center}
\caption[]{
$\Ga(\hbb)$, $\Ga(\htautau)$ and $\Ga(\hcc)$ are shown as a function
of $\Mh$. The Higgs-propagator corrections have 
been evaluated at the one- and at the \twol\ level. The QED, 
gluon  and gluino contributions are included. The other parameters are
$\mu = -100 \gev$, $M_2 = \msusy$, $\mgl = 500 \gev$, $\Ab = \At$, 
$\Tb = 3, 40$. The result is given in the no-mixing
scenario.
}
\label{fig:Ghbbttcc}
\end{figure}

\smallskip
In \reffi{fig:Ghbbttcc} the three decay rates $\Ga(\hbb)$,
$\Ga(\htautau)$ and $\Ga(\hcc)$ are shown as a function of $\Mh$.
The results are given in the no-mixing scenario for $\msusy = 500 \gev$
and $\Tb = 3, 40$. 

\begin{figure}[ht!]
\begin{center}
\mbox{
\psfig{figure=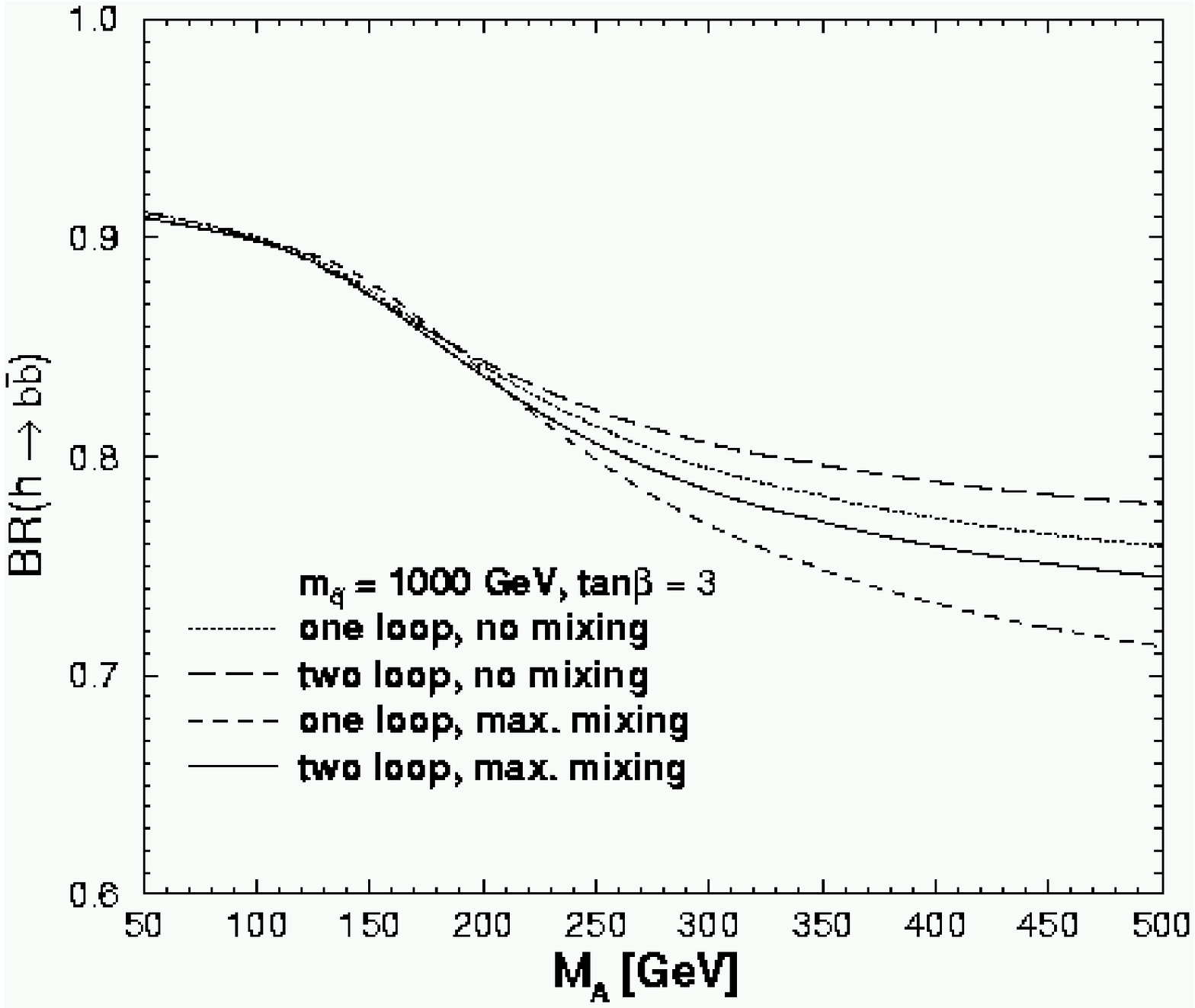,width=7cm,height=7cm}}
\hspace{1.5em}
\mbox{
\psfig{figure=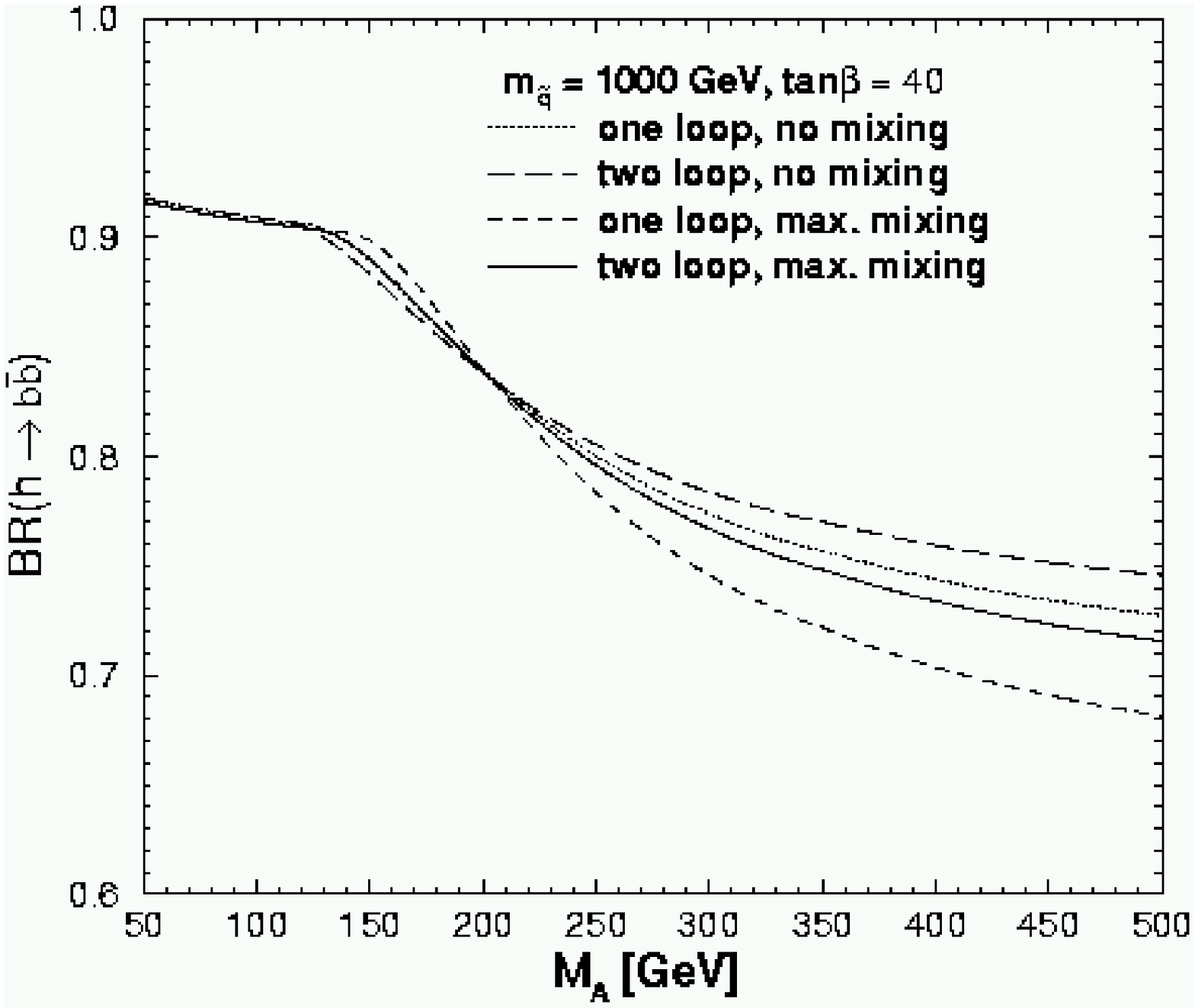,width=7cm,height=7cm}}
\end{center}
\begin{center}
\mbox{
\psfig{figure=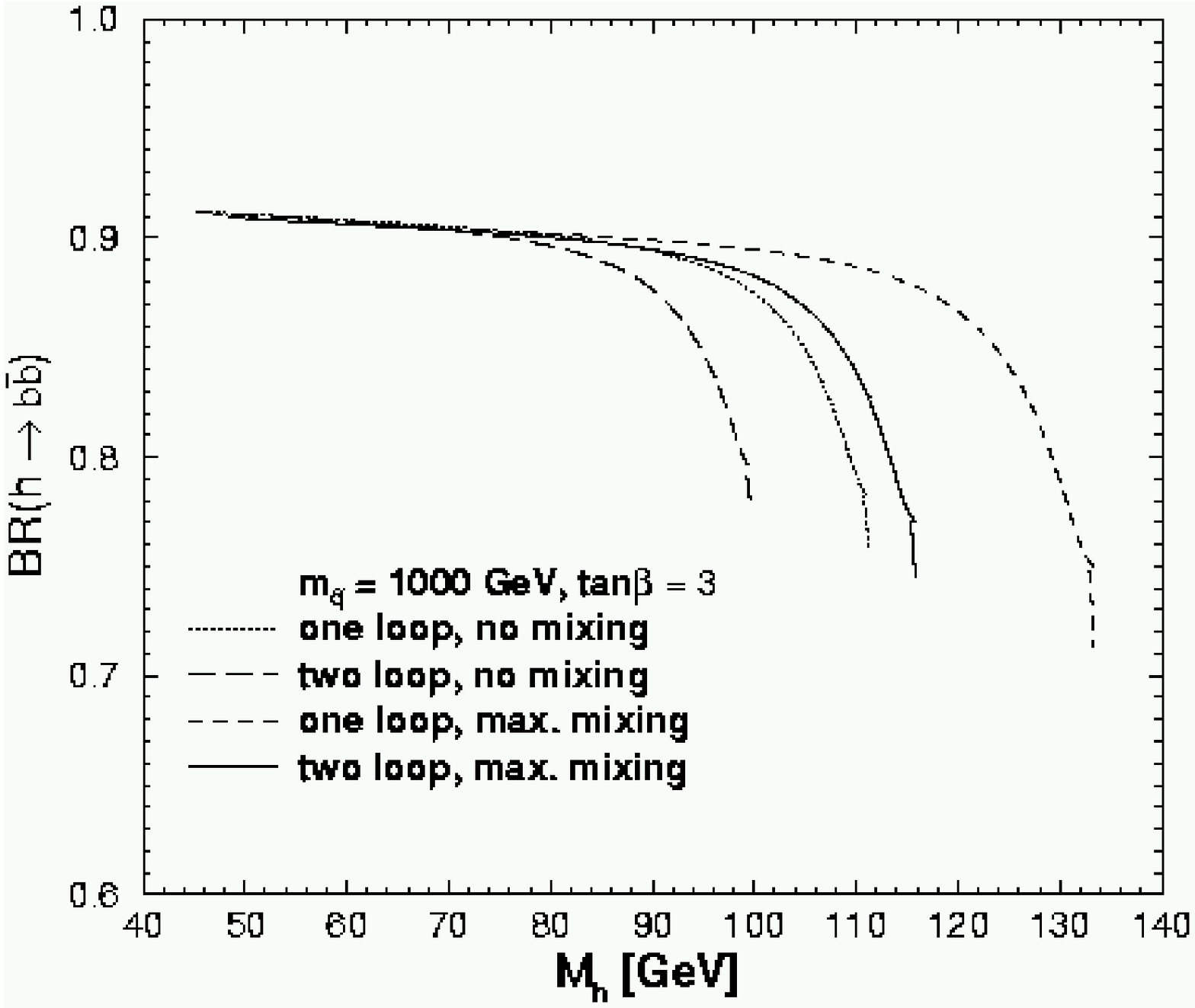,width=7cm,height=7cm}}
\hspace{1.5em}
\mbox{
\psfig{figure=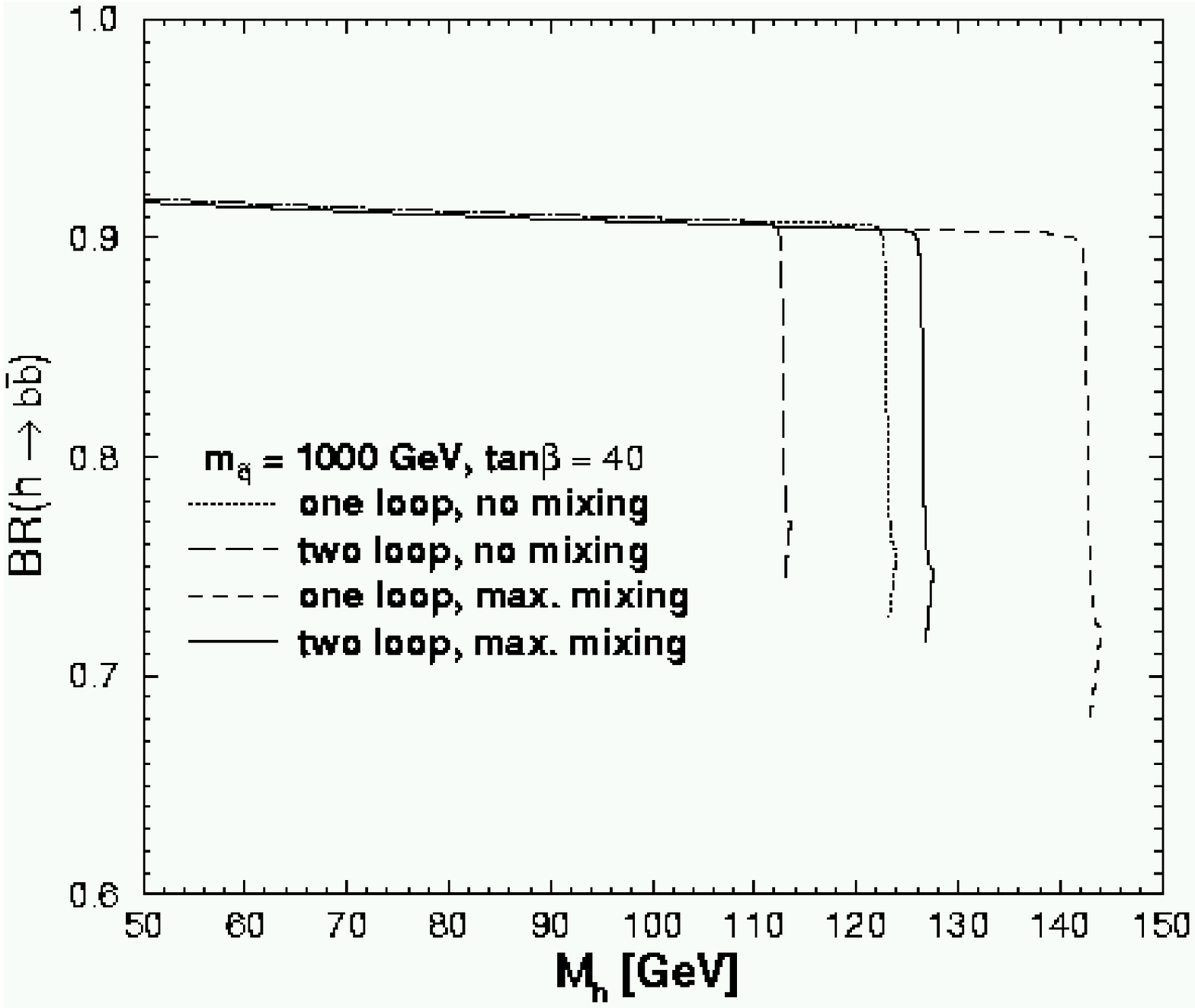,width=7cm,height=7cm}}
\end{center}
\caption[]{
$\br(\hbb)$ is shown as a function of $\MA$ and $\Mh$ 
for the same settings as in
\reffi{fig:Ghbb}. The QED, gluon and gluino contributions are included.
}
\label{fig:BRhbb}
\end{figure}

In the low $\tb$ scenario
$\Ga(\hbb)$ and $\Ga(\htautau)$ are lowered at the \twol\ level, while
$\Ga(\hcc)$ is increased. The decay rate for $\hbb$ is about one and
two orders of magnitude larger compared to the ones of $\htautau$ and
$\hcc$, respectively. In the large $\tb$ scenario the shifts are
again dominated by the kinematical effect from the different values
of $\Mh$ at the one- and \twol\ level.
In the maximal-mixing case, which is not plotted here, we find
qualitatively the same behavior. 

\smallskip
We now turn to the effects of the \twol\ corrections to the branching
ratios. 
For not too large values of $\Mh$, $\Gtot$ is strongly dominated
by $\Ga(\hbb)$. For large values of $\Mh$ the decay into gluons
becomes more relevant. 
In \reffi{fig:BRhbb} we show the branching ratio $\br(\hbb)$ as a
function of $\Mh$ and $\MA$. For values of $\MA \gsim 250 \gev$
there is a non-negligible difference between \onel\ and \twol\ order,
where at the \twol\ level the branching ratio is slightly enhanced. 
Compared in terms of $\Mh$ there is nearly no change for small values
of $\Mh$ in the low and in the high $\tb$ case. 
Here $\br(\hbb)$ is changed by less than about 1\%, see \reffi{fig:BRhbb}. 
$\br(\htautau)$ is
increased by less than about 2\%.
$\br(\hcc)$ can be increased at the
\twol\ level by ${\cal O}(50\%)$, but remains numerically relatively
small. For $\tb = 40$ the
main difference arises at the endpoints of the spectrum, again due to
the fact that different Higgs boson masses can be obtained at the
\onel\ and at the \twol\ level. For $\tb = 3$, however, also several
GeV below the kinematical endpoints there is a sizable effect on
$\br(\hbb)$. 
Thus, in the experimentally allowed region of $\Mh$, the \twol\
corrections can have an important effect on $\br(\hbb)$.


\subsection{Vanishing decay rate for $\hbb$}
\label{subsec:hbbzero}

The search for the Higgs boson, especially at \epem\ colliders, often
relies on $b$ tagging, 
since on the one hand the lightest $\cp$-even 
Higgs boson decays dominantly into $b\bar{b}$, and on the other hand
$b$~tagging can be performed with high efficiency. For some
combinations of parameters, however, $\Ga(\hbb)$ can become very small
and thus $\br(\hbb)$ can approach zero as a consequence of large
Higgs-boson propagator corrections or large gluino vertex-corrections,
making Higgs boson search 
possibly very difficult for these parameters. 
Higgs boson searches then have to rely on flavor-independent decay modes.
In order to have reliable
predictions for these regions of parameter space a full calculation of
the \onel\ $hb\bar b$~vertex
corrections, including all $\oa$ contributions, is
necessary~\cite{hfff1l}.


\subsubsection{Effects of \twol\ propagator corrections}

We first demonstrate the effect of the \twol\ propagator corrections on
the values of the parameters, especially of $\MA$, for which
$\br(\hbb)$ goes to zero. We 
also show the impact of the inclusion of the momentum dependence of
the Higgs boson self-energies that is often neglected in 
phenomenological analyses of the decays of the lightest $\cp$-even
Higgs boson.

\begin{figure}[ht!]
\begin{center}
\mbox{
\psfig{figure=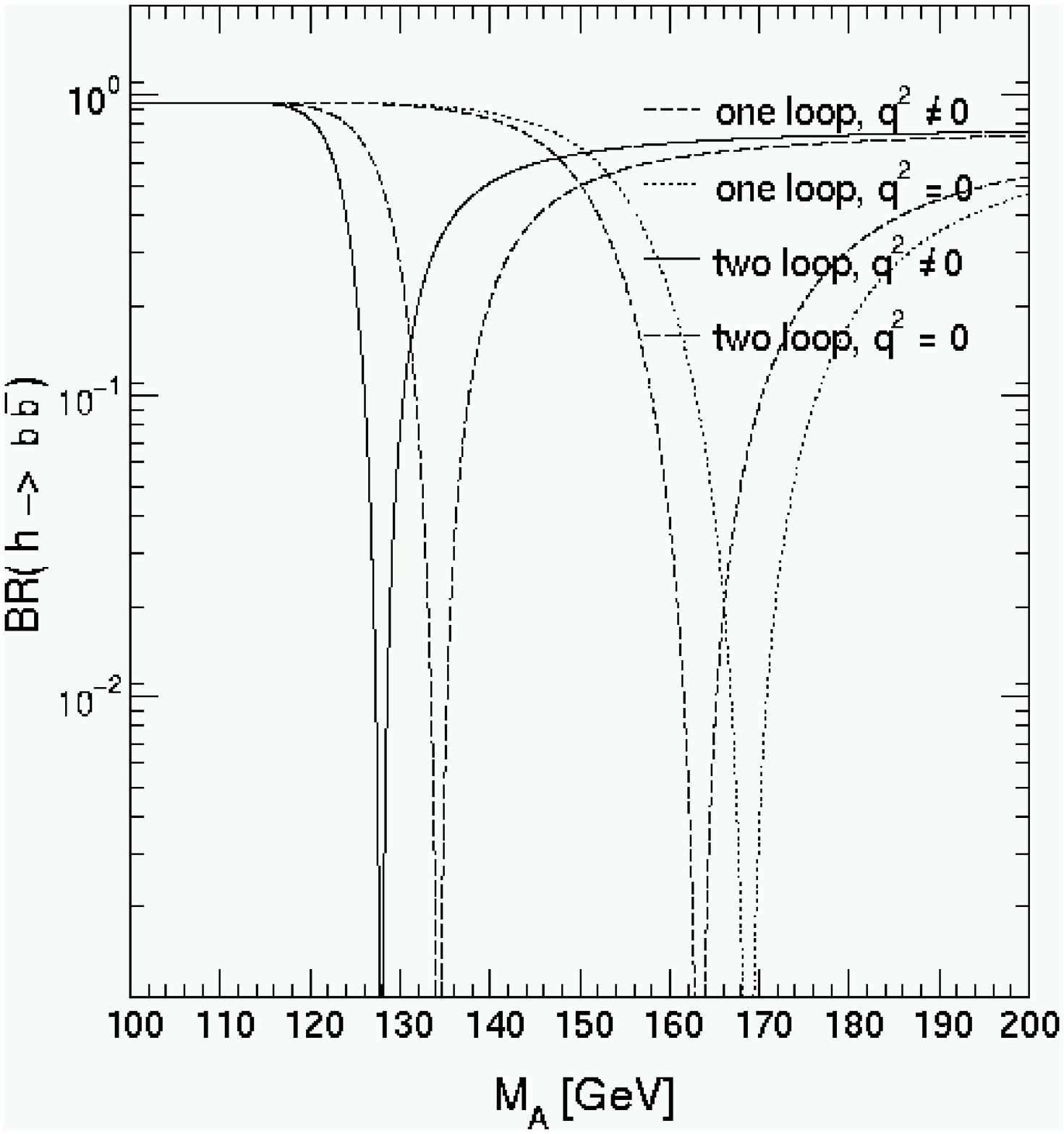,width=10cm,height=8cm}}
\end{center}
\caption[]{
$\br(\hbb)$ is shown as a function
of $\MA$. The Higgs boson self-energies are evaluated at the \onel\
and at the \twol\ level with and without momentum dependence (see
\refeq{hffzeroexternalmomentum}). The other parameters are
$\tb = 25$, $\msusy = 500 \gev$, $\mgl = 400 \gev$, $M_2 = 400 \gev$,
$\Xt = 400 \gev$, $\Ab = \At$, $\mu = -1000 \gev$. 
}
\label{fig:brhbbzero}
\end{figure}

In \reffi{fig:brhbbzero} $\br(\hbb)$ is shown as a function
of $\MA$. The Higgs boson self-energies are evaluated at the \onel\
and at the \twol\ level with and without momentum dependence (see
\refeq{hffzeroexternalmomentum}). The other parameters are
$\tb = 25$, $\msusy = 500 \gev$, $\mgl = 400 \gev$, $M_2 = 400 \gev$,
$\Xt = 400 \gev$, $\Ab = \At$, $\mu = -1000 \gev$. The inclusion of
the \twol\ propagator corrections shifts the $\MA$ value for which 
$\br(\hbb)$ becomes very small by about $-35 \gev$. The
inclusion of the momentum dependence of the Higgs boson self-energies
induces another shift of about $-6 \gev$. In order to have reliable
phenomenological predictions for the problematic $\MA$ values the
\twol\ corrections as well as the inclusion of the momentum dependence
is necessary. 
Note that the inclusion of the gluino vertex corrections as well as
the purely weak vertex corrections can also have a large impact on the
critical $\MA$ values.


\subsubsection{Effects of \order{\als\al_b}}

We now turn to the effects of the additional contributions induced by
\order{\als \al_b} corrections and the corresponding resummation as
discussed in \refse{subsec:recentHO}. 
We recall the tree-level couplings of the lightest $\cp$-even Higgs 
boson to the up-type and down-type SM fermions:
\BE
\Ga_h^{\,u} = \frac{i \,e \,m_u}{2 \,\sw \,\MW} \, \frac{-\Ca}{\Sbe}\,,  
\hspace{2cm}
\Ga_h^{\,d} = \frac{i\, e \,m_d}{2 \,\sw\, \MW}  \, \frac{\Sa}{\Cb}\,.
\label{treecouplings}
\end{equation}
The main source (besides the SM QCD corrections and the Higgs boson
propagator corrections) modifying the
$hb\bar b$ and $h\tau^+\tau^-$ vertices are the corrections modifying
the relation between the bottom quark or $\tau$~lepton mass and the
corresponding Yukawa couplings.


The potentially large corrections to the $\Ga_h^{\,d}$ couplings
come from the $\tb$-enhanced threshold effects in the relation
between the down-type fermion mass and the corresponding Yukawa
coupling~\cite{deltamb1}, already mentioned in
\refse{subsec:evaloaas}. 
A simple way to take into account these
effects is to employ the effective Lagrangian formalism of
\citeres{deltamb,deltamb2}, see \refse{subsec:recentHO}, 
where the coupling of the lightest Higgs
boson to down-type fermions (expressed through the fermion mass) 
can be written as
\BE
\left(\Ga_h^{\,d}\right)_{\rm eff} 
= \frac{i\, e \, m_d}{2 \,\sw\, \MW}  \, \frac{\sin \aeff}{\Cb}\,
\left[1 - \frac{\Delta m_d}{1+\Delta m_d}\,(1+\cot\aeff \,\cot\beta)\right]\, ,
\label{effcoupling}
\end{equation}
where $\Delta m_d$ contains (as described in \refse{subsec:recentHO})
the $\tb$-enhanced terms, and other
subleading (i.e.\ non $\tb$-enhanced) corrections have been
omitted. In the case of the coupling to the bottom quarks, the leading
contributions to $\Delta \mb$ are of \order{\als} and \order{\alt},
coming from diagrams with sbottom--gluino and stop--chargino loops,
respectively. In the case of the $\tau$ leptons, the leading
contributions are of \order{\alpha_{\tau}}, coming from
sneutrino--chargino loops. The terms containing $\Delta m_d$ in
\refeq{effcoupling} may be relevant for large $\tb$ and moderate
values of $\MA$. When $\MA$ is much bigger than $\MZ$, the product
$\cot\aeff \,\cot\beta$ tends to $-1$, and the SM limit is again
recovered.

\begin{figure}[ht!]
\begin{center}
\hspace{-.4cm}
\epsfig{figure=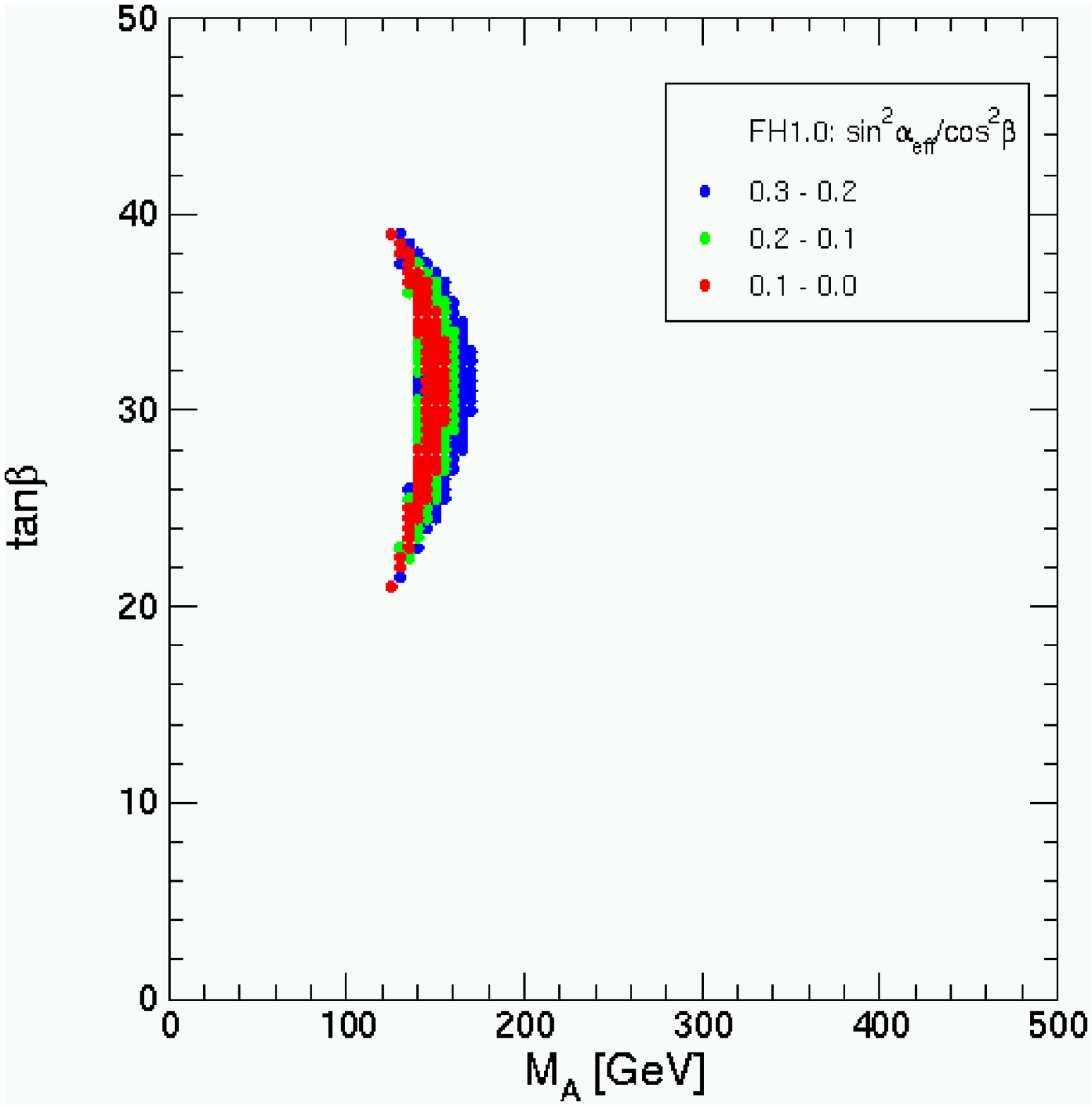,width=8cm,height=8cm}
\hspace{2mm}
\epsfig{figure=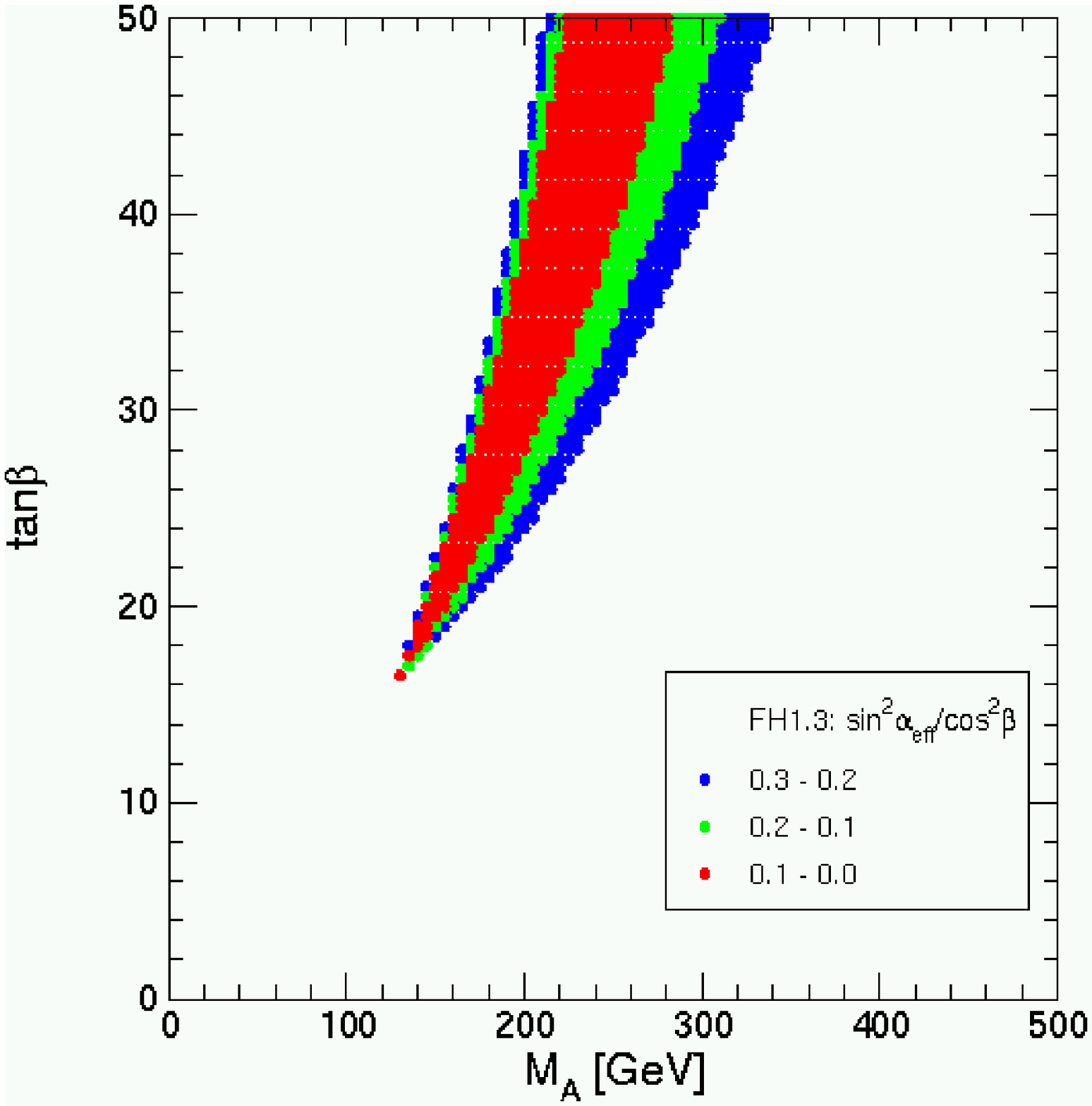,width=8cm,height=8cm}\\[3.0em]
\epsfig{figure=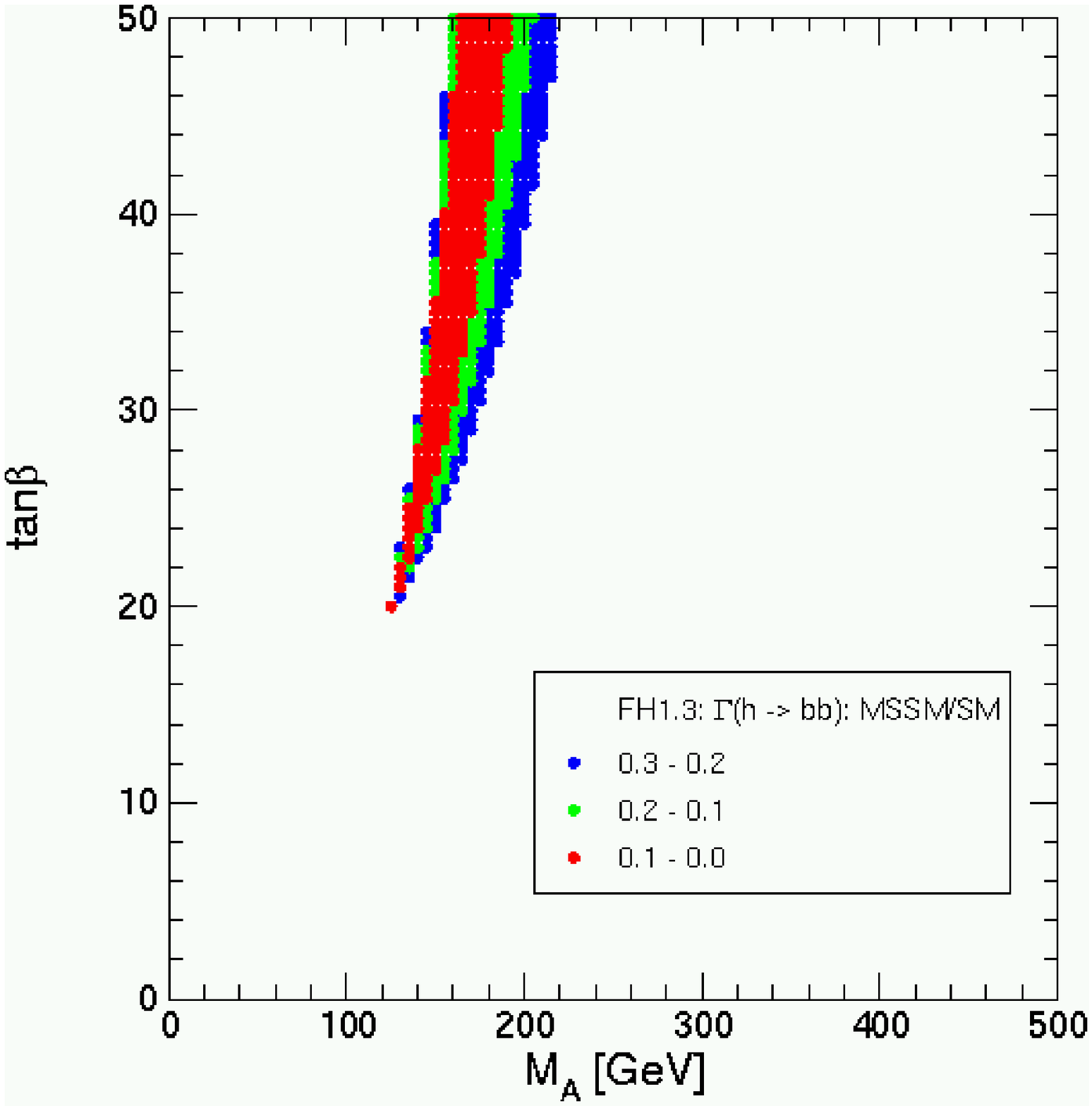,width=8cm,height=8cm}
\end{center}
\vspace{-0.5cm}
\caption{Regions of significant suppression of the coupling of $h$ to 
down-type fermions in the $\MA$--$\tb$-plane within the ``small $\aeff$''
benchmark scenario. The upper left plot shows the ratio 
$\sin^2 \aeff/\CQb$ as evaluated with \fh1.0 (i.e.\ without 
\order{\alb\als} and \order{\alb (\als\tb)^n} corrections), while the
upper right plot shows the same quantity as evaluated with \fh1.3 (i.e.\
including these corrections). The plot in the lower row shows the ratio
$\Gamma(h \rightarrow b\bar{b})_{\rm MSSM}
\,/\,\Gamma(h\rightarrow b\bar{b})_{\rm SM}$, i.e.\ the partial width
for $h \to b \bar b$ normalized to its SM value, where the other term
in \refeq{effcoupling} and further genuine loop corrections are taken 
into account (see text).
}
\label{fig:smallaleff}
\end{figure}

The effects of the higher-order corrections to the couplings of 
the lightest Higgs boson to the down-type fermions appear very
pronounced in the ``small $\aeff$''~scenario~\cite{benchmark},
see the Appendix.

In the upper row of \reffi{fig:smallaleff} we show the ratio 
$\sin^2 \aeff/\CQb$ in the $\MA$--$\tb$ plane, evaluated in the 
``small $\aeff$'' scenario. As explained above, replacing $\alpha$ by
$\aeff$ in the tree-level couplings of \refeq{treecouplings} 
gives rise to regions in the $\MA$--$\tb$ plane where the effective 
coupling of the lightest Higgs boson to the down-type fermions is
significantly suppressed with respect to the Standard Model. The region
of significant suppression of $\sin^2 \aeff/\CQb$ as evaluated 
without the inclusion of the sbottom corrections
beyond one-loop order, is shown in the upper left plot.
The upper right plot shows the corresponding result as evaluated with
\fhto, where the \order{\alb\als} and \order{\alb (\als\tb)^n}
corrections as well as the new \order{\alt^2} ones are included.
The new corrections are seen to have a drastic impact on the region
where $\sin^2 \aeff/\CQb$ is small. While without the new corrections a 
suppression of 70\% or more occurs only in a small area of
the $\MA$--$\tb$-plane for $20 \lsim \tb \lsim 40$ and $100 \gev \lsim
\MA \lsim 200 \gev$, the region where $\sin^2 \aeff/\CQb$ is very small
becomes much larger once these corrections are included. It now reaches
from $\tb \gsim
15$ to $\tb > 50$, and from $\MA \gsim 100 \gev$ to $\MA \lsim 350
\gev$. The main reason for the change is that the one-loop
\order{\alb} corrections to the Higgs-boson mass matrix, which for large
$\tb$ would prevent $\aeff$ from going to zero, are heavily suppressed
by the resummation of the \order{\alb (\als\tb)^n} corrections in the
bottom Yukawa coupling. This kind of suppression depends strongly on
the chosen MSSM parameters, and especially on the sign of $\mu$. 

In order to interpret the physical impact of the effective coupling
shown in the upper row of \reffi{fig:smallaleff}, the $\Delta m_d$ terms
in \refeq{effcoupling} as well as further genuine loop corrections
occurring in the $\hbb$ process have to be taken into account as
described in \refse{subsec:evaloaas}.
The effect of these contributions can be seen from the plot in the lower
row of \reffi{fig:smallaleff}, where the ratio $\Gamma(h \rightarrow
b\bar{b})_{\rm MSSM} \,/\,\Gamma(h\rightarrow b\bar{b})_{\rm SM}$ is
shown, which has been evaluated by including all terms of
\refeq{effcoupling} as well as all further corrections given in
\refeq{decayampresummed}. 
The region where the partial width 
$\Gamma(h \rightarrow b\bar{b})$ within the MSSM is suppressed compared
to its SM value is seen to be somewhat reduced and 
shifted towards smaller values of $\MA$ as compared to the region where
$\sin^2 \aeff/\CQb$ is small.


\subsection{Effects of the gluino vertex corrections}
\label{subsec:gluinoeffect}

In this subsection we focus on the effect of the gluino-exchange
contribution to the $hf\bar{f}$ vertex corrections in a more general
way, i.e.\ for parameters where $hf \bar f$ does not go to zero.

\begin{figure}[ht!]
\begin{center}
\mbox{
\psfig{figure=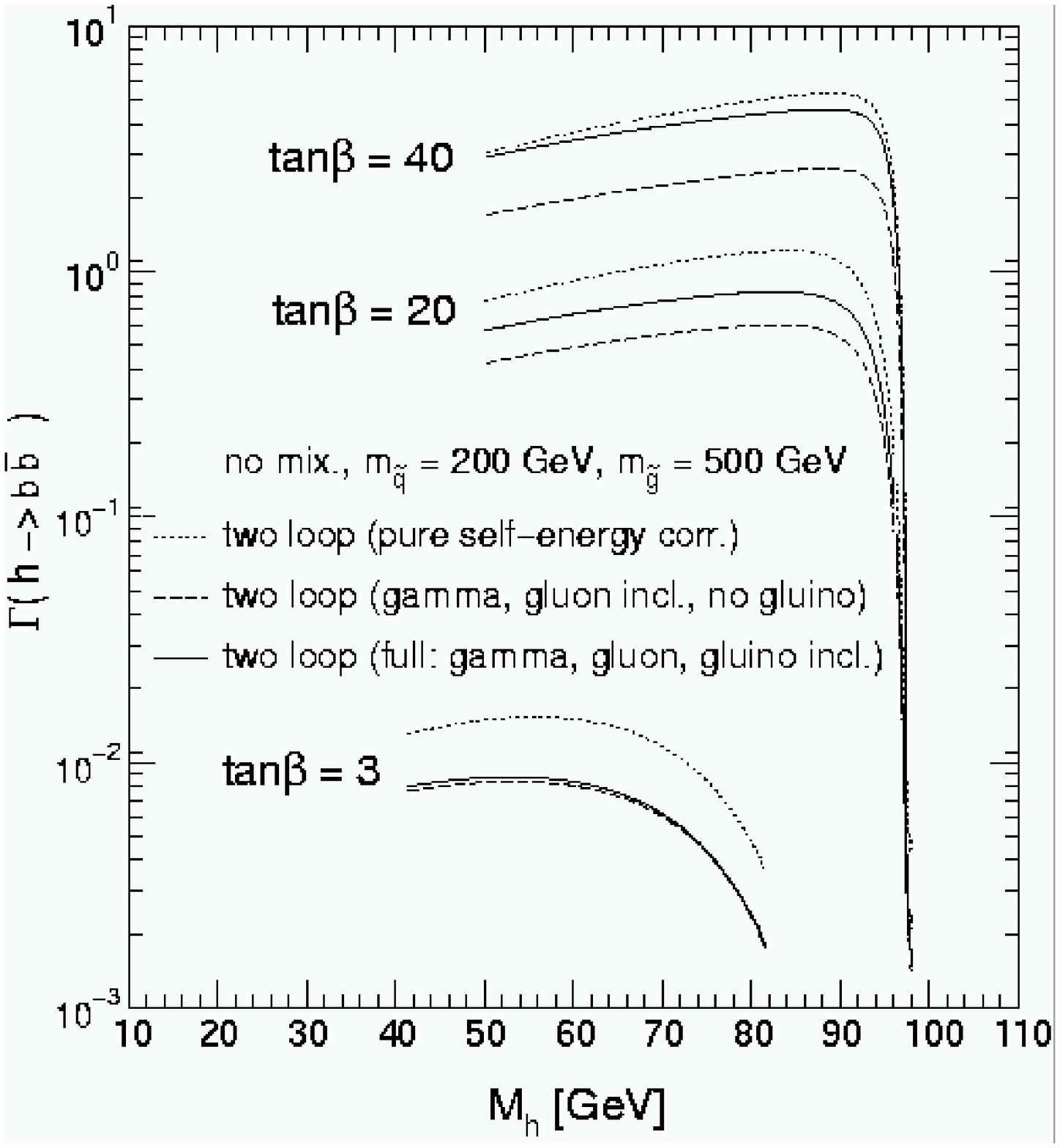,width=7cm,height=7cm}}
\hspace{1.5em}
\mbox{
\psfig{figure=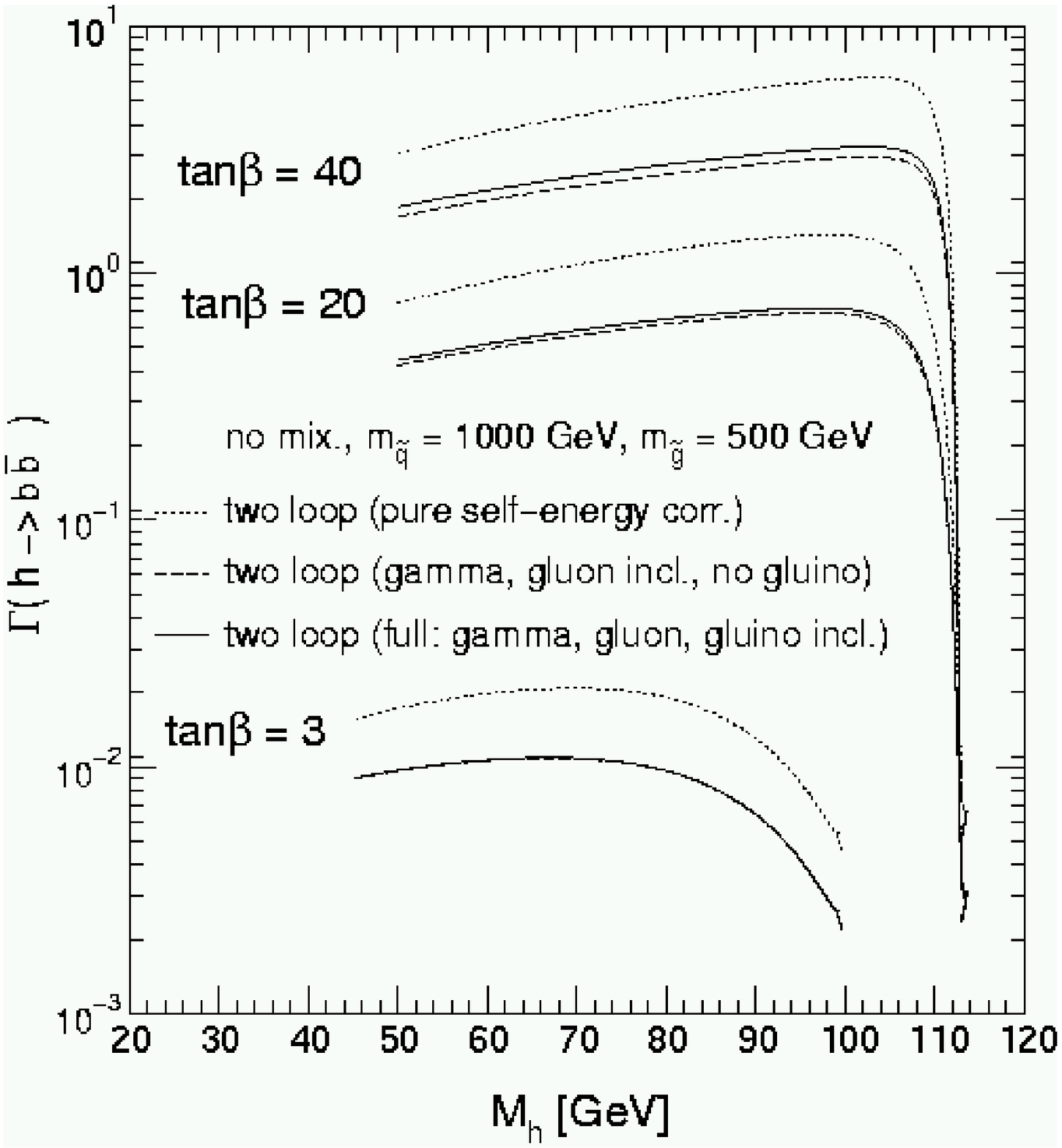,width=7cm,height=7cm}}
\end{center}
\caption[]{
$\Ga(\hbb)$ is shown as a function
of $\Mh$ for three values of $\tb$. The Higgs-propagator corrections have 
been evaluated at the \twol\ level in the no-mixing scenario.
The dotted curves shows the results containing only the pure 
self-energy corrections. The results given in the 
dashed curves in addition contain the QED correction and the gluon-exchange
contribution. The solid curves show the full result, including
also the gluino correction. The other parameters are
$\mu = -100 \gev$, $M_2 = \msq$, $\Ab = \At$.
}
\label{fig:Ghbbgluino}
\end{figure}

\begin{figure}[ht!]
\vspace{3em}
\begin{center}
\mbox{
\psfig{figure=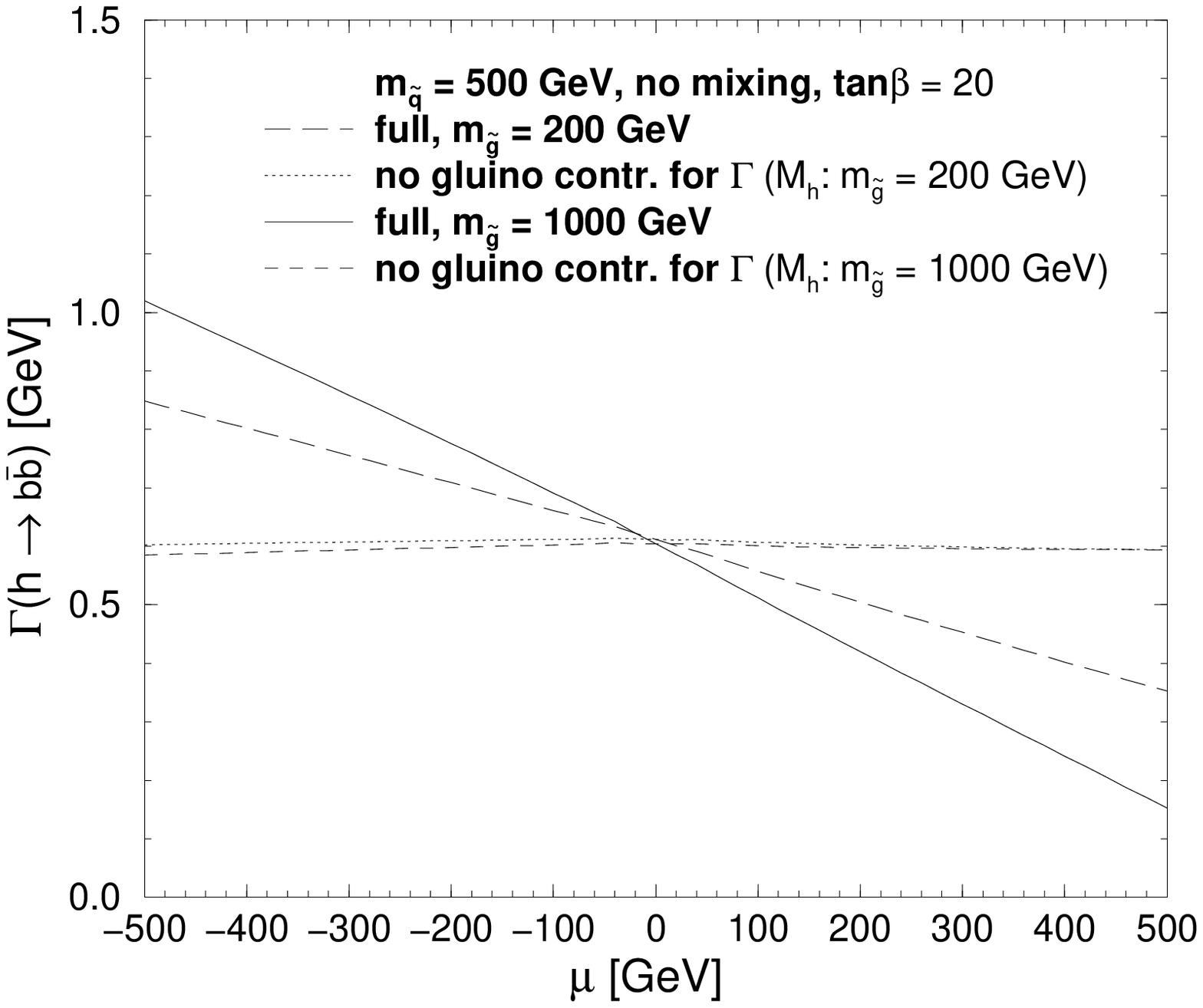,width=7cm,height=7cm}}
\hspace{1.5em}
\mbox{
\psfig{figure=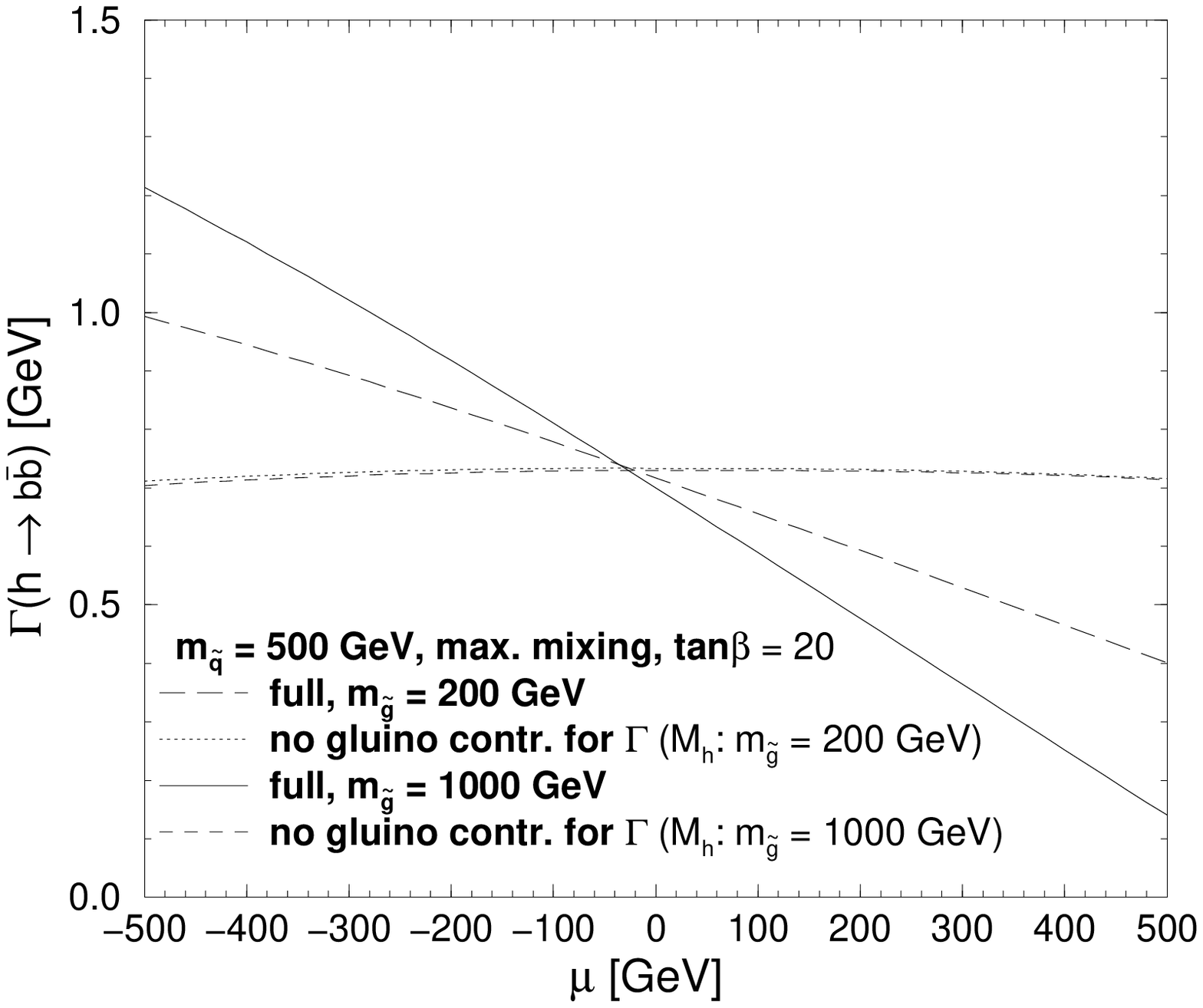,width=7cm,height=7cm}}
\end{center}
\caption[]{
$\Ga(\hbb)$ is shown as a function
of $\mu$ for $\Tb = 20$ and two different values of $\mgl$ in the
no-mixing and the maximal-mixing scenario.
The other parameters are
$\msq = 500 \gev$, $M_2 = 500 \gev$, $\Ab = \At$, $\MA = 100 \gev$.
}
\label{fig:Ghbbgluinomue}
\end{figure}

\reffi{fig:Ghbbgluino} shows $\Ga(\hbb)$ in three steps of accuracy:
the dotted curves contain only the pure self-energy correction, the
dashed curves contain in addition the QED and the gluon-exchange
correction. The solid curves show the full results including also the
gluino-exchange correction. The results are shown for $\Xt = 0$ (no mixing),
$\mu = -100 \gev$, 
$\mgl = 500 \gev$, $\msq = 200, 1000 \gev$, in the left and right part
of \reffi{fig:Ghbbgluino}, respectively. For $\tb$, three values have
been chosen: $\tb = 3, 20, 40$.

The left plot of \reffi{fig:Ghbbgluino} corresponds to a small soft
SUSY-breaking scale, $\msq = 200 \gev$. The effect of the gluon
contribution is large and negative, the effect of the
QED correction is small. For this combination of $\mgl$, $\msq$
and $\mu$ the effect of the gluino correction is large and positive, as
can be seen from the transition from the dashed to the solid
curves. For $\Tb = 40$ it nearly compensates the gluon effect, for 
$\tb = 20$ it amounts up to 20\% of the gluonic correction, while for 
$\tb = 3$ the gluino-exchange contribution is negligible. Note that we
have chosen a relatively small value of $\mu$, $\mu = -100 \gev$. For
larger values of $|\mu|$ even larger corrections can be obtained. Hence
neglecting the gluino-exchange correction in the large $\tb$
scenario can lead to results which
deviate by 50\% from the full $\oas$ calculation (see also
\refse{subsubsec:decaywidthgluino}). 
The right plot of \reffi{fig:Ghbbgluino} corresponds to $\msq = 1000 \gev$. 
The gluino-exchange effects are still visible, but much smaller
than for $\msq = 200 \gev$. The same observation has already been
made in \citere{mhiggsf1lC}.
For $\Xt = 2 \msq$ (maximal mixing) we find qualitatively the same behavior
for the gluino-exchange corrections as for $\Xt = 0$. 

In \reffi{fig:Ghbbgluinomue} the pure gluino-exchange effect is
shown as a function of $\mu$. This effect increases with rising%
\footnote{
This is correct for all values of $\mgl$ considered in this work. A
maximal effect is reached around $\mgl \approx 1500 \gev$.
The decoupling of the gluino takes place only for very large values,
$\mgl \gsim 5000 \gev$.
}
$\mgl$ and $|\mu|$, where for
negative (positive) $\mu$ there is an enhancement (a decrease) in
$\Ga(\hbb)$. The size of the gluino-exchange contribution also depends
on $\MA$, where larger effects correspond to smaller values of $\MA$,
see also \citere{mhiggsf1lC}. The small difference between the curves where the
decay rate has been calculated without gluino contribution is due to
the variation of $\Mh$ induced by different values of $\mgl$ which
enters at $\oaas$.
\reffi{fig:Ghbbgluinomue} demonstrates again
that neglecting the gluino contribution in the fermion decay rates can
yield (strongly) misleading results.

\bigskip
As a final remark, one should note that
the gluino exchange contribution has only a relatively small impact on  
$\br(\hbb)$. However, it can have a large influence on
$\br(\htautau)$ or $\br(\hWW)$. Both branching ratios
are expected to be measurable at the per-cent level, see
\refta{tab:br}. 
While the Higgs-propagator contributions are universal corrections
that affect $\Ga(\hbb)$ and $\Ga(\htautau, \hWW)$ in the same way (i.e.\ the
influence on the effective coupling is the same in both cases), the
gluino corrections, 
which influence only $\Ga(\hbb)$, can lead to a different behavior of
the decay widths. This has been analyzed in the mSUGRA, GMSB and AMSB
scenario in \citeres{asbs2,ehow2}.


\subsection{The $\aeff$-approximation}
\label{subsec:hffeffect}

\begin{figure}[ht!]
\begin{center}
\mbox{
\psfig{figure=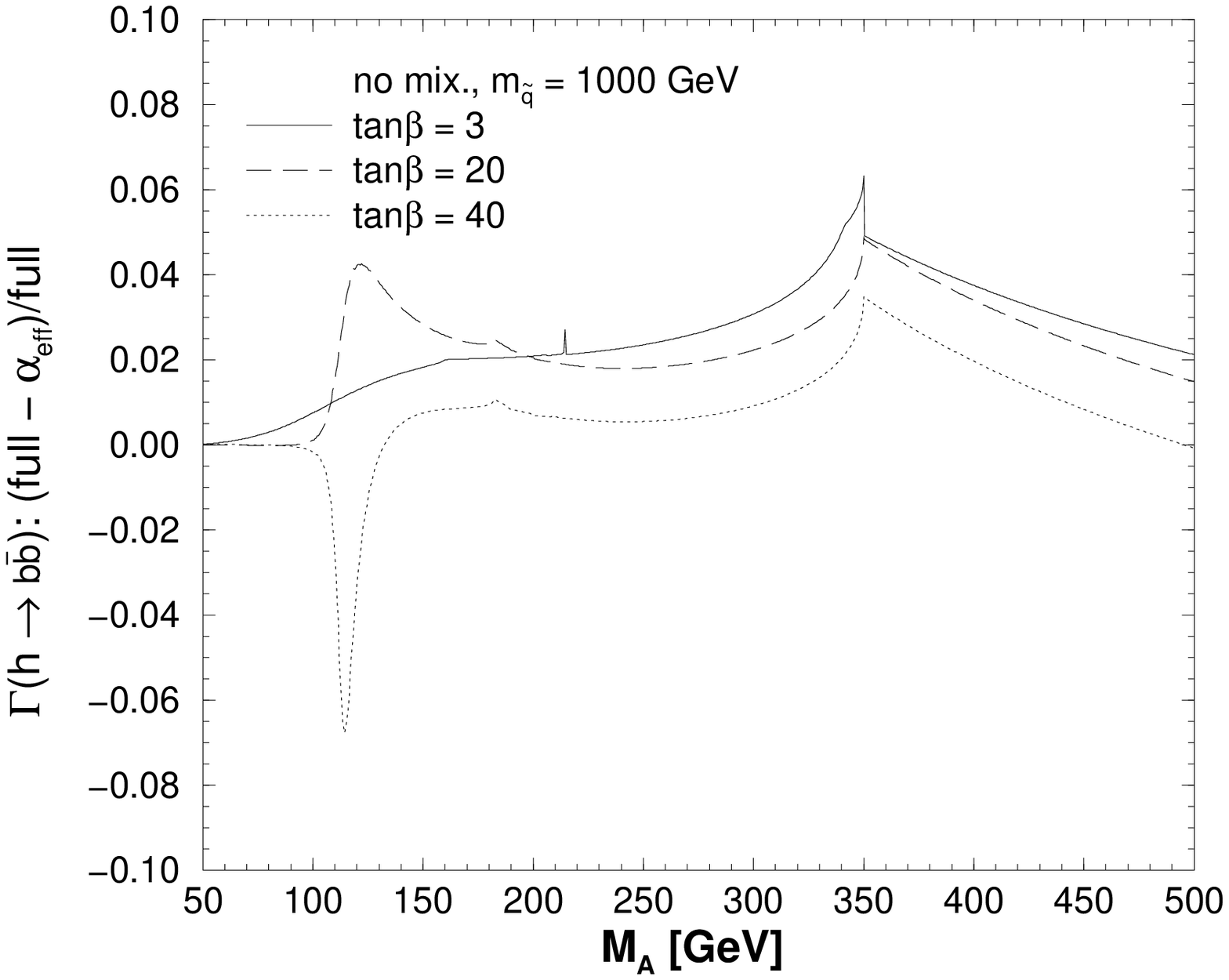,width=7cm,height=6.5cm}}
\hspace{1.5em}
\mbox{
\psfig{figure=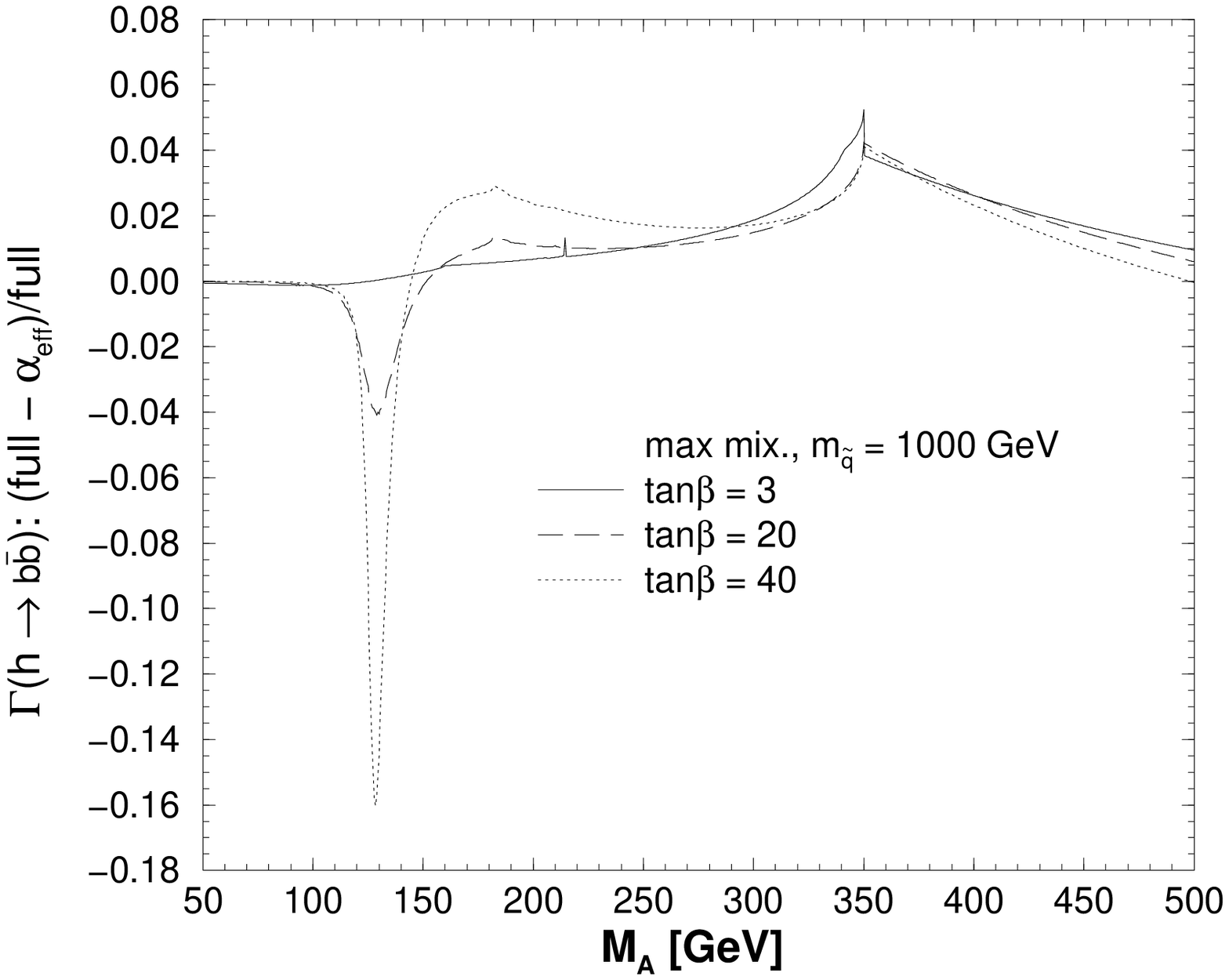,width=7cm,height=6.5cm}}
\end{center}
\caption[]{
$\De\Ga(\hbb) = 
(\Ga^{\rm full}(\hbb) - \Ga^{\aeff}(\hbb))/\Ga^{\rm full}(\hbb)$ 
is shown as a function of $\MA$ for three values of $\tb$.
The QED, gluon- and gluino-contributions are neglected here. 
The other parameters are 
$\mu = -100 \gev$, $M_2 = \msq$, $\mgl = 500 \gev$, $\Ab = \At$, 
$\Tb = 3, 20, 40$. $\Xt$ has been set to $\Xt = 0$ (no mixing) or
$\Xt = 2 \msq$ (maximal mixing). 
}
\label{fig:RCGhbbaeff}
\end{figure}

Finally, we investigate the quality of the
$\aeff$-approximation. In \reffi{fig:RCGhbbaeff} we display the
relative difference between the full result~(\ref{hsefull}) and the
$\aeff$ result, where the external momentum of the Higgs self-energies
has been neglected, see \refeq{hseq2zero}.
The relative difference 
$\De\Ga(\hbb) = 
(\Ga^{\rm full}(\hbb) - \Ga^{\aeff}(\hbb))/\Ga^{\rm full}(\hbb)$ 
is shown as a function of $\MA$ for $\msq = 1000 \gev$ and for three values
of $\tb$ in the no mixing and the maximal mixing scenario. 
Large deviations
occur only in the region $100 \gev \lsim \MA \lsim 150 \gev$, especially
for large $\tb$. In this region of parameter space the values of $\Mh$
and $\MH$ are very close to each other. This results in a high
sensitivity to small deviations in the Higgs boson self-energies
entering the Higgs-boson mass
matrix~\refeq{deltaalpha}. 
Another source of differences between the full and the approximate
calculation is the threshold $\MA = 2\,\mt = 350 \gev$ in the \onel\
contribution, originating from the top-loop diagram in the $A$
self-energy, see \citere{hff} for more details.
Here the deviation can amount up to 6\%.
This threshold is only present in the pure on-shell renormalization
(see also \refse{subsec:recentHO}), which has been used for these
plots. Making use of the hybrid \drbar/on-shell renormalization (as
incorporate in the latest version of \fh), these thresholds are
absent.

\begin{figure}[ht!]
\begin{center}
\mbox{
\psfig{figure=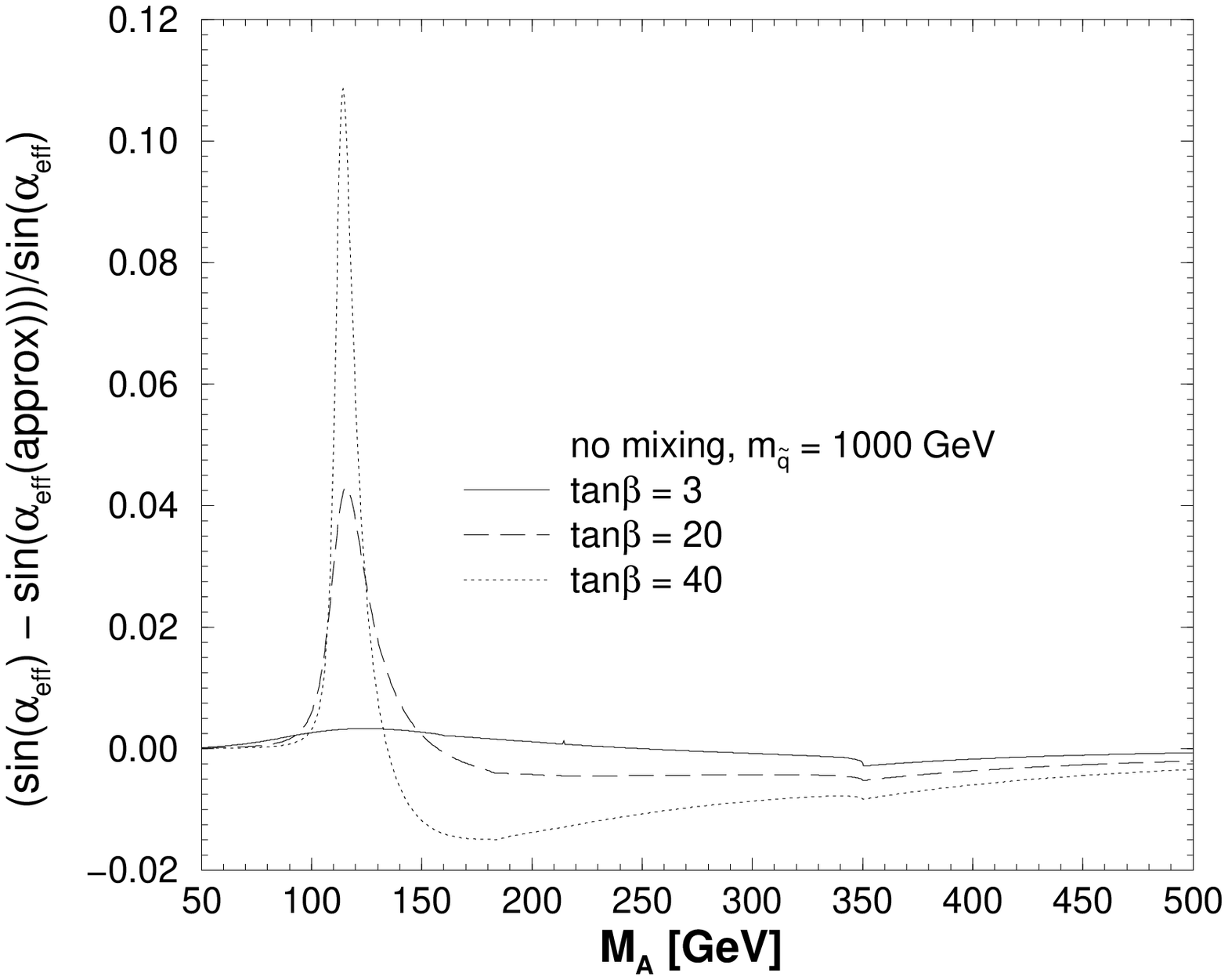,width=7cm,height=6.5cm}}
\hspace{1.5em}
\mbox{
\psfig{figure=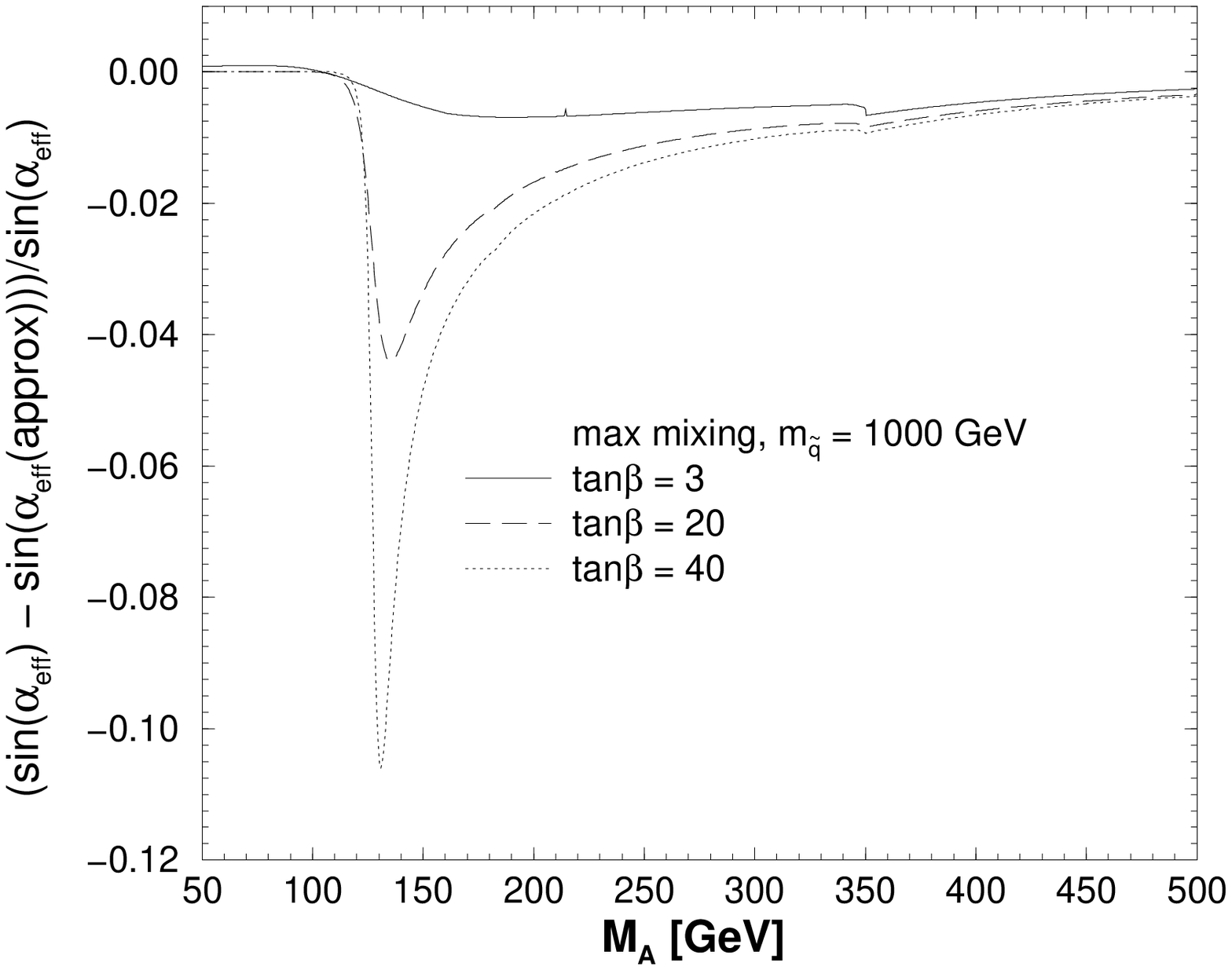,width=7cm,height=6.5cm}}
\end{center}
\caption[]{
The relative difference $(\sin\aeff - \sin\aeffapprox)/\sin\aeff$  
(see \refeqs{hseq2zero} and~(\ref{hselle})) is shown as a function
of $\MA$ for three values of $\tb$ for $\Xt = 0$ (no mixing) and 
$\Xt = 2 \msq$ (maximal mixing). The other parameters are chosen as in
\reffi{fig:RCGhbbaeff}. 
}
\label{fig:RCsaeff}
\end{figure}

In \reffi{fig:RCsaeff} we compare the $\aeff$ result (\ref{hseq2zero})
with the $\aeffapprox$ result (\ref{hselle}), where the
Higgs boson self-energies have been approximated by the compact
analytical expression obtained in~\citere{mhiggslle}.
Figure~\ref{fig:RCsaeff}
displays  the relative difference in the effective mixing angles,
$(\sin\aeff - \sin\aeffapprox)/\sin\aeff$. Via \refeq{ampeffhbb} 
$\Saeff$ directly determines 
the decay width $\Ga(\hbb)$. The result is shown for 
$\msq = 1000 \gev$, for three values of $\tb$ in the minimal and the
maximal mixing scenario.
Apart from the region around
$\MA \approx 120 \gev$ (compare \reffi{fig:RCGhbbaeff}) both effective
angles agree better than 3\% with each other. 

Concerning the comparison of the $\aeff$-approximations in terms of
$\Mh$ (which is not plotted here), 
due to the neglected external momentum or the neglected subdominant
one- or \twol\ terms, $\Mh$ receives a slight shift.
Besides this kinematical effect, 
the decay rate is approximated rather well for most of the $\Mh$ values: 
independently of $\msq$, the differences stay mostly below 2--4\%, 
for the no-mixing case as well as for the maximal-mixing case.
Only at the endpoints of the spectrum, due to the different
Higgs-boson mass values, the difference is not negligible.

%


\newpage

\chapter{Conclusions}

We have reviewed the current status of higher-order corrections in the
MSSM Higgs sector. This comprises the corrections to the Higgs boson
masses in the real and complex MSSM, the existing corrections to Higgs
boson production cross sections at the LC as well as corrections to
Higgs boson decays to SM fermions. 

Concerning Higgs boson masses and couplings in the MSSM with real
parameters, the higher-order corrections are known to be huge. The
corrections described in this report lower the intrinsic error due to
unknown higher-order corrections down to
about $\sim 3 \gev$ for the lightest Higgs boson mass. Likewise a
precision at the per-cent level for the couplings can be reached. In
the case of the MSSM with complex parameters, uncertainties of more
than $\sim 5 \gev$ for the lightest Higgs boson mass have been
eliminated. The effects of the full one-loop corrections have been
shown to be substantial for all Higgs boson masses and couplings. 
We have investigated the higher-order contributions to Higgs boson
production channels at the LC and found corrections of up to 
$\sim 20\%$ for the channel \eetohZ. In the $WW$~fusion channel
especially the non-decoupling behavior of the heavy $\cp$-even Higgs
boson has been investigated. Taking loop corrections into account the
reach of the LC could be extended by up to $\sim 200 \gev$. Concerning
the Higgs boson decays to SM fermions, we have shown that the
corrections arising in the Feynman-diagrammatic approach are
subtantial and have to be taken into account in an accurate analysis. 
This holds especially for gluino-loop corrections for the decay $\hbb$.

To summarize, in all investigated cases the higher-order corrections are
substantial and must not 
be neglected in phenomenological analyses. Once a MSSM-like Higgs
boson will have been discovered, and the corresponding quantities like
masses, couplings and production cross sections will be measured at
the LHC and (hopefully) furtheron at the LC, it will be mandatory to
have even more precise evaluations at hand. In view of the complexity
of these calculations in the past, and the time needed to perform
them, further higher-order corrections will have to be evaluated
continously in the future. Only then the required precision will be
available at the same time as the experimental data will be determined
from the collider experiments. 

At the same time it will be necessary not only to improve on the
higher-order corrections but also on the reduction of the parametric
uncertainties due to the imperfect knowledge of the input
parameters. This is especially true for the top quark mass. Here
certainly the LC precision will be necessary to match even the
anticipated LHC precision in $\Mh$. 

The tasks for theorist described above are huge. However, it still 
seems to be possible to achieve the required calculations,
provided that there is enough financial and man power
support from the whole community~\cite{ypp}. Then (but only then) it
can be possible 
that theoretical and experimental high-energy physics has a bright
future. 

\addcontentsline{toc}{chapter}{Appendix}
\begin{appendix}
\chapter*{Appendix}
\label{chap:appendix}

For our numerical results, the following values of 
the SM parameters are used (all other quark and lepton masses are
negligible):
\BE 
\begin{aligned}
G_F &= 1.16639\times 10^{-5}, &\quad
m_\tau &= 1.777 \gev, \\
\MW &= 80.450 \gev, &
\mt &= 174.3 \gev, \\
\MZ &= 91.1875 \gev, &
\mb &= 4.25 \gev, \\
\Gamma_Z &= 2.4952 \gev, &
m_c &= 1.5 \gev .
\end{aligned} 
\end{equation}

For our numerical evaluation we mostly rely on four benchmark
scenarios that have been defined in \citere{benchmark} for Higgs boson
searches at hadron colliders and beyond.
The four benchmark scenarios are
(more details can be found in \citere{benchmark}) 
\begin{itemize}

\item
the ``$\mhmax$'' scenario, which
yields a maximum value of $\Mh$ for given $\MA$ and $\tb$,
\BE
\begin{aligned}
{}& \mt = 174.3 \gev, \quad \msusy = 1 \tev, \quad
\mu = 200 \gev, \quad M_2 = 200 \gev, \\
{}& \Xt = 2\, \msusy, \quad
\Atau = \Ab = \At, \quad \mgl = 0.8\,\msusy\,,
\end{aligned}
\label{mhmax}
\end{equation}

\item
the ``no-mixing'' scenario, with no mixing in the $\Stop$~sector,
\BEA
{}&& \mt = 174.3 \gev, \quad \msusy = 2 \tev, \quad
\mu = 200 \gev, \quad M_2 = 200 \gev, \non \\
{}&& \Xt = 0, \quad
\Atau = \Ab = \At, \quad \mgl = 0.8\,\msusy\,,
\label{nomix}
\EEA

\item
the ``gluophobic-Higgs'' scenario, with a suppressed $ggh$ coupling,
\BEA
{}&& \mt = 174.3 \gev, \quad \msusy = 350 \gev, \quad
\mu = 300 \gev, \quad M_2 = 300 \gev, \non \\
{}&& \Xt = -750 \gev, \quad
\Atau = \Ab = \At, \quad \mgl = 500 \gev\,,
\label{gluophobicH}
\EEA

\item
the ``small-$\aeff$'' scenario, with possibly reduced decay rates for
$\hbb$ and $\htautau$,
\BEA
{}&& \mt = 174.3 \gev, \quad \msusy = 800 \gev, \quad
\mu = 2.5 \, \msusy, \quad M_2 = 500 \gev, \non \\
{}&& \Xt = -1100 \gev, \quad 
\Atau = \Ab = \At, \quad \mgl = 500 \gev\,.
\label{smallaeff}
\EEA

\end{itemize}
As explained above, for the sake of simplicity, $\msusy$ is chosen as a
common soft SUSY-breaking parameter for all three generations.

\end{appendix}

\newpage
\clearpage
\subsection*{Acknowledgements}

I thank all my collaborators with whom I have worked on the various
calculations presented here. These are 
O.~Brein,
M.~Carena,
G.~Degrassi,
M.~Frank,
H.~Haber,
T.~Hahn,
W.~Hollik,
J.~Rosiek,
P.~Slavich,
C.~Wagner,
and 
G.~Weiglein.
Furthermore, I thank 
S.~Dawson, 
S.~Dittmaier,
U.~Nierste,
H.~Rzehak, 
and 
M.~Spira
for helpful discussions.
I am also gratefull to
S.~Dittmaier, 
W.~Hollik
and 
G.~Weiglein
for a critical reading of the manuscript.
Finally, I thank I.~Campos for inspirational as well as W.~Lagavulin and
W.~Laphroigh for spiritual encouragement.
This work has been supported by the European Community's Human
Potential Programme under contract HPRN-CT-2000-00149 Physics at
Colliders. 

\newpage
\addcontentsline{toc}{chapter}{Bibliography}




\end{document}